\documentclass[11pt,a4paper,openright,twoside]{report}
\usepackage[dvips]{graphicx}
\usepackage{fancyheadings}
\usepackage{float}
\usepackage{subfigure}
\renewcommand{\chaptermark}[1]{\markboth{#1}{}}
\renewcommand{\sectionmark}[1]{\markright{ \thesection\ \  #1}}
\usepackage{rotating}
\rotdriver{dvips}

\begin{document}


\thispagestyle{empty}
\vspace*{-2.5cm}
\begin{center}
\centerline{\includegraphics[width=5cm]{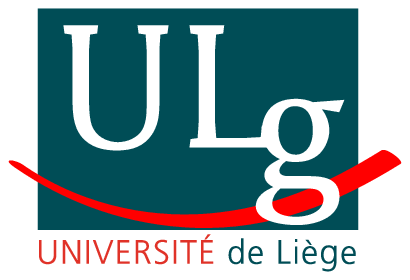}}
\Large
Facult\'e des Sciences\large\ \ \  
\end{center}

\vspace*{3.8cm}

\begin{center}

\Huge   \sc The Nucleon-Nucleon Interaction in a Chiral Constituent Quark Model  \\

\vspace*{3.8cm}
\LARGE  \sc Daniel Bartz
\end{center}

\vspace*{4.8cm}

\begin{center}
\large  
Th\`ese pr\'esent\'ee en vue de l'obtention du \\ Grade de Docteur en Sciences Physiques \\

\vspace*{1.2cm}

\large Mars 2002
\end{center}
\newpage
\thispagestyle{empty} 
\ 
\newpage


\newpage
\thispagestyle{empty} 
\ 
\ \\

\ \\

\ \\

\begin{flushright}
To Lorena
\end{flushright}
\newpage
\thispagestyle{empty} 
\ 
\newpage
\chapter*{Acknowledgments}
\thispagestyle{empty}

\vspace{3.5cm}
\ 

First and foremost, I wish to express my deep gratitude to my supervisor Fl. Stancu for giving me inspiration and guidance in the research work contained in this thesis. I am very grateful to her for prompt help whenever I needed it.  Her professional and human support was so fundamental that it is really hard to estimate what a chance I had meeting her. I would also like to thank her especially for careful readings of all the manuscripts.
\\

Many thanks to J.-P. Jeukenne and J. Cugnon, who offered me the possibility to start, and to achieve this thesis.
\\

I should not forget the University of Li\`ege and the {\it Institut Interuniversitaire des Sciences Nucl\'eaires} (IISN) for funding this work and conference appearances. Thus, providing me with the opportunity to interact with a wide scientific community.
\\

I am also indebted to all the members of the group of fundamental theoretical physics for the scientific atmosphere. I owe special acknowledgments to that crazy friend and colleague Michel Mathot for all of its ``good'' advices, special thanks also to the Others...
\\
\newpage
\thispagestyle{empty} 
\ 
\newpage

\setcounter{page}{1}


\chapter{Introduction}\label{introductionchapter}

\vspace{3.5cm}
\ 

The strong interaction between two nucleons is the basic ingredient of nuclear physics. Since about 1980 there have been many efforts to derive the nuclear force from the under\-lying theory of strong interactions, the quantum chromodynamics (QCD). However the basic equations of QCD are known for their difficulties to obtain a comprehensive solution at all energies. Indeed, the QCD is often split into two parts : the high-energy and the low-energy regimes. In the high-energy region, where the coupling constant of QCD becomes very small, leading to the asymptotic freedom, a perturbative treatment is applicable, allowing for a satisfactory description of many experimental data, such as the deep inelastic lepton-hadron scattering. On the other hand, due to its non-perturbative character in the low energy regime, as for example the nucleon or the nucleon-nucleon interaction, QCD cannot be solved exactly. Moreover, in this energy region new characteristics of QCD become important such as the confinement of quarks and gluons and the spontaneous breaking of chiral symmetry. This renders difficult the treatment of QCD using its own degrees of freedom. A first hope could come from lattice calculations. However, certainly more computer power and more deve\-lopment of sophisticated algorithms are still necessary. Consequently, several so-called QCD inspired models have been used to study the nucleon properties and to derive the nucleon-nucleon interaction, where the original degrees of freedom of QCD are replaced by effective degrees of freedom.
\\

Among all the available models, in this thesis we are concerned with a particular cons\-tituent quark model. Our choice goes to a new potential model : the Goldstone boson exchange (GBE) or the chiral constituent quark model, where the interaction between quarks is due to a pseudoscalar meson exchange (Goldstone boson). This model incorporates the chiral symmetry breaking effect of QCD, the importance of which becomes more and more clear to nuclear physicists.
\\

The GBE model reproduces well the baryon properties and it is interesting to find out if the nucleon-nucleon problem, as a system of six quarks, can be described as successfully. A strong motivation of using this model is that it contains all necessary ingredients for des\-cribing the nucleon-nucleon interaction. Its hyperfine interaction, which is flavor and spin dependent, besides a short-range part, essential for baryon spectroscopy, also contains a long-range part of Yukawa type, which is necessary to describe the long-range behaviour of the nucleon-nucleon interaction. This allows a fully consistent description of the nucleon-nucleon interaction and avoids the extra addition of meson exchange at the nucleon level, as imposed in the one-gluon exchange models, or of meson exchange between quarks, overlapped on the gluon exchange, like in the so-called hybrid models.
\\

Within the GBE model, besides one-meson exchanges, one can naturally include two or more pseudoscalar meson exchanges. The incorporation of two-meson exchanges which gives rise to a spin independent central component, simulated by a $\sigma$-meson exchange, could in principle explain the middle-range attraction of the nuclear forces. Apart from that, together with the one-pion exchange, two correlated pions bring a contribution to the tensor part of the quark-quark interaction, required to explain the binding of the deuteron and the behaviour of the $^3S_1$ phase shift.
\\

The best known way to study the nucleon-nucleon (NN) interaction as the interaction between two composite particles is via the resonating group method (RGM), or the closely related, the generator coordinate method (CGM). Here we chose the RGM to treat the nucleon-nucleon problem as a system of six interacting quarks. An important feature of this method is that the exchange of quarks between nucleons can be treated exactly. This leads to important non-local effects in the NN potential and allows the study of both bound and scattering states of the NN system.
\\

This thesis is divided in six chapters and six appendices. The work presented here begins with a short historical overview of the nucleon-nucleon interaction studies. The review presented in Chapter \ref{microchapter} starts from the approach beginning with Yukawas's theory in the thirties and ends with the most modern considerations where quark structure and chiral symmetry are taken into account. In particular we shortly introduce the three best known potential models : the one-gluon exchange model (OGE), the hybrid model and the GBE model. To our knowledge, our work is the first study of the NN interaction within the GBE model and based on RGM.
\\

Chapter \ref{gbechapter} is entirely devoted to the detailed study of the GBE model. The origin and its foundations are discussed in details. Results and in particular its success in reproducing spectra of light baryons is presented. We also underline how the GBE model provides a promising basis for further investigations, such as the nucleon-nucleon interaction. We also note that the long-range part of the quark-quark interaction which does not play an important role in the baryon spectra, leaves room for more adjustment of the GBE parametrization when studying the nucleon-nucleon interaction.
\\

The application of the GBE interaction to the study of six-quark systems starts in Chapter \ref{preliminarychapter}. Our first goal is to find out if the GBE interaction can explain the short-range repulsion of a two-nucleon system. That chapter constitutes in fact a first step of our investigation towards finding out whether or not the GBE model can give rise to a short-range repulsion in the nucleon-nucleon interaction. To do so we use the Born-Oppenheimer approximation and consider two different ways of constructing six-quark states : the cluster model basis and the molecular orbital basis. We show why the molecular basis is more appropriate to the derivation of the nucleon-nucleon interaction. Also we show how the details of the GBE model parametrization does not affect the results qualitatively. Finally, we propose the introduction of a scalar-meson exchange interaction between quarks which can simulate the missing middle-range attraction.
\\

However the results obtained in Chapter \ref{preliminarychapter} represent only a preliminary study, inasmuch as we derive only a static local potential. A dynamical treatment of the nucleon-nucleon interaction is developed in Chapter \ref{rgmchapter} where we apply the RGM to study the NN interaction. This method allows us to calculate bound and scattering states of the NN system in taking into account the compositeness of the nucleon. One of the concerns of Chapter \ref{rgmchapter} is to find out the role played by GBE interaction and the quark interchange due to the antisymmetrization, on the short-range NN interaction. In that chapter we first present  the resonating group method in the context of the NN problem. Next we apply it to the study of the relative s-wave scattering of two nucleons. The influence of the coupled channels $NN-\Delta\Delta-CC$ in RGM calculations is presented. Then, searching for an agreement of the $^1S_0$ phase shift with experimental data, we propose to explicitly include a $\sigma$-meson exchange which can provide an important amount of middle-range attraction. In order to reproduce the experimental data for the $^3S_1$ phase shift, we have to introduce the tensor part of hyperfine interaction potential which can provide the binding of the two nucleons.
\\

Our concluding remarks are gathered together in the last chapter. Many of the analytical details related to Chapter \ref{rgmchapter} are given in the Appendices \ref{appendixCOUPLED}-\ref{appendixTENSOR}.
\newpage
\thispagestyle{empty} 
\ 



\chapter{Microscopic Description of the Nucleon-Nucleon Interaction}\label{microchapter}

\vspace{3.5cm}

\section{Historical overview}
\

The nucleon-nucleon (NN) interaction is one of the most important key issues in nuclear physics. That is why so much effort has been devoted to the study of this interaction. Already in the thirties, Yukawa postulated the existence of a new particle, called later the pion, which explained the basic force between two nucleons. If this simple model was very good for a long-range interaction, of the order of two fermis, more complicated approaches were needed to reproduce the shorter range interaction. Knowing that the one-pion exchange potential (OPEP) plays the dominant role in the long-range region, there are however some ambiguities in the medium- and short-range parts of the potential. The nucleon-nucleon scattering data suggest a strong repulsive interaction at short-range as seen from the high-energy region $E_{lab} \geq 250$ MeV in NN scattering experiments. Accordingly, phenomenological potentials with a repulsive core have been introduced, see for example Reid \cite{REI68}. Later on, in the spirit of meson theory, besides the pion exchange, one had included two-pion exchanges for the intermediate region (between one and two fermis) as well as vector mesons exchanges, for example the isoscalar $\omega$ and the isovector $\rho$ meson, which are responsible for the short-range repulsion. This approach led to the more sophisticated potentials, like the Paris \cite{LAC80} or the Bonn \cite{MAC87} potential, which are able to describe the nucleon-nucleon scattering in a pretty good way. However in the new high-precision potentials as for example the Nijmegen potential \cite{STO93}, in order to get a $\chi^2/$datum close to 1, tens of parameters (about 45) were needed. If these potentials are employed or have been employed with success as a basic ingredient in the study of nuclear many-body problems, this amount of parameters clearly indicates the presence of some lack of knowledge.
\\

Moreover it is well known today that the nucleon itself has a finite size and an internal structure. The nucleon structure has been usually taken into account in terms of a nucleon form factor when electromagnetic and weak properties of the nucleon were discussed. In the nucleon-nucleon problem, the finite-size effect should also be taken into account. This has been achieved by introducing a form factor for the meson-nucleon vertex. Again, this form factor is phenomenological and has been adapted by comparison with the experiment. The current status of phenomenological or semi-phenomenological models can be found for example in Ref. \cite{MAC01}.
\\

Since about 1980, a second line of approach, at a more fundamental level, has been developed and this is the line we adopt here. This is because it is currently believed that quantum chromodynamics (QCD) is the fundamental theory of the strong interaction, where coloured quark fields interact with coloured gluon fields through a local gauge-invariant coupling \cite{CLO79}. Indeed the interaction range of the $\omega$ meson is about 0.2 fm which means that the nucleon internal structure has a crucial role in the short-range interaction because the root mean square radius of the proton is about 0.8 fm indicating the overlap of both nucleons. In this context the nucleon is then considered as a composite system of quarks and gluons described by QCD. Recent research of the NN interaction has been conducted on this more basic level. But the problem with QCD is that it is quite impossible to apply it directly to a nuclear problem such as the baryon-baryon interaction. That is why many quark mo\-dels inspired by QCD have been used in trying to describe the nucleon itself as well as the more complicated problem of the nucleon-nucleon interaction or even other nuclear problems.
\\

Three of the most important characteristics of QCD are clearly the colour confinement, the asymptotic freedom and the chiral symmetry. That is the basis of many well known models to describe the internal structure of hadrons. There are relativistic models such as the MIT bag model \cite{CHO74} or the soliton bag model of Friedberg and Lee \cite{FRI78}. In the MIT bag model, one postulates an infinitely sharp transition at the bag radius $R$. Inside the bag the quarks interact weakly. Outside there is no quark, it is the true QCD vacuum. This confining boundary condition breaks the chiral symmetry. It is the same in the Friedberg and Lee model where the confinement is realized by introducing a scalar field coupled to quarks. This scalar field is supposed to simulate all non-perturbative effects of QCD. In order to restore chiral symmetry other models such as the chiral bag \cite{BRO79} and the cloudy bag \cite{THE80} have been proposed. The conservation of the axial current is then ensured by the introduction of new fields. On the other hand there are non-relativistic or semi-relativistic (see later) models based on potentials \cite{ISG77}. In these models an important difference with respect to relativistic models comes from the phenomenology used to introduce the quark confinement \cite{BAR97}. In the bag models, confinement is introduced at the bag surface by imposing boundary conditions or introducing additional fields. In the potential models the confinement is described in terms of interactions between quarks. In all these models we then have a hadron constructed from confined quarks interacting between themselves via a hyperfine interaction. Presently the hyperfine interaction is either due to one-gluon exchange (OGE, see Section \ref{MICOGEsection}) or to meson exchange (GBE, see Section \ref{MICGBEsection}) or to a mixture of them (hybrid model, see Section \ref{MIChybridsection}). Note however that models containing only meson fields to describe baryons, like the Skyrme model, have also been proposed \cite{SKY61,WIT83}.
\\

Many hadron properties have been described with success by such models. It was then natural to extend the studies to the nucleon-nucleon interaction and the hadron in general, within quark models. The first attempts from Liberman \cite{LIB77} (potential model) and DeTar \cite{DET78} (bag model) did not really succeed. They obtained a soft core. This is today understood because they both used an adiabatic approximation and neglected important symmetry states. Harvey \cite{HAR81} performed a calculation of the Liberman type but with a more realistic Hamiltonian and including the missing symmetry states. He showed that indeed, both the mixing of $\Delta\Delta$ and a hidden colour states into the two-nucleon wave function or alternatively some states of higher spatial symmetry had dramatic effects. In other words this means that before Harvey only the state of spatial symmetry $[6]_O$ was used. Harvey proved the importance of the $[42]_O$ state in the nucleon-nucleon wave function describing the $L=0$ relative motion.
\\

Based on a constituent quark one-gluon exchange model, and still in an adiabatic approach, Maltman and Isgur \cite{MAL83} were later able to describe reasonably well some properties of the nucleon-nucleon system as {\it e. g.} the middle-range attraction. However their work remains nowadays under controversy \cite{OKA80}. Moreover, the long-range part of the nucleon-nucleon potential was described by the pion exchange at the level of the nucleon itself. This picture implies that something is not understood. The origin of this assumption is however very clear now : for such a range we are in a small momentum transfer region where the perturbative approach of QCD does not work. Maltman and Isgur, as well as others, avoid the problem simply by neglecting what happens at large distances at the quark level and add meson exchange at the nucleon level. We shall see in the following how these drawbacks have been removed.
\\

Actually in the nucleon-nucleon studies based on quark models there are three major steps : 1) the choice of the quark-quark interaction such as to properly describe the nucleon spectrum, 2) the truncation of the Hilbert space, as discussed above, which means the choice of the most important six quark states used in diagonalizing the Hamiltonian and 3) the treatment of the six-quark system itself, as a many body problem.
\\

So far we have presented studies based on a static treatment of the six-quark system, namely the adiabatic approximation (or the Born-Oppenheimer approximation). This is a well known approximation where one defines a separation distance between two composite clusters and calculates the interaction potential by assuming that the internal degrees of freedom are frozen. Note that if this separation distance is identified with the relative coordinate between the two clusters one obtains the Born-Oppenheimer approximation.
\\

It is in the beginning of the eighties, with the works of Oka, Yazaki {\it et al.} \cite{OKA80} in the framework of the resonating group methods (RGM) and of Harvey {\it et al.} \cite{HAR84} based on the generator coordinate method (GCM) that a dynamical treatment of the nucleon-nucleon interaction has been considered. These methods first introduced by Wheeler in 1937 \cite{WHE37} are appropriate to treat the interaction between two composite systems and have been first applied to the $\alpha-\alpha$ scattering\footnote{See the review of microscopic methods for the interaction between complex nuclei in the works of Saito \cite{SAI77}, Horiuchi \cite{HOR77} and Kamimura \cite{KAM77}. In particular the generator coordinate method and the resonating group method are explained in details.}. They allow to calculate both scattering and bound states. Restricting to quark degrees of freedom and non-relativistic kinematics, the resonating group method can straightforwardly be applied to the study of the baryon-baryon interaction in the quark model. The phase shifts calculated by Oka and Yazaki, for the first time, demons\-trated the existence of a short-range repulsion in the nucleon-nucleon potential as due to the one gluon exchange hyperfine quark-quark interaction. Another important issue is the introduction of non-local effects in the potential.
\\

But let us return to step 1 : the model itself. Nowadays, there are three kinds of constituent quark models based on a quark-quark interaction potential. The oldest ones are based on the exchange of one gluon between quarks. These models are called one-gluon exchange model or {\it OGE Models}. These models do not take into account the chiral aspect of QCD and in particular the spontaneous breaking of chiral symmetry. Therefore, in the nucleon-nucleon problem, they are unable to describe the middle- and long-range interaction. In order to avoid this last difficulty, one can include a pion exchange interaction at the quark level in addition to the OGE interaction. This approach is at the basis of the so-called {\it Hybrid Model}. Recently Glozman and Riska \cite{GLO96a} proposed a model where they dropped the gluon exchange hyperfine interaction leaving in the Hamiltonian only a pseudoscalar meson exchange. The main argument in doing so is that at the energy considered here (of the order of 1 GeV) the fundamental degrees of freedom of QCD, gluons and current quarks, have to be replaced by effective degrees of freedom such as the so-called Goldstone bosons (pseudoscalar mesons) and constituent quarks, due to the spontaneous breaking of chiral symmetry (see Section \ref{GBEchiralsection}). In this sense for light hadrons, there is no need of quark-gluon interaction anymore. Moreover the long-range NN problem is resolved automatically because of the presence of these Goldstone bosons. Such models are called Goldstone boson exchange models or {\it GBE Models}. More details on these three models will be presented in the following.
\\

Moreover, in quark models one has used either a non-relativistic or a relativistic kinematic. But considering the size of baryons, which is about 0.5 - 1.0 fm, it seems that the non-relativistic treatment can hardly be justified because it leads to velocities of the order of $v \sim c$. Surprisingly, in the framework of non-relativistic models, the agreement with expe\-riment was good for many observables. However semi-relativistic approaches have also been proposed. This means that in the potential model, the kinetic term $\frac{p^2}{2m}+m^2$ is replaced by $\sqrt{p^2+m^2}$. If for some observables this description is of crucial importance, the treatment of the relative motion of two baryons being essentially non-relativistic in the energy range where data are well established, the study of nucleon-nucleon interaction in a non-relativistic approach should be the first task to achieve.
\\

Non-relativistic models are all based on a Hamiltonian formed of a kinetic term and a confinement potential on the one hand, and a hyperfine spin-dependent term describing a short-range quark-quark interaction inside hadrons on the other hand. In other terms, the Hamiltonian is generically written as

\begin{equation}\label{MIChamilgeneric}
H = T + V_{Conf}+ V_{Hyp}
\end{equation}
\ 

\noindent where the kinetic term is obviously defined by

\begin{equation}\label{MICkinetic}
T=\sum_i^N m_i+\sum_i^N \frac{p^2_i}{2m_i}-K_G
\end{equation}
\ 

\noindent where $N$ is the number of quarks in the system, three for baryons, six for the nucleon-nucleon system and so on, $m_i$ are the quark masses, $p_i$ their momenta and $K_G$ is the kinetic energy of the center of mass. The confinement two-body term contains a colour operator and has the form

\begin{equation}\label{MICconf}
V_{Conf}=\sum_{i<j}^N V_{Conf}(r_{ij})=\sum_{i<j}^N \left( -\frac{3}{8} \lambda_i^c \lambda_j^c  \right) V_{conf}(r_{ij})
\end{equation}
\ 

\noindent where $\lambda^c$ are the colour $SU_C(3)$ generators. This form of confinement is mainly inspired by lattice QCD where the quark-quark interaction responsible for confinement comes from the chromoelectric part of the gluon field and simulates the non-linear aspects of QCD. However QCD lattice calculations lead to rather contradictory results. Some \cite{TAK01} give support to the Y-type flux tube picture {\it i. e.} a genuine three-body force, while others \cite{BAL01} favor the $\Delta$ ansatz {\it i. e.} a pair-wise interaction as in Eq. (\ref{MICconf}). In the following we shall use Eq. (\ref{MICconf}) with

\begin{equation}
V_{conf}(r)=V_0 + C r.
\end{equation}
\ 

\noindent where $V_0$ is a global constant introduced to adjust the position of the entire spectrum.
\\

The last term of Eq. (\ref{MIChamilgeneric}), $V_{Hyp}$ defines the model used in hadron spectroscopy. In this thesis we shall use the GBE model with its hyperfine interaction potential. However to really understand the advantages (and the disadvantages) of this model, in the following three sections we shall give a brief review of the three nowadays versions of constituent quark models mentioned above.
\\

\section{The OGE Model}\label{MICOGEsection}
\

In this section we present the hyperfine interaction potential introduced in Eq. (\ref{MIChamilgeneric}) for the one-gluon exchange model. This model is based on the idea that at short-range perturbative QCD theory can be applied, which implies gluon exchange between quarks in the lowest order. The validity of the perturbative method depends on the value of the strong interaction coupling constant $\alpha_s$. In these potential models the value of $\alpha_s$ is generally greater than one, which indicates a contradiction with perturbative QCD. Good agreement with experiment is however obtained for hadron spectra \cite{ISG77}.
\\

In the case of a system of quarks where the interaction is of a gluon exchange nature, the hyperfine interaction is similar to that of quantum electrodynamics (QED) for photon exchange (the Breit-Fermi interaction)

\begin{equation}\label{MICOGE}
V_{Hyp}^{OGE} =\sum_{i<j}^N V_{hyp}(r_{ij})=\sum_{i<j}^N (\alpha_e q_i q_j +\alpha_s \sum_c \frac{\lambda_i^c}{2}\frac{\lambda_j^c}{2})\ S_{ij}
\end{equation}
\ 

\noindent where $\alpha_e$ and $\alpha_s$ are the QED and QCD coupling constants respectively. The charge of the quark $i$ is given by $q_i$. But in the following we shall neglect the electromagnetic contribution because $\alpha_e << \alpha_s$. The form of the $S_{ij}$ operator given by de R\`ujula {\it et al.} \cite{RUJ75} is 

\begin{equation}\label{MICSij}
S_{ij}=S_{ij}^0+S_{ij}^{SS}+S_{ij}^T+S_{ij}^{SO}
\end{equation}

\noindent with

\begin{equation}\label{MICSij0}
S_{ij}^0=\frac{1}{r_{ij}}-\frac{1}{2m_im_j}(\frac{\vec{p}_i \cdot \vec{p}_j}{r_{ij}}-\frac{(\vec{r}_{ij}\cdot{\vec{p}_i})(\vec{r}_{ij}\cdot{\vec{p}_j})}{r_{ij}^3})-\frac{\pi}{2} (\frac{1}{m_i^2}+\frac{1}{m_j^2}) \delta^3(\vec{r}_{ij}),
\end{equation}
\ 

\begin{equation}\label{MICSijSS}
S_{ij}^{SS}=-\pi \frac{8\vec{S}_i\cdot \vec{S}_j}{3 m_i m_j}\ \delta^3(\vec{r}_{ij}),
\end{equation}
\\

\begin{equation}\label{MICSijT}
S_{ij}^T=-\frac{3(\vec{S}_i \cdot \vec{r}_{ij})(\vec{S}_j \cdot \vec{r}_{ij})}{m_i m_j r_{ij}^5}) + \frac{\vec{S}_i \cdot \vec{S}_j}{m_i m_j r_{ij}^3}
\end{equation}

and

\begin{eqnarray}\label{MICSijSO}
S_{ij}^{SO} &=& \frac{1}{r_{ij}^3}  \left[ \frac{1}{2m_i^2}(\vec{r}_{ij} \times \vec{p}_i) \cdot \vec{S}_i - \frac{1}{2m_j^2}(\vec{r}_{ij} \times \vec{p}_j) \cdot \vec{S}_j \right. \nonumber \\
 &&\left.+ \frac{1}{m_i m_j}\left( (\vec{r}_{ij} \times \vec{p}_i) \cdot \vec{S}_j - (\vec{r}_{ij} \times \vec{p}_j) \cdot \vec{S}_i \right) \right] .
\end{eqnarray}
\\

In this model there are three different kinds of terms : a spin-spin term $S_{ij}^{SS}$ proportional to a delta-function, therefore operating only on quark pairs with zero angular momentum, a tensor term $S_{ij}^T$ contributing only for non-zero angular momenta and a spin-orbit term $V_{ij}^{SO}$. The Coulomb type term in $S_{ij}^0$, proportional to $r^{-1}$, is often integrated in the confinement potential and $S_{ij}^{SO}$ as well as $S_{ij}^T$ terms are usually neglected due to their small contribution to baryon spectra.
\\

Thus in practical applications the hyperfine interaction reduces to its spin-spin part. However in order to test the role of the tensor term, Isgur and Karl \cite{ISG77} for example, took it in the form

\begin{equation}\label{MICisgur}
V_{Hyp}^{OGE} \simeq \sum_{i<j}^N \alpha_s \left\{ S_{ij}^{SS} + S_{ij}^T \right\} \sum_c  \frac{\lambda_i^c}{2}\frac{\lambda_j^c}{2}
\end{equation}
\\

The characteristic feature of the operator of Eq. (\ref{MICisgur}) is the presence of both colour and spin operators. The presence of these operators is believed to explain the hadron spectroscopy and the short-range part of the nucleon-nucleon interaction.
\\

But the first and most important drawback of this model is its difficulty to reproduce the correct order of the lowest part of the nucleon and $\Delta$ spectra and in particular the position of the {\it Roper} resonance relative to the first negative-parity states. In the baryon-baryon problem, the difficulties come from the absence of meson exchange which are not included in the Hamiltonian (\ref{MICOGE}). The latter point is the origin of the hybrid models but it is with the proposal of Goldstone boson exchange type models that a correct ordering in both strange and non-strange baryons has been achieved.
\\

\section{The Hybrid Model}\label{MIChybridsection}
\

The second category of models used in the study of the nucleon-nucleon interaction, but less extensively in hadron spectroscopy, contains those models where in addition to one-gluon exchange, quarks interact also via a pseudoscalar and a scalar meson exchange. In these hybrid models the short-range repulsion of the nucleon-nucleon interaction is still attributed to one-gluon exchange but the middle- and long-range attraction is due to meson exchanges between quarks. The hyperfine part of the Hamiltonian (\ref{MIChamilgeneric}) associated to these models has a supplementary term coming from pion exchange and $\sigma$-meson exchange

\begin{equation}\label{MICfullhybrid}
V_{Hyp}^{hybrid} = V_{Hyp}^{OGE} + V^{OPE}_{Hyp} + V^{\sigma}_{Hyp}.
\end{equation}
\\

The {\it OPE} stands for one-pion exchange and is written as

\begin{equation}\label{MIChybrid}
V^{OPE}_{Hyp} = \sum_{i<j}^N V_{Hyp}^{OPE}(r_{ij}) \simeq \sum_{i<j} \alpha_{\pi} \vec{\tau}_i\cdot \vec{\tau}_j \ S_{ij}^{OPE}
\end{equation}
\ 

\noindent where $\alpha_{\pi}$ is the pion-quark coupling constant, $S_{ij}^{OPE}$ is given this time by a form associated to the pseudoscalar mesons. The detailed forms of $S_{ij}^{OPE}$ and $V^{\sigma}_{Hyp}$ can be found in \cite{KUS91}.
\\

\section{The GBE Model}\label{MICGBEsection}
\ 

More recently, a third category of models was proposed where the quark-quark interaction, besides confinement, is due entirely to meson exchanges between quarks. This is the chiral constituent quark model (or Goldstone boson exchange model) proposed by Glozman and Riska in Ref. \cite{GLO96a}. If a quark-pseudoscalar meson coupling is assumed, in a non-relativistic limit, one obtains a quark-meson vertex proportional to $\vec{\sigma}\cdot \vec{q}\ \lambda^f$ with $\vec{\sigma}$ the Pauli matrices, $\vec{q}$ the momentum of the meson and $\lambda^f$ the Gell-Mann flavour matrices. This generates a pseudoscalar meson exchange interaction between quarks which is spin and flavour dependent. Its schematic form, used in Section \ref{GBEmodelsection} of the next chapter, is

\begin{equation}\label{MICGBE}
V_{Hyp}^{GBE} = - \sum_{i<j} v(r_{ij}) \vec{\lambda}_i^f \vec{\lambda}_j^f\ \vec{\sigma}_i \cdot \vec{\sigma}_j
\end{equation}
\ 

\noindent where $v(r_{ij})$ is a radial form, independent of $f$ in the $SU(3)$ invariant limit. The explicit form, proposed initially by Glozman, Papp and Plessas \cite{GLO96b} for the spin-spin part of the GBE interaction, is

\begin{eqnarray}\label{MICglozman}
V_{Hyp}^{GBE}(r_{ij}) &=&
\left\{\sum_{f=1}^3 V_{\pi}(r_{ij}) \lambda_i^f \lambda_j^f \right. \nonumber \\
&+& \left. \sum_{f=4}^7 V_{\rm K}(r_{ij}) \lambda_i^f \lambda_j^f
+V_{\eta}(r_{ij}) \lambda_i^8 \lambda_j^8
+V_{\eta^{\prime}}(r_{ij}) \lambda_i^0 \lambda_j^0\right\}
\vec\sigma_i\cdot\vec\sigma_j,
\end{eqnarray}
\ 

\noindent with

\begin{equation}\label{MICglozmanpot}
V_\gamma (\vec r_{ij})=\frac{g_\gamma^2}{4\pi}\frac{1}{3}\frac{1}{4m_i m_j}\{\mu_\gamma^2\frac{e^{-\mu_\gamma r_{ij}}}{ r_{ij}}-4\pi\ \delta (\vec r_{ij})\},  \hspace{5mm} (\gamma = \pi, K, \eta, \eta' ) \end{equation}
\ 

\noindent where $\mu_\gamma$ are the  meson masses and $g_\gamma^2/4\pi$ are the quark-meson coupling constants. Note that this form also contains the exchange of $\eta'$ mesons. The contact term has to be regularized. Two non-relativistic versions of this model are available. They differ from each other by the way the contact term has been regularized. Both versions will be used in our derivation of the nucleon-nucleon interaction. For this reason the GBE model and in particular its non-relativistic versions will be described in details in the next chapter. Baryon spectra, form factors and strong decays will be presented in order to show the reader the performances of this model.
\\

With Chapter \ref{preliminarychapter} we shall enter the core of the subject, namely the nucleon-nucleon interaction. That is the starting part of my personal contribution too. That chapter constitute a first step in our investigation of finding out whether or not the GBE model can give rise to a short-range repulsion in the the nucleon-nucleon interaction. We used two different approaches based on the Born-Oppenheimer approximation and taking into account Harvey's transformations. In Chapter \ref{rgmchapter} we study the NN interaction in a dynamical way using the resonating group method (RGM). We present results for the s-wave phase shifts in the nucleon-nucleon scattering. We also introduce some necessary extensions of the GBE model like the contribution of the $\sigma$-meson exchange interaction between quarks or the tensor force and discuss the comparison with the OGE results and experiment. In the last chapter, conclusions and outlook of this work will be presented. Details of the calculations have been gathered in appendices.



\chapter{The Goldstone Boson Exchange Model} \label{gbechapter}

\vspace{3.5cm}

\section{Introduction}
\

\begin{figure}[H]
\begin{center}
\includegraphics[width=13.5cm]{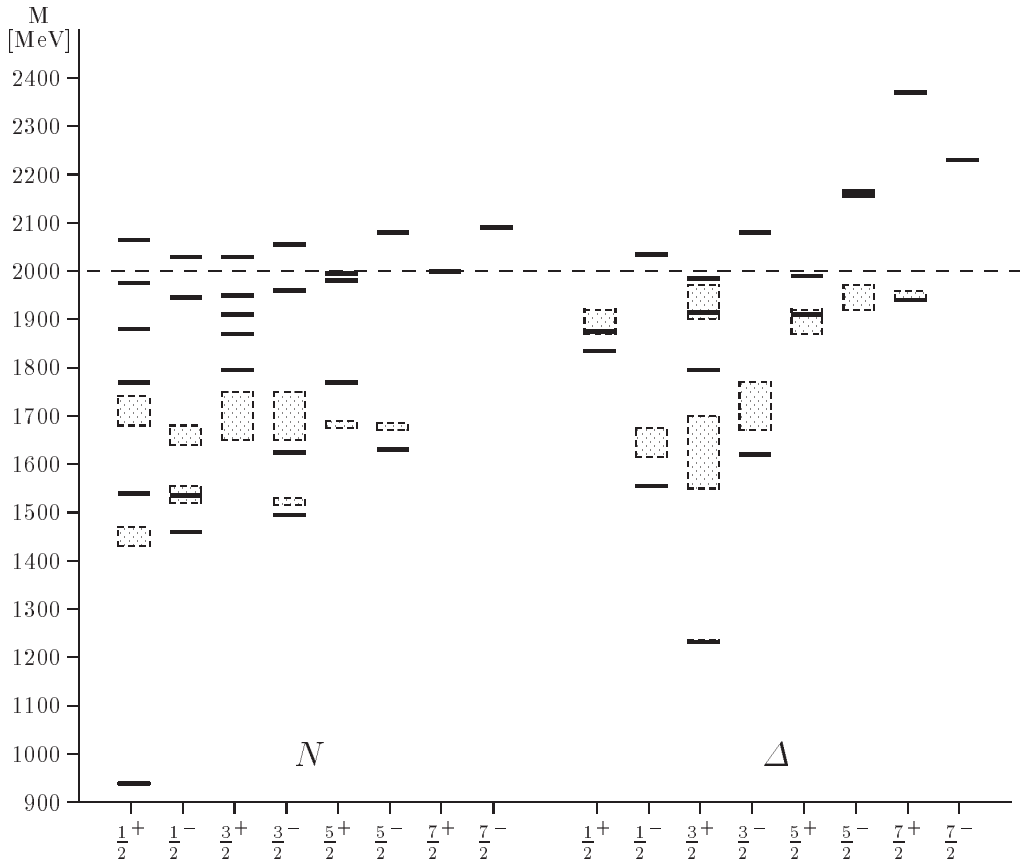}
\end{center}
\caption{\label{GBEisgurmasses}  Spectrum of the light baryons in the semi-relativistic version of the Capstick and Isgur model \cite{CAP86} where the contact term of Eq. (\ref{MICSijSS}) is smeared with the function $\delta(r) \simeq \frac{\sigma^3}{\pi^{3/2}} e^{-\sigma^2 r^2}$. The value of $\sigma$ is chosen to be equivalent to 0.1 fm. The dashed shade boxes give the experimental values in the commonly accepted Breit-Wigner fit. This figure has been extracted from Ref. \cite{THE01} }
\end{figure}

As it has been mentioned in the previous chapter, one-gluon exchange (OGE) models are motivated by perturbative QCD. However it is important to note that the justification of these models relies more on its successes than on its theoretical grounds. Indeed for OGE the only relevant energy scale is $\Lambda_{QCD} \approx 200$ MeV, the confinement scale. This point of view ignores the two-scale picture of Manohar and Georgi \cite{MAN84} in which the degrees of freedom should be the constituent quarks and the chiral meson fields below the chiral scale $\Lambda_{\chi} \approx 1$ GeV.  However, besides this theoretical drawback, OGE models have two other fundamental shortcomings. The first is the wrong prediction in the ordering of positive- and negative-parity states of baryons as compared to the experimental data. For example in Fig. \ref{GBEisgurmasses} Capstick and Isgur \cite{CAP86} gave the spectra of the light baryons $N$ and $\Delta$ in a semi-relativistic version of the one-gluon exchange model compared to the well accepted Breit-Wigner fit of experimental data. It is here important to note that this fit is not really unique in the sense that it relies on conventions which are sometimes different from one experimental group to another. Here we use the Particle Data Group (PDG) values.
\\

In OGE models, the Roper resonance $N_{1440}$ as well as its equivalent in the $\Delta$ spectrum, namely the $\Delta_{1600}$ resonance, are systematically above the negative-parity doublets, the $N_{1520}-N_{1535}$ and the $\Delta_{1620}-\Delta_{1700}$ respectively. It must be stressed that this problem can not be arranged by any specific parametrization of the interaction. There are two reasons for that. The first is the type of interaction itself, its dominant part being the spin-spin part of the Breit-Fermi interaction, given by

\begin{equation}\label{GBEcm}
V_{Hyp}^{OGE} \propto - \sum_{i<j} v(r_{ij}) \vec{\lambda}_i^c \vec{\lambda}_j^c\ \vec{\sigma}_i \cdot \vec{\sigma}_j = - \sum_{i<j} v(r_{ij}) O^{OGE}_{ij}
\end{equation}
\ 

\noindent where $\lambda^c$ are the $SU_C(3)$ Gell-Mann matrices and $\vec{\sigma}$ the $SU_S(2)$ Pauli matrices associated to the spin and $v(r)$ the regularized form of (\ref{MICSijSS}). If we look at the matrix elements of this interaction between two quarks with a definite colour-spin symmetry, we have

\begin{equation}\label{GBEOGE}
\left< [f]_C \times [f']_S : [f'']_{CS} |O^{cm}_{ij}|[f]_C \times [f']_S : [f'']_{CS} \right> = \left\{ \begin{array}{cl} 
\ 8 & [11]_C \times [11]_S : [2]_{CS},\\
-\frac{8}{3} & [11]_C \times [2]_S : [11]_{CS}.
\end{array} \right.
\end{equation}

According to (\ref{GBEOGE}), the immediate consequence is that the $V_{Hyp}^{OGE}$ contribution to the $\Delta^{[111]_{CS}}$ state, proportional to 8, is repulsive, while for the ground state nucleon, proportional to -8, it is attractive. The $\Delta$ is then heavier than $N$. But the {\it Roper} resonance $N_{1440}$ and $N_{1520}-N_{1535}$ doublets have the same $[21]_{CS}$ mixed symmetry as the ground state nucleon which indicates that the contribution of the spin-colour term is the same. Thus, because $N_{1440}$ belongs to $N=2$, and the lowest negative-parity states $N_{1520}-N_{1535}$ to $N=1$, they should lie about $\hbar \omega$ under the {\it Roper} resonance. Similar arguments can be used for the $\Delta$ case.
\\

A second reason for missing the level order comes from the potential shape of the spin independent part of the interaction. H\o gaasen and Richard \cite{HOG83} showed that if the Laplacian of a two-body potential $V$ is positive, $\Delta V > 0$, then $E(2S) > E(1P)$, {\it i. e.}, the first radial excitation comes above the first orbital excitation. This is the case for a harmonic oscillator or a linear confinement potential. As mentioned above, $V_{Hyp}^{OGE}$ cannot reverse this order, so that $N_{1440}$ appears above $N_{1520}-N_{1535}$ contrary to the experiment. Several solutions have been proposed, including the philosophy that a 100 MeV discrepancy on a particular level is not necessary a dramatic problem \cite{CAP86}. Another solution is the introduction in the potential of some terms with a negative Laplacian such as a scalar meson-exchange $V^{\sigma} \propto -e^{-\mu r}/r$. However, it has been showed \cite{STA00} that its strength is not sufficient to change the level order if a regularization term is added. But for $V^{\sigma} \propto -e^{-\mu r}/r$ alone, one can find a strength for which the order is reversed. Three-body forces have also been proposed as candidates to affect radial excitations of the nucleon, while producing no effect on states with mixed symmetry \cite{DES92}. The introduction of a specific spin-flavour dependence in the potential leads to an effective potential for negative-parity states which differs from the one governing the ground state and its radial excitations. This is in fact what occurs in the pseudoscalar meson exchange model of our interest, introduced first by Glozman and Riska \cite{GLO96a} and presented hereafter. In that model, the level ordering is definitely solved for non-strange and strange baryons simultaneously. For other considerations on level ordering in baryon spectra see the review of Richard \cite{RIC01} about few-body problems in hadron spectroscopy and references there.
\\

Another problem associated with the OGE interaction is the presence of large spin-orbit forces. In actual calculations the spin-orbit term coming from the Breit-Fermi interaction is simply dropped \cite{ISG79} because this term, given by Eq. (\ref{MICSijSO}) in the previous chapter, should give contributions which are not observed experimentally. Practitioners of the OGE models explained that a spin-orbit force due to the confining interaction cancels part of the spin-orbit (\ref{MICSijSO}). However the spin-orbit problem is still not clear in any constituent quark model and further investigations will be necessary in order to better understand the situation.
\\

In addition, indications against the dominant role of strong gluon exchange interactions at low energy are provided by QCD lattice calculations of Chu {\it et al.} \cite{CHU94}, Negele \cite{NEG99} and Liu {\it et al.} \cite{LIU99}. Moreover, the one-pion exchange potential alone between quarks appears naturally as an iteration of the instanton-induced interaction in the $t$ channel \cite{GLO00b}.
\\

As already mentioned, in our study of the NN system seen as a six-quark system, we shall use the GBE model. Its origin is thought to be in the spontaneous breaking of chiral symmetry \cite{GLO96a}. For this reason, in the following section we shall discuss in a more detailed way the consequence of this concept. Also, in this chapter some hadronic properties obtained in the framework of the Goldstone boson exchange model (GBE) will be presented. A description and a more detailed justification of the model is also given. Results on baryon spectra as well as form factors are analyzed. Decays properties are presented too. The very good agreement with experiment at the baryon level can be considered as the main motivation of this work where we apply the GBE model to the study of a baryon-baryon system and in particular to the nucleon-nucleon interaction.
\\

\section{A note on chiral symmetry}\label{GBEchiralsection}
\

In this section we shall present in more details few aspects of chiral symmetry. On a simple example the spontaneous breaking of this symmetry will be introduced as well as the consequences it produces.
\\

A chiral symmetric Lagrangian is a Lagrangian where not only the vector current but also the axial current are conserved under the following global chiral transformation :

\begin{equation}\label{GBEchiral}
\Psi^{'}_a = e^{-i\gamma_5 \vec{\varepsilon}\cdot\frac{\vec{\tau}}{2}}\Psi_a.
\end{equation}
\

Generally this transformation should act in the $SU_F(3)$ flavour space but here, for simplicity, we restrict to $SU_I(2)$, {\it i. e.} we consider the isospin space only. In Eq. (\ref{GBEchiral}), $\vec{\varepsilon}$ is a three-component vector defining a rotation angle in the isospin space. A Lagrangian with massless fermions, hence a massless QCD Lagrangian is invariant under this transformation. However we know that even the lighter quarks have a non-zero current mass (of about 5 MeV). Therefore if we introduce a mass term of the form $M \bar{\Psi} \Psi$ in the Lagrangian, the chiral symmetry is lost. This slight symmetry breaking due to the quark masses is the basis of the partial conserved axial current hypothesis (PCAC) and we shall see in the next example what are the important implications of this symmetry breaking.
\\

\subsection{Example of a chirally symmetric Lagrangian}
\

Let us consider a simple Lagrangian, as introduced by Gell-Mann and L\'evy \cite{GEL60}. This Lagrangian contains a fermionic isodoublet $\Psi$ of zero mass, a scalar field $\sigma$ and a pseudoscalar isovector field $\vec{\pi}$

\begin{eqnarray}\label{GBElagrangian}
{\cal{L}}&=& \frac{i}{2} ( \bar{\Psi} \gamma^{\mu} (\partial_{\mu} \Psi) -  (\partial_{\mu}\bar{\Psi}) \gamma^{\mu} \Psi ) + g \bar{\Psi}(\sigma + i \vec{\tau}\cdot\vec{\pi} \gamma_5 )\Psi \nonumber \\
&&+\frac{1}{2} (\partial_{\mu} \sigma)^2 + \frac{1}{2} (\partial_{\mu} \vec{\pi})^2 - C^2(\sigma^2 + \vec{\pi}^2 - A)^2
\end{eqnarray}
\ 

\noindent where $C$ and $A$ are real and $g$ is the boson-quark coupling constant. Note that the colour dependence is not written explicitly. This Lagrangian is invariant under the chiral transformation (\ref{GBEchiral}).
\\

This can be easily shown by using an infinitesimal chiral transformation, under which the fields become

\begin{eqnarray}\label{GBEfields}
\Psi  & \rightarrow & \left(1-i \varepsilon_\alpha \gamma_5 \frac{\tau^\alpha}{2} \right) \Psi,  \nonumber \\
\sigma  & \rightarrow & \sigma - \vec{\pi} \cdot \vec{\varepsilon}, \nonumber \\
\vec{\pi}  & \rightarrow & \vec{\pi} + \sigma\ \vec{\varepsilon}.
\end{eqnarray}
\ 

It turns then out that the vector and axial currents given by

\begin{eqnarray}\label{GBEcurrents}
\vec{j}^\mu &=&   \bar{\Psi} \gamma^{\mu} \frac{\vec{\tau}}{2}\Psi + \vec{\pi} \times (\partial^\mu \vec{\pi}) \nonumber \\
\vec{j}^\mu_5 &=& \bar{\Psi} \gamma^{\mu} \gamma_5 \frac{\vec{\tau}}{2}\Psi + \sigma (\partial^\mu \vec{\pi}) -  (\partial^\mu \sigma) \vec{\pi}
\end{eqnarray}
\ 

\noindent are conserved.
\\

One can rewrite the chiral rotation Eq. (\ref{GBEchiral}) in an alternative way \cite{STA02} as

\begin{equation}\label{GBEchiralRL}
\Psi^{'}_a = e^{-\frac{i}{2}(1+\gamma_5) \vec{\varepsilon}\cdot\frac{\vec{\tau}}{2}}\ e^{\frac{i}{2}(1-\gamma_5) \vec{\varepsilon}\cdot\frac{\vec{\tau}}{2}}\Psi_a.
\end{equation}
\ 

\noindent {\it i. e.} as a product of a right ($R$) and a left ($L$) transformation. Due to (\ref{GBEchiralRL}) one can say that the chiral rotation form the $SU_R(2) \times SU_L(2)$ direct product group. This implies that the Lagrangian (\ref{GBElagrangian}) is invariant under $SU_R(2) \times SU_L(2)$ transformations. It is useful to introduce such a language because the chiral transformations written as in (\ref{GBEchiral}) do not form a group. The reason is that the operators $\gamma_5 \tau_i\ (i=1,2,3)$ do not form a closed Lie algebra. Actually because of the property $\gamma_5^2=1$ one has

\begin{equation}\label{GBEnoLie}
\left[\gamma_5 \tau_i, \gamma_5 \tau_j \right] = i \varepsilon_{ijk} \tau_k
\end{equation}
\ 

\noindent {\it i. e.} the operator on the right hand side is different from those in the left hand side, so we do not have a Lie algebra. The situation is however different for the $R$ (and $L$) transformations. For example one has

\begin{equation}\label{GBEsuRL}
\left[\frac{1+\gamma_5}{2} \tau_i, \frac{1+\gamma_5}{2} \tau_j \right] = i  \varepsilon_{ijk} \frac{1+\gamma_5}{2}\tau_k.
\end{equation}
\ 

This is an $SU_R(2)$ algebra with generators $\frac{1+\gamma_5}{2} \vec{\tau}_i \ (i=1,2,3)$. Similarly, the generator of the $SU_L(2)$ algebra are $\frac{1-\gamma_5}{2} \vec{\tau}_i$. In practice one works with the transformation (\ref{GBEchiral}) but keeps in mind (\ref{GBEchiralRL}).
\\

The symmetry $SU_L(2) \times SU_R(2)$ can be seen explicitly in the Lagrangian (\ref{GBElagrangian}). In particular $\sigma^2+\vec{\pi}^2$ is a chiral invariant term and expresses the fact that the fields $\sigma, \pi_i\ (i=1,2,3)$ have the same mass equal to $m=\sqrt{4|A|C^2}$.
\\

\subsection{Spontaneous breaking of the chiral symmetry}\label{spontaneoussubsection}
\

Using again the Lagrangian density (\ref{GBElagrangian}), we can see that if the value of $A$ is positive, the minimum of the self-interaction potential is not located at $(0,0)$ but at $(<\sigma>,<\vec{\pi}>) \neq (0,0)$. However classical solutions may be considered as mean field approximations. Classical minima may then be understood as the vacuum states of the considered fields. If we get this minimum back to the origin we shall see that quarks could acquire a mass. Of course in the example considered here, this mass is related to the parameters $A$ and $C$.
\\

Let us first look at the Hamiltonian associated to the bosonic part of the Lagrangian. This Hamiltonian density gets the form

\begin{equation}\label{GBEboson}
{\cal{H}} = \frac{1}{2}(\dot{\vec{\pi}}^2 + \dot{\sigma}^2) + \frac{1}{2}(\nabla \sigma)^2 + \frac{1}{2}(\nabla \vec{\pi})^2 + C^2(\sigma^2 + \vec{\pi}^2 - A^2)^2.
\end{equation}
\ 

For the ground state of the Hamiltonian (\ref{GBEboson}), two situations can occur
\begin{itemize}
\item $A\leq 0$

In this case the minimum of the potential is obviously located at $(\sigma,\vec{\pi})=(0,0)$ as seen on the left part of Fig. \ref{GBEspontaneous}. We can then expand the solutions around this point. The mesonic fields have the same mass, and the fermionic field have no mass. The symmetry is here explicit and is known as the Wigner-Weyl mode.
\\

\begin{figure}[H]
\begin{center}
\includegraphics[width=15cm]{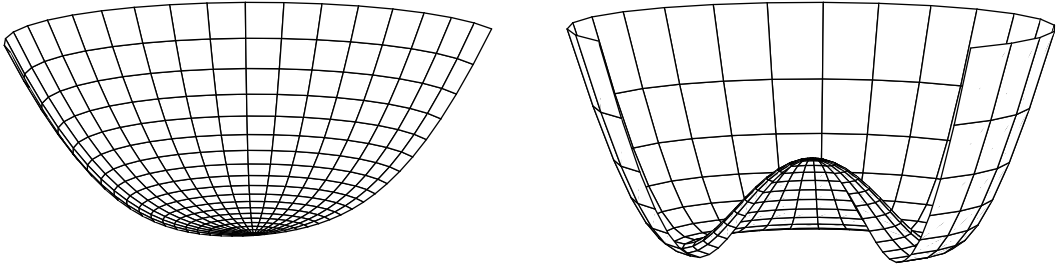}
\end{center}
\caption{\label{GBEspontaneous} Effective potential $C^2\ (\sigma^2 + \vec{\pi}^2 - A^2)^2$ without ($A\leq 0$, left) and with ($A>0$, right) spontaneous breaking of symmetry.}
\end{figure}

\item $A>0$

Here the potential has an infinity of minima for $\sigma$ and $\vec{\pi}$ corresponding to the equation $\sigma^2 + \vec{\pi}^2 = A$, called the chiral circle. The situation is illustrated on the right part of Fig. \ref{GBEspontaneous}. If we choose to keep $<\vec{\pi}> = 0$ then $<\sigma> = \pm \sqrt{A}$. Let us choose the positive square root. Noting that these values represent the mean field in the vacuum state, to get them at the origin we need to do the following transformation :

\begin{eqnarray}
\vec{\pi}&\rightarrow&\vec{\pi} \nonumber \\
\sigma&\rightarrow& \sigma - <\sigma>=\sigma - \sqrt{A}.
\end{eqnarray}
\ 

In this situation the total Lagrangian (\ref{GBElagrangian}) density becomes

\begin{eqnarray}\label{GBElagrangian2}
{\cal{L}}&=& \frac{i}{2} ( \bar{\Psi} \gamma^{\mu} (\partial_{\mu} \Psi) -  (\partial_{\mu}\bar{\Psi}) \gamma^{\mu} \Psi ) + g \bar{\Psi}(\sigma + i \vec{\tau}\cdot\vec{\pi} \gamma_5 )\Psi - M \bar{\Psi}\Psi \nonumber \\
&&+\frac{1}{2} (\partial_{\mu} \sigma)^2 + \frac{1}{2} (\partial_{\mu} \vec{\pi})^2 - \frac{1}{2} m_\sigma^2 \sigma^2 - \frac{1}{2} m_\pi^2 \vec{\pi}^2 \nonumber\\
&&- C^2(\sigma^2 + \vec{\pi}^2)^2 -4\sqrt{A}C^2 \sigma(\sigma^2+\vec{\pi}^2)
\end{eqnarray}
where
\begin{eqnarray}
M&=&-g\sqrt{A},\\
m_\pi^2&=&0,\\
m_\sigma^2&=&8AC^2.
\end{eqnarray}
\ 

We then see that the quark field acquires a mass and the masses of the mesonic fields are not equal anymore. The pion field is still massless but the $\sigma$ field acquired a mass $m_\sigma$. It is because $\sigma$ and $\vec{\pi}$ do not have the same mass and fermions have nonzero mass that we talk about hidden or spontaneously broken symmetry. The massless $\pi$ boson is called a Goldstone boson. Finally note that the Lagrangian is still chirally symmetric. In this case the symmetry is realized in the Nambu-Goldstone (hidden) mode.
\end{itemize}
\ 

\subsection{PCAC}\label{PCACsection}
\

Now we study the concept of partial conservation of the axial current (PCAC). In order to understand it we shall introduce a term in the Lagrangian density (\ref{GBElagrangian}) which breaks chiral symmetry. We take it in the form $c\ \sigma$. The Lagrangian then becomes

\begin{eqnarray}\label{GBElagrangian3}
{\cal{L}}&=& \frac{i}{2} ( \bar{\Psi} \gamma^{\mu} (\partial_{\mu} \Psi) -  (\partial_{\mu}\bar{\Psi}) \gamma^{\mu} \Psi ) + g \bar{\Psi}(\sigma + i \vec{\tau}\cdot\vec{\pi} \gamma_5 )\Psi \nonumber \\
&&+\frac{1}{2} (\partial_{\mu} \sigma)^2 + \frac{1}{2} (\partial_{\mu} \vec{\pi})^2 - C^2(\sigma^2 + \vec{\pi}^2 - A)^2 + c\ \sigma.
\end{eqnarray}
\ 

With this density one finds that the axial current and its four-divergence are given by

\begin{eqnarray}\label{GBEcurrentsPCAC}
\vec{j}^\mu_5 &=& \bar{\Psi} \gamma^{\mu} \gamma_5 \frac{\vec{\tau}}{2}\Psi + \sigma (\partial^\mu \vec{\pi}) -  (\partial^\mu \sigma) \vec{\pi} \nonumber \\
\partial_\mu \vec{j}^\mu_5 &=& -c\ \vec{\pi}.
\end{eqnarray}
\ 

Chiral symmetry is then explicitly broken when $c\neq 0$.
\\

If we look now after the minimum of this new potential, and choosing again $<\vec{\pi}>=0$, we get the following equation

\begin{equation}\label{GBEsigmaPCAC}
\left. \frac{\partial \left( C^2(\sigma^2+\vec{\pi}^2-A)^2-c\ \sigma \right)}{\partial \sigma} \right|_{\vec{\pi}=0} = 4C^2\sigma(\sigma^2-A)-c=0
\end{equation}
\\

Hence the ground state values of $\sigma$ satisfies

\begin{equation}\label{GBEsigmaGround}
<\sigma>^2-A= \frac{c}{4C^2<\sigma>}
\end{equation}
\\

Introducing a new field $\sigma \rightarrow \sigma - <\sigma>$ in the Lagrangian density and using $A$ as given by (\ref{GBEsigmaGround}) we get

\begin{eqnarray}\label{GBElagrangian4}
{\cal{L}}&=& \frac{i}{2} ( \bar{\Psi} \gamma^{\mu} (\partial_{\mu} \Psi) -  (\partial_{\mu}\bar{\Psi}) \gamma^{\mu} \Psi ) + g \bar{\Psi}(\sigma + i \vec{\tau}\cdot\vec{\pi} \gamma_5 )\Psi - M \bar{\Psi}\Psi \nonumber \\
&&+\frac{1}{2} (\partial_{\mu} \sigma)^2 + \frac{1}{2} (\partial_{\mu} \vec{\pi})^2 - \frac{1}{2} m_\sigma^2 \sigma^2 - \frac{1}{2} m_\pi^2 \vec{\pi}^2 \nonumber\\
&&+ C^2(\sigma^2 + \vec{\pi}^2)^2 +4C^2\sigma (\sigma^2+\vec{\pi}^2)<\sigma> + \frac{c^2}{16C^2<\sigma>^2}-c<\sigma>
\end{eqnarray}
where
\begin{eqnarray}
M&=&-g<\sigma>\label{GBEpcacM},\\
m_\pi^2&=&\frac{c}{<\sigma>}\label{GBEpcacmpi},\\
m_\sigma^2&=&\frac{c}{<\sigma>}+8C^2<\sigma>^2.\label{GBEpcacmsigma}
\end{eqnarray}
\ 

We see that the pion and $\sigma$ field acquire again different masses. Introducing (\ref{GBEpcacM}) and (\ref{GBEpcacmpi}) in (\ref{GBEpcacmsigma}) we obtain a relation which links the masses of the three fields

\begin{equation}
m_\sigma^2=m_\pi^2+\frac{8C^2}{g^2}M^2
\end{equation}
\\

Using Eq. (\ref{GBEpcacM}) we can eliminate $<\sigma>$ in (\ref{GBEpcacmpi}) to get

\begin{equation}
c=-\frac{M}{g}m_\pi^2
\end{equation}
\ 

Replacing this expression of $c$ in the divergence (\ref{GBEcurrentsPCAC}) we have

\begin{equation}
\partial_\mu \vec{j}^\mu_5 =m_\pi^2\frac{M}{g} \vec{\pi}
\end{equation}
\ 

From the matrix element of the weak axial vector current, related to pion decay, one can obtain
\begin{equation}
M= g f_\pi
\end{equation}
\ 

\noindent where $f_\pi$ is the pion decay constant. This lead to

\begin{equation}\label{GBEpcacrelation}
\partial_\mu \vec{j}^\mu_5 =f_\pi m_\pi^2 \vec{\pi}
\end{equation}
\ 

The above equation is referred to as PCAC. It implies that the axial current is almost conserved because the pion mass is small. Therefore the nonzero pion mass expresses the amount of explicit breaking of chiral symmetry. The PCAC relation (\ref{GBEpcacrelation}) connects the weak current $\vec{j}^\mu_5$ and the strong interacting pion field and has important experimental implications. Its application to physical observables is made through low energy theorems as {\it e. g.} the Goldberger-Treiman relation

\begin{equation}\label{GBEgtrelation}
f_\pi g_{\pi N}= m_N g^A_{\pi N}
\end{equation}
\ 

\noindent where $g_{\pi N}$ is the pion-nucleon coupling constant, $m_N$ the nucleon mass and $g^A_{\pi N}$ the axial vector coupling constant. The coupling constant of the quark with the pion $g_{\pi q}$ can be derived from an analog of the relation (\ref{GBEgtrelation}) at the quark level. The pion-quark coupling constant is a necessary ingredient of the GBE model. We shall extend the Goldberger-Treiman relation to the pion-quark coupling

\begin{equation}\label{GBEgtrelationq}
f_\pi g_{\pi q}= m_q g^A
\end{equation}
\ 

\noindent where $g_{\pi q}$ is the pion-(constituent)quark coupling constant, $m_q$ the $q$ quark mass ($q=u,d,s$) and $g^A$ the associated axial vector coupling constant. Weinberg has shown \cite{WEI90} that the constituent quarks have the bare unit axial coupling constant ($g^A=1$) and no anomalous magnetic moment. One thus obtains the relation

\begin{equation}\label{GBEgqn}
g_{\pi q}=\frac{3}{5}\frac{m_q}{m_N}g_{\pi N}
\end{equation}
\ 

The factor $\frac{3}{5}$ above comes from the spin-isospin matrix element when we consider the pion-nucleon interaction as the interaction between the pion and 3 constituent quarks. With $\frac{g^2_{\pi N}}{4\pi}=14.2$ one has $\frac{g^2_{\pi q}}{4\pi}=0.67$.
\\

Finally, it is interesting to note that if $A$ is chosen to be negative in Eq. (\ref{GBElagrangian3}), the ground state is given by $(<\sigma>,<\vec{\pi}>)=(0,0)$ which means that both fields have the same mass.
\\

Let us discuss the meaning of the spontaneous breaking of symmetry. We could assume that an effective QCD Lagrangian at zero temperature has a form similar to that described above with $A>0$. Since the ground state is not at the center, one of the fields will have a nonzero value. Usually this is the $\sigma$ field because it carries the vacuum quantum numbers. In quark language, this means that we expect to have a finite scalar quark condensate $<\bar{q}q>\neq 0$. In this way, pionic excitations are similar to small rotations of the ground state in the valley of right part of the Fig. \ref{GBEspontaneous}. Pion mass should then be zero as seen in the previous example. Excitations in the radial direction correspond to a perturbation of the $\sigma$ field and therefore are massive. Note that this does not break any symmetry and we are in total agreement with the Goldstone mode introduced above.
\\

The importance of the chiral symmetry for strong interactions was realized early \cite{PAG75}. This symmetry, which is almost exact in the light $u$ and $d$ flavour sector is however only approximate in QCD when strangeness is included because of the large mass of the $s$ quark (see Table \ref{GBEquarks}). Nevertheless even in 3-flavour QCD the current quark masses may, in a first approximation, be set to zero (the chiral limit) and their deviation from zero treated as a perturbation. The small finite masses of the current quark are however very important for the finite masses of the mesons. In the chiral limit all members of the pseudoscalar octet would have zero mass and we recover the situation depicted on the right part of the Fig. \ref{GBEspontaneous}.
\\

Indeed vacuum QCD contains $<q\bar{q}>$ states with a nonzero value as mentioned above. Shifman {\it et al.} \cite{SHI79} gives the approximate following values to the quark condensates

\begin{equation}\label{GBEshifman}
<u\bar{u}> \approx <d\bar{d}>\approx <s\bar{s}> \approx - (0.25\ {\rm GeV})^3
\end{equation}
\ 

\noindent which show that we are in the presence of spontaneous symmetry breaking of vacuum QCD. The relation from Gell-Mann-Oakes-Renner \cite{GEL68} relating the pseudoscalar mesons masses to the quark condensates (\ref{GBEshifman}) shows how the mesons acquire a mass. For example for pions we have

\begin{eqnarray}\label{GBEpion}
<m^2_{\pi^0}> = \frac{-1}{f^2_{\pi}}(m_u <u\bar{u}>+m_d <d\bar{d}>) \nonumber \\
<m^2_{\pi^{\pm}}> = \frac{-1}{f^2_{\pi}}\frac{m_u + m_d}{2}(<u\bar{u}>+<d\bar{d}>)
\end{eqnarray}
\ 

\noindent where $f_\pi$ is the pion decay constant already introduced and $m_u \approx m_d$ are the current mass of the quarks given in Table \ref{GBEquarks}. Analog relations also exist for other pseudoscalar mesons. Another important remark from Table \ref{GBEquarks} is the difference between $u$ and $s$ current mass. In the following we shall see that this is the basis of the explicit breaking of the $SU_F(3)$ flavour symmetry.
\\

In our work we shall deal with a non-relativistic Hamiltonian were the fundamental current quarks of QCD are replaced by the so-called constituent quarks, also called valence quarks. Table \ref{GBEquarks} also shows the most accepted values of constituent quark masses.
\\

\begin{table}[H]
\centering

\begin{tabular}{|l|cccccc|}

\hline
mass (GeV)                  & $u$          & $d$           & $s$          & $c$          & $b$         & $t$  \\

\hline
\hline

current \cite{GRO00}    & 0.001-0.005  & 0.003-0.009  & 0.075-0.170  & 1.15-1.35 & 4.0-4.4  & 160.8-179.4       \\
constituent & $\sim$ 0.330  & $\sim$ 0.330  & $\sim$ 0.500  & $\sim$ 1.2 & $\sim$ 4.2  & $\sim$ 175 \\

\hline

\end{tabular}
\caption{Commonly accepted values of the different flavour for current and constituent quark masses}\label{GBEquarks}

\end{table}

We note that the more quarks are heavy, the smaller is the difference between current and constituent quark masses. Therefore a non-relativistic description of heavy particles is entirely justified. However non-relativistic models are also used to describe light flavour baryons, mesons and other composite systems and the results obtained are very close to the experimental data. One of the reason for this is the possibility to easily extract the center of mass motion in a non-relativistic model.
\\

\section{The GBE model}\label{GBEmodelsection}
\

As already pointed out before, Manohar and Georgi \cite{MAN84} showed that there are two dif\-ferent scales in three flavour QCD. The first one, $\Lambda_{QCD} \approx 200$ MeV characterizes confinement and then gives more or less the size of a baryon. At the other one, $\Lambda_{\chi} \approx 1$ GeV the spontaneous breaking of chiral symmetry occurs and hence at distances beyond 0.2 fm dynamical constituent quark masses as well as Goldstone bosons (mesons) appear. The constituent quarks are particles with internal complex structure and the mesons are the chiral fields.
\\

Now looking at the confirmed states of the nucleon for example, one can split the spectrum in a low energy part where states are well separated and without nearby parity partners and a high energy part with an increasing number of near parity doublets. A natural interpretation of this feature is that the approximate chiral symmetry of QCD is realized in the hidden Nambu-Goldstone mode at low excitation and in the explicit Wigner-Weyl mode at high excitation. In an $SU_F(3)$ flavour QCD, the spontaneous breaking of chiral symmetry leads then to the existence of an octet of low mass pseudoscalar mesons. The $U(1)$ anomaly decouples the $\eta'\ SU_F(3)$-singlet from the original nonet. In line with these considerations one can conclude that below the chiral symmetry spontaneous breaking scale, a baryon should be considered as a system of three constituent quarks with an effective quark-quark interaction that is formed of a central confining part and a chiral part where the interaction between the constituent quarks is mediated by the octet of pseudoscalar mesons.
\\

The simplest representation of the most important component of the chiral interaction in the $SU_F(3)$ invariant limit is

\begin{equation}\label{GBEgbe}
V_{\chi} \propto - \sum_{i<j} v(r_{ij}) \vec{\lambda}_i^f \vec{\lambda}_j^f\ \vec{\sigma}_i \cdot \vec{\sigma}_j = - \sum_{i<j} v(r_{ij}) O^{\chi}_{ij}
\end{equation}
\ 

\noindent where $\lambda^f_i$ are the $SU_F(3)$ Gell-Mann matrices and $v(r)$ a smeared version of the $\delta$-function dominating at short range. Note the contrast with Eq. (\ref{GBEcm}) which contains $\lambda^c_i$ instead of $\lambda^f_i$. Because of the flavour dependent factor $ \vec{\lambda}_i^f \vec{\lambda}_j^f$ the interaction (\ref{GBEgbe}) will lead to correct ordering of the positive- and negative-parity states in the baryon spectra in all strange and non-strange sectors. By contrast with the OGE model for the nucleon or the $\Delta$, to some appropriate strength, the chiral interaction between the constituent quarks shifts the lowest positive-parity state in the $N=2$ band below the negative parity states in the $N=1$ band. The question arises then about what happens in the spectrum of the $\Lambda$ where experimental values give the first negative-parity state below the $N=2$ positive-parity state. Let us analyze the symmetry structure of the operator $O^{\chi}_{ij}$ to show that the $\Lambda$ ordering is reproduced as well as that of the non-strange baryon spectra.
\\

If we look at the matrix elements of $O^{\chi}_{ij}$ of Eq. (\ref{GBEgbe}) between two quarks with a definite flavour-spin symmetry, we have

\begin{equation}
\left< [f]_F \times [f']_S : [f'']_{FS} |O^{\chi}_{ij}|[f]_F \times [f']_S : [f'']_{FS} \right> = \left\{ \begin{array}{cl} 
\frac{4}{3}   & [2]_F \times [2]_S : [2]_{FS},\\
\ 8           & [11]_F \times [11]_S : [2]_{FS},\\
\ -4          & [2]_F \times [11]_S : [11]_{FS},\\
-\frac{8}{3} & [11]_F \times [2]_S : [11]_{FS}.
\end{array} \right.
\end{equation}
\ 

Consequently, because $v(r_{ij})$ is positive (smeared $\delta$-function), one finds the two following important properties : 1) the chiral interaction is attractive in symmetrical flavour-spin pairs and repulsive in anti-symmetrical ones, and 2) among the flavour-spin symmetrical pairs, the flavour anti-symmetrical ones experience a much larger attraction than the flavour symme\-trical ones. We find that the states $N_{1440}$ and $\Delta_{1600}$ which belong to the $N=2$ band are lowered in mass much more than the states $N_{1520}-N_{1535}$ and $\Delta_{1620}-\Delta_{1700}$ because of their symmetry structure as shown in Table (\ref{GBEtablesym}). The spacial symmetry of the state is indicated by the Young pattern $[f]_O$ and the angular momentum by $L$. The $[f]_F$, $[f]_S$ and $[f]_{FS}$ Young patterns denote the flavour, spin and combined flavour-spin symmetries, respectively. The totally antisymmetric colour state $[111]_C$, which is common to all the baryon states is suppressed in the notation.
\\

The possibility to reproduce the $\Lambda$ spectrum is based on the fact that the $\Lambda(1405)-\Lambda(1520)$ is lowered nearly as much as $\Lambda(1600)$, as seen in Table \ref{GBEtablesym} and thus the ordering implied by a confining interaction of oscillator or linear type is maintained. However this chiral attraction is generally not enough to reproduce precisely the experimental value of $\Lambda(1405)-\Lambda(1520)$.
\\

\begin{table}[H]
\centering

\begin{tabular}{|l|l|c|}

\hline
Physical states $J^\pi\ B(MeV)$                           &   $L[f]_O[f]_{FS}[f]_F[f]_S$       & $\left< O^{\chi}_{ij} \right>$\\

\hline
\hline

$\frac{1}{2}^+ N(939)$                                      &  $0[3]_O[3]_{FS}[21]_F[21]_S$    & -14\\
$\frac{1}{2}^+ N(1440)$                                     &  $0[3]_O[3]_{FS}[21]_F[21]_S$    &-14\\
$\frac{1}{2}^- N(1535) - \frac{3}{2}^- N(1520)$             &  $1[21]_O[21]_{FS}[21]_F[21]_S$  &-2\\
\hline
$\frac{3}{2}^+ \Delta(1232)$                                &  $0[3]_O[3]_{FS}[3]_F[3]_S$      &-4\\
$\frac{3}{2}^+ \Delta(1600)$                                &  $0[3]_O[3]_{FS}[3]_F[3]_S$      &-4\\
$\frac{1}{2}^- \Delta(1620) - \frac{3}{2}^- \Delta(1700)$   &  $1[21]_O[21]_{FS}[3]_F[21]_S$   &4\\
\hline
$\frac{1}{2}^+ \Lambda(1115)$                               &  $0[3]_O[3]_{FS}[21]_F[21]_S$    &-14\\
$\frac{1}{2}^- \Lambda(1405) - \frac{3}{2}^- \Lambda(1520)$ &  $1[21]_O[21]_{FS}[111]_F[21]_S$ &-8\\
$\frac{1}{2}^+ \Lambda(1600)$                               &  $0[3]_O[3]_{FS}[21]_F[21]_S$    &-14\\

\hline

\end{tabular}
\caption{Structure of the lowest states for $N$, $\Delta$ and $\Lambda$. The notation in the first column is $J^\pi\ B$ with $B$ the considered baryon, $J$ the total angular momentum and $\pi$ the parity. Experimental masses are given in brackets. The last column gives the matrix elements of the $O^{\chi}_{ij}$ operator for the considered state.}\label{GBEtablesym}

\end{table}
\ 

\subsection{The parametrization of the pseudoscalar exchange interaction}
\

In a simple effective chiral symmetric description of the baryon the coupling of the quarks and the pseudoscalar Goldstone bosons in an exact $SU_F(3)$ symmetry will have the form

\begin{equation}\label{GBEsimplelagrangian}
{\cal L} \sim i g \bar{\Psi}\gamma_5 \vec{\lambda}^f\cdot \vec{\phi} \Psi.
\end{equation}
\ 

\noindent where $\Psi$ is the fermion constituent quark field operator, $\vec{\phi}$ the octet boson field operator and $g$ the pseudoscalar boson-fermion coupling constant. A coupling of this form in a non-relativistic reduction produces a spin- and flavour-dependent interaction potential between the constituent quarks containing a spin-spin and a tensor part

\begin{equation}\label{GBEsimplepotential}
V_{\chi}(\vec{r}_{ij}) =  \vec{\lambda}^f_i \cdot  \vec{\lambda}^f_j \left\{ V^{SS}(\vec{r}_{ij}) \vec{\sigma}_i\cdot \vec{\sigma}_j + V^T(\vec{r}_{ij}) S^T_{ij}\right\}
\end{equation}
\ 

\noindent where $S^T_{ij}$ given by Eq. (\ref{MICSijT}) is the tensor operator between the quarks $i$ and $j$. Note in particular that the GBE interaction to lowest order does not lead to any spin-orbit force, in contrast to the OGE interaction. But a spin-orbit interaction will appear to second order (two correlated pions).
\\

Up to now we have considered only exact $SU_F(3)$ flavour symmetry. However, in reality this symmetry is broken. Indeed with the same $SU_F(3) \times SU_S(2)$ symmetry, the lowest $N$ and $\Lambda$ states have different experimental energies. In the $SU_F(3)$ symmetric limit the constituent quark masses would be equal ($m_u=m_d=m_s$), the pseudoscalar octet would be degenerate and the meson-constituent quark coupling constant would be flavour independent. In a broken $SU_F(3)$ symmetry the pseudoscalar fields, {\it i. e.} the pion, kaon and $\eta$ mesons have a different interaction with the quarks because of the different constituent quark masses ($m_u\neq m_d\neq m_s$), different meson masses  ($m_\pi\neq m_K \neq m_\eta$) and different meson-quark coupling constants ($g_{\pi q} \neq g_{Kq} \neq g_{\eta q}$). Note however that the very small mass difference between the $u$ and $d$ quark is neglected in the following. Thus if $SU_F(3)$ is broken, the flavour-dependent part of the potential is then split into

\begin{eqnarray}\label{GBEpotsu3broken}
V_{\chi}(\vec{r}_{ij}) &\rightarrow& \sum_{a=1}^3  \lambda^a_i \cdot  \lambda^a_j V_\pi(\vec{r}_{ij}) + \sum_{a=4}^7  \lambda^a_i \cdot  \lambda^a_j V_K(\vec{r}_{ij}) + \lambda^8_i \cdot  \lambda^8_j V_\eta(\vec{r}_{ij}) + \lambda^0_i \cdot  \lambda^0_j V_{\eta'}(\vec{r}_{ij}) \nonumber \\ 
\end{eqnarray}
where $V(\vec{r}_{ij})$ is either the spin-spin part $ V^{SS}(\vec{r}_{ij})$ or the tensor part $V^T(\vec{r}_{ij})$ of the interaction potential. The last term is the pseudoscalar singlet exchange potential. Goldstone bosons manifest themselves in the octet of pseudoscalar meson ($\pi, K, \eta$). In the large-$N_C$ limit, when axial anomaly vanishes \cite{WIT79}, the spontaneous breaking of chiral symmetry $U_L(3) \times U_R(3) \rightarrow U_V(3)$ implies a ninth Goldstone boson \cite{COL80}, which corresponds to the flavour singlet $\eta'$. Under real conditions, for $N_C=3$, a certain contribution from the flavour singlet remains and the $\eta'$ must thus be included in the GBE interaction.
\\

In a non-relativistic reduction the coupling (\ref{GBEsimplelagrangian}) will give rise to a Yukawa interaction between constituent quarks so that the meson exchange potential in Eqs. (\ref{GBEpotsu3broken}) becomes

\begin{equation}\label{GBEmodel}
V_{\gamma}^{SS}(\vec{r}_{ij}) = \frac{g^2_\gamma}{4\pi}\frac{1}{12m_i m_j} \left\{ \mu^2_\gamma \frac{e^{-\mu_\gamma r_{ij}}}{r_{ij}} -4\pi \delta(\vec{r}_{ij}) \right\}
\end{equation}
\ 

for the spin-spin part and
\ 

\begin{equation}\label{GBEmodeltensor}
V_{\gamma}^{T}(\vec{r}_{ij}) = \frac{g^2_\gamma}{4\pi}\frac{1}{12m_i m_j} \left\{ \mu^2_\gamma + \frac{3 \mu_\gamma}{r_{ij}} + \frac{3}{r^2_{ij}}  \right\} e^{-\mu_\gamma r_{ij}}
\end{equation}
\ 

\noindent for the tensor part, with $\gamma$ standing for $\pi,K,\eta$ and $\eta'$. The quark and meson masses are given by $m_i$ and $\mu_\gamma$ ($\gamma=\pi,K,\eta,\eta'$), respectively.
\\

In a chiral constituent quark model, the total Hamiltonian thus consists of the sum of a kinetic term (\ref{MICkinetic}), a confinement term (\ref{MICconf}) and the chiral interaction (\ref{GBEpotsu3broken})

\begin{equation}\label{GBEhamiltonian}
H = \sum_i^N m_i+\sum_i^N \frac{p^2_i}{2m_i}-K_G + \sum_{i<j}^N  -\frac{3}{8}\lambda_i^c \lambda_j^c\ V_{conf}(r_{ij}) + \sum_{i<j}^N V_{\chi}(\vec{r}_{ij})
\end{equation}
\ 

\noindent where $K_G$ is the center of mass kinetic energy operator.
\\

The chromoelectric confinement interaction is taken in a linear form with a strength parameter $C$ identified to be the string tension

\begin{equation}\label{GBEconf}
V_{Conf} = \sum_{i<j}^N  -\frac{3}{8}\lambda_i^c \lambda_j^c\ C\ r_{ij}
\end{equation}
\ 

This represents a very good approximation of the regular $Y$-shape flux tube picture for the 3-body force with a string configuration. We shall see later that in a semi-relativistic GBE model there is a confinement strength of the order of $C\approx 2.3$ fm$^{-2}$. We note that this value appears to be quite realistic, as it is consistent both with Regge trajectories slopes and also with the string tension extracted in lattice QCD. However because the confinement mechanism is still very difficult to understand, the choice of this confinement parametrization is only effective and should be interpreted as a phenomenological potential gathering all non-linear effect of QCD.
\\

The potentials (\ref{GBEmodel}) and (\ref{GBEmodeltensor}) are strictly applicable only for point-like particles. Since one deals with structured particles of finite extension, namely the constituent quarks and the pseudoscalar mesons, one must regularize the short-range interaction. In the following we shall use two different parametrizations detailed in the next subsections.
\\

\subsection{Model I}
\

Eq. (\ref{GBEmodel}) contains both the traditional long-range Yukawa potential as well as a $\delta$-function term. It is the latter that is of crucial importance for baryon physics. But this form is strictly valid only for point-like particles. It must be smeared out however, as the constituent quarks and pseudoscalar mesons have a finite size and in addition the boson fields in a chiral Lagrangian should in fact satisfy a nonlinear equation. In the Model I described in this section it is assume that 1) tensor force is neglected and 2) at distances $r<r_0$, where $r_0$ can be related to the constituent quark and pseudoscalar meson sizes, there is no chiral boson exchange interaction as this is the region of perturbative QCD with the original QCD degrees of freedom. The interactions at these very short distances are supposed not to be essential for the low energy properties of baryons. Consequently a two parameter representation for the $\delta$-function term was chosen \cite{GLO96b}

\begin{equation}\label{GBEdeltaI}
4\pi\ \delta({\vec{r}_{ij}})\rightarrow \frac{4}{\sqrt{\pi}}\ \alpha^3 e^{-\alpha^2(r-r_0)^2}.
\end{equation}
\ 

Following the arguments above one should also cutoff the Yukawa part of the GBE for $r<r_0$. In order to avoid any cutoff parameter a step function is used at $r=r_0$ so that the total spin-spin part of the interaction is given by

\begin{equation}\label{GBEMODELI}
V_{\gamma}(r_{ij})= \frac{g_\gamma^2}{4\pi}\frac{1}{12 m_i m_j} \left\{ \mu_\gamma^2 \frac{e^{-\mu_\gamma r_{ij}}}{r_{ij}}\ \theta(r-r_0) -  \frac{4}{\sqrt{\pi}}\ \alpha^3 e^{-\alpha^2(r-r_0)^2} \right\}
\end{equation}

\begin{table}[H]
\centering

\begin{tabular}{|cccc|}

\hline
$m_{u,d}$ & $\mu_\pi$ & $\mu_\eta$ & $\mu_{\eta'}$      \\

\hline
\hline

340       & 139     & 547        & 958 \\

\hline

\end{tabular}
\caption{The {\it a priori} determined parameters of the GBE Model I and Model II (MeV).}\label{GBEtableMODELIa}

\end{table}

Table \ref{GBEtableMODELIa} gathers the physical meson masses and the $u$ and $d$ constituent quark masses used as parameters in this model. Note that the quark masses are consistent with nucleon magnetic moments. The pion-quark coupling constant can be extracted from the phenomenological pion-nucleon coupling as $g_8^2/4\pi=0.67$ as seen in Section \ref{PCACsection}. For simplicity and to avoid additional free parameters, the same coupling constant is assumed for the coupling between the $\eta$ meson and the constituent quark. This is in the spirit of unbroken $SU_F(3)$ symmetry. However, for the flavour-singlet $\eta'$, a different coupling $g_0^2/4\pi$ is taken as the $\eta'$ decouples from the pseudoscalar octet due to the $U_A(1)$ anomaly. Lacking a phenomenological value, $g_0^2/4\pi$ is treated as a free parameter. The three other free parameters are presented in Table \ref{GBEtableMODELIb}. They have been adjusted to describe in the best way all the lowest states of the $N$ and $\Delta$ spectra. In the next section these spectra will be presented.
\\
\begin{table}[H]
\centering

\begin{tabular}{|cccccc|}

\hline
$V_0$ (MeV) & $C$ (fm$^{-2}$) & $g_8^2/4\pi$ & $g_0^2/4\pi$ & $r_0$ (fm) & $\alpha$ (fm$^{-1}$)  \\

\hline
\hline

0           & 0.474           & 0.67         & 1.206        & 0.43       & 2.91                 \\

\hline

\end{tabular}
\caption{Free parameters of the Hamiltonian \cite{GLO96b} in Model I.}\label{GBEtableMODELIb}

\end{table}
\ 

\subsection{Model II}
\

The regularization of the short-range interaction in the second version, called Model II, is related to the property of a vanishing volume integral of the pseudoscalar meson exchange interaction. In the the Model I this is not the case. However, in a non-relativistic reduction in momentum space we have $V(\vec{q}=0)=0$ which implies that the volume integral ({\it i. e.} the Fourier transform at $\vec{q}=0$) of the Goldstone boson exchange interaction in configuration space should vanish : $\int d\vec{r}\ V(\vec{r})=0$. In order to design a parametrization that meets this requirement on makes use of the common Yukawa-type smearing of the $\delta$-function. This leads to a meson-exchange potential with the spatial dependence

\begin{equation}\label{GBEMODELII}
V_{\gamma}(r_{ij})= \frac{g_\gamma^2}{4\pi}\frac{1}{12 m_i m_j} \left\{ \mu_\gamma^2 \frac{e^{-\mu_\gamma r_{ij}}}{r_{ij}} -  \Lambda_\gamma^2 \frac{e^{-\Lambda_\gamma r_{ij}}}{r_{ij}} \right\}
\end{equation}
\ 

\noindent involving the cutoff parameters $\Lambda_\gamma$. If we employ the phenomenological values for the different meson masses $\mu_\gamma$, different cutoff parameter $\Lambda_\gamma$ corresponding to each meson exchange should also be allowed : with a larger meson mass, $\Lambda_\gamma$ should also increase. In the attempt to keep the number of free parameters as small as possible, one has to avoid fitting each individual cutoff parameter $\Lambda_\gamma$. A linear dependence on the meson mass was then chosen

\begin{equation}\label{GBEcutallMODELII}
\Lambda_\gamma = \Lambda_0 + \kappa \mu_\gamma,
\end{equation}
\ 

\noindent which contains only two parameters $\Lambda_0$ and $\kappa$. In order to reproduce the baryon spectra, the coupling constant $g_8^2/4\pi$ cannot be taken $0.67$ as before and has to be adjust as another free parameter. In fact, in a non-relativistic approach, there is no strict constraint to keep $g_8^2/4\pi$ equal to 0.67, a value deduced from $\pi-N$ coupling. Anyway, allowing a value of $g_8^2/4\pi$ different from 0.67 is certainly justified since in a non-relativistic constituent quark model the parameters must be considered as effective. However in the semi-relativistic version of the Model II $g_8^2/4\pi=0.67$ will be again recovered \cite{GLO98}. As compared to the previous parametrization of the quark-quark interaction of Model I, one must also employ an additional constant $V_0$ different from $0$ in the confinement potential in order to make the nucleon ground-state level match the experimental value. The {\it a priori} determined parameters are the same as in Model I and are presented in Table \ref{GBEtableMODELIa}. The remaining free parameters of Model II are given in the Table \ref{GBEtableMODELIIb} below. 
\\

\begin{table}[H]
\centering

\begin{tabular}{|cccccc|}

\hline
$V_0$ (MeV) & $C$ (fm$^{-2}$) & $g_8^2/4\pi$ & $g_0^2/4\pi$ & $\Lambda_0$ (fm$^{-1}$) & $\kappa$  \\

\hline
\hline

-112        & 0.77            & 1.24         & 2.77         & 5.82                     & 1.34      \\

\hline

\end{tabular}
\caption{Free parameters of the Hamiltonian in Model II \cite{GLO97a}.}\label{GBEtableMODELIIb}

\end{table}
\ 

\section{Spectra} \label{GBEspectra}
\

In this section we shall present results obtained by the Graz group in different versions of the GBE model. They used two completely different approaches to calculate the three-quark bound-state levels. One is to solve the Fadeev 3-body equations, the other is based on a stochastic variational method. The $qq$ potential $V_{Conf}+V_\chi$ represents the dynamical input into the 3-body Hamiltonian. The Faddeev equations were solved along the method of Ref. \cite{PAP96} designed for an efficient resolve of any 3-body bound-state problem. It has already been successfully employed in atomic and nuclear problems. The Graz group has carefully checked the accuracy of the results for all baryon levels. In particular they have ensured convergence with respect to all dynamical ingredients. In the most extensive calculations, namely the higher excited states, they went up to including as many as 20 angular-momentum-spin-isospin channels. All these numbers have been cross-checked with the recent stochastic variational method of Ref. \cite{VAR95}.
\\

\begin{figure}[H]
\begin{center}
\includegraphics[width=14.5cm]{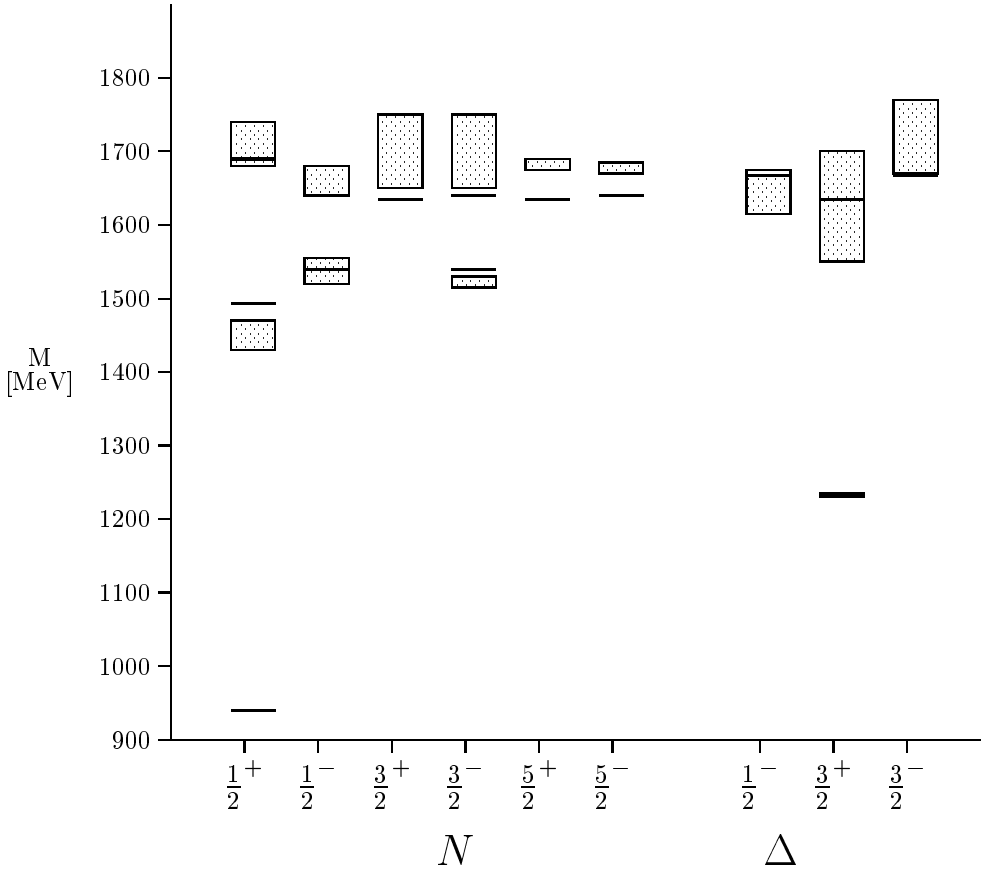}
\end{center}
\caption{\label{GBEgbe1nonmasses} Spectra of $N$ and $\Delta$ in the non-relativistic GBE Model I (from Ref. \cite{GLO96b}).}
\end{figure}

In Fig. \ref{GBEgbe1nonmasses} we show their results for the non-relativistic Model I, mentioned in the previous section. It is clear that the whole set of lowest $N$ and $\Delta$ states is quite correctly reproduced. In the most unfavorable cases, deviations from experimental values do not exceed 3 \%. In addition, all level orderings are correct. Most impressive is the correct level ordering of the positive- and negative-parity states : the {\it Roper} $N$(1400) resonance lies below the pair $N$(1535)-$N$(1520) of negative-parity states. The same is true in the $\Delta$ spectrum. Thus a long-standing problem of baryon spectroscopy is now definitely resolved. Note that here no spin-orbit or tensor force is included, therefore the fine-structure splittings in the $LS$-multiplets are not introduced. We shall see later that in the new versions of the model the tensor force is taken into account.
\\

\ 

\begin{figure}[H]
\begin{center}
\includegraphics[width=14cm]{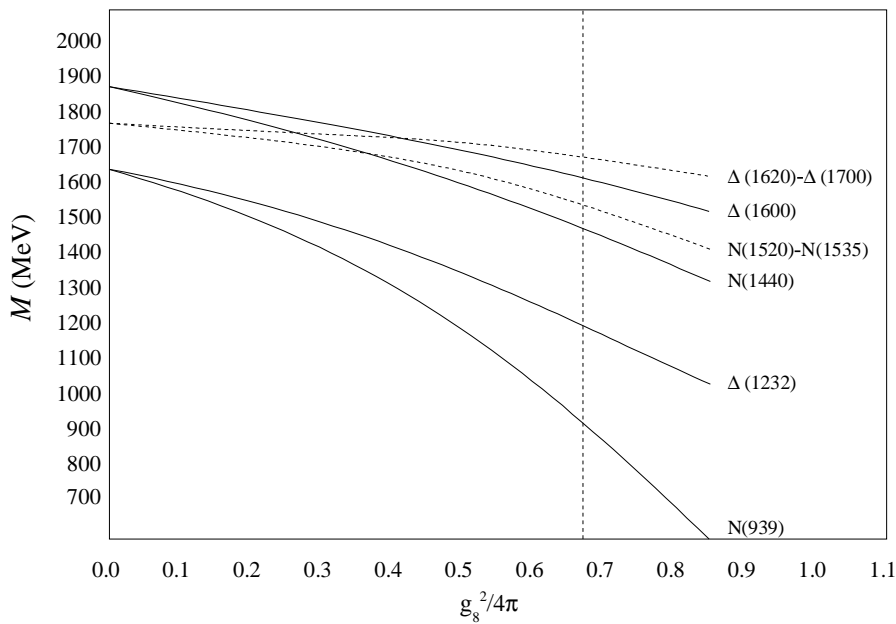}
\end{center}
\caption{\label{GBEcoupling1} Level shifts of lowest $N$ and $\Delta$ states as a function of the coupling constant $g_8^2/4\pi$ in Model I of the GBE interaction \cite{GLO96b}. Solid and dashed lines correspond to positive- and negative-parity states, respectively (from Ref. \cite{GLO96b}).}
\end{figure}

\begin{figure}[H]
\begin{center}
\includegraphics[width=14cm]{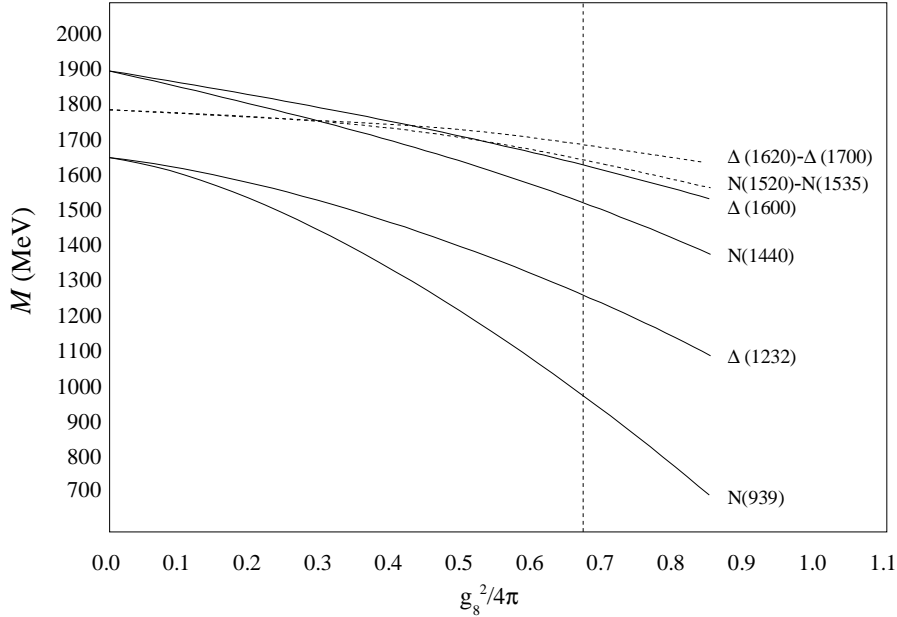}
\end{center}
\caption{\label{GBEcoupling1harmonic} Same as Fig. \ref{GBEcoupling1} but for $s^3$ harmonic oscillator wave function approximation (from Ref. \cite{BAR97}).}
\end{figure}

It is instructive to learn how the GBE interaction affects the energy levels when it is switched on and its strength (coupling constant) is gradually increased as shown in Fig. \ref{GBEcoupling1} for the Model I. Starting out from the case with confinement only, one observes that the degeneracy of states is removed and the inversion of the ordering of positive- and negative-parity states is achieved, both in the $N$ and $\Delta$ excitations after the coupling constant reaches the value 0.67. The reason for this behavior lies in the flavour dependence of the GBE interaction and with $SU_F(3)$ symmetry breaking. It is interesting also to compare the results of Fig. \ref{GBEcoupling1} with our results based on a simple one parameter variational solution. The value of the parameter $\beta$ is found from the condition

\begin{equation}
\frac{\partial}{\partial \beta} <N|H|N>=0
\end{equation}
\ 

\noindent where $|N>$ is a $s^3$ harmonic oscillator wave function with $\beta$ the harmonic oscillator size. This ansatz is very satisfactory because $<N|H|N>$ takes a minimal value of 970 MeV at $\beta=0.437$ fm, {\it i. e.} only 30 MeV above the actual value in the dynamical three-body calculation. We used the same approximation for the $\Delta$ ground state and we found it at 1272 MeV with $\beta=0.511$ fm. Note that the $\Delta$-$N$ splitting is also correct. Fig. \ref{GBEcoupling1harmonic} shows the change of levels with the coupling constant in a simple variational method based on the harmonic oscillator. This variational solution will be used in the study of the nucleon-nucleon interaction. That is why it is important that the results are close to exact calculations. Note however that the single harmonic description is not enough in order to describe higher baryon excitations and in particular the negative-parity states, even if correct ordering is obtained for $N$ and $\Delta$ separately.

\ 

Authors of Ref. \cite{GLO96b} have also calculated root-mean-square radii for the nucleon and the $\Delta$. They obtained $\sqrt{<r_N^2>}=0.465$ fm and $\sqrt{<r_\Delta^2>}=0.54$ fm. Our results in the harmonic approximation are very close to those of Ref. \cite{GLO96b}. For the nucleon the axial {\it r.m.s.} radius is $\sqrt{<r_{\rm axial}^2>}=0.68$ fm and the proton charge {\it r.m.s.} radius is $\sqrt{<r_{\rm p}^2>}=0.862$ fm. However it is clear that GBE results must be smaller, as both these phenomenological values include effects from the finite size of the constituent quarks and from meson-exchange currents.

\begin{figure}[H]
\begin{center}
\includegraphics[width=14cm]{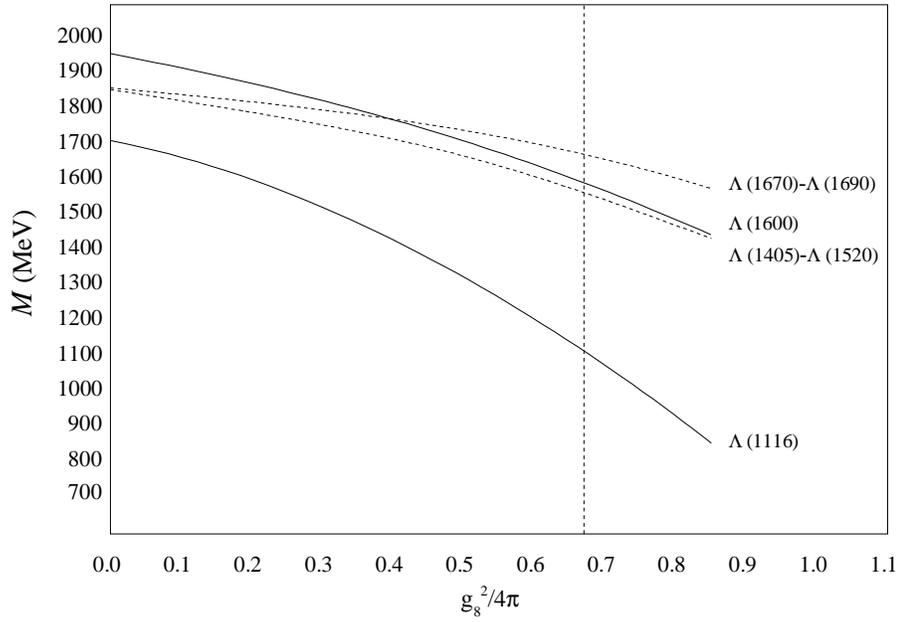}
\end{center}
\caption{\label{GBEcouplinglambda} Same as Fig. \ref{GBEcoupling1} but for $\Lambda$ lowest positive- (solid lines) and negative-parity (dashed lines) levels \cite{GLO97a}.}
\end{figure}

It is interesting to note that a good description is also achieved for the strange baryons ($\Lambda , \Sigma , \Xi , \Omega$) without changing or adding a single parameter. But of course one must take into account the kaon exchange in (\ref{GBEpotsu3broken}). The largest deviation in the strange baryon spectra  is of the order of 150 MeV ($\approx 10$ \%) for $\Lambda(1405)$. Else, the discrepancies are smaller than a few percents. Anyhow, the level ordering is in all cases right. This fact is most remarkable because for strange baryons the experimental ordering of positive- and negative-parity states is opposite as compared to the $N$ and $\Delta$ spectra. A flavour-independent quark-quark interaction, such as one-gluon exchange, is not able to account for this distinction. The flavour-dependence is naturally included in the GBE model. Again, it is the symmetry of the chiral interaction (\ref{GBEpotsu3broken}) which accounts for this property. The change of the lowest part of the $\Lambda$ spectrum with the increase of the coupling constant is shown in Fig. \ref{GBEcouplinglambda}. One can see that the flavour-singlet negative-parity state remains the first excitation above the positive-parity ground state. Contrary to $N$, the first negative-parity state of $\Lambda$ is strongly influenced by the chiral interaction (see Table \ref{GBEtablesym}) but unfortunately is not lowered enough.
\\

\begin{figure}[H]
\begin{center}
\includegraphics[width=14cm]{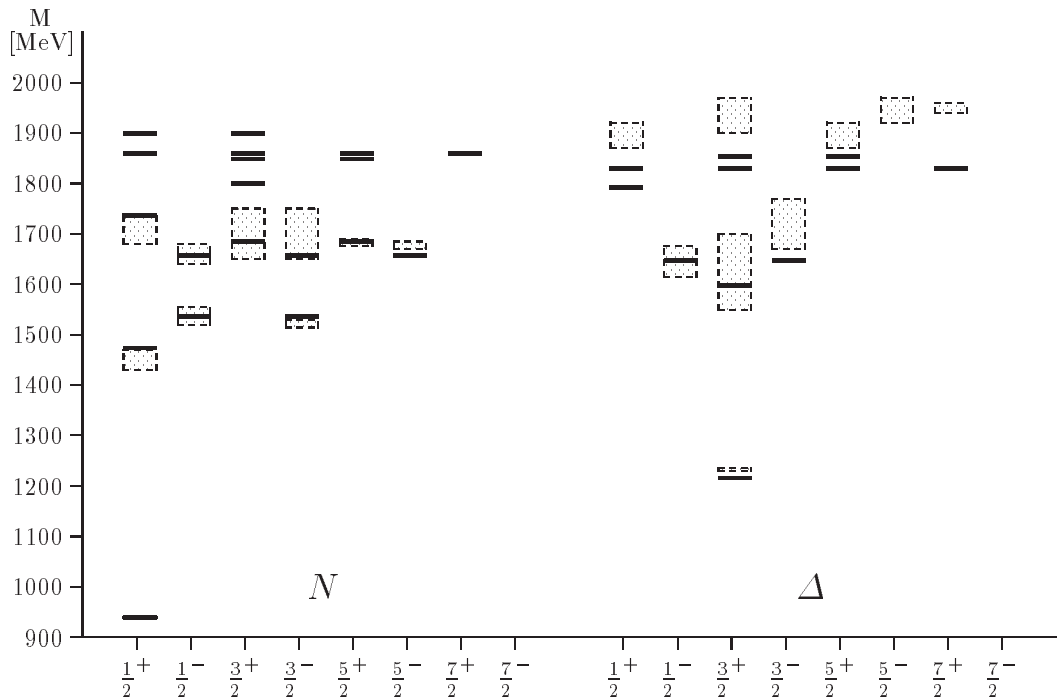}
\end{center}
\caption{\label{GBEgbe2nonmasses} Spectra of $N$ and $\Delta$ in the non-relativistic GBE Model II (from Ref. \cite{THE01}).}
\end{figure}

The parametrization of Model I has the unwanted property that the volume integral of the chiral potential does not vanish as it should be for a pseudoscalar exchange interaction with a finite meson mass. That is why the Model II has been proposed. With the parametrization of Model II a similar description of the baryon spectra is achieved. The $N$ and $\Delta$ spectra are reproduced in Fig. \ref{GBEgbe2nonmasses}. Strange spectra have the same quality that in Model I.
\\

Finally, it is of first importance to note that the parametrization of Table \ref{GBEtableMODELIIb} is only one of several possible choices. Clearly further constraints on the parameters would be welcome. They could come from strong or electromagnetic decay studies or the nucleon-nucleon interaction.

\ 
\ 

In the next chapter we will study the NN interaction in the two non-relativistic versions of the GBE interaction, namely Model I and Model II. We mention however below the recent developments of the GBE constituent quark model, in a semi-relativistic version of the model. This is based on the following three-quark Hamiltonian

\begin{equation}
H = \sum_{i=1}^N \sqrt{\vec{p}_i^2 + m_i^2} + V_{Conf} + V_{Hyp}.
\end{equation}
\ 

Here the relativistic form of the kinetic-energy operator is employed with $\vec{p}_i$, the three-momenta and $m_i$ the masses of the constituent quarks. The dynamical parts, namely the confinement and the hyperfine interaction, are chosen with exactly the same form as that of Model II but with other parameter values. These values are gathered in Table \ref{GBEtablesemiMODELIa} and \ref{GBEtablesemiMODELIIb} for the {\it a priori} fixed and free parameters, respectively \cite{GLO98}.
\\

\begin{table}[H]
\centering

\begin{tabular}{|cccccc|}

\hline
$m_{u,d}$ & $m_s$ & $\mu_\pi$ & $\mu_K$ & $\mu_\eta$ & $\mu_{\eta'}$      \\

\hline
\hline

340   & 500   & 139 & 494  & 547        & 958 \\

\hline

\end{tabular}
\caption{The {\it a priori} determined parameters of the semi-relativistic Model II (MeV) \cite{GLO98}.}\label{GBEtablesemiMODELIa}

\end{table}

\begin{table}[H]
\centering

\begin{tabular}{|cccccc|}

\hline
$V_0$ (MeV) & $C$ (fm$^{-2}$) & $g_8^2/4\pi$ & $g_0^2/4\pi$ & $\Lambda_0$ (fm$^{-1}$) & $\kappa$  \\

\hline
\hline

-416        & 2.33            & 0.67         & 0.898         & 2.87                     & 0.81      \\

\hline

\end{tabular}
\caption{Free parameters of the Hamiltonian in the semi-relativistic Model II \cite{GLO98}.}\label{GBEtablesemiMODELIIb}

\end{table}

\begin{figure}[H]
\begin{center}
\includegraphics[width=13.5cm]{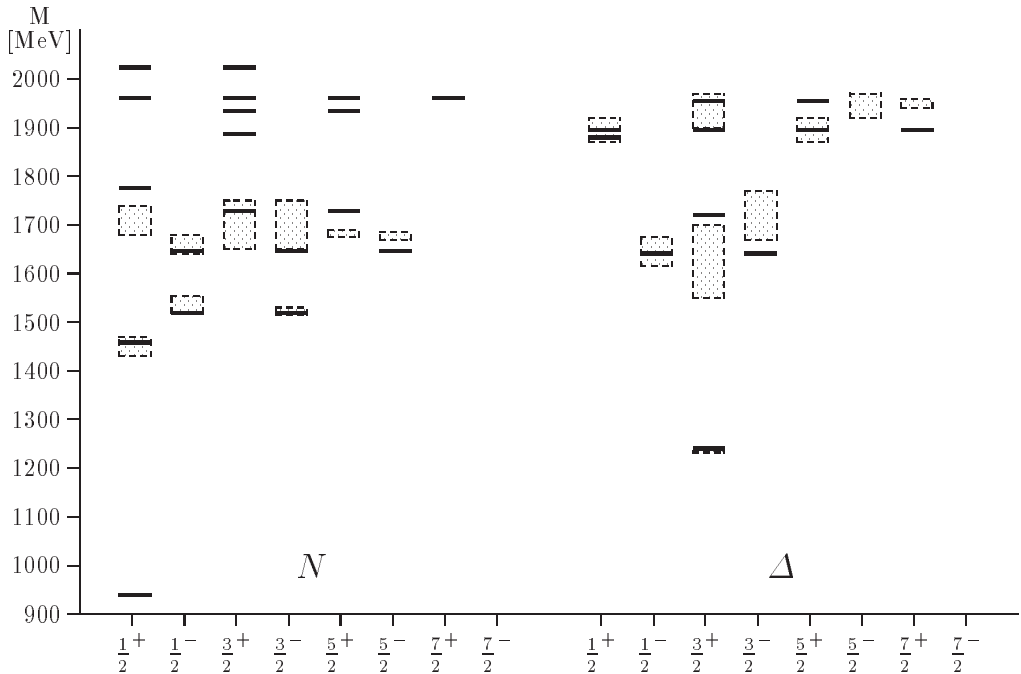}
\end{center}
\caption{\label{GBEgbe2semimasses} Spectra of the $N$ and $\Delta$ in the semi-relativistic Model II of the GBE.}
\end{figure}

The same remark about the sets of the parameters fitting the spectra has to be mentioned here : the parameters are only one choice out of many possible sets that could lead to a similar quality description of the baryon spectra. The baryon spectra alone do not guarantee a unique determination of the model parameters. Nevertheless, it is rewarding to find the present parameter values of reasonable magnitude. For example the confinement strength is comparable with the string tension extracted from lattice calculations and it is also consistent with the slopes of Regge trajectories. The strength of the coupling constant for all the octet mesons is extracted from the phenomenological pion-nucleon coupling constant as discussed in Section \ref{PCACsection} and has been considered as a fixed parameter in the calculations of Ref \cite{GLO98}. In this way one can say that in practice this model involves only five free parameters.

\begin{figure}[H]
\begin{center}
\includegraphics[width=11.5cm]{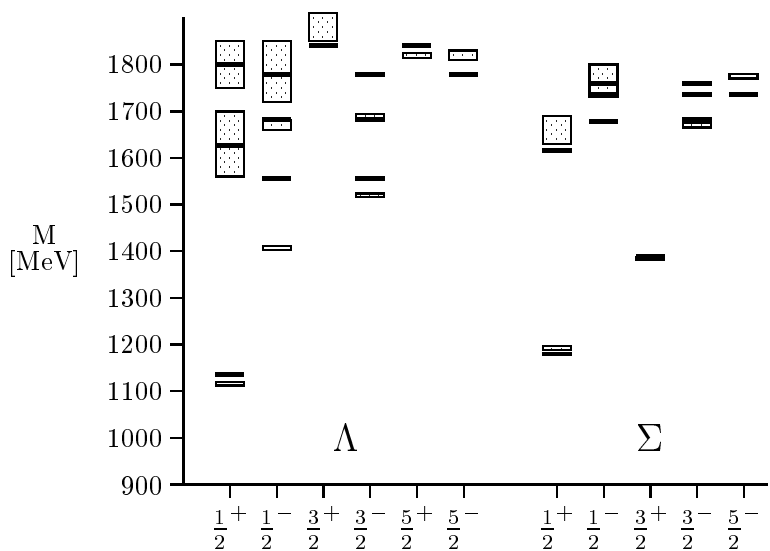}
\end{center}
\caption{\label{GBElambdasigma} Spectra of the $\Lambda$ and $\Sigma$ in the semi-relativistic Model II \cite{GLO98}.}
\end{figure}

\begin{figure}[H]
\begin{center}
\includegraphics[width=11.5cm]{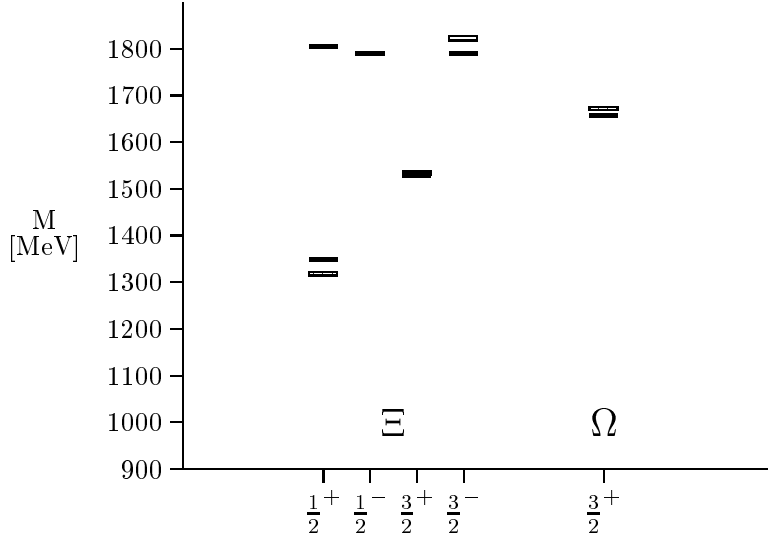}
\end{center}
\caption{\label{GBExiomega} Spectra of the $\Xi$ and $\Omega$ in the semi-relativistic Model II \cite{GLO98}.}
\end{figure}

The results obtained with the semi-relativistic version of the GBE Model II are presented in Figs. \ref{GBEgbe2semimasses}-\ref{GBExiomega} for the light non-strange and strange spectra. In particular we note the right level ordering of the {\it Roper} resonance and the first negative-parity states in the $N$ spectrum. However, in the $\Delta$ spectrum, the GBE semi-relativistic parametrization predicts the $\Delta_{1600}$ $\frac{3}{2}^+$ to be heavier than the lowest negative-parity states, contrary to the non-relativistic parametrization. This is the main qualitative difference between spectra employing a non-relativistic and a semi-relativistic kinematics for light or strange flavours.
\\

The difference is not so surprising if one looks at Table \ref{GBEtablesym} which implies that the chiral interaction is less powerful in $\Delta$ than in $N$. In practice the spin-flavor matrix element of $V_\chi$ for $\Delta_{1600}$ is about three times smaller than the corresponding matrix element for the {\it Roper} resonance. Moreover, the confinement in the semi-relativistic case is much larger than the confinement in the non-relativistic case. This means that level spacing is higher in the semi-relativistic case. For the $\Delta_{1600}$, which has a larger spatial extension than the {\it Roper} resonance, the short-range hyperfine interaction is then not enough to reproduce the correct ordering. On the other side, the small value of the confinement strength in the non-relativistic approach is not really consistent with the values commonly accepted for the QCD string constant, $C\approx 2.5$ fm$^{-2}$. But as we have already mentioned, in a non-relativistic approach the parameters have to be considered as effective.

\begin{figure}[H]
\begin{center}
\includegraphics[width=11cm]{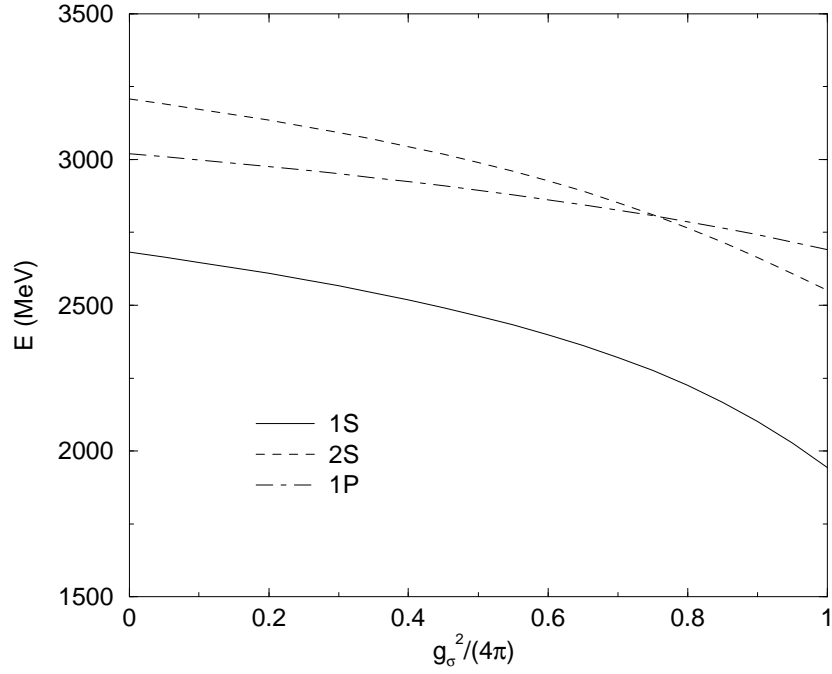}
\end{center}
\caption{\label{GBEscalarinfinity} Energies of the first three eigenvalues of Hamiltonian (\ref{GBEhamilscalar}) for increasing values of the coupling constant $g_{\sigma}^{2}/(4\pi)$ in the limit $\Lambda \rightarrow \infty$ \cite{STA00}.}
\end{figure}

\begin{figure}[H]
\begin{center}
\includegraphics[width=11cm]{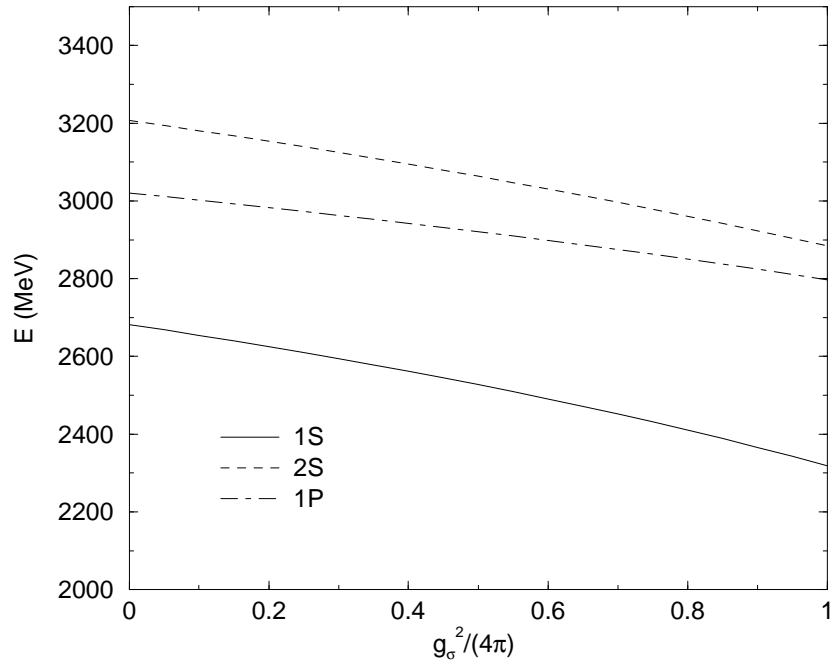}
\end{center}
\caption{\label{GBEscalar2} Same as Fig. \ref{GBEscalarinfinity} but for a finite cut-off $\Lambda = 2$ GeV \cite{STA00}.}
\end{figure}

Although the models considered up to now are thought to be a consequence of the spontaneous chiral symmetry breaking, the chiral partner of the pion, the $\sigma$ meson is not considered explicitly. One can think that its contribution is already included in the parameters of the spin independent part of the Hamiltonian. But let us consider it explicitly as in the following semi-relativistic Hamiltonian

\begin{equation}\label{GBEhamilscalar}
H=\sum_{i=1}^N \sqrt{\vec{p}_i^2 + m_i^2} + V_0 + \sum_{i=1}^N C r_{ij} - \frac{g_{\sigma}^2}{4 \pi} \left\{ \frac{e^{-\mu_{\sigma} r}}{r_{ij}} - \frac{e^{-\Lambda_\sigma r}}{r_{ij}}\right\}.
\end{equation} 
\ 

\noindent where $m_i=340$ MeV, $C=2.534$ fm$^{-2}$ which is very close to the semi-relativistic GBE parametrization and $\mu_\sigma=600$ MeV. The coupling constant $\frac{g_\sigma^2}{4 \pi}$ and the regularization parameter $\Lambda_\sigma$ are taken as variable parameters. Note that this interaction is attractive whenever $\mu_\sigma < \Lambda_\sigma$. In Fig. \ref{GBEscalarinfinity}-\ref{GBEscalar2} we present the eigenvalues of the $1S$, $2P$ and $2S$ states as a function of $\frac{g_{\sigma}^2}{4 \pi}$ obtained in Ref. \cite{STA00} for two different values of $\Lambda_\sigma$.
\\

Let us look at the limit $\Lambda_\sigma \rightarrow \infty$ first. In this case one can see in Fig. \ref{GBEscalarinfinity} that the mass difference between the radially excited state and the ground state remains practically constant as a function of the coupling constant while the mass difference between the radially and orbitally excited states decreases with $\frac{g_\sigma^2}{4 \pi}$ until it becomes negative for $\frac{g_\sigma^2}{4 \pi} > 0.75$. This is precisely the desired behavior of reproducing the correct order of the experimental spectrum as it was achieved with the GBE interaction but this time coming from a potential whose Laplacian is negative in a region around the origin.

\ 

\ 

Realistically, one expected $\Lambda_\sigma$ to be finite. The results are shown in Fig. \ref{GBEscalar2} for $\Lambda_\sigma = 2$ GeV. In this case the conclusion is different : the crossing effect of $2S$ with $1P$ disappears because when a regularizing term with a finite $\Lambda$ is subtracted from the attractive Yukawa-type term this leads to a $\Delta V \geq 0$ contribution at small values of $r$. Indeed, as shown in Fig. \ref{GBEscalar2} the level spacings are much less sensitive to the strength of the coupling constant. In this case the $\sigma$-exchange potential leads in practice to a global shift of the whole spectrum which can be compensated by the constant $V_0$ in (\ref{GBEhamilscalar}).
\\

\begin{figure}[H]
\begin{center}
\includegraphics[width=14cm]{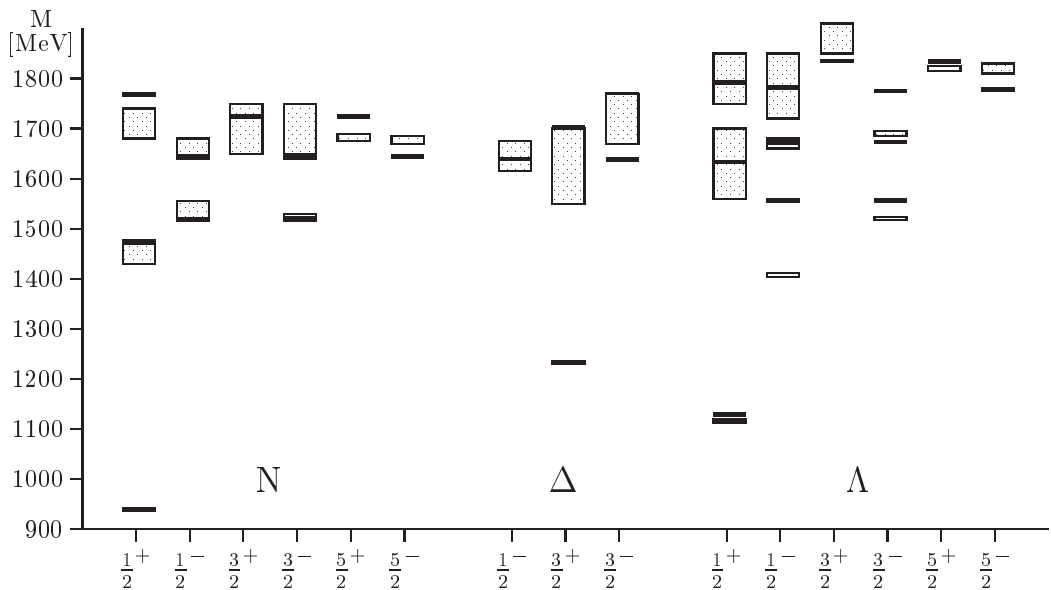}
\end{center}
\caption{\label{GBEgbetensormasses} Spectra of the $N$, $\Delta$ and $\Lambda$ in the extended semi-relativistic Model II with pseudoscalar, vector and scalar exchanges \cite{WAG99,WAG00,WAG01}.}
\end{figure}

We now present the extension of the GBE chiral quark model \cite{WAG99,WAG00,WAG01} beyond the previous pseudoscalar exchange version, which considered the spin-spin hyperfine interaction only. In this extended model the tensor part has also been introduced because the pseudoscalar exchange gives rise to a spin-spin component but to a tensor component as well. The inclusion of the multiple GBE has also been considered by the introduction of vector and scalar meson exchanges. The vector meson nonet exchange interaction has central, spin-spin, tensor and spin-orbit components. The scalar singlet $\sigma$-meson exchange comes with only central and spin-orbit forces. Detailed of the parametrization can be found in Refs. \cite{WAG99,WAG00,WAG01}. The spectra of the $N$, $\Delta$ and $\Lambda$ up to an energy of $E\approx 1800$ MeV are shown in Fig. \ref{GBEgbetensormasses}. The various levels are well reproduced in rather good agreement with the experimental data. All essential features of the GBE model are present, in particular the correct level orderings of positive- and negative-parity states are reproduced. Furthermore, the extremely small splittings of equal-parity multiplets existing in the experimental data are also well described even though all tensor force components are now included in the hyperfine interaction. The reason lies in the fact that the individually large tensor force contributions from the pseudoscalar and vector meson exchanges practically cancel each other.
\\

The extended GBE chiral quark model provides a promising basis for a number of further investigations. The addition of vector and scalar exchanges to the pseudoscalar exchange and the inclusion of all force components turns out to be essential in properties such as the electromagnetic nucleon form factors, the hadronic decays of baryon resonances, and also in the derivation of the baryon-baryon interactions.
\\

\section{Form Factors}
\

In the previous section we showed the remarkable success of the GBE model in reprodu\-cing the detailed features of the lowest part of the excitation spectra of light non-strange and strange baryons. Baryon spectroscopy, however, is only a first, though quite demanding test of low-energy models of QCD. Furthermore, constituent quark models should also provide a comprehensive description of other hadron phenomena, such as electromagnetic properties, resonance decays, etc. More stringent tests of any constituent quark model consist in the proton and neutron electromagnetic form factors, $G_E$ and $G_M$ observed in elastic electron-nucleon scattering. Further important constraints are furnished by the nucleon weak form factors, {\it i. e.} the axial form factor $G_A$ and the induced pseudoscalar form factor $G_P$. They reflect the structure of the nucleons as probed by an axial vector field in processes such as beta decay, muon capture and pion production. In contrast to the electromagnetic form factors, the weak form factors involve a combination of the proton and neutron wave functions. This provides another test for the nucleon ground state obtained from the GBE eigensolutions. In this section we present the recent covariant results obtained by Boffi {\it et al.} \cite{BOF01} for all elastic electroweak nucleon form factors.
\\

\begin{table}[H]
\centering

\begin{tabular}{|l|cccc|}

\hline
                             &  Experience                   & PFSA  & NRIA  & Conf. \\

\hline
\hline

$r_p^2$ [fm$^2$]             & 0.780(25) \cite{MEL00}         & 0.81  & 0.10  &  0.37 \\
$r_n^2$ [fm$^2$]             & -0.113(7) \cite{KOP95}         & -0.13 & -0.01 & -0.01 \\
$\mu_p$ [n. m.]              & 2.792847337(29) \cite{GRO00}   & 2.7   & 2.74  &  1.84 \\
$\mu_n$ [n. m.]              & -1.91304270(5) \cite{GRO00}    & -1.7  & -1.82 & -1.20 \\
$\sqrt{<r_A^2>}$ [fm]        & 0.635(23) \cite{LIE99}         & 0.53  & 0.36  &  0.43 \\
$g_A$                        & 1.255$\pm$ 0.006 \cite{GRO00}  & 1.15  & 1.65  &  1.29 \\

\hline

\end{tabular}
\caption{Proton and neutron charge radii as well as magnetic moments and nucleon axial radius as well as axial charge. Predictions of the GBE model in PFSA (third column), in NRIA (fourth column), and with the confinement interaction only (last column).}\label{GBEemradii}

\end{table}
\ 

The calculation are performed in a covariant form using the point form approach to the relativistic quantum mechanics \cite{DIR49}. In the point form the four-momentum operators $P^{\mu}$ containing all the dynamics commute with each other and can be simultaneously diagonalized. All other generators of the Poincar\'e group are not affected by interactions. In particular because the Lorentz generators do not contain any interaction terms, the theory is manifestly covariant. Moreover the electromagnetic current operator $J^{\mu}(x)$ can be written in such a way that it transform as an irreducible tensor operator under the Poincar\'e group. Thus the electromagnetic form factors can be calculated as reduced matrix elements of such an irreducible tensor operator in the Breit frame. The same procedure can be applied to the axial current. Once $G_A$ is known, $G_P$ can be extracted from the longitudinal part of the axial current in the Breit frame. The current operator is a single-particle current operator for point-like constituent quarks. In the literature it is called point form spectator approximation (PFSA).

\begin{figure}[H]
\begin{center}
\includegraphics[width=7.69cm]{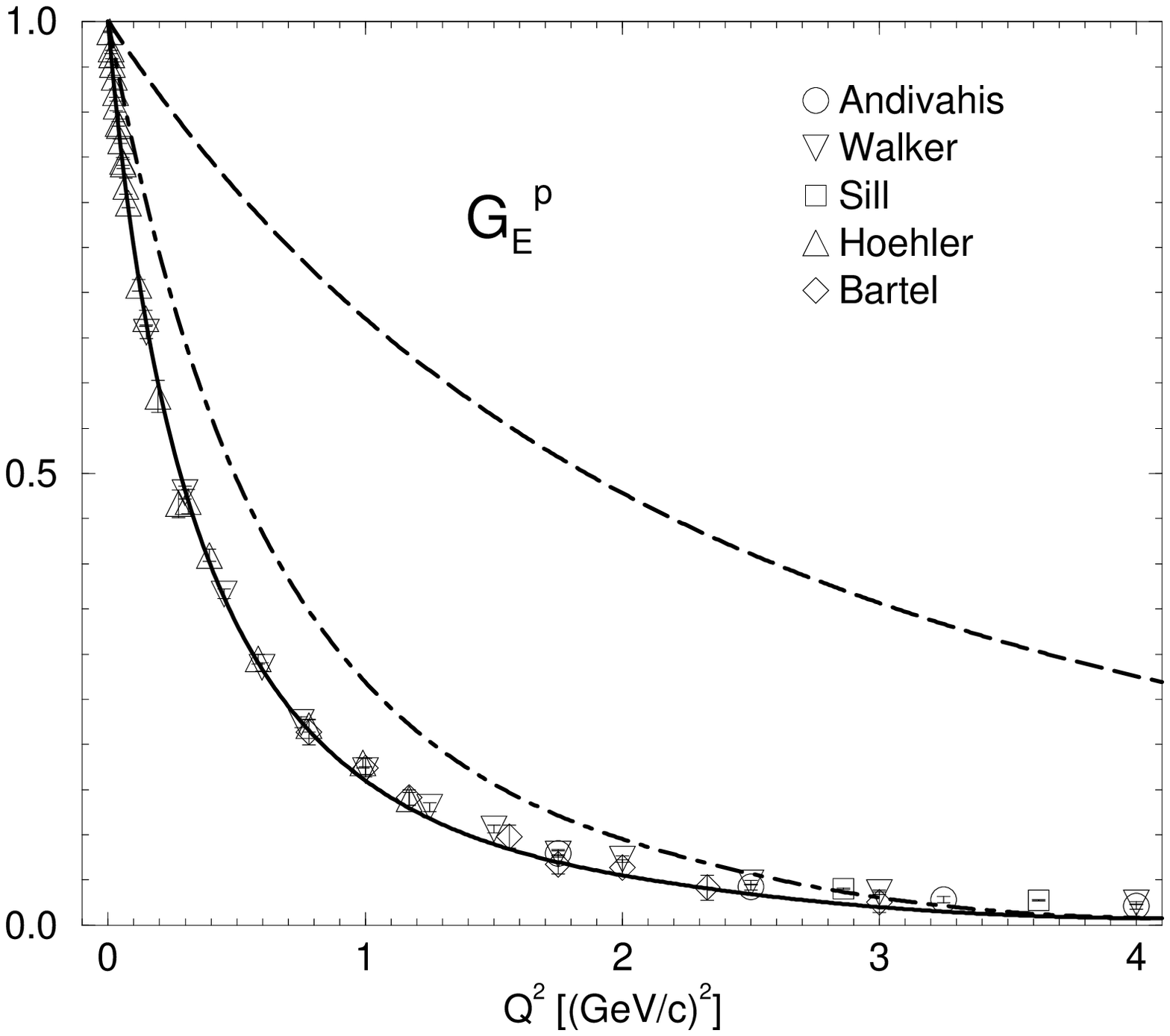}
\includegraphics[width=7.69cm]{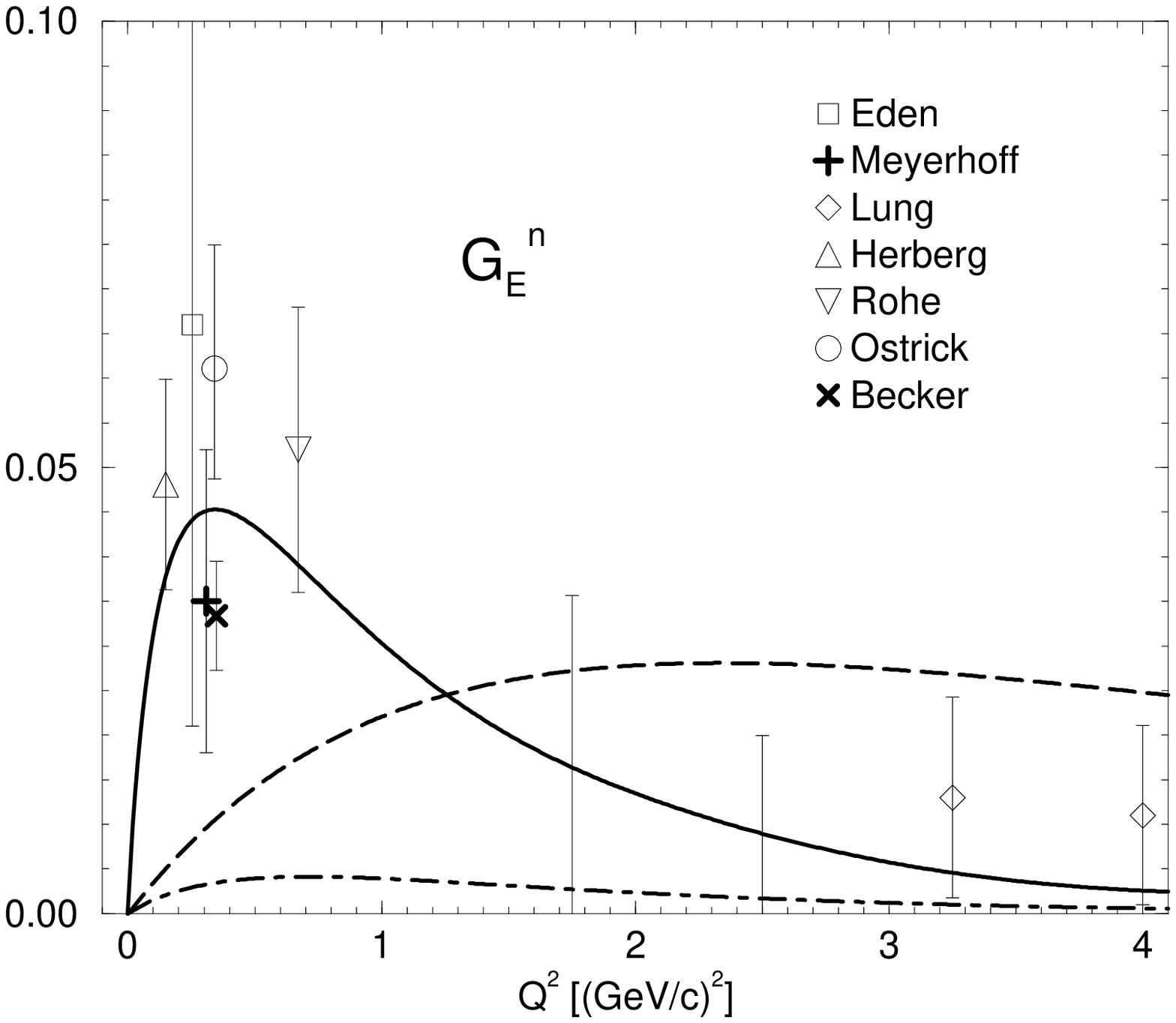}
\includegraphics[width=7.69cm]{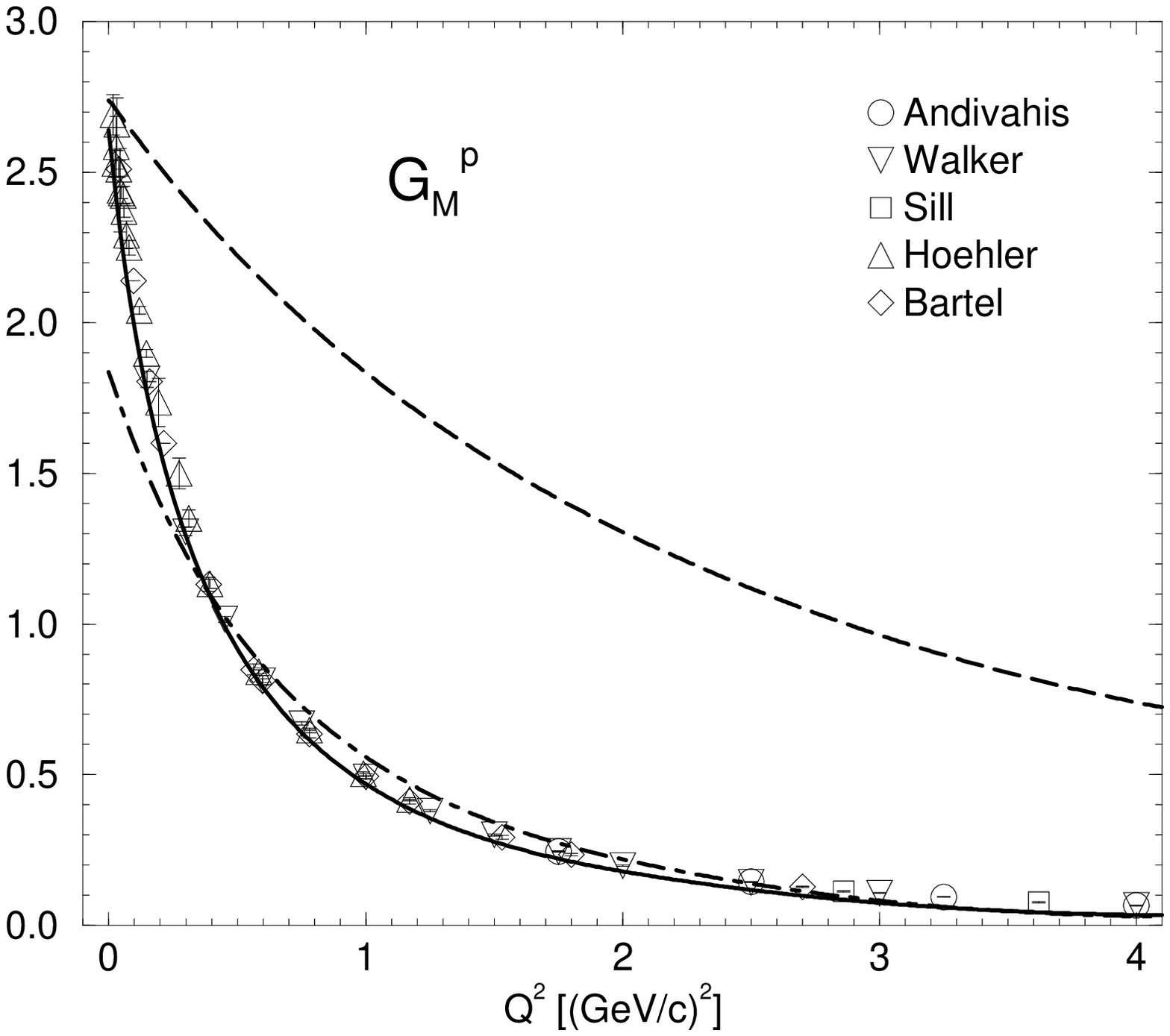}
\includegraphics[width=7.69cm]{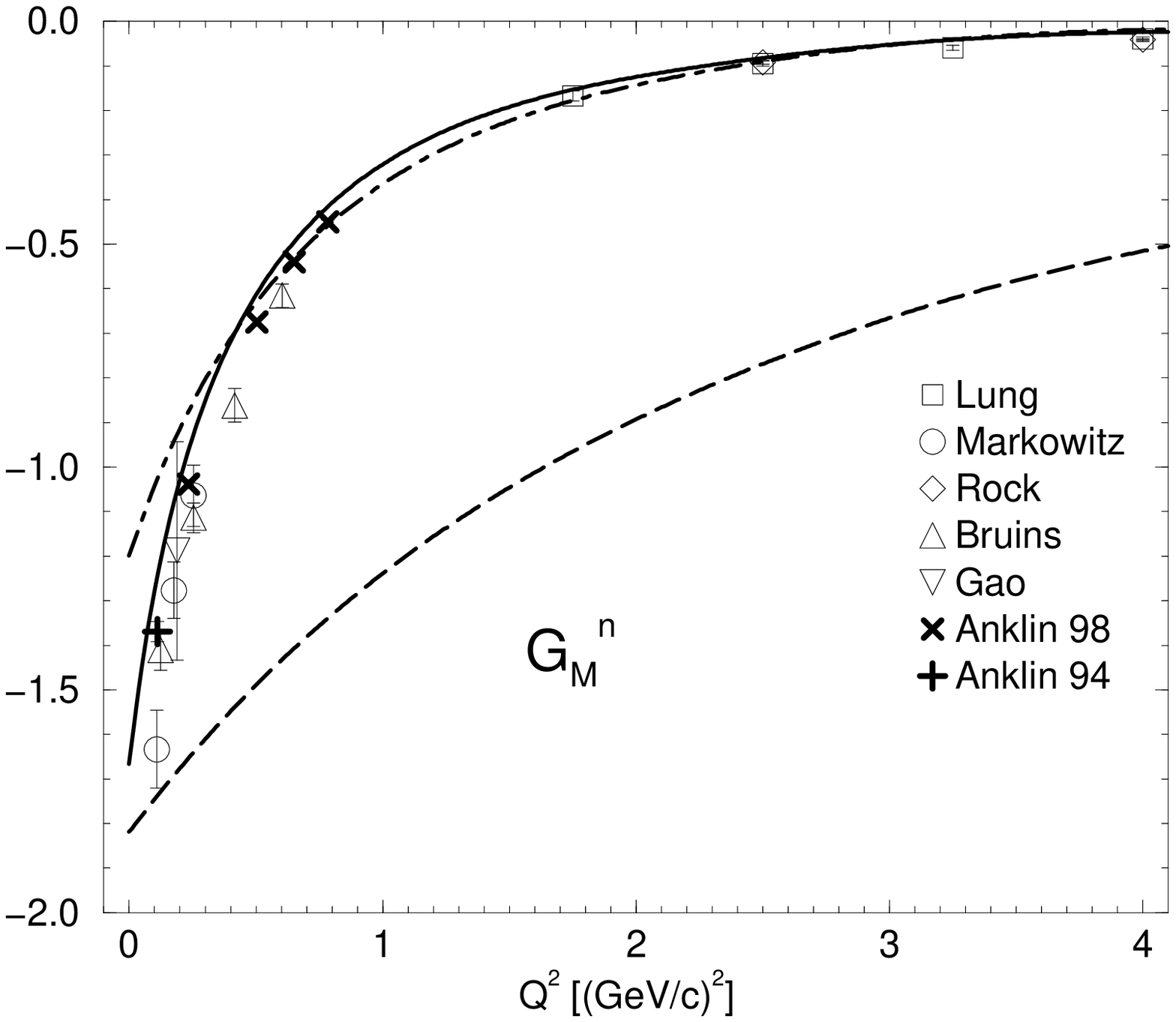}
\end{center}
\caption{\label{GBEgempf} Proton (left) and neutron (right) electric (upper) and magnetic (lower) form factors as predicted by the GBE model in PFSA (solid lines). A comparison is given to the results in NRIA (dashed) and the case with the confinement interaction only (dashed-dotted). The experimental data are from Ref. \cite{AND94} (from Ref. \cite{WAG01b}).}
\end{figure}
\ 

The prediction \cite{WAG01b} of the GBE for the nucleon electromagnetic form factors are shown in Fig. \ref{GBEgempf}. Their properties at zero momentum transfer are reflected by the charge radii and magnetic moments given in Table \ref{GBEemradii}. The input into the calculations consists only in the proton and neutron three-quark wave functions as produced by the ground state of the GBE Hamiltonian \cite{GLO98}. One observes that an extremely good description of both the proton and neutron electromagnetic structure is achieved. Relativity plays a major role here. For comparison Wagenbrunn {\it et al.} \cite{WAG01b} also showed results for the form factors when calculated in non-relativistic impulse approximation (NRIA), {\it i. e.} with the standard non-relativistic form of the current operator and without Lorentz boosts applied to the nucleon wave functions. Evidently there is no way of describing the nucleon electromagnetic form factors in a non-relativistic theory if quarks are considered as point-like.
\\

In order to get an idea of the role of the GBE hyperfine interaction in the form factors, Wagenbrunn {\it et al.} have considered the case with the confinement potential only. In addition to differences in the wave functions, the nucleons now also have a larger mass of $m_N^{\rm conf}=1353$ MeV. This different mass is very important for the behavior of the form factors for low $Q^2$ and is essentially responsible for the corresponding results given in the last column of Table \ref{GBEemradii}. Shifting the nucleon mass artificially to $m_N=939$ MeV would change the charge radii and magnetic moments in the following way: $r^2\to r^2(m_N^{\rm conf}/m_N)^2$ and $\mu\to\mu(m_N^{\rm conf}/m_N)$. As a result the proton charge radius as well as the magnetic moments of both the proton and the neutron would then already be very close to the values obtained with the full interaction. Only the neutron charge radius would still remain much too small, due to the absence of the mixed-symmetry component in the wave function for the case with the confinement potential only. Though the mixed-symmetry component brought about by the hyperfine interaction is rather small, it turns out to be most essential for reproducing the neutron charge radius in a reasonable manner. It is then evident that $G_E^n$ is essentially driven by the combined effects of small mixed-symmetry components in the neutron wave function which are induced only by the hyperfine interaction and Lorentz boosts.
\\

\begin{figure}[H]
\begin{center}
\includegraphics[width=10cm]{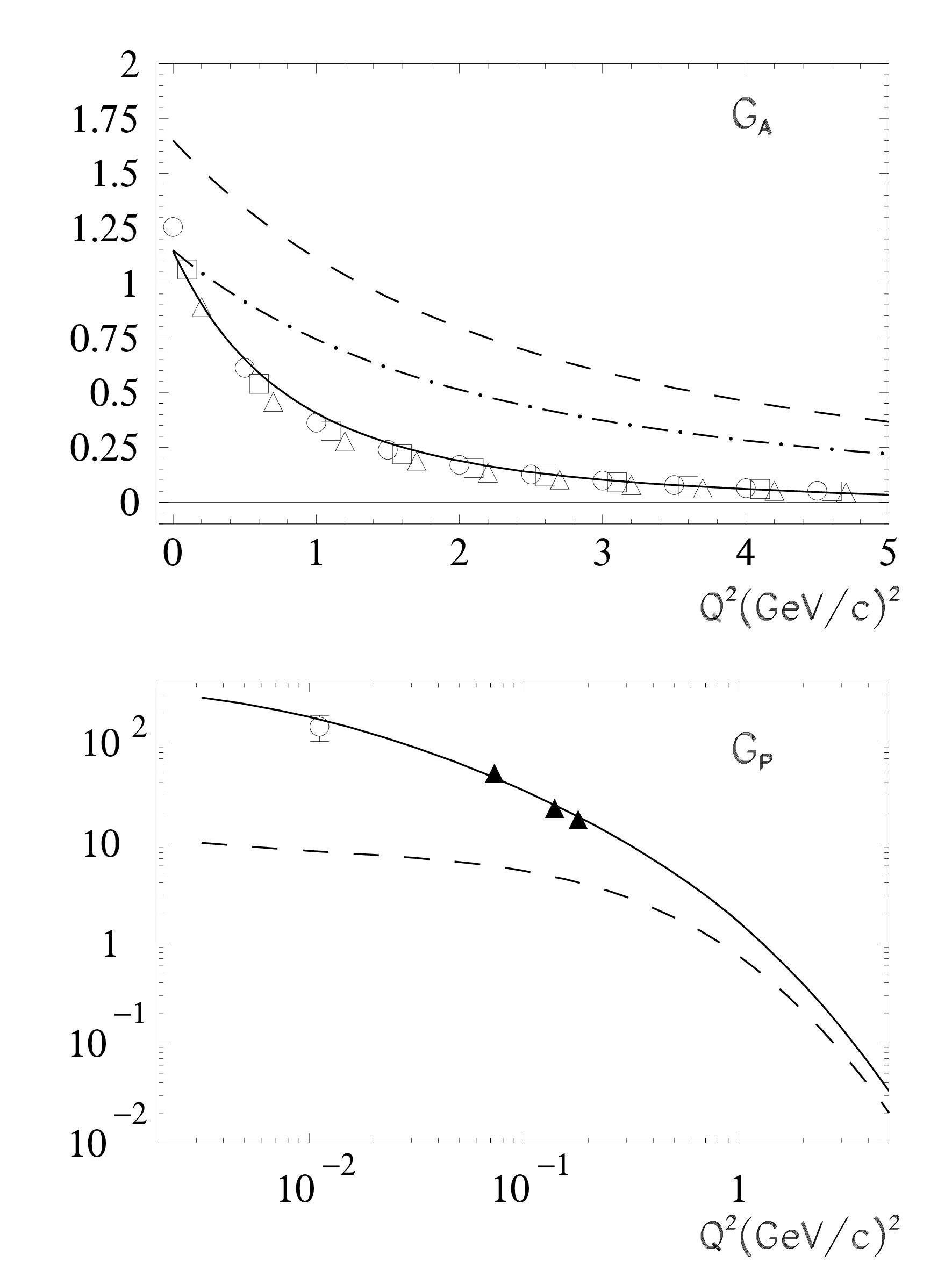}
\end{center}
\caption{\label{GBEappf} Nucleon axial and induced pseudoscalar form factors $G_A$ and $G_P$, respectively (from Ref. \cite{BOF01}). The PFSA predictions of the GBE model are always represented by solid lines. In the top panel a comparison is given with the NRIA results (dashed) and to the case with a relativistic current operator but no boosts included (dot-dashed); experimental data are shown assuming a dipole parametrization with the axial mass value $M_A$ deduced from pion electroproduction (world average : squares, Mainz experiment {\protect\cite{LIE99}} : circles) and from neutrino scattering {\protect\cite{KIT83}} (triangles). In the bottom panel the dashed line refers to the calculation of $G_P$ without any pion-pole contribution. Experimental data are from Ref. {\protect\cite{BAR81}}.}
\end{figure}

The nucleon axial form factor $G_A$ and the induced pseudoscalar form factor $G_P$ are shown in Fig. \ref{GBEappf} and the axial radius $\sqrt{<r_A^2>}$ as well as the axial charge $g_A$ are given in Table \ref{GBEemradii}. In the top panel of Fig. \ref{GBEappf} the $G_A$ predictions of the GBE model in PFSA are compared to experimental data, which are presented assuming the common dipole parametrization with the axial charge $g_A=1.255\pm0.006$, as obtained from $\beta$-decay experiments \cite{GRO00}, and three different values for the nucleon axial mass $M_A$

\begin{equation}\label{GBEaxialdipole}
G_A(Q^2)=\frac{g_A}{\left( 1+\frac{Q^2}{M_A2} \right)^2}
\end{equation}
\ 

Again a remarkable agreement of the theory and experiment is detected; only at $Q^2=0$ does the PFSA calculation underestimate the experimental value of $g_A$ and, consequently, also the axial radius. In contrast, both the NRIA results and also the results from a calculation with a relativistic axial current but no boosts on the wave functions fall tremendously short. Again the inclusion of all relativistic effects, in order to produce a covariant result, appears most essential.
\\

The PFSA predictions of the GBE model for the induced pseudoscalar form factor $G_P$ also fall readily on the available experimental data.  For this result the pion-pole term occurring in the axial current turns out to be most important, especially at low $Q^2$. This is clearly seen by a comparison of the solid curve in the lower panel of Fig. \ref{GBEappf} with the results obtained without the pion-pole term. It follows that at least for low $Q^2$ values the role of pions is essential. It is also remarkable that the agreement of the PFSA predictions with experiment is obtained by using the same value of the quark-pion coupling  constant $g_{\pi q}^2/4\pi=0.67$ as employed in the GBE model of Ref. \cite{GLO98}.

In summary, the chiral constituent quark model based on GBE dynamics predicts all observables of the electroweak nucleon structure in a consistent manner. The covariant results calculated in the framework of the point form relativistic quantum mechanics always fall rather close to the available experimental data. This indicates that a quark model using the proper low-energy degrees of freedom may be capable of providing a reasonable description also of other dynamical phenomena in addition to the baryon spectra.
\\

\section{Decays}
\

In this section we shall present some new investigations done by Theu{\ss}l {\it et al.} \cite{THE01} in the description of $\pi$ and $\eta$ decays of $N$ and $\Delta$ resonances within different constituent quark mo\-dels. A comparison between the modern experimental database \cite{GRO00} and the results obtained in the GBE model (with or without tensor force) and the OGE model, in a relativistic or semi-relativistic description will be presented. In that work, two approaches were considered for the derivation of the strong interaction transition operator.
\\

\begin{figure}[H]
\begin{center}
\includegraphics[width=11.5cm]{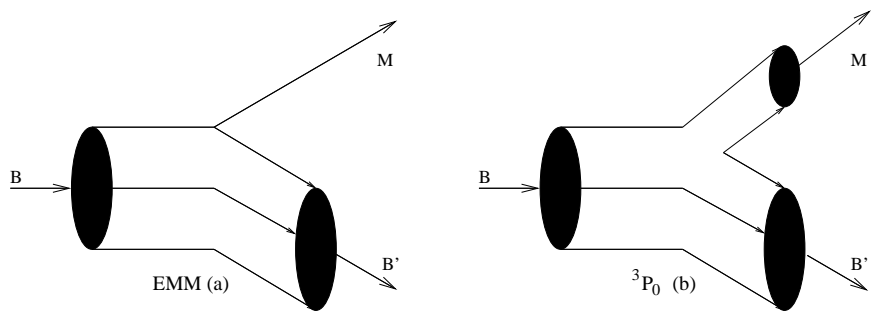}
\end{center}
\caption{\label{GBEeem-3P0} Decay of a baryon B by the emission of a point-like meson M (a) and by the emission of a composite meson (b).}
\end{figure}

The first approach is the elementary emission model (EEM) where the decay takes place through the emission of a point-like meson by one of the quarks of the baryon as shown in Fig. \ref{GBEeem-3P0} (a). The baryons are therefore considered as composite objects, whereas the meson is treated as a point-like particle that couples to the quark. As a consequence, the interaction potential between quarks intervenes only through the baryon wave functions in the initial and final states. The process should therefore be a good test for the quality of the wave function provided by the constituent quark model. However the performance of the EEM in describing strong decays of non-strange baryon resonances is far from being good which give rise to serious doubts whether this decay mechanism, namely the EEM, used in the calculations, is really appropriate.

It seems then natural to investigate the influence of the most important model simplification of the EEM : the point-like nature of the emitted meson. Indeed in strong-interaction physics, in particular with light quark flavours, the extended size of any hadron should not be negligible. This feature is taken into account in decay models that describe the transition process by the creation of an additional quark-antiquark pair. The corresponding models are generally referred to as quark-pair creation models. In the simplest version, one assumes that all the quarks in the initial hadron are spectators, so that the $q\bar{q}$-pair is necessarily created from the vacuum which is flavour and colour singlet and has zero momentum and total angular momentum $J^{PC}=0^{++}\ (L=1,S=1)$. These quantum numbers have given the name to the simplest and most popular model of hadronic transitions : the $^3P_0$ model. In the following, we shall present the results obtained in a modified version of the $^3P_0$ model \cite{CAN96} including a relativistic boost effect. In this model there are two important ingredients. The first is a parameter $\gamma$ representing the probability for a $q\bar{q}$-pair creation anywhere in the vacuum. This parameter is adjusted in order to reproduce the $\Delta_{1232} \rightarrow N\pi$ or $N_{1535} \rightarrow N\eta$ decay width. The second ingredient is the shape of the meson wave function. Theu{\ss}l {\it et al.} \cite{THE01b} used both a Yukawa- and a Gaussian-like wave function with an extension $r_\pi$ for the pion and $r_\eta$ for the eta meson. They showed that the specific form have only minor influence on the prediction of the decays widths. However the size of the meson is important even if the introduction of a finite size meson wave function does not lead to a striking improvement in predictions for decay widths, except in few cases as {\it e. g.} the {\it Roper} resonance.
\\

\subsection{The $\pi$ decays}
\

The results for the partial widths of the $\pi$ decay modes of the $N$ and $\Delta$ resonances are shown in Table \ref{GBEgausspi}. All values presented here have been calculated with a Gaussian-type parametrization of the meson wave function. For the baryons, the theoretical masses have been used as predicted by the different constituent quark models given in the table heading. All the decay widths can then be considered as genuine predictions of the models along the modified $^3P_0$ model.
\\

\begin{table}[H]
\centering

\begin{tabular}{|cccccccc|}
\hline

\rule[-3mm]{0mm}{7mm}$N^*$ & $J^{\pi}$ & \multicolumn{6}{c|}{$\Gamma(N^*\rightarrow N\pi$) [MeV]} \\

\hline
\hline

\rule[-2mm]{0mm}{6mm}&&  GBE SR & GBE SR T & GBE NR & OGE SR & OGE NR & Exp. \\

\hline
\hline

\rule[-2mm]{0mm}{7mm}$N_{1440} $      &$ \frac{1}{2}^+$  & $517$  & $646$ & $258$ & $1064$ & $161$ & $(227\pm18)^{+70}_{-59}$   \\
\rule[-2mm]{0mm}{0mm}$N_{1710} $      &$ \frac{1}{2}^+$  & $54$   & $87$  & $14$  & $202$  & $8$   & $(15\pm5)^{+30}_{-5}$      \\
\rule[-2mm]{0mm}{0mm}$\Delta_{1232}$  &$\frac{3}{2}^+$   & $120$  & $120$ & $120$ & $120$  & $120$ & $(119\pm1)^{+5}_{-5}$      \\
\rule[-2mm]{0mm}{0mm}$\Delta_{1600}$  &$ \frac{3}{2}^+$  & $43$   & $87$  & $34$  & $174$  & $14$  & $(61\pm26)^{+26}_{-10}$    \\
\rule[-2mm]{0mm}{0mm}$N_{1520} $      &$ \frac{3}{2}^-$  & $131$  & $146$ & $161$ & $108$  & $168$ & $(66\pm6)^{+9}_{-5}$       \\
\rule[-2mm]{0mm}{0mm}$N_{1535} $      &$ \frac{1}{2}^-$  & $336$  & $294$ & $75$  & $462$  & $109$ & $(67\pm15)^{+55}_{-17}$    \\
\rule[-2mm]{0mm}{0mm}$N_{1650} $      &$ \frac{1}{2}^-$  & $53$   & $176$ & $5$   & $87$   & $8$   & $(109\pm26)^{+36}_{-3}$    \\
\rule[-2mm]{0mm}{0mm}$N_{1675} $      &$ \frac{5}{2}^-$  & $34$   & $61$  & $35$  & $40$   & $52$  & $(68\pm8)^{+14}_{-4}$      \\
\rule[-2mm]{0mm}{0mm}$N_{1700} $      &$ \frac{3}{2}^-$  & $6$    & $64$  & $6$   & $7$    & $9$   & $(10\pm5)^{+3}_{-3}$       \\
\rule[-2mm]{0mm}{0mm}$\Delta_{1620} $ &$ \frac{1}{2}^-$  & $26$   & $22$  & $3$   & $41$   & $5$   & $(38\pm8)^{+8}_{-6}$       \\
\rule[-2mm]{0mm}{0mm}$\Delta_{1700} $ &$ \frac{3}{2}^-$  & $28$   & $41$  & $29$  & $20$   & $38$  & $(45\pm15)^{+20}_{-10}$    \\
\rule[-2mm]{0mm}{0mm}$N_{1680} $      &$ \frac{5}{2}^+$  & $85$   & $112$ & $85$  & $149$  & $313$ & $(85\pm7)^{+6}_{-6}$       \\
\rule[-2mm]{0mm}{0mm}$N_{1720} $      &$ \frac{3}{2}^+$  & $377$  & $577$ & $100$ & $689$  & $238$ & $(23\pm8)^{+9}_{-5}$       \\

\hline

\multicolumn{2}{|c}{\rule[-2mm]{0mm}{6mm} $\gamma$}     & 15.365 & 16.525 & 14.635& 18.015 & 11.868&                           \\

\hline

\end{tabular}

\caption{Decay widths of baryon resonances for the GBE and OGE constituent quark models both in non-relativistic (NR) and semi-relativistic (SR) parametrizations (from Refs. \cite{THE01,THE01b}). A Gaussian-type meson wave function with $r_{\pi}=0.565$ fm was used along with a modified $^3P_0$ decay model. Experimental data are from Ref. \protect\cite{GRO00}.}\label{GBEgausspi}

\end{table}
\ 

Table \ref{GBEgausspi} also allows a comparison of the theoretical results to experimental data as compiled by the Particle Data Group (PDG) \cite{GRO00}. For the latter there arise two kinds of uncertainties : first, the total decay width of each resonance is given by a central value and a lower and upper bound. Second, the partial decay width has its own uncertainty. In Table \ref{GBEgausspi} we quote the value for the $\pi$ decay widths deduced from the central value of the total width and first add the uncertainty from the partial decay width itself (numbers inside the brackets in the last column). Then we indicate also the range of the total decay width by an upper and lower bound. We understand that the total uncertainty in a partial decay width must be estimated by combining both types of uncertainties (inherent separately in the total and partial widths).
\\

The OGE model chosen here is the model of Bhaduri, Cohler and Nogami \cite{BHA81} where the hyperfine interaction consists of a Coulomb term, a linear confinement and a flavour-independent spin-spin interaction where the parameters have been determined from a fit of the baryon spectra. The GBE model is the Model II presented in the previous section.
\\

Analyzing the results in details we note that for the $N_{1440}$ $\frac{1}{2}^+$ resonance the semi-relativistic (SR) GBE prediction is obviously too large, whereas the pertinent non-relativistic (NR) result lies within the experimental error bars. The SR OGE result overshoots the experiment by far, its non-relativistic version is also much smaller than the SR one and lies just at the lower end of the experimental error bar. The results for the next $\frac{1}{2}^+$ excitation of the nucleon, the $N_{1710}$, show a similar relative pattern as the ones for the {\it Roper} resonance, though all the values are smaller by about an order of magnitude. The fact that for each case, $N_{1440}$ and $N_{1710}$, the predictions of the SR parametrizations of both the OGE and GBE models exceed by far their NR counterparts can be readily understood because of the higher momentum components present in the SR parametrizations, as compared to the NR ones. In case of the OGE SR this effect is enhanced by a phase space that is much too large and which is due to the bad prediction of the resonance energy.
\\

For the $N_{1720}$ $\frac{3}{2}^+$ resonance the results again have similar characteristics, with the SR cases drastically overshooting the experimental data. Here, however, none of the NR versions can come close to the rather small experimental width. This problem may hint to a wrong symmetry assignment (or a strong mixing) of this state. Only for the $N_{1680}$ $\frac{5}{2}^+$ resonance the GBE model produces correct results, both in its SR and NR versions. In this case the results from both variants of the OGE constituent quark model are again too high.
\\

For the negative-parity $N_{1535}$ $\frac{1}{2}^-$ resonance the SR results are also much too high, whereas the predictions from the NR versions agree with experiment. For the $N_{1650}$ $\frac{1}{2}^-$ the situation is just reversed. Most disappointingly, in all instances the widths of the $N_{1535}$ resonance are larger than the ones of $N_{1650}$, contrary to experiment, where the $N_{1535}$ width appears to be smaller or is at most as large as the $N_{1650}$ width. Regarding the $L=1$, $S=\frac{3}{2}$ multiplet $N_{1650}-N_{1675}-N_{1700}$, one notes that the SR parametrizations give approximately the correct ratios of these widths, as it is expected from the corresponding spin-isospin matrix elements. These features are not found for the NR parametrizations due to the exceedingly small value of the $N_{1650}$ width. Concerning the negative-parity $N$ excitations, it is interesting to note that certain resonances are more sensible to the different parametrizations than others.
\\

The decay widths for the $\Delta$ resonances are practically all correct for the SR GBE model. In case of the other models the one or the other shortcoming appears.
\\

\subsection{The $\eta$ decays}
\

Table \ref{GBEgausseta} gives the results for $\eta$ decays. Theu{\ss} {\it et al.} use the same spatial part for the meson wave function as for $\pi$ decays but the constant $\gamma$ is adjusted so as to reproduce the $\eta$ decay width of the $N_{1535}$ resonance. Furthermore, they use an unmixed flavour wave function for the $\eta$ meson, {\it i. e.} a pure flavour octet state. For non-strange decays as regarded here, a possible mixing would only influence the normalization of this wave function, which can effectively be absorbed into the coupling constant $\gamma$.

\begin{table}[H]
\centering

\begin{tabular}{|cccccccc|}
\hline

\rule[-3mm]{0mm}{7mm}$N^*$ & $J^{\pi}$ & \multicolumn{6}{c|}{$\Gamma(N^*\rightarrow N\eta$) [MeV]} \\

\hline
\hline

\rule[-2mm]{0mm}{6mm}&&  GBE SR & GBE SR T & GBE NR & OGE SR & OGE NR & Exp.        \\

\hline
\hline

\rule[-2mm]{0mm}{7mm}$N_{1440} $      &$ \frac{1}{2}^+$  &        &       &        & 6     & 10   &       \\
\rule[-2mm]{0mm}{0mm}$N_{1710} $      &$ \frac{1}{2}^+$  & $26$   & $15$  & $4$    & $50$  & $10$ &       \\
\rule[-2mm]{0mm}{0mm}$N_{1520} $      &$ \frac{3}{2}^-$  & $0$    & $0$   & $0$    & $0$   & $0$  &       \\
\rule[-2mm]{0mm}{0mm}$N_{1535} $      &$ \frac{1}{2}^-$  & $64$   & $64$  & $64$   & $64$  & $64$ & $(64\pm19)^{+76}_{-15}$  \\
\rule[-2mm]{0mm}{0mm}$N_{1650} $      &$ \frac{1}{2}^-$  & $113$  & $33$  & $68$   & $140$ & $94$ & $(10\pm5)^{+4}_{-1}$     \\
\rule[-2mm]{0mm}{0mm}$N_{1675} $      &$ \frac{5}{2}^-$  & $2$    & $2$   & $4$    & $3$   & $5$  &  \\
\rule[-2mm]{0mm}{0mm}$N_{1700} $      &$ \frac{3}{2}^-$  & $0$    & $0$   & $1$    & $1$   & $1$  &  \\
\rule[-2mm]{0mm}{0mm}$N_{1680} $      &$ \frac{5}{2}^+$  & $0$    & $0$   & $1$    & $2$   & $6$  &  \\
\rule[-2mm]{0mm}{0mm}$N_{1720} $      &$ \frac{3}{2}^+$  & $15$   & $15$  & $11$   & $30$  & $25$ &  \\

\hline

\multicolumn{2}{|c}{\rule[-2mm]{0mm}{6mm} $\gamma$} & 5.929 & 4.80 & 6.682 & 6.572 & 4.937 & \\

\hline

\end{tabular}

\caption{Same than in Fig. \ref{GBEgausspi} but for $N^*\rightarrow N\eta$, $r_{\eta}=0.565$ fm (from Refs. \cite{THE01,THE01b}).}\label{GBEgausseta}

\end{table}
\ 

The $\eta$ widths of the {\it Roper} resonance $N_{1440}$ for the GBE parametrizations (NR as well as SR) are rigorously zero, since in both cases the theoretically predicted masses lie below the $\eta$ threshold, in accordance with experiment. For the OGE parametrizations, the decay $N_{1440}\rightarrow N\eta$ is possible, the corresponding widths remain rather small, however.
\\

In all, there are four resonances predicted with considerable branching ratios in the $\eta$ decay channel. Only for the $N_{1535}$ and $N_{1650}$ resonances one can compare the theoretical predictions to experiment, since these are the only ones with an experimental width assigned by the PDG \cite{GRO00}. The relative magnitudes of the experimental decay widths in both of these cases are missed by all theoretical models. This is again reminiscent of the EEM, where a similar effect is found.
\\

In addition to $N_{1535}$ and $N_{1650}$, also the widths of the $N_{1710}$ and the $N_{1720}$ resonances come out appreciably large. The PDG does not quote any experimental data for these states. This does not necessarily mean that their widths are vanishing or too small to be measured. It may simply be the case that experimental ambiguities do not allow yet for a reliable determination. In fact, there are single partial-wave analyzes that assign an appreciable $\eta$ decay width, for example, also to the $N_{1710}$, see Ref. \cite{BAT95}.
\\

From the presented results it is still difficult to draw definite conclusions about the qua\-lity of the  wave functions stemming from different quark models. In fact, the various decay widths seem to be more determined by the choice of the SR or NR parametrizations rather than by the use of either type of dynamics, GBE or OGE. One might expect that the decays of these resonances are quite sensitive to the tensor force in the quark-quark interaction. But the inclusion of these force components does not really improve the description of both $N\pi$ and $N\eta$ decays for these resonances in the SR GBE as shown in Tables \ref{GBEgausspi} and \ref{GBEgausseta} in the column GBE SR T. Thus, the description of strong decays of baryon resonances within present constituent quark models is not yet fully satisfactory. The reasons for the persisting difficulties may on the one hand reside in the baryon wave functions. On the other hand one must realize that the $^3P_0$ decay model may also fall short as it is based on intuitive grounds and lacks a firm theoretical foundation. A consistent microscopic description of the strong-decay processes thus remains a challenging task. One can think of a number of improvements to be done. For example, the proper inclusion of relativistic effects appears mandatory. The encouraging work, which is on the way, by Plessas {\it et al.} \cite{PLE02}, within the point form formalism presented in the previous section indicates a serious improvement of the pion decays of $N$ and $\Delta$. In particular the width of the {\it Roper} resonance acquires a reasonable magnitude.
\\

\section{Conclusion}
\

In this chapter we discussed the theoretical foundation and phenomenological implications of the GBE hyperfine interaction introduced in Chapter \ref{microchapter}. The early quark model was successful in classifying hadrons and describing some gross properties of their spectra but no firm evidence on the dynamics of the valence quarks was achieved. Even when motivated by QCD, the concept of one-gluon exchange was introduced as a hyperfine interaction between confined constituent quarks, we have seen that a number of delicate problems remained unsolved. In that context one has not been able to explain the correct level ordering of the first positive-parity and the first negative-parity excited states in both light- and strange-baryons spectra. This shortcoming essentially stem from the fact that the implications of the spontaneous breaking of chiral symmetry presented in Section \ref{GBEchiralsection} are not properly taken into account in such models.  As a consequence of this symmetry breaking, baryons are to be considered as systems of three constituent quarks that interact by Goldstone boson exchange, the pseudoscalar mesons, and are subject to confinement.
\\

The remarkable successes of the GBE quark-quark interaction in reproducing the spectra has been shown as coming from the particular symmetry introduced through the spin-flavor operator $\vec{\sigma}_i\cdot\vec{\sigma}_j\ \vec{\lambda}_i\cdot\vec{\lambda}_j$ and by the short-range part of the interaction with a proper sign. Several parametrizations of this interaction have been presented in this chapter, both within a relativistic and a non-relativistic kinematic. However, we have seen from the spectra calculated by Glozman {\it et al.} \cite{GLO96b}-\cite{GLO98} and presented in this chapter, that the quality of the results depends only a little on the parametrization interaction. Extended versions of the GBE model, necessary for further investigations, were also presented.
\\

However, baryon spectra is only a first test of a low-energy QCD modelization. That is why, to show the performances of this model, we presented some stringent tests of the GBE interaction where, apart from its Hamiltonian eigenvalue, the wave function were analyzed. First, we discussed the predictions of observables related to the electroweak nucleon structure in the framework of the point form approach to the relativistic quantum mechanics. Theoretical predictions fall quite close to the available experimental data if the GBE dynamics is used in a semi-relativistic version.
\\

We have then looked at the strong decays widths in the $^3P_0$ decay model, and in particular the pion decay of $N$ and $\Delta$ resonances. Poor agreement has been obtained as compared to the experimental data, the widths seem to be more determined by the choice of the kinetic parametrization (relativistic or non-relativistic) rather than by the type of dynamics (GBE or OGE). The origin of these discrepancies may come either from the baryon wave function or from the $^3P_0$ decay model considered in the derivation of the results. Anyhow, promising works under progress within a point form approach seem to indicate that it is the formalism which is responsible of the difficulties to reproduce experiment.
\\

In all, the GBE chiral quark model provides a promising basis for a number of further investigations. In particular the middle- and long-range parts of the quark-quark interaction is not really constrained by the observables at the baryon level, as for example the $\sigma$-meson exchange introduced in Section \ref{GBEspectra}. This leaves room for more adjustment of the parametrization. It is then natural to extend the study of the GBE interaction to systems composed of more than three quarks. Our first interest here is to find out if the GBE interaction can explain the short-range repulsion of a two-nucleon system. This will be the main purpose of the following chapter.
\newpage
\thispagestyle{empty} 
\ 
\newpage


\chapter{Preliminary Studies}\label{preliminarychapter}

\vspace{3.3cm}
\ 

Following the successes of the baryon description reviewed in the previous chapter, it is quite natural to extend the application of the GBE model to other systems. The present chapter is a first exploratory step towards calculating the NN interaction starting from a system of six interacting quarks. The first arising question is whether or not the chiral constituent quark model presented in the previous chapter is able to produce a short-range repulsion in the NN system. For this purpose, we diagonalize the corresponding Hamiltonian in a harmonic oscillator basis containing up to two excitations quanta. Using the Born-Oppenheimer (adiabatic) approximation, we obtain an effective internucleon potential between nucleons from the difference between the lowest eigenvalue of the six-quark Hamiltonian and two times the nucleon (three-quark) mass calculated in the same model.
\\

First we compute the adiabatic potential at zero separation distance only. We use a cluster model single particle basis. The result indicates the presence of a short-range repulsion. We then repeat these calculations in an orbital molecular basis. We still obtain a repulsion at zero separation. Furthermore we extend our calculations to any separation distance $Z$ between two nucleons in both the cluster and molecular basis. The resulting potential indicates the presence of a repulsive core of about 1 fm which is quite reasonable. The potential obtained in the cluster and the molecular bases are very similar. However at short-range, due to a larger Hilbert space, the molecular basis improves slightly the results.
\\

Both Model I and Model II, presented in the previous chapter, have been used in the derivation of the adiabatic potential. The results are very similar. However some differences appear, in particular in the shape of the potential. In Model II, the repulsion is higher at small $Z$ but has a shorter range as compared to that of Model I.
\\

Finally we show that the introduction of a scalar-exchange interaction at the quark level, considered as due to two pion exchanges, provides a middle-range attraction.

\section{The GBE Hamiltonian}
\ 

In this section we briefly recall the GBE models \cite{GLO96a,GLO96b} used in the study of the NN interaction. In both cases, the Hamiltonian reads
 
\begin{equation}
H= 6m + \sum_i \frac{\vec{p}_{i}^{2}}{2m} - \frac {(\sum_i \vec{p}_{i})^2}{12m} + \sum_{i<j} V_{conf}(r_{ij}) + \sum_{i<j} V_\chi(r_{ij})
\label{PREham}
\end{equation}
\ 

\noindent where $m$ is the constituent quark mass and $r_{ij} = | \vec{r}_i - \vec{r}_j |$ is the interquark distance.
\\
 
The confining interaction is
 
\begin{equation}
V_{conf}(r_{ij}) = -\frac{3}{8}\lambda_{i}^{c}\cdot\lambda_{j}^{c} \left( C\, r_{ij} + V_0 \right)
\label{PREconf}
\end{equation}
\ 

\noindent where $\lambda_{i}^{c}$  are the  $SU(3)$-color matrices and $C$ and $V_0$ are parameters given below.
\\

The spin-spin component of the GBE interaction between the constituent quarks $i$ and $j$ reads

\begin{eqnarray}
V_\chi(\vec r_{ij})
&=&
\left\{\sum_{a=1}^3 V_{\pi}(\vec r_{ij}) \lambda_i^a \lambda_j^a \right.
\nonumber \\
&+& \left. \sum_{a=4}^7 V_{\rm K}(\vec r_{ij}) \lambda_i^a \lambda_j^a
+V_{\eta}(\vec r_{ij}) \lambda_i^8 \lambda_j^8
+V_{\eta^{\prime}}(\vec r_{ij}) \lambda_i^0 \lambda_j^0\right\}
\vec\sigma_i\cdot\vec\sigma_j,
\label{PREvchi}
\end{eqnarray}
\ 

\noindent where $\lambda^a, a=1,...,8$ are the flavor Gell-Mann matrices and $\lambda^0 = \sqrt{2/3}~{\bf 1}$, where $\bf 1$ is the $3\times3$ unit matrix. Thus the interaction (\ref{PREvchi}) includes $\pi$, $K$, $\eta$ and $\eta'$ exchanges. While the $\pi$, $K$, $\eta$  mesons are the Goldstone bosons of the spontaneously broken $SU(3)_L \times SU(3)_R \rightarrow SU(3)_V$ chiral symmetry, the $\eta'$ (flavor singlet) is not {\it a priori} a Goldstone boson. But it becomes a Goldstone boson in the large $N_c$ limit (see Chapter \ref{gbechapter}). This is an argument to include it in the model.  For systems formed of $u$ and $d$ quarks only, the $K$-exchange does not contribute. It is the case here because we deal only with nucleons.
\\

In the simplest case, when both the constituent quarks and mesons are point-like particles and the boson field satisfies the linear Klein-Gordon equation, one has the following spatial dependence for the meson-exchange potentials \cite{GLO96a}

\begin{equation}
V_\gamma (\vec r_{ij})=\frac{g_\gamma^2}{4\pi}\frac{1}{3}\frac{1}{4m^2}\{\mu_\gamma^2\frac{e^{-\mu_\gamma r_{ij}}}{ r_{ij}}-4\pi\delta (\vec r_{ij})\},  \hspace{5mm} (\gamma = \pi, K, \eta, \eta' )
\label{PREpoint} \end{equation}
\ 

\noindent where $\mu_\gamma$ are the  meson masses and $g_\gamma^2/4\pi$ are the quark-meson coupling constants given below.
\\

Eq. (\ref{PREpoint}) contains both the traditional long-range Yukawa potential as well as a $\delta$-function term. It is the latter that is of crucial importance for baryon spectroscopy since it has a proper sign to provide the correct hyperfine splittings in baryons. Since one deals with structured particles (both the constituent quarks and pseudoscalar mesons) of finite extension, one must smear out the $\delta$-function in (\ref{PREpoint}). The Graz group has proposed several parametrizations of the model, some presented in the previous chapter, in order to improve the description of the baryon spectra. Here we use only two non-relativistic versions of the model as justified below. We shall show that the regularized $\delta$-function term is crucial in describing the short-range repulsion in the NN potential. But as we shall see in this preliminary studies, the details of this regularization do not really matter. In Ref. \cite{GLO96b} a smooth Gaussian term has been employed instead of the $\delta$-function

\begin{equation}
4\pi \delta(\vec r_{ij}) \Rightarrow \frac {4}{\sqrt {\pi}}\alpha^3 \exp(-\alpha^2(r-r_0)^2). \label{PREcontact}
\end{equation}
\ 

\noindent where $\alpha$ and $r_0$ are adjustable parameters. In this version of the model the parameters of the Hamiltonian (\ref{PREham}) are \cite{GLO96b}

$$\frac{g_{\pi q}^2}{4\pi} = \frac{g_{\eta q}^2}{4\pi} = 0.67,\ \frac{g_{\eta ' q}^2}{4\pi} = 1.206,$$
$$m_{u,d} = 340 \, {\rm MeV},~~   C= 0.474 \, {\rm fm}^{-2},$$
$$\mu_{\pi}=139\ {\rm MeV},\ \mu_{\eta}=547\ {\rm MeV},\ \mu_{\eta'}=958\ {\rm MeV},$$
\begin{equation}\label{PREparam1}
r_0 = 0.43 \, {\rm fm}, ~\alpha = 2.91 \, {\rm fm}^{-1},~ V_0= 0 \ {\rm MeV}.
\end{equation}
\\

The Hamiltonian (\ref{PREham}) with the parameters (\ref{PREparam1}) provides a very satisfactory description of the low-lying $N$ and $\Delta$ spectra in a fully dynamical non-relativistic 3-body calculation \cite{GLO96b}. In the following, we shall call this version the Model I. 
\\

The other parametrization, called here Model II, is given in Ref. \cite{GLO97a}

\begin{equation}\label{PREmodelII}
V_{\gamma}(r) = \frac{g_{\gamma}^2}{4\pi} \frac{1}{12m_im_j} \{ \mu_{\gamma}^2 \frac{e^{- \mu_{\gamma} r}}{r} - \Lambda_{\gamma}^2 \frac{e^{- \Lambda_{\gamma} r}}{r} \} ,
\end{equation}
\ 

\noindent where $\Lambda_{\gamma}=\Lambda_0+\kappa \mu_{\gamma}$ and the parameters
$$\frac{g_{\pi q}^2}{4\pi} = \frac{g_{\eta q}^2}{4\pi} = 1.24,\ \frac{g_{\eta' q}^2}{4\pi} = 2.7652,$$
$$m_{u,d}=340\ {\rm MeV},\ \ C=0.77\ {\rm fm}^{-2},$$
$$\mu_{\pi}=139\ {\rm MeV},\ \mu_{\eta}=547\ {\rm MeV},\ \mu_{\eta'}=958\ {\rm MeV},$$
\begin{equation}\label{PREparam2}
\ \ \Lambda_0=5.82\ {\rm fm}^{-1},\  \kappa = 1.34,\  V_0=-112\ {\rm MeV}.
\end{equation}
\ 

\noindent Certainly more fundamental studies are required to understand this second term and attempts are being made in this direction. The instanton liquid model of the vacuum (for a review see \cite{SCH98}) implies point like quark-quark interactions.  To obtain a realistic description of the hyperfine interaction this interaction has to be iterated in the t-channel \cite{GLO99}. The t-channel iteration admits a meson exchange interpretation \cite{RIS99}.
\\

At present we restricted to use an $s^3$ harmonic oscillator wave function for the nucleon in the NN problem. The parametrization (\ref{PREparam1}) is especially convenient for this purpose since it allows to use the $s^3$ as a variational ansatz. Otherwise the structure of N should be more complicated. With an $s^3$ ansatz the average $<N|H|N>$ takes a minimal value of 969.6 MeV at a harmonic oscillator parameter value of $\beta=0.437$ fm \cite{PEP97}, i.e. only 30 MeV above the actual value in the dynamical 3-body calculation. In this way one satisfies one of the most important constraint for the microscopic study of the NN interaction : the nucleon stability condition (see next chapter) \cite{OKA84}

\begin{equation}
\frac {\partial}{\partial \beta} < N | H | N > = 0 .
\label{PREstab} \end{equation}

The other condition, the qualitatively correct $\Delta-N$ splitting, is also satisfied \cite{PEP97}.
\\

We keep in mind, however, that a non-relativistic description of baryons cannot be completely adequate. In principle it would be better to use a parametrization of the GBE interaction as given in \cite{GLO98} based on a semi-relativistic Hamiltonian. Within the semi-relativistic description of baryons \cite{GLO97c} the parameters extracted from the fit to baryon masses become considerably different and even the representation of the short-range part of GBE has a different form. Within a semi-relativistic description the simple $s^3$ wave function for the nucleon is not adequate anymore. All this suggests that the description of the nucleon based on the parameters (\ref{PREparam1}) or (\ref{PREparam2}) and an $s^3$ wave function is only effective. Since here we study only qualitative effects, related to the spin-flavor structure and sign of the short-range part of GBE interaction, we consider the present non-relativistic parametrization as a reasonable framework.
\\

\section{A qualitative analysis at the Casimir operator level}\label{sectionCasimir}
\ 

In order to have a preliminary qualitative insight it is convenient first to consider a schematic model which neglects the radial dependence of the GBE interaction. In this model the short-range part of the GBE interaction between the constituent quarks is approximated by \cite{GLO96a}

\begin{equation}
 V_{\chi} = - C_{\chi} \sum_{i<j}  \lambda_{i}^{f} . \lambda_{j}^{f} \vec{\sigma}_i . \vec{\sigma}_j , \label{PREopFS}
\end{equation}
\ 

\noindent where $\lambda^{f}$ with an implied summation over f (f=1,2,...,8) and $\vec{\sigma}$ are the quark flavor Gell-Mann and spin matrices respectively. The minus sign of the interaction (\ref{PREopFS}) is related to the sign of the short-range part of the pseudoscalar meson-exchange interaction (which is opposite to that of the Yukawa potential tail), crucial for the hyperfine splittings in baryon spectroscopy as seen in Chapter \ref{microchapter}. The constant $C_{\chi}$ can be determined from the $\Delta-N$ splitting. For that purpose one only needs the spin (S), flavor (F) and flavor-spin (FS) symmetries of the $N$ and $\Delta$ states, identified by the corresponding partitions [f] associated with the groups $SU(2)_S, SU(3)_F$ and $SU(6)_{FS}$

\begin{eqnarray}
|N> &=& |s^3 [3]_{FS} [21]_F [21]_S > , \\
|\Delta> &=& |s^3 [3]_{FS} [3]_F [3]_S > .
\end{eqnarray}
\ \\
Then the matrix elements of the interaction (\ref{PREopFS}) are \cite{GLO96a}

\begin{eqnarray}
<N | V_{\chi} | N> &=& -14\ C_{\chi} , \\
<\Delta | V_{\chi} | \Delta> &=& -4\ C_{\chi} .
\end{eqnarray}
\ \\
Hence $E_{\Delta}-E_N = 10 C_{\chi}$, which gives $C_{\chi} = 29.3$MeV, if one uses the experimental value of 293 MeV for the $\Delta - N$ splitting.
\\

To see the effect of the interaction (\ref{PREopFS}) in the six-quark system, the most convenient is to use the coupling scheme called FS, where the spatial $[f]_O$ and color $[f]_C$ parts are coupled together to $[f]_{OC}$, and then to the $SU(6)_{FS}$ flavor-spin part of the wave function in order to provide a totally antisymmetric wave function in the OCFS space \cite{HAR81}. The antisymmetry condition requires $[f]_{FS} = [\tilde{f}]_{OC}$, where $[\tilde{f}]$ is the conjugate of $[f]$.
\\

The color-singlet 6q state is $[222]_C$. Assuming that N has a $[3]_O$ spatial symmetry, there are two possible states $[6]_O$ and $[42]_O$ compatible with the S-wave relative motion in the NN system \cite{NEU77}. The flavor and spin symmetries are $[42]_F$ and $[33]_S$ for $^1S_0$ and $[33]_F$ and $[42]_S$ for $^3S_1$ channels. Applying the inner product rules of the symmetric group for both the $[f]_O \times [f]_C$ and $[f]_F \times [f]_S$ products one arrives at the following 6q antisymmetric states associated with the $^3S_1$ and $^1S_0$ channels \cite{HAR81,STA96} : $| [6]_O [33]_{FS} >$, $| [42]_O [33]_{FS} >$, $| [42]_O [51]_{FS} >$, $| [42]_O [411]_{FS} >$, $| [42]_O [321]_{FS} >$, $| [42]_O [2211]_{FS} >$.
\\
 
Then the expectation values of the GBE interaction (\ref{PREopFS}) for these states  can be easily calculated in terms of the Casimir operators eigenvalues of the groups $SU(6)_{FS}$, $SU(3)_F$ and $SU(2)_S$ using the following formula

\begin{equation}
<\sum_{i<j} \lambda_i . \lambda_j \vec{\sigma}_i . \vec{\sigma}_j> = 4 C_{2}^{SU(6)} - 2 C_{2}^{SU(3)} - \frac{4}{3} C_{2}^{SU(2)} - 8N
\end{equation}
\ 

\noindent where $N$ is the number of particles, here $N=6$, and $C_{2}^{SU(n)}$ is the Casimir operator eigenvalues of $SU(n)$ which can be derived from the expression

\begin{eqnarray}
C_{2}^{SU(n)} = \frac{1}{2} [f_1'(f_1'+n-1) + f_2'(f_2'+n-3) + f_3'(f_3'+n-5) \nonumber \\
+f_4'(f_4'+n-7) + ... + f_{n-1}'(f_{n-1}'-n+3) ] - \frac{1}{2n}(\sum_{i=1}^{n-1} f_i')^2
\label{casimir}
\end{eqnarray}
\ 

\noindent where $f_i'= f_i-f_n$, for an irreducible representation given by the partition $[f_1,f_2,...,f_n]$.
\\

The corresponding matrix elements are given in Table \ref{PREexpectation}, from where one can see that, energetically, the most favorable configuration is $[51]_{FS}$ for $V_\chi$. This is a direct consequence of the general rule that at short-range and with fixed spin and flavor, the more ``symmetric" a given FS Young diagram is, the more negative is the expectation value of (\ref{PREopFS}). The difference in the potential energy between the configuration $[51]_{FS}$ and $[33]_{FS}$ or $[411]_{FS}$ is of the order

\begin{equation}
\begin{array}{ccccc}
<[33]_{FS} | V_{\chi} | [33]_{FS}> & - & <[51]_{FS} | V_{\chi} | [51]_{FS}> & = & \\
<[411]_{FS} | V_{\chi} | [411]_{FS}> & - & <[51]_{FS} | V_{\chi} | [51]_{FS}> & = &24\ C_{\chi}\\
\end{array}
\end{equation}
\ 

\noindent and using $C_{\chi}$ given above one obtains approximately 700 MeV for both the $SI = 10$ and $01$ sectors.

\begin{table}[H]
\centering

\begin{tabular}{|l|c|c|c|c|}
\hline
& \multicolumn{2}{c|}{$I=0, S=1$} & \multicolumn{2}{c|}{$I=1, S=0$} \\
$[f]_o [f]_{FS}$  & $<V_{\chi}>$ & $<V_{cm}>$ & $<V_{\chi}>$ & $<V_{cm}>$ \\

\hline
\hline

$[6]_o [33]_{FS}$  & -28/3 & 8/3 & -8 & 8 \\
$[42]_o [33]_{FS}$  & -28/3 & -26/9 & -8 & -4/3 \\
$[42]_o [51]_{FS}$  & -100/3 & 16/9 & -32 & 16/9 \\
$[42]_o [411]_{FS}$  & -28/3 & 20/9 & -8 & 44/9 \\
$[42]_o [321]_{FS}$  & 8/3 & -164/45 & 4 & 232/45 \\
$[42]_o [2211]_{FS}$  & 68/3 & -62/15 & 60 & 42/5 \\

\hline

\end{tabular}
\caption{Expectation values of the operators defined by Eqs. (\ref{PREopFS}) and (\ref{PREvcm}) for all compatible symmetries $[f]_O [f]_{FS}$ in the $SI=10$ and $01$ sector. $<V_{\chi}>$ is in units of $C_{\chi}$ and $<V_{cm}>$ in units of $C_{cm}$.}\label{PREexpectation}

\end{table}

In a harmonic oscillator basis containing up to $2\hbar\omega$ excitation quanta, there are two dif\-ferent 6q states corresponding to the $[6]_O$ spatial symmetry with removed center of mass motion. One of them, $|s^6 [6]_O>$, belongs to the $N=0$ shell, where $N$ is the number of excitation quanta in the system,  and the other, $ \sqrt{\frac {5}{6}} |s^52s [6]_O> - \sqrt{\frac {1}{6}} |s^4p^2 [6]_O>$, belongs to the $N=2$ shell. There is only one state with $[42]_O$ symmetry, the $|s^4p^2 [42]_O>$ state belonging to the $N=2$ shell. While here and below we use notations of the shell model it is always assumed that the center of mass motion is removed.
\\

The kinetic energy $KE$ for the $|s^4p^2 [42]_O>$ state is larger than the one for the $|s^6[6]_O>$ state by $KE_{N=2} - KE_{N=0} = \hbar \omega$. Taking $\hbar \omega\simeq 250$ MeV \cite{GLO96a}, and denoting the kinetic energy operator by $H_0$, we obtain

\begin{equation}
<s^6 [33]_{FS} | H_0 + V_{\chi} | s^6 [33]_{FS}> -  <s^4p^2 [51]_{FS} | H_0 + V_{\chi} | s^4p^2 [51]_{FS}> \simeq 453\ {\rm MeV}
\end{equation}
\ 

\noindent which shows that $[51]_{FS}$ is far below the other states of Table \ref{PREexpectation}. For simplicity, here we have neglected a small difference in the confinement potential energy between the above configurations.
\\

This qualitative analysis suggests that in a more quantitative study, where the radial dependence of the GBE interaction is taken into account, the state $| s^4p^2 [42]_O [51]_{FS}>$ will be highly dominant and, due to the important lowering of this state by the GBE interaction with respect to the other states, the mixing angles with these states will be small. This is indeed the case, it will be proved in the following.
\\

Table \ref{PREexpectation} and the discussion above indicate that for the diagonalization of a realistic Hamiltonian as (\ref{PREham}) the following most important configurations should be taken into account

\begin{equation}
\begin{array}{cll}
|1> &=&| s^6 [6]_O [33]_{FS} > \\
|2> &=&| s^4p^2 [42]_O [33]_{FS} > \\
|3> &=&| s^4p^2 [42]_O [51]_{FS} > \\
|4> &=&| s^4p^2 [42]_O [411]_{FS} > \\
\end{array}
\label{PREbasis}
\end{equation}
\ \\

As already mentioned, one also has to include the $ \sqrt{\frac {5}{6}} |s^52s [6]_O> - \sqrt{\frac {1}{6}} |s^4p^2 [6]_O>$ state which belongs to the $N=2$ shell if one takes up to two excitation quanta in the system. Stancu {\it et al.} \cite{STA97} have extended the basis (\ref{PREbasis}) by adding the configuration $| (\sqrt{5/6} s^{5}2s - \sqrt{1/6}s^4p^2  [6]_o [33]_{FS} >$ in their calculations and showed that it practically does not change much the result at $Z=0$. That is why we shall neglect it in the treatment given below.
\\

Now we want to give a rough estimate of the interaction potential of the NN system at zero separation distance between nucleons. We calculate this potential in the Born-Oppenheimer (or adiabatic) approximation defined as

\begin{equation}
V_{NN}(Z) = <H>_Z - <H>_{\infty}
\label{PREborn}
\end{equation}
\ 

\noindent where $Z$ is a collective coordinate which is the separation distance between the two  $s^3$ nucleons, $<H>_Z$ is the lowest expectation value of the Hamiltonian describing the 6q system at fixed $Z$ and $<H>_{\infty} = 2 m_N$ for the NN problem, i.e. the energy of two well separated nucleons. As above, we ignore the small difference between the confinement energy of $<H>_{Z=0}$ and $<H>_{\infty}$. That this difference is small follows from the $\lambda_{i}^{c}. \lambda_{j}^{c}$ structure of the confining interaction and from the identity

\begin{equation}
<[222]_c | \sum_{i<j}^{6} \lambda_{i}^{c} . \lambda_{j}^{c} | [222]_c> = 2 <[111]_c | \sum_{i<j}^{3} \lambda_{i}^{c} . \lambda_{j}^{c} | [111]_c> .
\end{equation}
\ 

This identity shows that if the space parts $[6]_O$ and $[3]_O$ contain the same single particle state, for example an s-state, then the difference is identically zero.
\\

It has been shown by Harvey \cite{HAR81} that when the separation $Z$ between two $s^3$ nucleons approaches zero, then only two types of 6q configurations survive: $|s^6 [6]_O>$ and $|s^4p^2 [42]_O>$. Thus in order to extract an effective NN potential at zero separation between nucleons in the adiabatic approximation one has to diagonalize the Hamiltonian in the basis $|1>-|4>$. For the rough estimate below we take only the lowest configuration $|3>$. One then obtains

$$<s^4p^2 [42]_O [51]_{FS} | H_0 + V_{\chi} | s^4p^2 [42]_O [51]_{FS}> - 2 <N | H_0 + V_{\chi} | N> =$$

\begin{equation}
\left\{
\begin{array}{ccccc}
(-100/3 + 28)C_{\chi} + 7/4 \hbar \omega & = & 280 \mbox{ MeV}, & \mbox{if} & $SI=10$ \\ 
(-32 + 28)C_{\chi} + 7/4 \hbar \omega & = & 320 \mbox{ MeV}, & \mbox{if} & $SI=01$ \\
\end{array}
\right.
\label{PREestim}
\end{equation}
\\

At this stage it is useful to compare the nature of the short-range repulsion generated by the GBE interaction  to that produced by the OGE interaction.
\\

In the constituent quark models based on OGE the situation is more complex. Table \ref{PREexpectation} helps in summarizing the situation there. The OGE interaction is described by the simplified chromomagnetic interaction

\begin{equation}
V_{cm} = - C_{cm} \sum_{i<j} \lambda_{i}^{c} . \lambda_{j}^{c} \vec{\sigma}_i . \vec{\sigma}_j
\label{PREvcm}
\end{equation}
in units of the  constant $C_{cm}$ (the constant $C_{cm}$ can also be determined from the $\Delta - N$ splitting to be $C_{cm}\simeq 293/16$ MeV).
\\

The expectation values of (\ref{PREvcm}) can be easily obtained in the CS scheme with the help of Casimir operator formula above and can be transformed to the FS scheme by using the unitary transformations from the CS scheme to the FS scheme. All these transformations are given in Ref. \cite{STA97}.
\\

The chromomagnetic interaction pulls the configuration $|s^4p^2[42]_O [42]_{CS}>$ down to become approximately degenerate with $|s^6 [6]_O [222]_{CS}>$ which is pulled up. In a more detailed calculation with explicit radial dependence of the chromomagnetic interaction as well as with a Coulomb term the configuration $|s^6 [6]_O>$ is still the lowest one \cite{OKA84,OBU88}. With the model (\ref{PREvcm}) the $\hbar \omega$ should be about 500 MeV. Thus in the Born-Oppenheimer approximation we can roughly estimate an effective interaction given by the OGE model through the difference

\begin{equation}
< s^6 [6]_O [222]_{CS}| H_0 + V_{cm}| s^6[6]_O [222]_{CS}> - 2 < N | H_0 + V_{cm} | N >
\end{equation}

\begin{equation}
= \left\{ \begin{array}{ccc} \frac {56}{3}C_{cm} +  3/4 \hbar \omega = 717\ {\rm MeV}& \mbox{if} & SI = 10 \\ 
24C_{cm} +  3/4 \hbar \omega = 815\ {\rm MeV} & \mbox{if} & SI = 01 \end{array}\right.
\end{equation}
\\

We conclude that both the GBE and OGE models imply effective repulsion at short-range.
\\

The matrix elements of the Hamiltonian (\ref{PREham}) are calculated in the basis (\ref{PREbasis}) by using the fractional parentage technique described in Refs. \cite{HAR81} and also applied in Ref. \cite{STA97}. A program based on Mathematica \cite{WOL96} has been created for this purpose. In this way every six-body matrix element reduces to a linear combination of two-body matrix elements of either symmetric or antisymmetric states for which Eqs. (3.3) of Ref. \cite{GLO96a} can be used to integrate in the flavor-spin space.
\\

In dealing with $n$ particles the matrix elements of a symmetric two-body operator between totally (symmetric or) antisymmetric states $\psi_n$ and $\psi_n'$ reads

\begin{equation}
<\psi_n | \sum_{i<j} V_{ij} | \psi_n' > = \frac{n(n-1)}{2} <\psi_n | V_{n-1,n} | \psi_n' >
\end{equation}
\\

The matrix elements of $V_{n-1,n}$ are calculated by expanding $\psi_n$ and $\psi_n'$ in terms of products of antisymmetric states of the first $n-2$ particles $\psi_{n-2}$ and of the last pair $\phi_2$

\begin{equation}
\psi_n = \sum_{\alpha \beta} P_{\alpha\beta} \psi_{n-2}(\alpha) \phi_2(\beta)
\end{equation}
\ 

\noindent with $\alpha, \beta$ denoting the possible structures of $\psi_{n-2}$ and $\phi_2$ and $P_{\alpha\beta}$ the products of cfp coefficients in the orbital, spin-flavor and color space states. In practical calculations, the color space cfp coefficients are not required. The orbital cfp are taken from Ref. \cite{STA88}, Tables 1 and 2 by using the replacement $r^4l^2 \rightarrow s^4p^2$ and $r^5l \rightarrow s^5p$. The trivial ones are equal to one. The flavor-spin cfp for $SI=10$ are identical to the $\bar{K}$-matrices of Table 1 Ref. \cite{STA89} with $[42]_S [33]_F$ in the column headings. For $SI=01$ they are the same as for $SI=10$ but the column headings is $[42]_F [33]_S$ instead of $[42]_S [33]_F$ as above, and this is due to the commutativity of inner products of $S_n$ (see for example Ref. \cite{STA96}). The cfp used in the OC coupling are from Ref. \cite{STA88} Table 3, for $[42]_O \times [222]_C \rightarrow [3111]_{OC}$ and Table 5 of Ref. \cite{STA89} for $[42]_O \times [222]_C \rightarrow [222]_{OC}$ and $[42]_O \times [222]_C \rightarrow [21111]_{OC}$.
\\

In this way, after decoupling all degrees of freedom one can integrate out in the color, spin and flavor space. The net outcome of this algebra is that any six-body matrix element becomes, as we mentioned above, a linear combination of two-body orbital matrix elements, $<V_{\pi}>, <V_{\eta}>$ and $<V_{\eta'}>$. The coefficients of $<V_{\pi}>$ are the same for $SI=10$ and $01$, but the coefficients of $<V_{\eta}>$ are usually different. In both cases the coefficients of $<V_{\eta'}>$ are two times those of $<V_{\eta}>$. We found that the two-body GBE matrix elements satisfy the relations $<V_{\pi}> \simeq <V_{\eta}>$ and $<V_{\eta'}> \simeq 2<V_{\pi}>$. As an example, in Table \ref{PREOME} we show the matrix elements obtained for $SI=10$. Similar results are obtained in Model II. Except for $<ss | V_\gamma |pp>$, they are all negative, i.e. carry the sign of Eq. (\ref{PREopFS}).
\\

The six-quark states are constructed from the following single particle harmonic oscillator states
\begin{eqnarray}
|s> & = & \pi^{-3/4} \beta^{-3/2} \exp{(-r^2/2\beta^2)} \label{PREsstate}\\
|p>_m & = & 8^{1/2} 3^{-1/2} \pi^{-1/4} \beta^{-5/2} r \exp{(-r^2/2\beta^2)} \, Y_{1m} \label{PREpstate}
\label{PREsinglestate}
\end{eqnarray}

\begin{table}[H]
\centering

\begin{tabular}{|l|c|c|c|}
\hline
Two-body matrix elements & $\gamma=\pi$ & $\gamma=\eta$ & $\gamma=\eta'$ \\

\hline
\hline

$<ss | V_\gamma | ss>$ & -0.108357 & -0.104520 & -0.189153 \\
$<ss | V_\gamma | (pp)_{L=0}>$ & 0.043762 & 0.042597 & 0.076173 \\
$<sp | V_\gamma | sp>$ & -0.083091 & -0.079926 & -0.145175 \\
$<(pp)_{L=0} | V_\gamma | (pp)_{L=0}>$ & -0.081160 & -0.078594 & -0.142205 \\
\hline

\end{tabular}
\caption{All one-meson exchange two-body matrix elements (in GeV) for the sector $SI=10$ evaluated at $\beta = 0.437$ fm in the framework of the Model I. The remaining matrix element is $<sp | V_\gamma | ps> = - <ss | V_\gamma |(pp)_{L=0}>/\sqrt{3}$.}\label{PREOME}

\end{table}

All orbital two-body matrix elements of the confining potential $V_{conf}=C r$ required in theses calculation can be obtained analytically. They are

\begin{eqnarray}
<ss | V_{conf} | ss> & = & \sqrt{\frac{2}{\pi}} 2 C \beta \nonumber \\
<sp | V_{conf} | sp> & = & \sqrt{\frac{2}{\pi}} \frac{7 C \beta}{3} \nonumber \\
<sp | V_{conf} | ps> & = & -\sqrt{\frac{2}{\pi}}  \frac{ C \beta}{3} \nonumber \\
<ss | V_{conf} | (pp)_{L=0}> & = & -\sqrt{3} <sp | V_{conf} | ps> \nonumber \\
<(pp)_{L=0} | V_{conf} | (pp)_{L=0}> & = & \sqrt{\frac{2}{\pi}}  \frac{5 C \beta}{2}
\end{eqnarray}
\ 

Finally, the kinetic energy matrix elements can be calculated as above, by writing the relative kinetic energy operator as a two-body operator

\begin{equation}
T = \sum_i \frac{p_{i}^{2}}{2m} - \frac{1}{12} (\sum_i \vec{p}_i)^2 = \sum_{i<j} T_{ij}
\end{equation}
with

\begin{equation}
T_{ij} = \frac{1}{12m} (p_{i}^{2} + p_{j}^{2}) - \frac{1}{6m} \vec{p_i} .\vec{p_j}
\end{equation}
\\

Alternatively we can use an universal formula for the kinetic energy of harmonic oscillator states

\begin{equation}\label{PREKEHO}
KE = \frac{1}{2} [N + \frac{3}{2} (n-1) ] \hbar \omega + \frac{3}{4} \hbar \omega
\end{equation}
\ 

\noindent where $N$ is the number of quanta and $n$ the number of particles. The last term is the kinetic energy of the center of mass.
\\

\section{Hamiltonian diagonalization in the cluster model at $Z=0$}\label{sectionCluster0}
\ 

In Tables \ref{PREBOdiag01} and \ref{PREBOdiag10} we present our results obtained from the diagonalization of the Hamiltonian (\ref{PREham}) in the basis $|1> -\ |4>$ of (\ref{PREbasis}). We use the definition of the effective NN potential as given by the Born-Oppenheimer approximation (\ref{PREborn}) at zero separation between nucleons. All diagonal matrix elements and eigenvalues presented in Tables \ref{PREBOdiag01} and \ref{PREBOdiag10} are given relative to two-nucleon threshold, which means that the quantity $2<N|H|N>=1939$ MeV has always been subtracted. In the second column we present the diagonal matrix elements for all the states listed in the first column. In the third column we present all the eigenvalues obtained from the diagonalization of a $4\times4$ matrix. In the fourth column  the amplitudes of all components of the ground state are given. In agreement with the qualitative results above, one can see that the expectation value of the configuration $|s^4p^2 [42]_O [51]_{FS}>$ given in column 2 is much lower than all the other ones, and in particular it is about 1.4 GeV below  the expectation value of the configuration $|s^6 [6]_O [33]_{FS}>$.
\\

The substantial lowering of the configuration $|s^4p^2 [42]_O [51]_{FS}>$ relative to the other ones implies that this configuration is by far the most important component in the ground state eigenvector. The last column of Tables \ref{PREBOdiag01} and \ref{PREBOdiag10} implies that the probability of this configuration is 98\% both for $SI=10$ and $SI=01$. As a consequence, the lowest eigenvalue is only about 40 MeV lower than the expectation value of the configuration $|3>$.
\\

The main outcome is that $V_{NN}(Z=0)$ is highly repulsive in both $^3S_1$ and $^1S_0$ channels, the height being 0.915 GeV in the former case and 1.453 GeV in the latter one.

\begin{table}[H]
\centering

\begin{tabular}{|l|c|c|c|}
\hline

State & Diag. elem - 2 $m_N$ & Eigenvalues - 2 $m_N$ & Lowest state amplitudes\\

\hline
\hline

$|s^6[6]_O[33]_{FS} >$ & 2.346 & 0.915 & -0.10686\\
\hline
$|s^4p^2[42]_O[33]_{FS} >$ & 2.824 & 1.922 & 0.08922\\
\hline
$|s^4p^2[42]_O[51]_{FS} >$ & 0.942 & 2.956 & -0.98854\\
\hline
$|s^4p^2[42]_O[411]_{FS} >$ & 2.949 & 3.268 & 0.05843\\
\hline

\end{tabular}
\caption{Results of the diagonalization of the Hamiltonian (\ref{PREham}) for $SI=10$ in the GBE Model I. Column 1 - basis states, column 2 - diagonal matrix elements (GeV), column 3 - eigenvalues (GeV) in increasing order for a 4 x 4 matrix, column 4 - components of the lowest state. The results correspond to $\beta$ = 0.437 fm . The diagonal matrix elements  and the eigenvalues are relative to 2 $m_{N}$= 1939 MeV  }\label{PREBOdiag01}

\end{table}

In order to see that it is the GBE interaction which is responsible for the short-range repulsion, it is very instructive  to remove $V_\chi$ from the Hamiltonian (\ref{PREvchi}),  compute the ``nucleon mass" in this case, which turns out to be $m_N=1.633$ GeV at the harmonic oscillator parameter $\beta=0.917$ fm and diagonalize such a Hamiltonian again in the basis (\ref{PREbasis}). In this case the most important configuration is $|s^6 [6]_O [33]_{FS}>$. Subtracting from the lowest eigenvalue the ``two-nucleon energy" $2m_N=2\times1.633$ GeV one obtains $V^{no ~\chi}_{NN}(Z=0)=-0.197$ GeV. This soft attraction comes from the nonphysical color forces related to the pairwise confinement. These forces would not appear if the basis was restricted to the $|s^6>$ state only. If the spatially excited 3q clusters from the $s^4p^2$ configurations were removed the forces  would disappear and we would arrive at $V^{no ~\chi}_{NN}(Z=0)=0.$ Thus it is the GBE interaction which brings about 1 GeV repulsion, consistent with the previous discussion.

\begin{table}[H]
\centering

\begin{tabular}{|l|c|c|c|}
\hline

State & Diag. elem - 2 $m_N$ & Eigenvalues - 2 $m_N$ & Lowest state amplitudes\\

\hline
\hline

$|s^6[6]_O[33]_{FS} >$ & 2.990 & 1.453 & -0.10331\\
\hline
$|s^4p^2[42]_O[33]_{FS} >$ & 3.326 & 2.436 & 0.09371\\
\hline
$|s^4p^2[42]_O[51]_{FS} >$ & 1.486 & 3.557 & -0.98723\\
\hline
$|s^4p^2[42]_O[411]_{FS} >$ & 3.543 & 3.899 & -0.07694\\
\hline

\end{tabular}
\caption{Same as Table \ref{PREBOdiag01} but for $SI=01$.}\label{PREBOdiag10}
\end{table}

The effective repulsion obtained above implies a strong suppression of the $L=0$ relative motion wave function in the nucleon overlap region, as compared to the wave function of two well separated nucleons.
\\

There is another important mechanism producing additional effective repulsion in the NN system, which is related to the symmetry structure of the lowest configuration but not related to its energy relative to the NN threshold. This ``extra" repulsion, related to the ``Pauli forbidden state" \cite{SAI69},  persists if any of the configurations from the $|s^4p^2>$ shell becomes highly dominant \cite{NEU77}. Indeed, the NN phase shift calculated with a pure $[51]_{FS}$ state, which is projected ``by hands" (not dynamically) from the full NN state  in a toy model \cite{OKA84}, shows a behavior typical for  repulsive potentials. As a result the s-wave NN relative motion wave function has an almost energy independent node \cite{NEU75,NEU77}. A similar situation occurs in $^4He - ^4He$ scattering \cite{TAM65}. The only difference between this nuclear case and the NN system is that while in the former a configuration $s^8$ is indeed forbidden by the Pauli principle in eight-nucleon system, the configuration $s^6$ is allowed in a six-quark system, but is highly suppressed by dynamics, as it was discussed above. The existence of a strong effective repulsion, related to the energy balance in the adiabatic approximation, as in our case, suggests, however, that the amplitude of the oscillating NN wave function at short distance will be strongly suppressed. This is equivalent with a ``Pauli forbidden state effect''. In the OGE model this effect is absent because none of the $[42]_O$ states is dominant \cite{OKA84,OBU88,HAR84}.
\\

In the present section we have calculated an adiabatic NN potential at zero separation between nucleons in the framework of a chiral constituent quark model. Diagonalizing a Hamiltonian in a basis consisting of the most important 6q configurations in the nucleon overlap region, we have found a very strong effective repulsion of the order of 1 GeV in both $^3S_1$ and $^1S_0$ NN partial waves. Due to the specific flavor-spin symmetry of the Goldstone boson exchange interaction the configuration $|s^4p^2 [42]_O [51]_{FS}>$ becomes highly dominant at short-range. As a consequence, the projection of the full 6q wave function onto the NN channel should have a node at short-range in both $^3S_1$ and $^1S_0$ partial waves. The amplitude of the oscillation left to the node should be strongly suppressed as compared to the wave function of two well separated nucleons.
\\

Thus, within the chiral constituent quark model one has all the necessary ingredients to understand microscopically the NN interaction. There appears strong effective short-range repulsion from the same part of Goldstone boson exchange which also produces hyperfine splittings in baryon spectroscopy. The long-range attraction in the NN system is auto\-matically implied by the Yukawa part of pion exchange between quarks belonging to different nucleons. With this first encouraging result, it might be worthwhile to perform a more elaborate study of the NN system with the present quark model. This we be done in Chapter \ref{rgmchapter} in the resonating group method.
\\

\section{Six-quark states from molecular orbitals}
\ 

Here we follow closely Ref. \cite{STA87} where the use of molecular orbitals in the construction of six-quark states was originally proposed, instead of commonly used cluster model states as in the previous section. Let us denote by $Z$ the separation coordinate between the centers of the two clusters.  At finite $Z$, in the simplest cluster model basis, each of the six quarks is described by an orbital wave function represented by a Gaussian centered either at $Z/2$ or $-Z/2$. These non-orthogonal states are denoted by $R$ (right) and $L$ (left) respectively

\begin{equation}
R(\vec{r}) = \psi \left(\vec{r} - \frac{\vec{Z}}{2}\right) \hspace{8mm} \mbox{and}
\hspace{8mm} L(\vec{r}) =
\psi \left(\vec{r} + \frac{\vec{Z}}{2}\right).
\end{equation}
\ 

Alternatively, in a molecular basis we consider the two lowest states, $\sigma$ which is even and $\pi$ which is odd. These could be either the solutions of a static, axially and reflectionally symmetric independent particle model Hamiltonian (see for example \cite{KOE94}) or, as for the present purpose, can be constructed from $R$ and $L$ states.
\\

First we introduce pseudo-right and pseudo-left states $r$ and $l$ starting from the molecular orbitals $\sigma$ and $\pi$ as

\begin{equation} \label{PREpseudorl}
\left[ \begin{array}{c} r\\ l \end{array} \right] = 2^{-1/2}\, ( \sigma \pm \pi ) \hspace{5mm} {\rm{for\ all\ }}Z,
\end{equation}
\ 

\noindent where
\begin{eqnarray}
& <r|r> = <l|l> = 1 ,\,\, <r|l> = 0 .&
\end{eqnarray}
\ 

On the other hand, starting from the cluster model states, one can construct good parity, orthonormal states for all $Z$ by setting

\begin{equation}
\left[ \begin{array}{c} \sigma\\ \pi\end{array} \right] = [2( 1 \pm <R|L>)]^{-1/2} ( R \pm L ) ,
\end{equation}
\ 

\noindent which, introduced in Eq. (\ref{PREpseudorl}) gives

\begin{equation}
\left[ \begin{array}{c} r\\ l \end{array} \right] = \frac{1}{2} \left[ \frac{R+L}{(1+<R|L>)^{1/2}} \pm \frac{R-L}{(1-<R|L>)^{1/2}} \right].
\end{equation}
\ 

At $Z \rightarrow 0$ one has $\sigma \rightarrow s$ and $\pi \rightarrow p$ (with $m = 0,\pm1$), so that

\begin{equation} \label{PREZgoes0}
\left[ \begin{array}{c} r\\ l \end{array} \right] = 2^{1/2} (s \pm p) ,
\end{equation}
\ 

\noindent and at $Z \rightarrow \infty$ one has $r \rightarrow R$ and $l \rightarrow L$.
\\

From $(r,l)$ as well as from $(\sigma,\pi)$ orbitals one can construct six-quark states of required permutation symmetry. For the $S_6$ symmetries relevant for the NN problem the transformations between six-quark states expressed in terms of $(r,l)$ and $(\sigma,\pi)$ states are given in Table I of Ref. \cite{STA87}. This table shows that in the limit $Z \rightarrow 0$ six-quark states obtained from molecular orbitals contain configurations of type $s^np^{6-n}$ with $n = 0,1,...,6$. For example the $[6]_O$ state contains $s^6$, $s^6p^4$, $s^2p^4$ and $p^6$ configurations and the $[42]_O$ state associated to the $S$-channel contains $s^4p^2$ and $s^2p^4$ configurations. This is in contrast to the cluster model basis where $[6]_O$ contains only $s^6$ and $[42]_O$ only $s^4p^2$ configurations \cite{HAR81}. This suggests that the six-quark basis states constructed from molecular orbitals form a richer basis without introducing more single particle states. Here we examine its role in lowering the ground state energy of a six-quark system described by the Hamiltonian introduced in the next section.
\\

Using Table I of Ref. \cite{STA87} we find that the six-quark basis states needed for the $^3S_1$ or $^1S_0$ channels are

\begin{eqnarray}
\left.{\left|{33{\left[{6}\right]}_{O}{\left[{33}\right]}_{FS}}\right.}
\right\rangle\
& = & \frac{1}{4}\ \left.{\left| \left[{\sqrt {5}\ \left({{s}^{6}\ -\
{p}^{6}}\right)\ -\
\sqrt {3}\ \left({{s}^{4}{p}^{2}\ -\ {s}^{2}{p}^{4}}\right)}\right]\
{{\left[{6}\right]}_{O}{\left[{33}\right]}_{FS}}\right.}\right\rangle, \nonumber \\
\left.{\left|{33{\left[{42}\right]}_{O}{\left[{33}\right]}_{FS}}\right.}
\right\rangle\
& = & \sqrt {\frac{1}{2}}\ \left.{\left|{ \left[{{s}^{4}{p}^{2}\ -\
{s}^{2}{p}^{4}}\right]
{\left[{42}\right]}_{O}{\left[{33}\right]}_{FS}}\right.}\right\rangle, \nonumber \\
\left.{\left|{33{\left[{42}\right]}_{O}{\left[{51}\right]}_{FS}}\right.}
\right\rangle\
& = & \sqrt {\frac{1}{2}}\ \left.{\left|{ \left[{{s}^{4}{p}^{2}\ -\
{s}^{2}{p}^{4}}\right]
{\left[{42}\right]}_{O}{\left[{51}\right]}_{FS}}\right.}\right\rangle, \nonumber \\
\left.{\left|{33{\left[{42}\right]}_{O}{\left[{411}\right]}_{FS}}\right.}
\right\rangle\
& = & \sqrt {\frac{1}{2}}\ \left.{\left|{ \left[{{s}^{4}{p}^{2}\ -\
{s}^{2}{p}^{4}}\right]
{\left[{42}\right]}_{O}{\left[{411}\right]}_{FS}}\right.}\right\rangle, \nonumber \\
\left.{\left|{{42}^{+}{\left[{6}\right]}_{O}{\left[{33}\right]}_{FS}}\right.}
\right\rangle\
& = & {\frac{1}{4}} \sqrt {{\frac{1}{2}}}\ \left.{\left|{ \left[{{\sqrt
{15}}^{}\left({{s}^{6}\ +\ {p}^{6}}\right)\ -\ \left({{s}^{4}{p}^{2}\ +\
{s}^{2}{p}^{4}}\right)}\right]{\left[{6}\right]}_{O}{\left[{33}\right]}_{FS}}
\right.}\right\rangle, \nonumber \\
\left.{\left|{{42}^{+}{\left[{42}\right]}_{O}{\left[{33}\right]}_{FS}}\right.}
\right\rangle\
& = & \sqrt {\frac{1}{2}}\ \left.{\left|{ \left[{\left.{{s}^{4}
{p}^{2}}\right.\ +\
{s}^{2}{p}^{4}}\right]{\left[{42}\right]}_{O}{\left[{33}\right]}_{FS}}\right.}
\right\rangle, \nonumber \\
\left.{\left|{{42}^{+}{\left[{42}\right]}_{O}{\left[{51}\right]}_{FS}}\right.}
\right\rangle\
& = & \sqrt {\frac{1}{2}}\ \left.{\left|{ \left[{\left.{{s}^{4}
{p}^{2}}\right.\ +\
{s}^{2}{p}^{4}}\right]{\left[{42}\right]}_{O}{\left[{51}\right]}_{FS}}\right.}
\right\rangle, \nonumber \\
\left.{\left|{{42}^{+}{\left[{42}\right]}_{O}{\left[{411}\right]}_{FS}}\right.}
\right\rangle\
& = & \sqrt {\frac{1}{2}}\ \left.{\left|{ \left[{\left.{{s}^{4}
{p}^{2}}\right.\ +\
{s}^{2}{p}^{4}}\right]{\left[{42}\right]}_{O}{\left[{411}\right]}_{FS}}\right.}
\right\rangle, \nonumber \\
\left.{\left|{{51}^{+}{\left[{6}\right]}_{O}{\left[{33}\right]}_{FS}}\right.}
\right\rangle\
& = & \frac{1}{4}\ \left.{\left|{ \left[{\sqrt {3}\ \left.{\left({{s}^{6}\ - \
{p}^{6}}\right)}\right.\ +\ \sqrt {5}\ \left({{s}^{4}{p}^{2}\ -\
{s}^{2}{p}^{4}}\right)}\right]{\left[{6}\right]}_{O}{\left[{33}\right]}_{FS}}
\right.}\right\rangle,\nonumber \\
\label{PREorbitbasis}
\end{eqnarray}
where the notation $33$ and $mn^+$ in the left hand side of each equality above means $r^3\ell^3$ and $r^m\ell^n+r^n\ell^m$, respectively, as in Ref. \cite{STA87} (see also discussion below). Each wave function contains an orbital part ($O$) and a flavor-spin part ($FS$) which combined with the color singlet $[222]_C$ state gives rise to a totally antisymmetric state. We restricted the flavor-spin states to $[33]_{FS}$, $[51]_{FS}$ and $[411]_{FS}$ as in (\ref{PREbasis}) for the cluster model.
\\

Besides being poorer in $s^np^{6-n}$ configurations, as explained above, the number of basis states is smaller in the cluster model although we deal with the same $[f]_O$ and $[f]_{FS}$ symmetries and the same harmonic oscillator states $s$ and $p$ in both cases. This is due to the existence of three-quark clusters only in the cluster model states, while the molecular basis also allows configurations with five quarks to the left and one to the right, or {\it vice versa}, or four quarks to the left and two to the right or {\it vice versa}. At large separations these states act as ``hidden color" states but at zero separation they bring a significant contribution, as we shall see below.
\\

The matrix elements of the Hamiltonian (\ref{PREham}) are calculated in the basis (\ref{PREorbitbasis}) by using the fractional parentage technique. Another program based on Mathematica \cite{WOL96} has been created for this purpose. In this way every six-body matrix element reduces again to a linear combination of two-body matrix elements of either symmetric or antisymmetric states. Then the linear combinations contain orbital two-body matrix elements of the type $\left\langle{ss\left|{{V}_{\gamma}}\right|ss}\right\rangle$, $\left\langle{ss\left|{{V}_{\gamma}}\right|pp}\right\rangle$, $\left\langle{sp\left|{{V}_{\gamma}}\right|sp}\right\rangle$, $\left\langle{sp\left|{{V}_{\gamma}}\right|ps}\right\rangle$ and $\left\langle{pp\left|{{V}_{\gamma}}\right|pp}\right\rangle_{L\ =\ 0}$ where $\gamma = \pi$, $\eta$ or $\eta '$, as in (\ref{PREpoint}) or (\ref{PREmodelII}). Here we study the case $Z = 0$ for which harmonic oscillator states $s$ and $p$ are given by (\ref{PREsstate}) and (\ref{PREpstate}), respectively.
\\

The Hamiltonian (\ref{PREham}) in the six-quark basis (\ref{PREorbitbasis}) is then diagonalized and we calculate the NN interaction potential in the Born-Oppenheimer approximation given by Eq. (\ref{PREborn}).
\\

Here we study the case $Z = 0$, relevant for short separation distances between the nu\-cleons. Later in this chapter we will extend the calculations to any $Z$.
\\

In Tables \ref{PREBOstates01} and \ref{PREBOstates10} we present our results for $SI = 10$ and $01$ respectively, obtained from the diagonalization of $H$ with the parametrization of Model I \cite{BAR98}. From the diagonal matrix elements $H_{ii}$ as well as from the eigenvalues, the quantity $2m_N$ = 1939 MeV has been subtracted according to Eq. (\ref{PREborn}). Here $m_N$ is the nucleon mass calculated also variationally, with an $s^3$ configuration, as mentioned in the previous section. This value is obtained for a harmonic oscillator parameter $\beta$ = 0.437 fm \cite{PEP97}. For sake of comparison with Ref. \cite{STA97} we take same value of $\beta$ for the six-quark system as well.

\begin{table}[H]
\centering

\begin{tabular}{|l|c|c|c|}
\hline
State &$H_{ii}$ - 2 $m_N\,$ & Eigenvalues - 2 $m_N$ & Lowest state amplitudes\\

\hline
\hline

$|33[6]_O[33]_{FS} >$ & 2.616 & 0.718 & -0.04571\\
\hline
$|33[42]_O[33]_{FS} >$ & 3.778 & 1.667 & 0.02479\\
\hline
$|33[42]_O[51]_{FS} >$ & 1.615 & 1.784 & -0.31762\\
\hline
$|33[42]_O[411]_{FS} >$ & 2.797 & 2.309 & 0.04274\\
\hline
$|42^+[6]_O[33]_{FS} >$ & 3.062 & 2.742 & -0.07988\\
\hline
$|42^+[42]_O[33]_{FS} >$ & 2.433 & 2.784 & 0.12930\\
\hline
$|42^+[42]_O[51]_{FS} >$ & 0.850 & 3.500 & -0.93336\\
\hline
$|42^+[42]_O[411]_{FS} >$ & 3.665 & 3.752 & 0.00145\\
\hline
$|51^+[6]_O[33]_{FS} >$ & 2.910 & 4.470 & -0.01789\\
\hline

\end{tabular}
\caption{Results of the diagonalization of the Hamiltonian (\ref{PREham}) in the GBE Model I for $SI = 10$. Column 1 - basis states, column 2 - diagonal matrix elements (GeV), column 3 - eigenvalues (GeV) in increasing order, column 4 - lowest state amplitudes of components given in column 1. The results correspond to $\beta$ = 0.437 fm . The diagonal matrix elements $H_{ii}$ and the eigenvalues are relative to 2 $m_{N}$ = 1939 MeV (see text).}\label{PREBOstates01}

\end{table}

In both $SI = 10$ and $01$ cases the effect of using molecular orbitals is rather remarkable in lowering the ground state energy as compared to the cluster model value obtained in the four dimensional basis (\ref{PREBOdiag01} - \ref{PREBOdiag10}). Accordingly, the height of the repulsive core in the $^3S_1$ channel is reduced from 915 MeV in the cluster model basis to 718 MeV in the molecular orbital basis. In the $^1S_0$ channel the reduction is from 1453 MeV to 1083 MeV. Thus the molecular orbital basis is much better, inasmuch as the same two single particle states, $s$ and $p$, are used in both bases. 

\begin{table}[H]
\centering

\begin{tabular}{|l|c|c|c|}
\hline

State &$H_{ii}$ - 2 $m_N$\,& Eigenvalues - 2 $m_N$ & Lowest state amplitudes\\

\hline
\hline

$|33[6]_O[33]_{FS} >$ & 3.300 & 1.083 & -0.02976\\
\hline
$|33[42]_O[33]_{FS} >$ & 4.367 & 2.252 & 0.01846\\
\hline
$|33[42]_O[51]_{FS} >$ & 2.278 & 2.279 & -0.20460\\
\hline
$|33[42]_O[411]_{FS} >$ & 3.191 & 2.945 & -0.04729\\
\hline
$|42^+[6]_O[33]_{FS} >$ & 3.655 & 3.198 & -0.07215\\
\hline
$|42^+[42]_O[33]_{FS} >$ & 2.796 & 3.317 & 0.13207\\
\hline
$|42^+[42]_O[51]_{FS} >$ & 1.167 & 4.058 & -0.96531\\
\hline
$|42^+[42]_O[411]_{FS} >$ & 4.405 & 4.459 & -0.00081\\
\hline
$|51^+[6]_O[33]_{FS} >$ & 3.501 & 5.070 & -0.01416\\
\hline

\end{tabular}
\caption{Same as Table \ref{PREBOstates01} but for $SI = 01$.}\label{PREBOstates10}

\end{table}

The previous study performed in a cluster model basis indicated that the dominant configuration is associated to the symmetry $[42]_O[51]_{FS}$. It is the case here too and one can see from Tables \ref{PREBOstates01} and \ref{PREBOstates10} that the diagonal matrix element $H_{ii}$ of the state $|42^+[42]_O[51]_{FS} >$ is far the lowest one, so that this state is much more favored than $|33[42]_O[51]_{FS} >$ . As explained above, such a state represents a configuration with two quarks on the left and four on the right around the symmetry center. At $Z \rightarrow \infty$ its energy becomes infinite {\it i. e.} this state behaves as a hidden color state (see {\it e. g.} Ref. \cite{HAR81}) and it decouples from the ground state. But at $Z = 0$ it is the dominant component of the lowest state  with a probability of 87 \% for $SI=10$ and 93 \% for $SI=01$. The next important state is $|33[42]_O[51]_{FS} >$ with a probability of 10 \% for $SI=10$ and 4 \% for $SI=01$. The presence of this state will become more and more important with increasing $Z$. Asymptotically this state corresponds to a cluster model state with three quarks on the left and three on the right of the symmetry centre.

\begin{table}[H]
\centering

\begin{tabular}{|c|c|c|}
\hline

Energy &Cluster model &Molecular orbital\\
& $|s^4p^2[42]_O[51]_{FS}\rangle$ & $|42^+[42]_O[51]_{FS}\rangle$\\

\hline
\hline

$KE$ & 2.840 & 3.139\\
\hline
$V_{conf}$ & 0.385 & 0.364\\
\hline
$V_{\chi}$ & -2.384 & -2.754\\
\hline
$E$ & 0.841 & 0.749\\
\hline

\end{tabular}
\caption{Parts of the energy expectation values (GeV) of the dominant 6q state in the cluster model and the molecular orbital basis for $SI = 10$ in the GBE Model I.}\label{PREBOenergy01}

\end{table}

To have a better understanding of the lowering of the six-quark energy we present in Tables \ref{PREBOenergy01} and \ref{PREBOenergy10} the separate contribution of the kinetic energy $KE$, of the confinement $V_{conf}$ and of the GBE interaction $V_{\chi}$ to the dominant state in the cluster model $|s^4p^2[42]_O[51]_{FS}\rangle$ result and the dominant state in the molecular basis case respectively. Table  \ref{PREBOenergy01} corresponds to the $^3S_1$ channel and Table  \ref{PREBOenergy10} to the $^1S_0$ channel. We can see that $V_{conf}$ does not change much in passing from the cluster model to the molecular orbital basis. The kinetic energy $KE$ is higher in the molecular orbital basis which is natural because the $s^2p^4$ and $p^6$ configurations contribute with higher energies than $s^6$ and $s^4p^2$. Contrary, the contribution of the GBE interaction $V_{\chi}$ is lowered by several hundreds of MeV in both channels, so that $E = KE + V_{conf} + V_{\chi}$ is substantially lowered in the molecular orbital basis. This shows that the GBE interaction is more effective in the molecular orbital basis than in the cluster model basis. Note that $E$ differs from the value of the previous diagonal matrix elements by the additional quantity $6m - 2m_N$, where $m = m_u = m_d$.

\begin{table}[H]
\centering

\begin{tabular}{|c|c|c|}
\hline

Energy &Cluster model &Molecular orbital\\
& $|s^4p^2[42]_O[51]_{FS}\rangle$ & $|42^+[42]_O[51]_{FS}\rangle$\\

\hline
\hline

$KE$ & 2.840 & 3.139\\
\hline
$V_{conf}$ & 0.385 & 0.364\\
\hline
$V_{\chi}$ & -1.840 & -2.437\\
\hline
$E$ & 1.385 & 1.066\\
\hline

\end{tabular}
\caption{Same as Table \ref{PREBOenergy01} but for $SI=01$}\label{PREBOenergy10}

\end{table}

The practically identical confinement energy in both bases shows that the amount of Van der Waals forces, as discussed in \cite{STA97}, remains the same. However, the soft attraction brought in by the Van der Waals forces does not play an important role at short distances and in the adiabatic approximation it disappears at long distance as we will see in the next sections.
\\

For both $SI = 10$ and $01$ sectors we also searched for the minimum of $\langle H\rangle_{Z=0}$ as a function of the oscillator parameter $\beta$ in the Model I parametrization. For $SI = 10$ the minimum of 572 MeV has been reached at $\beta$ = 0.547 fm. For $SI = 01$ the minimum of 715 MeV was obtained at $\beta$ = 0.608 fm. These values are larger than the value of $\beta$ = 0.437 fm associated to the nucleon, which is quite natural because a six-quark system at equilibrium is a more extended object.

\begin{table}[H]
\centering

\begin{tabular}{|l|c|c|c|}
\hline
State &$H_{ii}$ - 2 $m_N\,$ & Eigenvalues - 2 $m_N$ & Lowest state amplitudes\\

\hline
\hline

$|33[6]_O[33]_{FS} >$ & 2.181 & 1.326 & -0.55253\\
\hline
$|33[42]_O[33]_{FS} >$ & 3.399 & 1.746 & 0.20484\\
\hline
$|33[42]_O[51]_{FS} >$ & 2.486 & 2.052 & -0.17413\\
\hline
$|33[42]_O[411]_{FS} >$ & 2.492 & 2.307 & 0.04526\\
\hline
$|42^+[6]_O[33]_{FS} >$ & 2.662 & 2.498 & -0.49115\\
\hline
$|42^+[42]_O[33]_{FS} >$ & 2.270 & 2.692 & 0.44882\\
\hline
$|42^+[42]_O[51]_{FS} >$ & 1.894 & 3.166 & -0.42124\\
\hline
$|42^+[42]_O[411]_{FS} >$ & 3.261 & 3.366 & 0.01596\\
\hline
$|51^+[6]_O[33]_{FS} >$ & 2.523 & 4.0153 & 0.00466\\
\hline

\end{tabular}
\caption{Results of the diagonalization of the Hamiltonian (\ref{PREham}) in the GBE Model II for $SI = 10$. Column 1 - basis states, column 2 - diagonal matrix elements (GeV), column 3 - eigenvalues (GeV) in increasing order, column 4 - lowest state amplitudes of components given in column 1. The results correspond to $\beta$ = 0.437 fm . The diagonal matrix elements $H_{ii}$ and the eigenvalues are relative to 2 $m_{N}$ = 1939 MeV (see text).}\label{PREBOstates01II}

\end{table}

Now, let us look at the results in the Model II. If we use the same $s^3$ description for $N$ with $|s>$ given by (\ref{PREsstate}), the variational parameter for the nucleon wave function is still $\beta=0.437$ fm.  Most of the results are very similar to those of Model I as we can see from Tables \ref{PREBOstates01II}-\ref{PREBOdiag10II}. Although $H_{ii}-2m_N$ associated with the $|42^+[42]_O[51]_{FS}>$ state remains the lowest one, this state is not anymore dominant. Both $|[6]_O[33]_{FS} >$ and $|[42]_O[33]_{FS} >$ states have significant contribution to the lowest eigenvector as seen in the last column of Table \ref{PREBOstates01II} for  $SI = 10$.

\begin{table}[H]
\centering

\begin{tabular}{|l|c|c|c|}
\hline

State &$H_{ii}$ - 2 $m_N$\,& Eigenvalues - 2 $m_N$ & Lowest state amplitudes\\

\hline
\hline

$|33[6]_O[33]_{FS} >$ & 2.407 & 1.392 & 0.43647\\
\hline
$|33[42]_O[33]_{FS} >$ & 3.548 & 1.847 & -0.19427\\
\hline
$|33[42]_O[51]_{FS} >$ & 2.690 & 2.154 & 0.17021\\
\hline
$|33[42]_O[411]_{FS} >$ & 2.511 & 2.357 & 0.00124\\
\hline
$|42^+[6]_O[33]_{FS} >$ & 2.849 & 2.561 & 0.46553\\
\hline
$|42^+[42]_O[33]_{FS} >$ & 2.286 & 2.905 & -0.51128\\
\hline
$|42^+[42]_O[51]_{FS} >$ & 1.895 & 3.322 & 0.51210\\
\hline
$|42^+[42]_O[411]_{FS} >$ & 3.483 & 3.580 & 0.002163\\
\hline
$|51^+[6]_O[33]_{FS} >$ & 2.684 & 4.238 & 0.04890\\
\hline

\end{tabular}
\caption{Same as Table \ref{PREBOstates01II} but for $SI = 01$.}\label{PREBOstates10II}

\end{table}

For comparison with Model I, we also show the diagonalization results in the cluster model basis in Tables \ref{PREBOdiag01II} and \ref{PREBOdiag10II} for $SI = 10$ and $SI = 01$, respectively. In this basis the lowest eigenvalue is higher by about 600 MeV for $SI = 10$ as compared to that in the Model I. Also the state $|[6]_O[33]_{FS} >$ appears with the largest probability, both in $SI = 10$ and $SI = 01$, contrary to the situation of Model I where $|[42]_O[51]_{FS} >$ is far the most dominant configuration.

\begin{table}[H]
\centering

\begin{tabular}{|l|c|c|c|}
\hline

State & Diag. elem - 2 $m_N$ & Eigenvalues - 2 $m_N$ & Lowest state amplitudes\\

\hline
\hline

$|s^6[6]_O[33]_{FS} >$ & 1.961 & 1.515 & 0.74272\\
\hline
$|s^4p^2[42]_O[33]_{FS} >$ & 2.553 & 2.022 & -0.47192\\
\hline
$|s^4p^2[42]_O[51]_{FS} >$ & 1.917 & 2.600 & 0.47073\\
\hline
$|s^4p^2[42]_O[411]_{FS} >$ & 2.596 & 2.890 & -0.06385\\
\hline

\end{tabular}
\caption{Results of the diagonalization of the Hamiltonian (\ref{PREham}) in the GBE Model II for $SI = 10$. Column 1 - basis states, column 2 - diagonal matrix elements (GeV), column 3 - eigenvalues (GeV) in increasing order for a 4 x 4 matrix, column 4 - components of the lowest state. The results correspond to $\beta$ = 0.437 fm . The diagonal matrix elements  and the eigenvalues are relative to 2 $m_{N}$= 1939 MeV  }\label{PREBOdiag01II}

\end{table}

\begin{table}[H]
\centering

\begin{tabular}{|l|c|c|c|}
\hline

State & Diag. elem - 2 $m_N$ & Eigenvalues - 2 $m_N$ & Lowest state amplitudes\\

\hline
\hline

$|s^6[6]_O[33]_{FS} >$ & 2.100 & 1.595 & 0.72712\\
\hline
$|s^4p^2[42]_O[33]_{FS} >$ & 2.626 & 2.082 & -0.50000\\
\hline
$|s^4p^2[42]_O[51]_{FS} >$ & 1.998 & 2.729 & 0.47036\\
\hline
$|s^4p^2[42]_O[411]_{FS} >$ & 2.705 & 3.022 & 0.00837\\
\hline

\end{tabular}
\caption{Same as Table \ref{PREBOdiag01II} but for $SI = 01$.}\label{PREBOdiag10II}

\end{table}

It is interesting to look also at the details of the most important states. The different values of the matrix elements are given in Tables \ref{PREBOenergy01IIM}-\ref{PREBOenergy10IIC}. Note that the values of $V_{Conf}$ given here do not include the parameter $V_0$ because it does not contribute in the adiabatic approach.

\begin{table}[H]
\centering

\begin{tabular}{|c|c|c|c|c|}
\hline

 & $|33[6]_O[33]_{FS}\rangle$ & $|42^+[6]_O[33]_{FS}\rangle$ & $|42^+[42]_O[33]_{FS}\rangle$ & $|42^+[42]_O[51]_{FS}\rangle$ \\

\hline
\hline

$KE$ & 2.541 & 2.890 & 3.139 & 3.139\\
\hline
$V_{conf}$ & 0.625 & 0.693 & 0.619 & 0.591\\
\hline
$V_{\chi}$ & -0.403 & -0.339 & -0.906 & -1.255\\
\hline
$E$ & 2.181 & 2.662 & 2.270 & 1.894 \\
\hline

\end{tabular}
\caption{Same as Table \ref{PREBOenergy01} for the molecular basis but in the GBE Model II.}\label{PREBOenergy01IIM}

\end{table}

\begin{table}[H]
\centering

\begin{tabular}{|c|c|c|c|c|}
\hline

 & $|33[6]_O[33]_{FS}\rangle$ & $|42^+[6]_O[33]_{FS}\rangle$ & $|42^+[42]_O[33]_{FS}\rangle$ & $|42^+[42]_O[51]_{FS}\rangle$ \\

\hline
\hline

$KE$ & 2.541 & 2.890 & 3.139 & 3.139\\
\hline
$V_{conf}$ & 0.625 & 0.693 & 0.619 & 0.591\\
\hline
$V_{\chi}$ & -0.177 & -0.152 & -0.889 & -1.253\\
\hline
$E$ & 2.407 & 2.849 & 2.288 & 1.895 \\
\hline

\end{tabular}
\caption{Same as Table \ref{PREBOenergy10} for the molecular basis but in the GBE Model II.}\label{PREBOenergy10IIM}

\end{table}

\begin{table}[H]
\centering

\begin{tabular}{|c|c|c|c|}
\hline

 & $|s^6[6]_O[33]_{FS}\rangle$ & $|s^4p^2[42]_O[33]_{FS}\rangle$ & $|s^4p^2[42]_O[51]_{FS}\rangle$ \\

\hline
\hline

$KE$ & 2.242 & 2.840 & 2.840\\
\hline
$V_{conf}$ & 0.634 & 0.681 & 0.625\\
\hline
$V_{\chi}$ & -0.334 & -0.386 & -0.967\\
\hline
$E$ & 1.961 & 2.553 & 1.917\\
\hline

\end{tabular}
\caption{Same as Table \ref{PREBOenergy01} for the cluster model but in the GBE Model II.}\label{PREBOenergy01IIC}

\end{table}

\begin{table}[H]
\centering

\begin{tabular}{|c|c|c|c|}
\hline

 & $|s^6[6]_O[33]_{FS}\rangle$ & $|s^4p^2[42]_O[33]_{FS}\rangle$ & $|s^4p^2[42]_O[51]_{FS}\rangle$ \\

\hline
\hline

$KE$ & 2.242 & 2.840 & 2.840\\
\hline
$V_{conf}$ & 0.634 & 0.681 & 0.625\\
\hline
$V_{\chi}$ & -0.195 & -0.314 & -0.886\\
\hline
$E$ & 2.100 & 2.626 & 1.998\\
\hline

\end{tabular}
\caption{Same as Table \ref{PREBOenergy10} for the cluster model but in the GBE Model II.}\label{PREBOenergy10IIC}

\end{table}

In this section we have calculated the NN interaction potential at zero separation distance between nucleons by treating NN as a six-quark system in a constituent quark model where the quarks interact via Goldstone boson (pseudoscalar meson) exchange. The orbital part of the six-quark states was constructed from molecular orbitals and the commonly used cluster model single particle states. The molecular orbitals posses the proper axially and reflec\-tionally symmetries and are thus physically more adequate than the cluster model states. Due to their orthogonality property they are also technically more convenient. Here we constructed molecular orbitals from harmonic oscillator $s$ and $p$ states. Such molecular orbitals are a very good approximation \cite{ROB87} to the exact eigenstates of a "two-center" oscillator, frequently used
in nuclear physics in the study of the nucleus-nucleus potential. The problem of calculating an NN potential is similar in many ways.
\\

We have shown that the upper bound of the ground state energy, and hence the height of the repulsive core in the NN potential, is lowered by about 200 MeV in the $^3S_1$ channel and by about 400 MeV in the $^1S_0$ channel with respect to the cluster model results. Hence using molecular orbitals is more efficient than working with a cluster model basis. A repulsive core of several hundred MeV is still present in both channels. Note also that the configurations $s^2p^4$ or $p^6$ introduced through the molecular orbitals might have an influence on the momentum distribution of the NN system as was discussed, for example, in \cite{KOE95} within the chromodielectric model.
\\

The following step will be to calculate the NN potential at $Z \neq 0$. The Yukawa potential tail in Eq. (\ref{PREpoint}) is expected to bring the required long-range attraction. An extra middle-range attraction will be also introduced. It can be considered as due to two correlated or uncorrelated pion exchanges.
\\

\section{NN interaction in the adiabatic approximation at any $Z$}
\ 

In this section we diagonalize the Hamiltonian \ref{PREham} in the six-quark cluster model basis and in the six-quark molecular orbital basis for values of the separation distance $Z$ up to 2.5 fm. Using in each case the lowest eigenvalue, denoted by $\langle H\rangle_Z$ we again define the NN interaction potential in the adiabatic (Born-Oppenheimer) approximation as

\begin{equation}
V_{NN}\left(Z\right) = \langle H\rangle_Z - 2m_N - K_{rel}
\label{PREadia}
\end{equation}
\ 

This is the same as the definition (\ref{PREborn}) but with $K_{rel}$ subtracted. The quantity $K_{rel}$ represents the relative kinetic energy of two 3q clusters separated at infinity

\begin{equation}
{K}_{rel}\ =\ {\frac{{3\hbar }^{2}}{4{m\beta }^{2}}}\ 
\end{equation}
\ 

\noindent where $m$ above and in the following designates the mass of the $u$ or $d$ quarks which is taken equal to 340 MeV. For the value of $\beta$ of both Model I and Model II this gives $K_{rel} = 0.448$ GeV.
\\

As above, $m_N$ is the nucleon mass obtained as a variational $s^3$ solution for a 3q system described by the Hamiltonian where the wave function has the form $\phi \propto \exp\left[{-\left( \rho^2+\lambda^2 \right)/2\beta^2}\right]$ with $\rho =\left(\vec{r}_1-\vec{r}_2\right)/\sqrt{2}$ and $\vec{\lambda} =\left(\vec{r}_1+\vec{r}_2-2\vec{r}_3\right)/\sqrt{6}$. The same value of $\beta$ as that obtained from minimizing  $m_N = \langle H\rangle_{3q}$ is also used for the 6q system. This is equivalent with imposing the ``stability condition" (\ref{PREstab}) which is of crucial importance in resonating group method calculations \cite{OKA84,SHI89}.
\ 

\begin{figure}[H]
\begin{center}
\includegraphics[width=13cm]{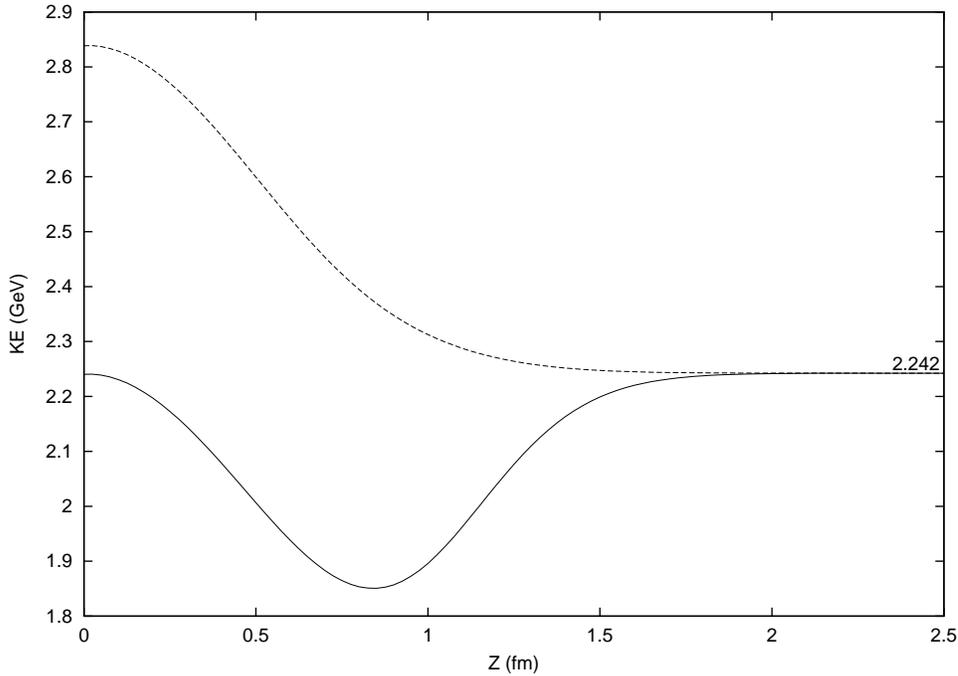}
\end{center}
\caption{\label{PREcluKE} The expectation value of the kinetic energy $\langle KE\rangle$ as a function of the separation distance $Z$ between two 3q clusters in the cluster model basis. The asymptotic value of 2.242 GeV , given by Eq. (\ref{PREKref}) is indicated. The full line corresponds to $| [6]_O \rangle$ and the dashed line to $| [42]_O \rangle $ states. (Model I)}
\end{figure}

\subsection{Cluster model basis}
\ 

First we show contributions to $V_{NN}$ coming from different parts of the Hamiltonian. In Fig. \ref{PREcluKE} we present the expectation value of the kinetic energy $\langle KE\rangle$ as a function of $Z$. One can see that for the state $\left|{\left.{{R}^{3}{L}^{3}{\left[{42}\right]}_{O}}\right\rangle}\right.$ it decreases with $Z$ but for the state $\left|{\left.{{R}^{3}{L}^{3}{\left[{6}\right]}_{O}}\right\rangle}\right.$ it first reaches a minimum at around $Z \cong$ 0.85 fm and then it tends to an asymptotic value equal to its value at the origin due to its $s^6$ structure. This value is

\begin{equation}
\langle KE\rangle_{Z=0} = \langle KE\rangle_{Z=\infty} = \frac{15}{4}~\hbar
\omega
\label{PREKref}
\end{equation}
\ 

\noindent where $\hbar \omega = \hbar^2/m\beta^2$. Actually this also the value for 6q states with $N=0$, as indicated by Eq. (\ref{PREKEHO}) minus $\frac{3}{4}\ \hbar \omega$ which is the center of mass motion. It is also the asymptotic value for all states, irrespective of their symmetry.
\\

\begin{figure}[H]
\begin{center}
\includegraphics[width=13cm]{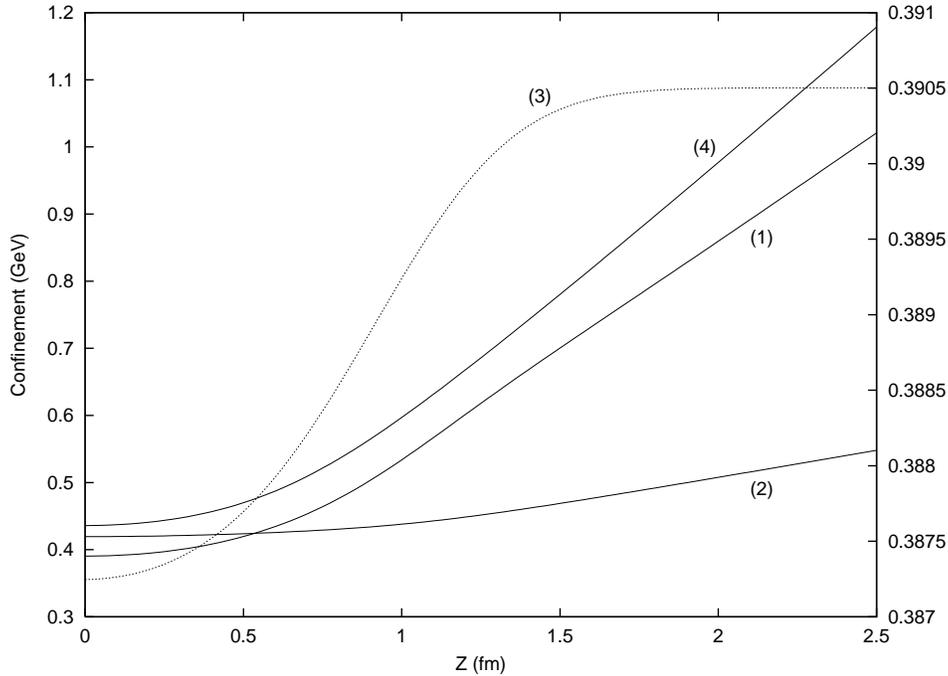}
\end{center}
\caption{\label{PREcluCONF} The cluster model basis. The expectation value of $V_{conf}$ of Eq. (\ref{PREconf}). The corresponding states are : (1) - $| [6]_O [33]_{FS} \rangle$, (2) - $| [42]_O [33]_{FS} \rangle$, (3) - $| [42]_O [51]_{FS} \rangle$, (4) - $| [411]_O [51]_{FS} \rangle$. Note that for curve (3) the scale is on the right hand side vertical line. (Model I)} 
\end{figure}

The diagonal matrix elements of the confinement potential are presented in Fig. \ref{PREcluCONF}. Beyond $Z >$ 1.5 fm one can notice a linear increase except for the $\left|{\left.{{R}^{3}{L}^{3}{\left[{42}\right]}_{O} {\left[{51}\right]}_{FS}}\right\rangle}\right.$ state where it reaches a plateau of 0.3905 GeV.
\\

In the following we present in detail the contributions to $V_{NN}(Z)$ of $V_{conf}$ and $V_{\chi}$ corres\-ponding to the GBE Model I. For Model II the results are qualitatively similar. We shall return to the GBE Model II at the end of this section.
\ \\

\begin{figure}[H]
\begin{center}
\includegraphics[width=13cm]{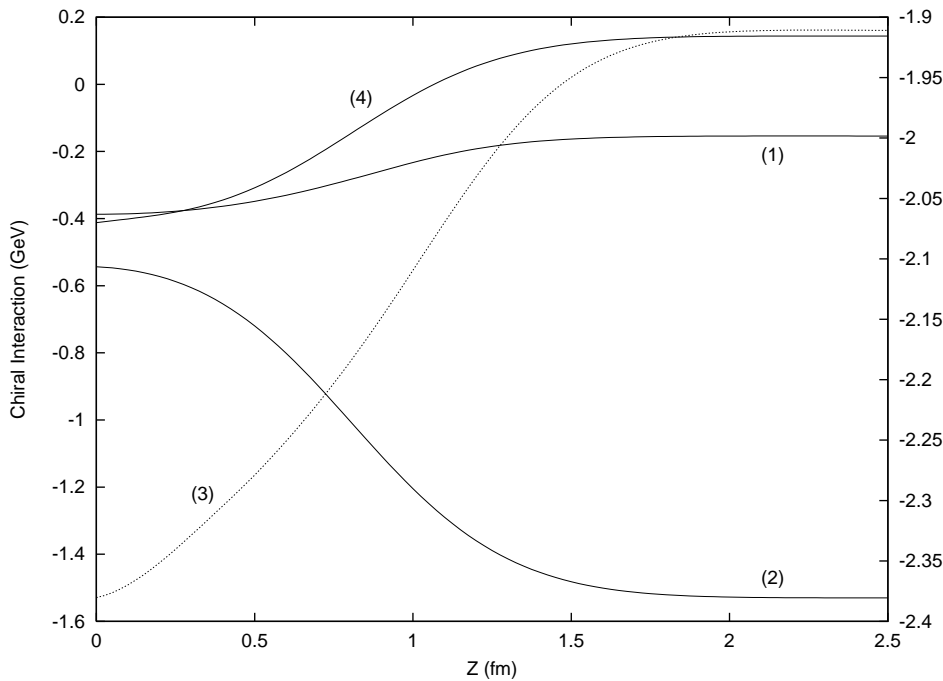}
\end{center}
\caption{\label{PREcluHYP10} The expectation value of the chiral interaction in the cluster model basis, Eq. (\ref{PREvchi}), for $SI= 10$. The curves are numbered as in Fig. \ref{PREcluCONF} and the scale for (3) is also on the right hand side vertical line. (Model I)}
\end{figure}

The diagonal matrix elements of the chiral interaction $V_{\chi}$ of the GBE Model I are exhibited in Fig. \ref{PREcluHYP10} for $SI=10$. At $Z=0$ one recovers the values obtained before. At $Z \rightarrow \infty$ the symmetries corresponding to baryon-baryon channels, namely $[51]_{FS}$ and $[33]_{FS}$, must appear with proper coefficients, as given by the following equation derived in Ref. \cite{HAR81}

\begin{equation}
{\psi }_{NN}\ =\ {\frac{1}{3}}\
\left|{\left.{{\left[{6}\right]}_{O}{\left[{33}\right]}_{FS}}\right\rangle}
\right.\
+\ {\frac{2}{3}}\
\left|{\left.{{\left[{42}\right]}_{O}{\left[{33}\right]}_{FS}}\right\rangle}
\right.\
-\ {\frac{2}{3}}\
\left|{\left.{{\left[{42}\right]}_{O}{\left[{51}\right]}_{FS}}\right\rangle}
\right.
\label{PREHarveyasymp}
\end{equation}
\\

The contribution due to these symmetries must be identical to the contribution of $V_{\chi}$ to two times the nucleon mass, also calculated with the Hamiltonian (\ref{PREham}). We have checked that this is indeed the case. In the total Hamiltonian the contribution of the $[411]_{FS}$ $V_{\chi}$ state tends to infinity when $Z \rightarrow \infty$. Then this state decouples from the rest which is natural because it does not correspond to an asymptotic baryon-baryon channel. It plays a role at small $Z$ but at large $Z$ its amplitude in the NN wave function vanishes, similarly to the ``hidden color" states.
\\

\begin{figure}[H]
\begin{center}
\includegraphics[width=13cm]{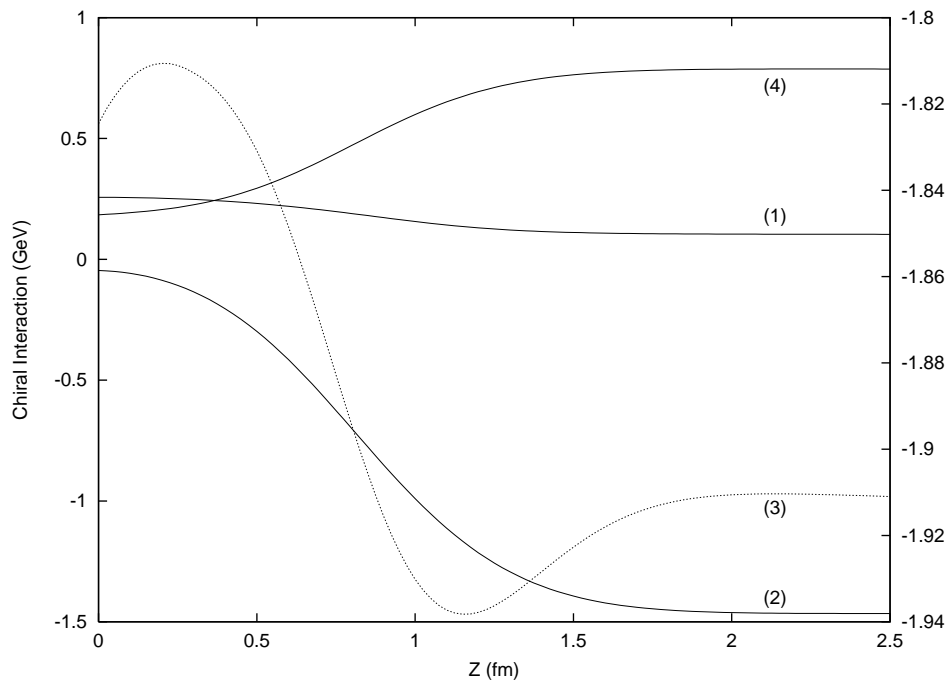}
\end{center}
\caption{The same as Fig. \ref{PREcluHYP10} but for $SI=01$.}\label{PREcluHYP01} 
\end{figure}

For comparison we also show in Fig. \ref{PREcluHYP01} the contribution of the chiral interaction for $SI= 01$ of the GBE Model I. For the states $|[6]_O [33]_{FS} >$, $|[42]_O [33]_{FS} >$ and $|[42]_O [411]_{FS} >$ it changes little with $Z$ but the state $|[42]_O [51]_{FS} >$ brings a contribution with a clear minimum at $Z=0$ before it reaches its maximum as the asymptotic value.
\ \\

\begin{figure}[H]
\begin{center}
\includegraphics[width=13cm]{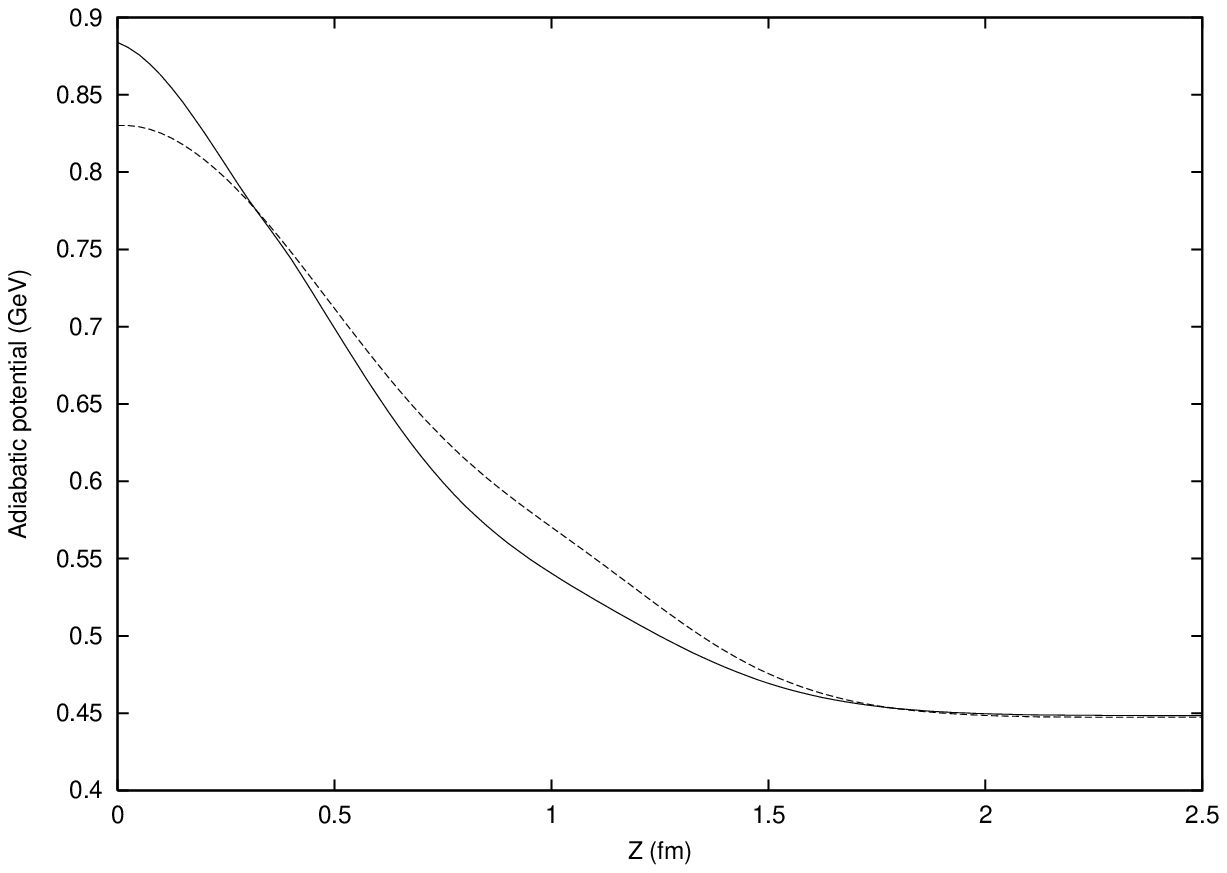}
\end{center}
\caption{\label{PREadiaCM10} Comparison of the adiabatic potential for $SI=10$, calculated in the cluster model basis (full curve) and the molecular orbital basis (dashed curve). (Model I)}
\end{figure}

\begin{figure}[H]
\begin{center}
\includegraphics[width=13cm]{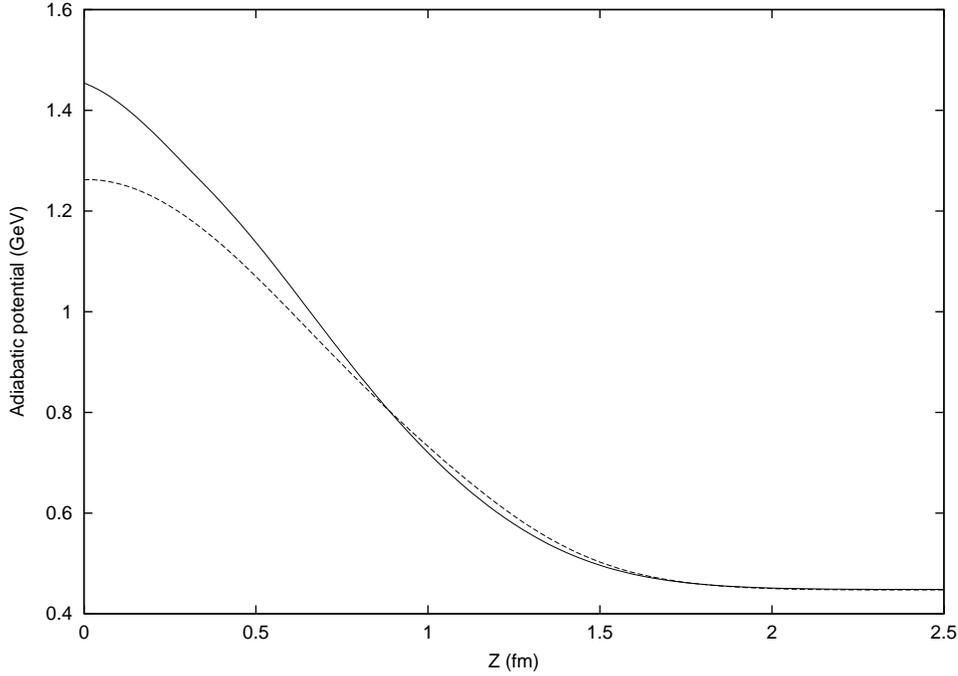}
\end{center}
\caption{\label{PREadiaCM01} Same as Fig. \ref{PREadiaCM10} but for $SI=01$.}
\end{figure}

The adiabatic potential drawn in Figs. \ref{PREadiaCM10} and \ref{PREadiaCM01} is defined according to Eq. (\ref{PREadia}) where $\langle H\rangle_Z$ is the lowest eigenvalue resulting from the diagonalization. Note however that $K_{rel}$ has not yet been subtracted so that $V\rightarrow 0.448$ GeV asymptotically instead of zero. Fig. \ref{PREadiaCM10} corresponds to $SI=10$ and Fig. \ref{PREadiaCM01} to $SI=01$. One can see that the potential is repulsive at any $Z$ in both sectors.

Our cluster model results can be compared to previous literature based on OGE models. A typical example for the $^3S_1$ and $^1S_0$ adiabatic potentials can be found in Ref. \cite{ELS84}. The results are similar to ours. There is a repulsive core but no attractive pocket. However, in our case, in either bases, the core is about twice higher at $Z$ = 0 and about 0.5 fm wider than in \cite{ELS84}.
\\

Another interesting remark is to see the difference in the $Z$ dependence of potential calculated in the basis 

\begin{eqnarray}
\psi_{NN} & = & {\frac{1}{3}}\ \left|{\left.{{\left[{6}\right]}_{O}{\left[{33}\right]}_{FS}}\right\rangle} \right.\ +\ {\frac{2}{3}}\ \left|{\left.{{\left[{42}\right]}_{O}{\left[{33}\right]}_{FS}}\right\rangle} \right.\ -\ {\frac{2}{3}}\ \left|{\left.{{\left[{42}\right]}_{O}{\left[{51}\right]}_{FS}}\right\rangle} \right.\nonumber \\
\psi_{\Delta \Delta} & = & \sqrt{{\frac{4}{45}}}\ \left|{\left.{{\left[{6}\right]}_{O}{\left[{33}\right]}_{FS}}\right\rangle} \right.\ +\ \sqrt{{\frac{16}{45}}}\ \left|{\left.{{\left[{42}\right]}_{O}{\left[{33}\right]}_{FS}}\right\rangle} \right.\ +\ \sqrt{{\frac{25}{45}}}\ \left|{\left.{{\left[{42}\right]}_{O}{\left[{51}\right]}_{FS}}\right\rangle} \right.\nonumber \\
\psi_{CC} & = & \sqrt{{\frac{4}{5}}}\ \left|{\left.{{\left[{6}\right]}_{O}{\left[{33}\right]}_{FS}}\right\rangle} \right.\ -\ \sqrt{{\frac{1}{5}}}\ \left|{\left.{{\left[{42}\right]}_{O}{\left[{33}\right]}_{FS}}\right\rangle} \right.
\label{PREHarveyasympbasis}
\end{eqnarray}
\ 

\noindent and the potential in the same basis but including coupling between different states. The transformation between the symmetry states $|[f]_O [f']_{FS} >$ and the physical states is

\begin{equation}
 \left( \begin{array}{c}
\psi_{NN} \\
\psi_{\Delta \Delta} \\
\psi_{CC} \end{array} \right) = {\bf A} \left( \begin{array}{c}
\left|{{\left[{6}\right]}_{O}{\left[{33}\right]}_{FS}}\right\rangle  \\
\left|{{\left[{42}\right]}_{O}{\left[{33}\right]}_{FS}}\right\rangle \\
\left|{{\left[{42}\right]}_{O}{\left[{51}\right]}_{FS}}\right\rangle  \end{array} \right)
\end{equation}
\ 

\noindent where {\bf A} depends on $Z$. In Harvey's transformation \cite{HAR81} {\bf A} is a constant matrix given by

\begin{equation}\label{PREAconstant}
 {\bf A} = \left( \begin{array}{ccc}
\frac{1}{3} & \frac{2}{3} & -\frac{2}{3}\\
 \sqrt{{\frac{4}{45}}} & \sqrt{{\frac{16}{45}}} & \sqrt{{\frac{25}{45}}}\\
\sqrt{{\frac{4}{5}}} & \sqrt{{\frac{1}{5}}} & 0
\end{array} \right)
\end{equation}
\ 

\begin{figure}[H]
\begin{center}
\includegraphics[width=12.5cm]{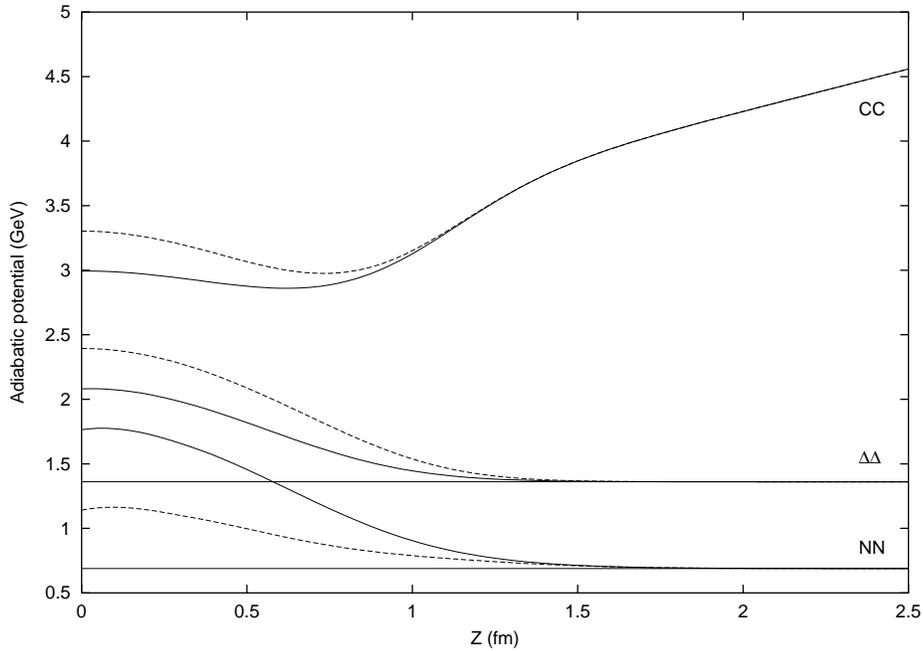}
\end{center}
\caption{\label{PREadiaVDW} The adiabatic potential in a $3\times 3$ basis cluster model in the coupled channel case (dashed curves) and in the pure $NN$, $\Delta \Delta$ and $CC$ basis (full curves) for the GBE Model I. The asymptotic values of the potential for $NN$ and $\Delta\Delta$ are represented by horizontal lines. These are $2m_N+K_{rel}$ and $2m_\Delta+K_{rel}$, respectively.}
\end{figure}

\begin{figure}[H]
\begin{center}
\includegraphics[width=12.75cm]{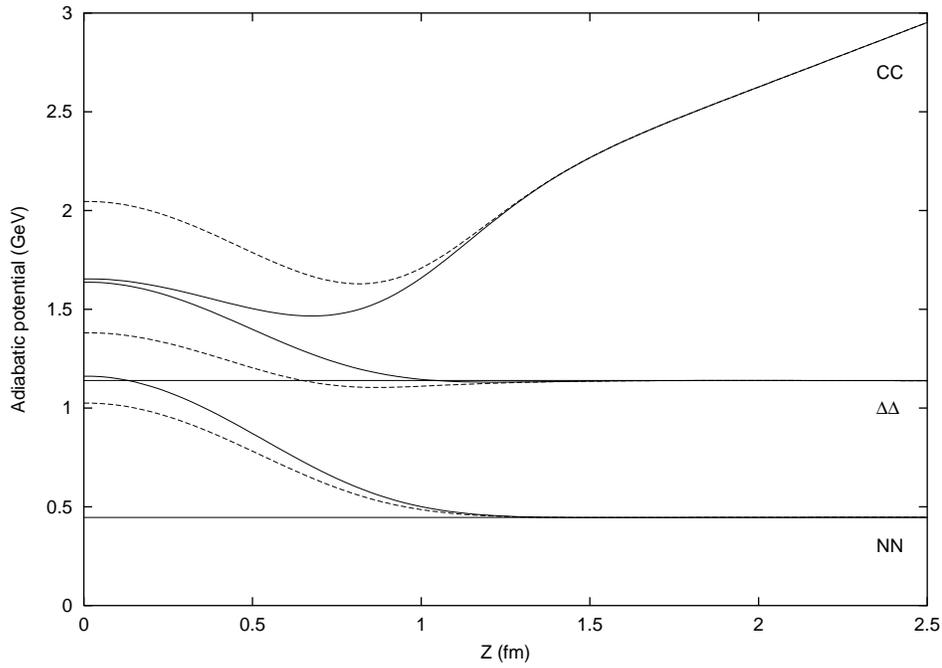}
\end{center}
\caption{\label{PREadiaVDW2} Same as Fig. \ref{PREadiaVDW} but in the GBE Model II.}
\end{figure}

In Fig. \ref{PREadiaVDW} and \ref{PREadiaVDW2} we show the adiabatic potential in the basis (\ref{PREHarveyasympbasis}) and in the coupled channel basis. Obviously the diagonalization lowers the potential. But the important point here is that the {\bf A}$(Z)$ matrix defined above depends only slightly on $Z$ and is close to the constant matrix (\ref{PREAconstant}). Note finally that the $Z$ dependence is more important in Model I than in Model II. This confirms Harvey's definition of the NN wave function.

\subsection{Molecular orbital basis}
\ 

\begin{figure}[H]
\begin{center}
\includegraphics[width=12.75cm]{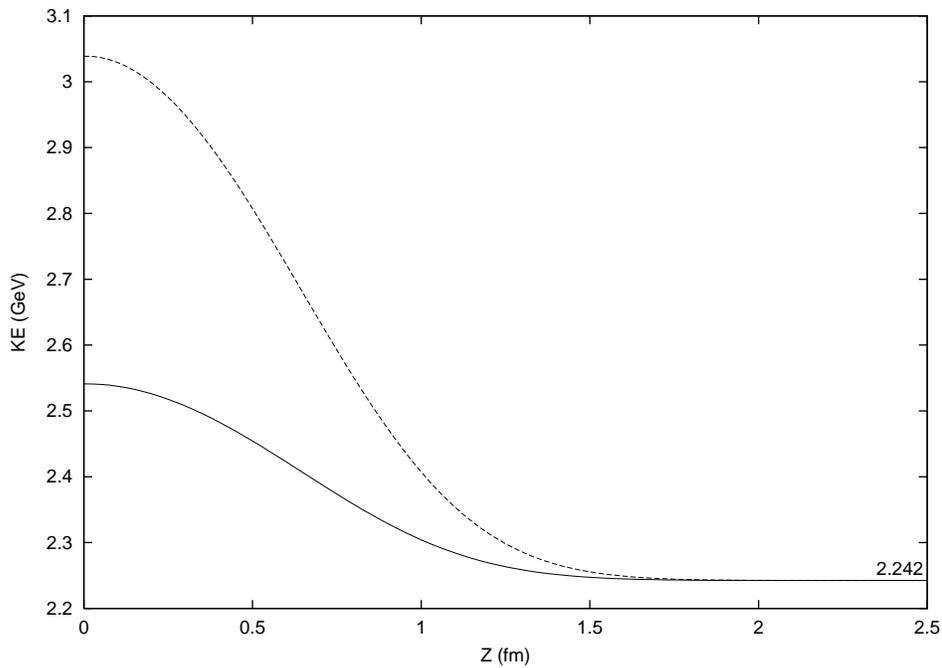}
\end{center}
\caption{\label{PREmolKE} The expectation value of the kinetic energy $\langle KE \rangle$ for the $| [6]_O [33]_{FS} \rangle$ (full curve) and $| 42^+ [42]_O [51]_{FS} \rangle$ (dashed curve) states for the GBE Model I in the molecular orbital basis. The latter is the most dominant state at $Z=0$ (see text).}
\end{figure}

In the molecular basis the diagonal matrix elements of the kinetic energy are similar to each other as decreasing functions of $Z$. As an illustration in Fig. \ref{PREmolKE} we show the expectation value of the kinetic energy $\langle KE\rangle$ corresponding to $\left|{\left.{33{\left[{6}\right]}_{O}{\left[{33}\right]}_{FS}}\right\rangle} \right.$ and to the most dominant state at $Z=0$, namely $\left|{\left.{42^+ {\left[{42}\right]}_{O}{\left[{51}\right]}_{FS}}\right\rangle}\right.$ (see Ref. \cite{BAR99a}). At finite $Z$ the kinetic energy of the latter is larger than that of the former because of the presence of the configuration $s^2 p^4$ with 50 \% probability while in the first state this probability is smaller as well as that of the $p^6$ configuration, as shown by the first and the seventh wave functions of Eqs. (\ref{PREorbitbasis}). The large kinetic energy of the state  $\left.{\left|{{42}^{+}{\left[{42}\right]}_{O}{\left[{51}\right]}_{FS}}\right.}\right\rangle\ $ is compensated by a large negative value of $\langle V_{\chi}\rangle$ so that this state becomes dominant at small $Z$ in agreement with Ref. \cite{BAR99a}.

\ 

The expectation values of the confinement potential increase with $Z$ becoming linear beyond $Z >$ 1.5 fm except for the state $\left|{\left.{33{\left[{42}\right]}_{O}{\left[{51}\right]}_{FS}}\right\rangle} \right.$ which gives a result very much similar to the cluster model state $\left|{\left.{R^3L^3{\left[{42}\right]}_{O}{\left[{51}\right]}_{FS}}\right \rangle} \right.$ drawn in Fig. \ref{PREcluCONF}. Such a behavior can be understood through the details given later in Subsection \ref{asympmolsection}. Due to the similarity to the cluster model results we do not show here $\langle V_{conf} \rangle$ explicitly for the molecular orbital basis.
\\

\begin{figure}[H]
\begin{center}
\includegraphics[width=13cm]{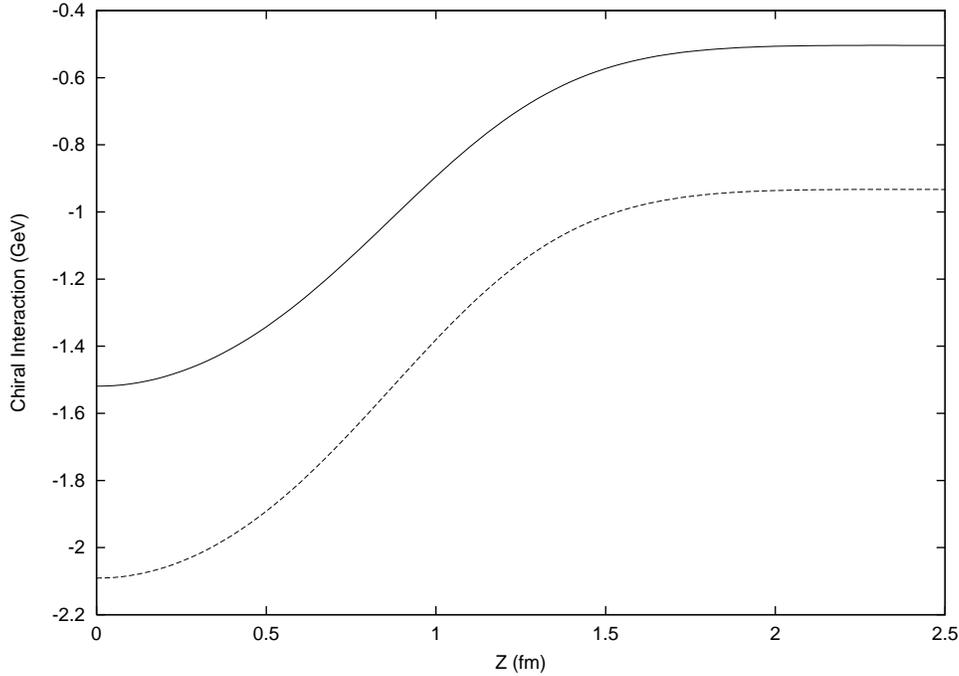}
\end{center}
\caption{\label{PREmolHYP} The expectation value of the chiral interaction for the GBE Model I in the molecular orbital basis for $| 42^+ [42]_O [51]_{FS} \rangle$ which is the most dominant state at $Z$ = 0. The dashed curve corresponds to $SI=10$ and the full curve to $SI=01$.}
\end{figure}

The expectation value of the chiral interaction either decreases or increases with $Z$,  depending on the state. In Fig. \ref{PREmolHYP} we illustrate the case of the $\left|{\left.{42^+{\left[{42}\right]}_{O}{\left[{51}\right]}_{FS}}\right\rangle}\right.$ state. both for $SI=10$ and $SI=01$ sectors in Model I. This state is the dominant component of ${\psi }_{NN}$ at $Z$ = 0 with a probability of 87 \% for $SI= 10$ and 93 \% for $SI = 01$ \cite{BAR99a}. With increasing $Z$ these probabilities decrease and tend to zero at $Z \rightarrow \infty$. In fact in the molecular orbital basis the asymptotic form of  ${\psi }_{NN}$ is also given by Eq. (\ref{PREHarveyasymp}) inasmuch as $r \rightarrow R$ and $l \rightarrow L$ as indicated below Eq. (\ref{PREZgoes0}).
\\

Adding together these contributions we diagonalize the Hamiltonian of Model I and use its lowest eigenvalue to obtain the NN potential according to the definition (\ref{PREadia}). The $SI=10$ and $SI=01$ cases are illustrated in Figs. \ref{PREadiaCM10} and \ref{PREadiaCM01} respectively, for a comparison with the cluster model basis. As shown before, at $Z=0$ the repulsion reduces by about 22 \% and 25 \% in the $^3S_1$ and $^1S_0$ channels respectively when passing from the cluster model basis to the molecular orbital basis. From Figs. \ref{PREadiaCM10} and \ref{PREadiaCM01} and  one can see that the molecular orbital basis has an important effect up to about $Z \approx$ 1.5 fm giving a lower potential at small values of $Z$. For $Z \approx 1$ fm it gives a potential larger by few tens of MeV than the cluster model potential. However there is no attraction at all in either case.
\\

\begin{figure}[H]
\begin{center}
\includegraphics[width=13cm]{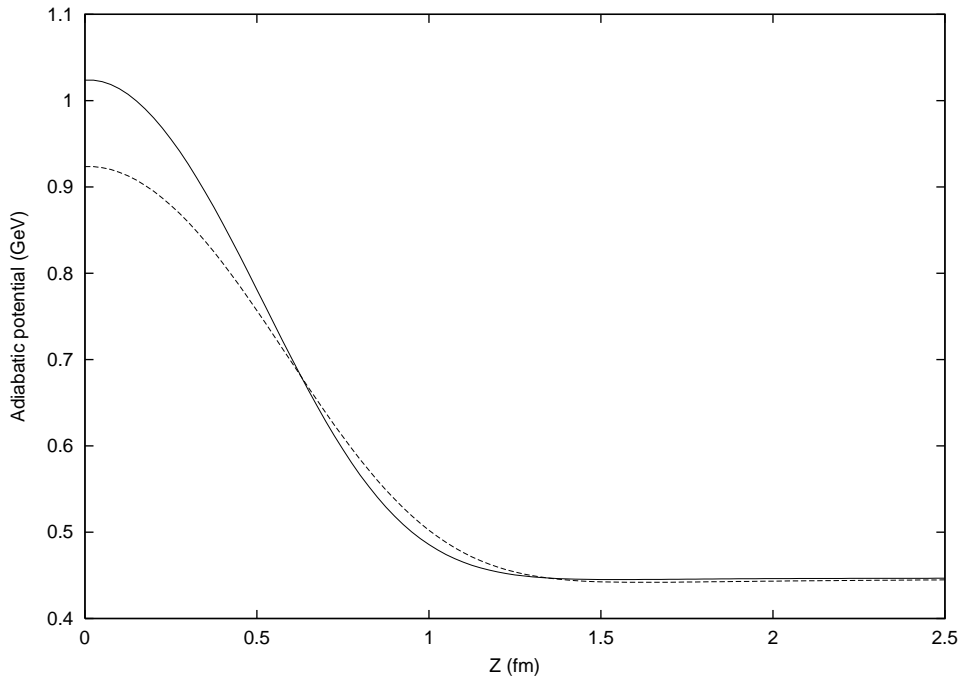}
\end{center}
\caption{\label{PREadiaCM10b} Comparison of the adiabatic potential for $SI=10$, calculated in the cluster model basis (full curve) and the molecular orbital basis (dashed curve) in the GBE Model II.}
\end{figure}

\begin{figure}[H]
\begin{center}
\includegraphics[width=13cm]{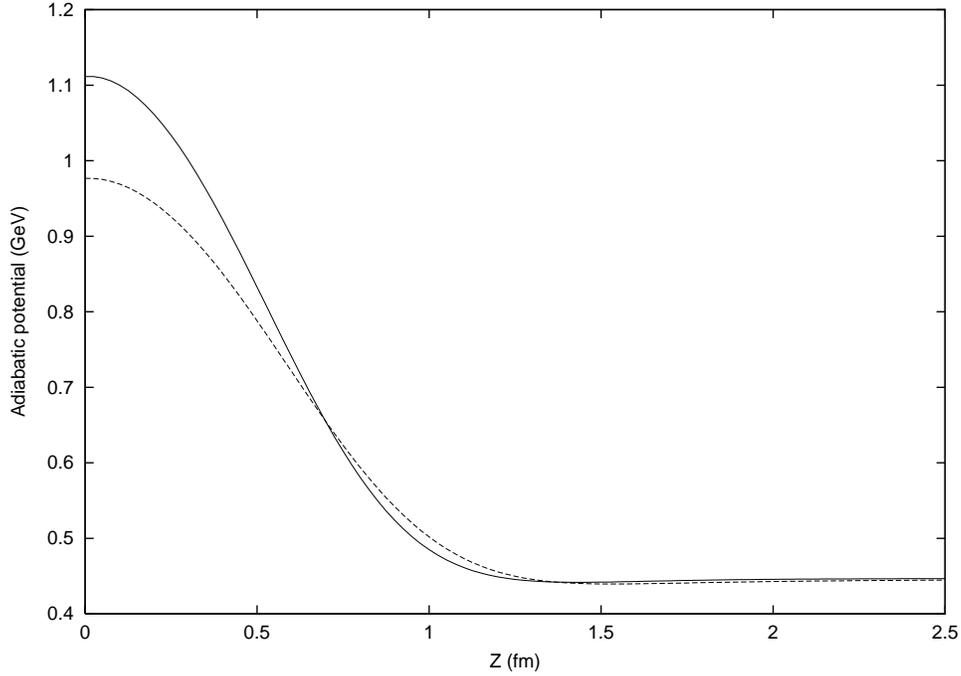}
\end{center}
\caption{\label{PREadiaCM01b} Same as Fig. \ref{PREadiaCM10b} but for $SI=01$.}
\end{figure}

The Model II gives similar results, the details of which we do not reproduce here. In Figs. \ref{PREadiaCM10b} and \ref{PREadiaCM01b} we give the final results of the adiabatic potential for $SI=10$ and $SI=01$, respectively. There is however a difference in the shape of the adiabatic potential. Indeed the repulsion at small $Z$ is more important in the GBE Model II and the range of the repulsive core is about 1.5 fm in the GBE Model I compared to 1 fm in the GBE Model II. This indicates that the parametrization of the GBE interactions influence the NN interaction potential.
\\

As already mentioned, by construction, the molecular orbital basis is richer at $Z$ = 0 \cite{STA87} than the cluster model basis. For this reason, at small $Z$ it leads to a lower potential than the cluster model basis. Within a truncated space this property may not hold beyond some value of $Z$. This could be a possible explanation of the fact that the molecular orbital result is higher than the cluster model result at $Z\approx 1$. However by an increase of the Hilbert space one can possibly bring the molecular potential lower again. In fact we choose the most important configurations from symmetry arguments \cite{STA97} based on Casimir operator eigenvalues. These arguments hold if the interaction is the same for all quarks in the coordinate space. This is certainly a better approximation for $Z$ = 0 than for larger values of $Z$. So it means that other configurations, which have been neglected, may play a role at $Z > 0.4$ fm. Then, if added, they could possibly lower the molecular basis result.

As defined before, the quantity $Z$ is the separation distance between two 3q clusters.  It represents the Jacobi relative coordinate between the two nucleons only for large $Z$.  There we view it as a generator coordinate and the potential we obtain represents the diagonal kernel appearing in the generator coordinate method (CGM) which can be related to the diagonal kernel of the resonating group method (see \cite{CVE83}). The resonating group method will be presented and applied in the next chapter. It leads to non local potentials.  However the adiabatic potentials, here obtained in the two bases, can be compared with each other in an independent and different way. On can introduce the quadrupole moment of the six-quark system
\begin{equation}
q_{20}=\sum_{i=1}^6 r_i^2\ Y_{20} (\hat{r}_i)
\end{equation}
\ 

\noindent and treat the square root of its expectation value

\begin{equation}
\langle Q\rangle=\langle \psi_{NN}|q_{20}|\psi_{NN}\rangle
\label{PREquadEQT}
\end{equation}
\ 

\noindent as a collective coordinate describing the separation between the two nucleons.  Obviously $\sqrt{\langle Q\rangle} \rightarrow Z $ for large $Z$.
\\

\begin{figure}[H]
\begin{center}
\includegraphics[width=12.5cm]{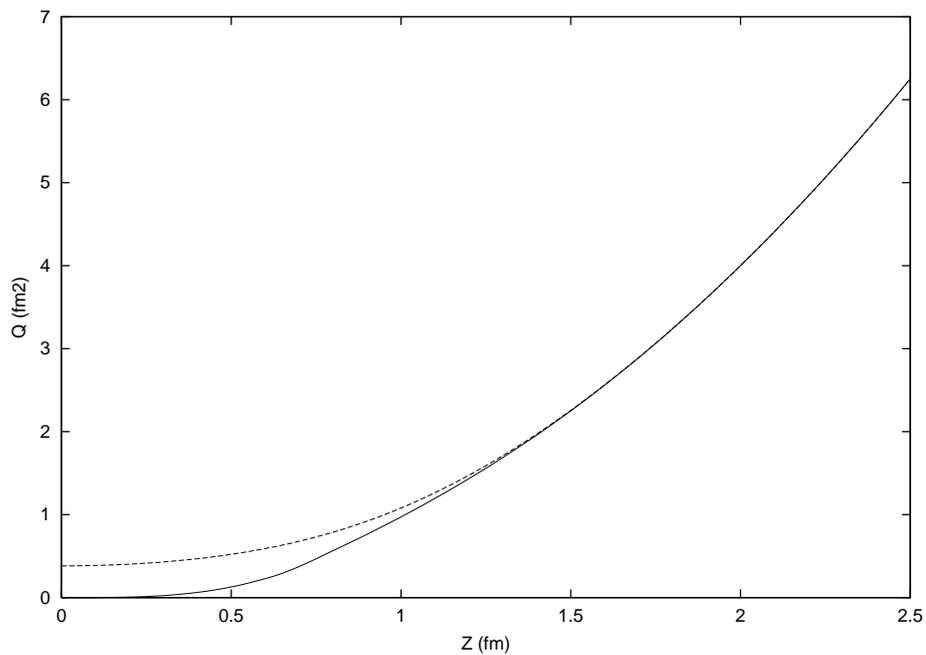}
\end{center}
\caption{\label{PREadiaQ} $\langle Q\rangle$ as a function of $Z$ with $\langle Q\rangle$ defined by Eq. (\ref{PREquadEQT}) and normalized such as to be identical to $Z^2$ at large $Z$. The full line corresponds to the cluster model basis and the dashed line to the molecular orbital basis. (Model I)}
\end{figure}

\begin{figure}[H]
\begin{center}
\includegraphics[width=12.5cm]{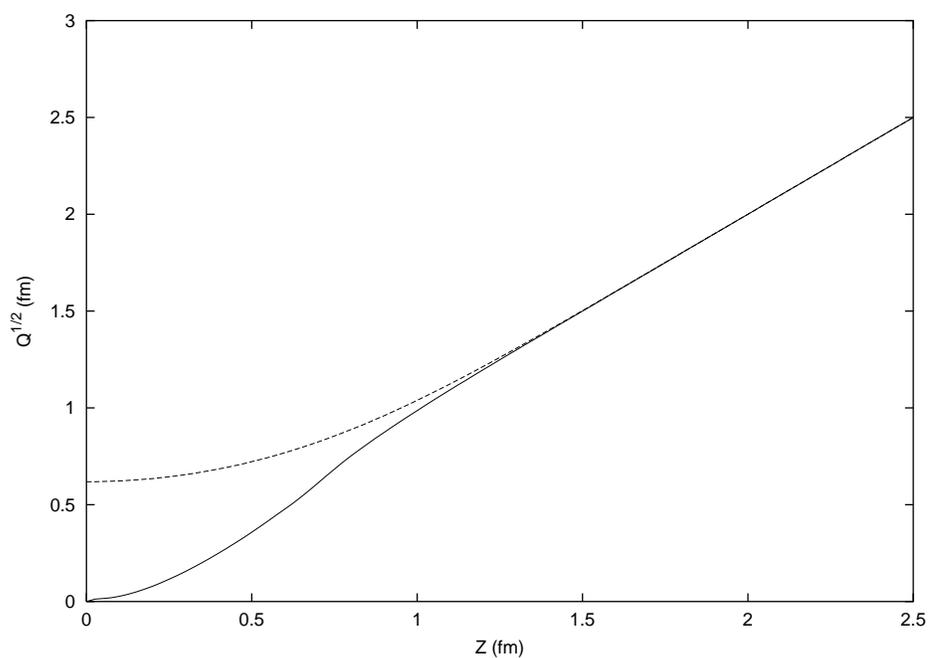}
\end{center}
\caption{\label{PREadiaSQ} $\sqrt{\langle Q\rangle}$ as a function of $Z$ with $\langle Q\rangle$ defined by Eq. (\ref{PREquadEQT}) and normalized as indicated in the text.  The full line corresponds to the cluster model basis and the dashed line to the molecular orbital basis. (Model I)}
\end{figure}

In Figs. \ref{PREadiaQ} and \ref{PREadiaSQ} we plot $\langle Q\rangle$ and $\sqrt{\langle Q\rangle}$ as a function of $Z$. The results are practically identical for $SI =10$ and $SI = 01$. Note that $\sqrt{\langle Q\rangle}$ is normalized such as to be identical to $Z$ at large $Z$. One can see that the cluster model gives $\sqrt{\langle Q\rangle}=0$ at $Z=0$, consistent with the spherical symmetry of the system, while the molecular basis result is $\sqrt{\langle Q\rangle}= 0.573 $ fm at $Z=0$, which suggests that the system acquires a small deformation in the molecular basis.  This also means that its {\it r.m.s.} radius is larger in the molecular basis.

\begin{figure}[H]
\begin{center}
\includegraphics[width=12.5cm]{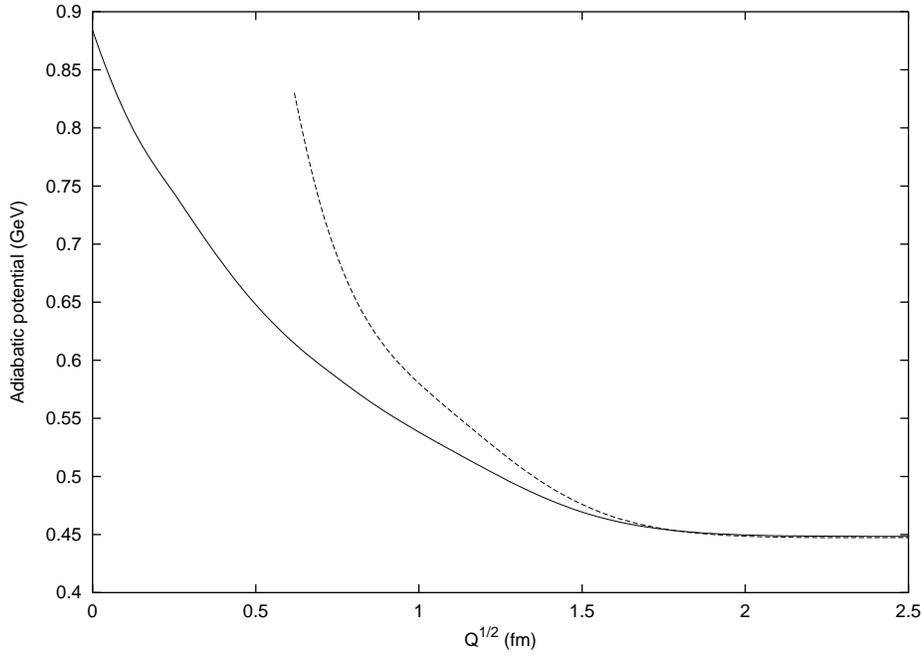}
\end{center}
\caption{\label{PREadiaSQ10} Adiabatic potential for $SI=10$ as a function of $\sqrt{\langle Q\rangle}$.  The full line is the cluster model result and the dashed line the molecular basis result. (Model I)}
\end{figure}

\begin{figure}[H]
\begin{center}
\includegraphics[width=12.5cm]{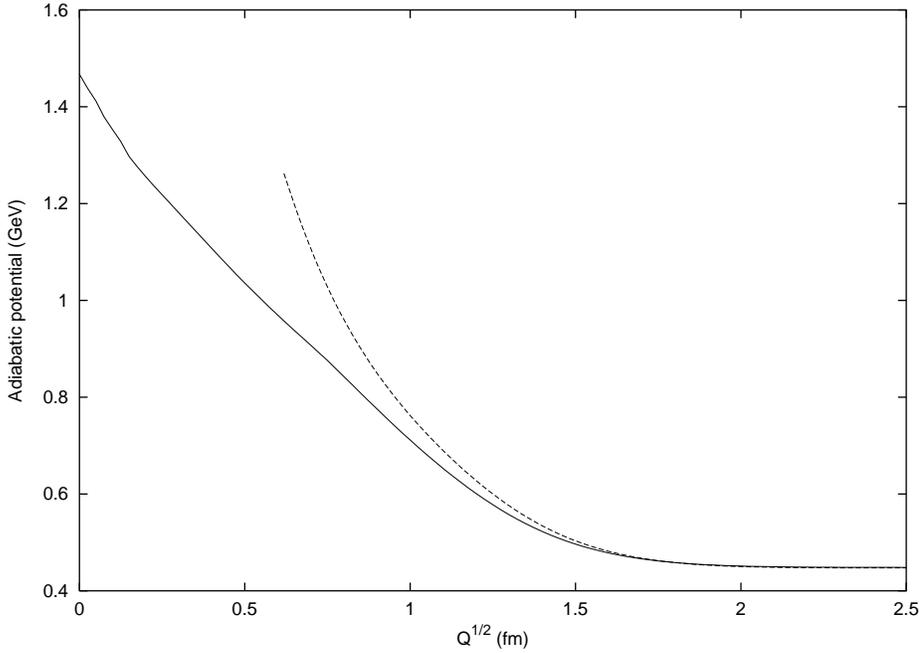}
\end{center}
\caption{\label{PREadiaSQ01} Same as Fig. \ref{PREadiaSQ10} but for $SI=01$.}
\end{figure}

In Figs. \ref{PREadiaSQ10} and \ref{PREadiaSQ01} we plot the adiabatic potentials as a function of $\sqrt{\langle Q\rangle}$ instead of $Z$, for $SI = 10$ and $01$ respectively. As $\sqrt{\langle Q\rangle}\neq 0 $ at any $Z$ in the molecular orbital basis, the corresponding potential is shifted to the right and appears above the cluster model potential at finite values of $\sqrt{\langle Q\rangle}$ but tends asymptotically to the same value.  The comparison made in Figs.  \ref{PREadiaSQ10} and \ref{PREadiaSQ01} is meaningful in the context of a Schr\"odinger type equation where the local adiabatic potential appears in conjunction with an ``effective mass'' depending on $\sqrt{\langle Q\rangle}$ also.

\

\section{A note on the asymptotic behavior in the molecular orbital basis}\label{asympmolsection}
\ 

Asymptotically, the molecular and the cluster basis must lead to the same result. In this section we study the behavior of the confinement potential in the molecular orbital basis at large separation distances $Z$ between the centers of two 3q clusters. As an example we consider the state $|42^+[42]_O [33]_{FS}\rangle$. Through the fractional parentage technique \cite{HAR81,STA96} the six-body matrix elements can be reduced to the calculation of two-body matrix elements. Using this technique and integrating in the color space one obtains

\begin{eqnarray}\label{PREasymptVCONF}
&&\langle 42^+[42]_O [33]_{FS} | V_{conf} | 42^+[42]_O [33]_{FS}\rangle = \frac{1}{40}~ [ 22 \langle \pi\pi |V| \pi\pi\rangle \nonumber \\ 
&&\hspace{3cm}\mbox{}+ 76 \langle \sigma\pi |V| \sigma\pi \rangle + 26 \langle \sigma\pi |V| \pi\sigma\rangle \nonumber \\ 
&&\hspace{3cm}\mbox{}- 58 \langle \pi\pi |V| \sigma\sigma \rangle + 22 \langle \sigma\sigma |V| \sigma\sigma\rangle ]
\end{eqnarray}
\ 

\noindent where the right-hand side contains two-body orbital matrix elements. According to Eq. (8) for $Z \rightarrow \infty$ one has

\begin{equation}
|\sigma\rangle \rightarrow \frac{1}{\sqrt{2}}~|R~+~L\rangle,\ \ |\pi\rangle \rightarrow \frac{1}{\sqrt{2}}~|R~-~L\rangle
\end{equation}

Replacing these asymptotic forms in the above equation one obtains matrix elements containing the states $|R\rangle$ and $|L\rangle$. Most of these matrix elements vanish asymptotically. The only surviving ones are

\begin{equation}
\langle R~R |V| R~R\rangle \rightarrow a~,\,\, \langle R~L |V| R~L\rangle \rightarrow b~Z
\end{equation}
\ 

\noindent where $a$ and $b$ are some constants. This brings us to the following asymptotic behavior of the matrix elements in the right-hand side of (\ref{PREasymptVCONF})

\begin{eqnarray}
\langle \sigma\sigma |V| \sigma\sigma\rangle \rightarrow (a~+~b~Z)/2 \nonumber\\
\langle \pi\pi |V| \pi\pi\rangle \rightarrow (a~+~b~Z)/2  \nonumber\\
\langle \sigma\pi |V| \sigma\pi \rangle \rightarrow (a~+~b~Z)/2 \nonumber\\
\langle \sigma\pi |V| \pi\sigma\rangle \rightarrow (a~-~b~Z)/2 \nonumber\\
\langle \pi\pi |V| \sigma\sigma \rangle \rightarrow (a~-~b~Z)/2
\end{eqnarray}
\ 

\noindent from which it follows that

\begin{equation}
\langle 42^+[42]_O [33]_{FS} | V_{conf} | 42^+[42]_O [33]_{FS}\rangle 
\rightarrow (11~a~+~19~b~Z)/10
\end{equation}
\ 

\noindent i. e. this matrix element grows linearly with $Z$ at large $Z$. In a similar manner one can show that in the confinement matrix element of the state $\left|{\left.{33{\left[{42}\right]}_{O}{\left[{51}\right]}_{FS}}\right\rangle} \right.$ the coefficient of the term linear in $Z$ cancels out so that in this case one obtains a {\it plateau} as in Fig. \ref{PREcluCONF}.
\ \\

\section{The middle-range attraction}
\ 

\begin{figure}[H]
\begin{center}
\includegraphics[width=13.5cm]{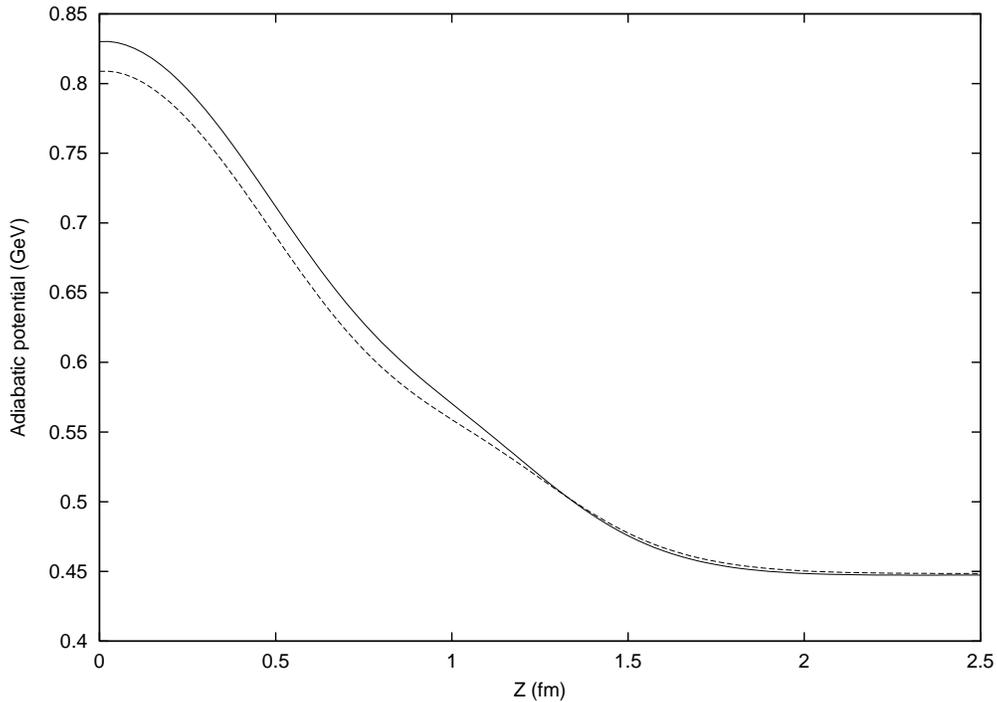}
\end{center}
\caption{\label{PREmolNoYuk10} The adiabatic potential in the molecular orbital basis for $SI = 10$ in the GBE Model I. The solid curve is the same as in Fig. \ref{PREadiaCM10}. The dashed curve is the result obtained by removing the Yukawa part of the quark-quark interaction (\ref{PREpoint}).}
\end{figure}

\begin{figure}[H]
\begin{center}
\includegraphics[width=13.5cm]{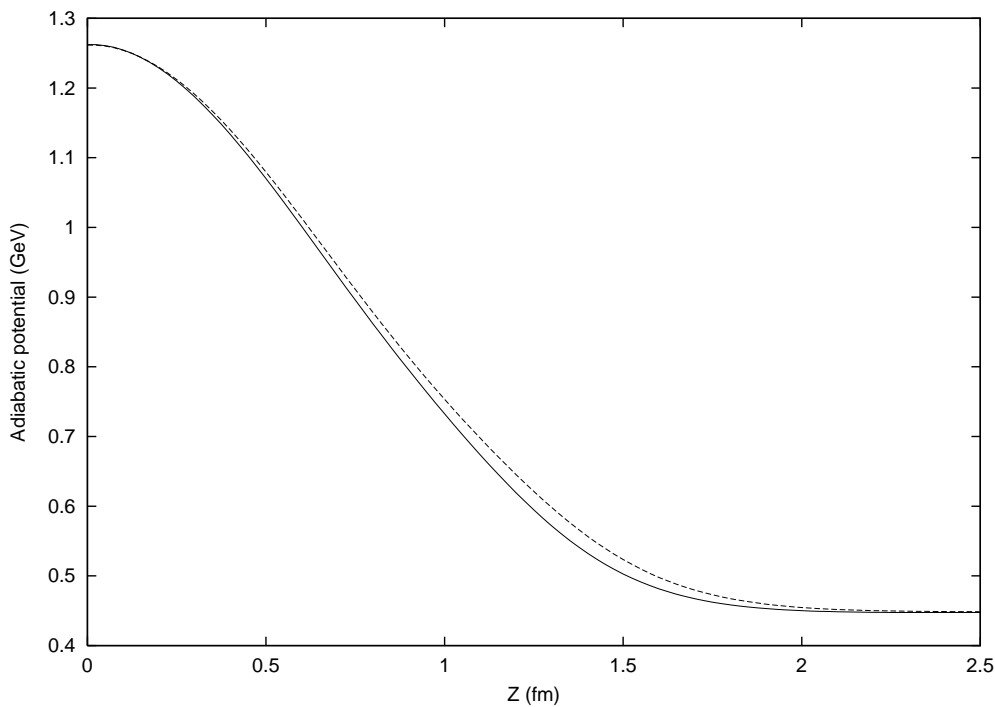}
\end{center}
\caption{\label{PREmolNoYuk01} The same as Fig. \ref{PREmolNoYuk10} but for $SI = 01$.}
\end{figure}

In principle we expected some attraction at large $Z$ due to the presence of the Yukawa potential tail in Eq. (\ref{PREpoint}). To see the net contribution of this part of the quark-quark interaction we repeated the calculations in the molecular orbital basis by completely removing the first term - the Yukawa potential part - in Eq. (\ref{PREpoint}). The result is shown in Fig. \ref{PREmolNoYuk10} for $SI=10$ and in  Fig. \ref{PREmolNoYuk01} for $SI=01$. One can see that beyond $ Z  \approx 1.3 $ fm the potential containing the contribution of the Yukawa potential tail is lower that the one with the Yukawa part removed. But the contribution of the Yukawa potential tail is very small, of the order of 1-2 MeV.  At small values of $Z$ the Yukawa part of (\ref{PREpoint})
contributes to increase the adiabatic potential because
it diminishes the attraction in the two body matrix elements.
\\

Therefore the potential is repulsive everywhere. The missing middle- and long-range attraction can in principle be simulated in a simple phenomenological way. For example, in Ref. \cite{OKA84} this has been achieved at the baryon level. Here we adopt a more consistent procedure assuming that besides the pseudoscalar meson exchange interaction there exists an additional scalar, $\sigma$-meson exchange interaction between quarks. This is in the spirit of the spontaneous chiral symmetry breaking mechanism (see Section \ref{spontaneoussubsection}) on which the GBE model is based. In that sense the $\sigma$-meson is the chiral partner of the pion and it should be considered explicitly.
\\

But in another language, once the one-pion exchange interaction between quarks is admitted, one can inquire about the role of at least two-pion exchanges. Recently it was found \cite{RIS99} that the two-pion exchange also plays a significant role in the quark-quark interaction. It enhances the effect of the isospin dependent spin-spin component of the one-pion exchange interaction and reduces up to canceling out its tensor component. Apart from that it gives rise to a spin independent central component, which averaged over the isospin wave function of the nucleon it produces an attractive spin independent interaction. These findings also support the introduction of a scalar ($\sigma$-meson) exchange interaction between quarks as an approximate description of the two-pion exchange loops.
\\

For consistency with the parametrization \cite{GLO96b} {\it i. e.} the Model I, we consider here a scalar quark-quark interaction of the form

\begin{equation}
V_\sigma (r)= \frac{g_\sigma^2}{4\pi}\frac{1}{12m_i m_j} \{\theta(r-r'_0)\mu_\sigma^2 \,\frac{e^{-\mu_\sigma r}}{ r}- \frac {4}{\sqrt {\pi}} \alpha'^3 \exp(-\alpha'^2(r-r'_0)^2)\}.
\label{PREsigmaI}
\end{equation}
\ 

\noindent where we fixed $\mu_\sigma$ = 675 MeV and $r'_0$, $\alpha'$ and the coupling constant $g^2_\sigma/4\pi$ are arbitrary parameters. In order to be effective at middle-range separation between nucleons we expect this interaction to have $r'_0 \neq r_0$ and $\alpha' \neq \alpha$. Note that the factor $1/m_i m_j$ has only been introduced for dimensional reasons. The value of $\mu_\sigma$ is close to the conventional value (600 MeV) of the $\sigma$-meson.
\\

We first looked at the baryon spectrum with the same variational parameters as before. The only modification is a shift of the whole spectrum which would correspond to taking $V_0 \approx - 60$ MeV in Eq. (\ref{PREconf}) instead of $V_0=0$ as in the parametrization (\ref{PREparam1}) of the GBE Model I.
\\

\begin{figure}[H]
\begin{center}
\includegraphics[width=13.5cm]{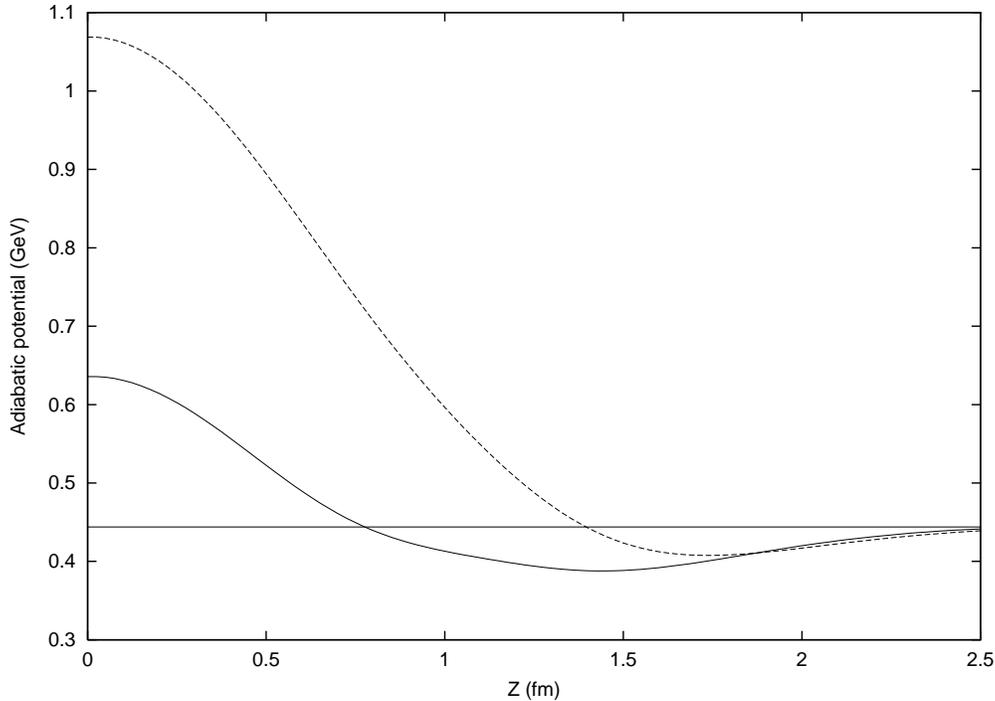}
\end{center}
\caption{\label{PREadiaMscal} The adiabatic potential in the molecular orbital basis for $SI = 10$ (full curve) and $SI = 01$ (dashed curve) with pseudoscalar + scalar quark-quark interaction of Model I.}
\end{figure}

For the 6q system we performed calculations in the molecular basis, which is more appropriate and easier to handle than the cluster model basis. We found that the resulting adiabatic potential is practically insensitive to changes in $\mu_\sigma$ and $r'_0$ but very sensitive to $\alpha'$. In Fig. \ref{PREadiaMscal} we show results for 

\begin{equation}
r'_0 = 0.86 \, { fm}, ~\alpha'  = 1.47 \, { fm}^{-1},~~ g^2_\sigma/4\pi = g^2_8/4\pi
\end{equation}
\\

\begin{figure}[H]
\begin{center}
\includegraphics[width=13.5cm]{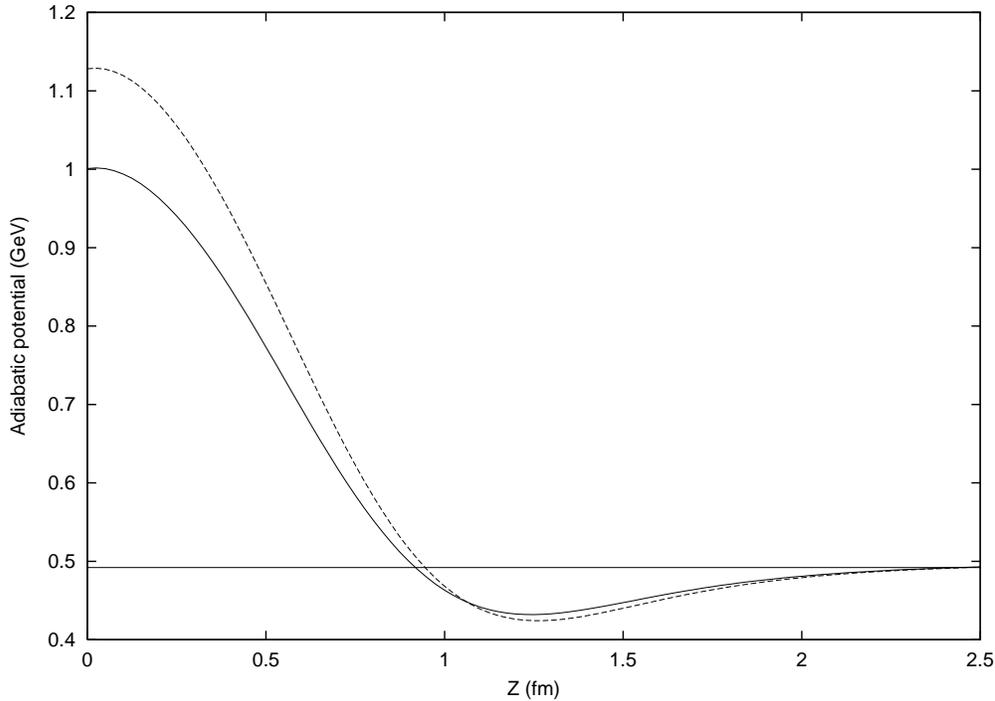}
\end{center}
\caption{\label{PREadiaMscalGl2} The adiabatic potential in the molecular orbital basis for $SI= 10$ (full curve) and $SI = 01$ (dashed curve) with pseudoscalar + scalar quark-quark interaction in the GBE Model II.}
\end{figure}

One can see that $V_\sigma$ produces indeed an attractive pocket, deeper for $SI=10$ than for $01$, as it may be for the NN system. The depth of the attraction depends essentially on $\alpha'$. The precise values of the parameters entering Eq. (\ref{PREsigmaI}) should be determined in further resonating group method calculations as discussed in the next chapter.  As mentioned above the Born-Oppenheimer potential is in fact the diagonal GCM kernel related to the RGM kernel (see Chapter \ref{rgmchapter}). Note that similar results are obtained with the Model II, as shown in Fig. \ref{PREadiaMscalGl2}. In this particular case, we take $V_{\sigma}$ as follows :

\begin{equation}\label{PRESIGMA}
V_{\sigma}=-\frac{g_{\sigma q}^2}{4\pi}~(\frac{e^{-\mu_{\sigma}r}}{r}-\frac{e^{-\Lambda_{\sigma}r}}{r})\ ,
\end{equation}
\ 

\noindent consistent with the form (\ref{PREmodelII}). Our first choice of the parameters is

\begin{equation}
\frac{g_{\sigma q}^2}{4\pi} = \frac{g_{\pi q}^2}{4\pi} = 1.24,~~~~~ \mu_{\sigma} = 600\ {\rm MeV}\ ,~~~~~\Lambda_{\sigma} = 830\ {\rm MeV}\ .
\end{equation}
\

The choice of these parameters with $\mu_\sigma=600$ MeV is given only for orientation. In the next chapter where the agreement with experiment is searched for the bound and the scattering states of the NN system, we make a more careful choice. The important point here is that the qualitative results are reasonable, like in the Model I.
\\

\ 

\section{Summary}
\ 

The present chapter represents the preliminary necessary step towards the study of the NN system. Here we have calculated the NN potential in the adiabatic approximation as a function of $Z$, the separation distance between the centers of the two 3q clusters. We used a constituent quark model where quarks interact via pseudoscalar meson exchange. The orbital part of the six-quark states was constructed either from cluster model or molecular orbital single particle states. The latter are more realistic, having the proper axially and reflec\-tionally symmetries. Also technically they are more convenient. We explicitly showed that they are important at small values of $Z$. In particular we found that the NN potential obtained in the molecular orbital basis has a less repulsive core than the one obtained in the cluster model basis. However none of the bases leads to an attractive pocket. We have simulated this attraction by introducing a $\sigma$-meson exchange interaction between quarks.
\\

To have a better understanding of the two bases we have also calculated the quadrupole moment of the 6q system as a function of $Z$.  The results show that in the molecular orbital basis the system acquires some small deformations at $Z=0$.  As a function of the quadrupole moment the adiabatic potential looks more repulsive in the molecular orbital basis than in the cluster model basis.  In this light one might naively expect that the molecular basis will lead to scattering phase shifts having a more repulsive behavior.
\\

Both Model I and Model II have been used in the derivation of the adiabatic potential. We showed that the results are very similar even if some small differences appear, in particular for the shape of the potential. Because the parametrization of the GBE Model II rely on more realistic grounds, we will not investigate further the Model I.
\\

The present calculations give us an idea about the size and shape of the hard core produced by the GBE interaction.  Except for small values of $Z$ the two bases give rather similar potentials. Taking $Z$ as a generator coordinate the following step is to perform a dynamical study. This is the goal of the next chapter where the resonating group method will be used provided phase shifts to be compared directly to the experiment.
\newpage
\thispagestyle{empty} 
\ 
\newpage


\chapter{The Resonating Group Method}\label{rgmchapter}

\vspace{3.4cm}
\

In Chapter \ref{preliminarychapter} we have used the adiabatic approximation to study the short-range repulsion of the NN system in the framework of the GBE model. However a dynamical treatment of the nucleon-nucleon interaction could not be considered with that method. In this chapter we shall extend the study of the previous chapter using the resonating group method (RGM) \cite{WHE37}, in order to calculate the nucleon-nucleon scattering phase shifts. Restricting to quark degrees of freedom and non-relativistic kinematics, the resonating group method which is appropriate to treat the interaction between two composite systems, can be straightforwardly applied to the study of the baryon-baryon interaction in the quark models. A very important aspect of the resonating group method is the introduction of non-local effects in the potential.
\\

In this chapter we shall limit ourselves to the use of only the GBE Model II in order to study the bound state and the low energy scattering of two nucleons. One of our concerns is to confirm the role of the quark exchange interaction induced by the antisymmetrization on the short-range repulsion. In order to achieve this goal the resonating group method appears to be particularly appropriate. In this chapter we shall first present the resonating group method in the framework of the NN problem for systems composed of quarks. We shall then apply it to the s-wave scattering of two nucleons. We shall show that the GBE model can reproduce the short-range repulsive interaction fairly well. In order to calculate the phase shifts of the NN scattering which can be compared with experiment, we have to include some middle-range attraction. In the way presented in the previous chapter, this middle-range attraction is provided by a $\sigma$-exchange potential, a scalar interaction at the quark level. Finally we incorporate the tensor part of the hyperfine interaction in order to describe the $^3S_1$ phase shifts.
\\

We shall show that the flavor-spin symmetry of the GBE model can induce a short-range repulsion comparable to that produced by the OGE model, due to its color-spin symmetry. A comparison between results obtained in these two models will be provided.

\section{The model}
\

In this chapter we shall focus our investigation on the Model II of the GBE Hamiltonian presented in the previous chapters. As already mentioned before, the reason of using the parametrization of the Model II \cite{GLO97a}, instead of Model I \cite{GLO96b}, as in the previous work \cite{STA97,BAR99a,BAR99b}, is that it is more realistic. Its volume integral, {\it i. e.} its Fourier transform at $\vec{q} = 0$, vanishes, consistently with the quark-pseudoscalar meson vertex proportional to $\vec{\sigma} \cdot \vec{q}\ \lambda^F$. In addition, this interaction does not enhance the quark-quark matrix elements containing $1p$ relative motion, as it is the case with the parametrization of the Model I \cite{GLO96b}. This point has been raised in Ref. \cite{STA99}.
\\

In order to eliminate all ambiguities, we recall explicitly the GBE Hamiltonian considered in this chapter

\begin{equation}\label{RGMhamiltonian}
H=\sum_i m_i+\sum_{i=1}{\frac{p^2_i}{2m_i}}-K_G+\sum_{i<j} V_{conf}(r_{ij})+\sum_{i<j} V_{\chi}(r_{ij})\ ,
\end{equation}
\ 

\noindent where $K_G$ is the kinetic energy of the center of mass. The linear confining interaction is

\begin{equation}
V_{conf}(r_{ij}) = -\frac{3}{8}\lambda_{i}^{c}\cdot\lambda_{j}^{c} \left( C\, r_{ij} + V_0 \right),
\label{RGMconf}
\end{equation}
\ 

\noindent and the spin-spin component of the GBE interaction in its $SU_F(3)$ form is

\begin{eqnarray}
V_\chi(\vec r_{ij})
&=&
\left\{\sum_{a=1}^3 V_{\pi}(\vec r_{ij}) \lambda_i^a \lambda_j^a \right.
\nonumber \\
&+& \left. \sum_{a=4}^7 V_{\rm K}(\vec r_{ij}) \lambda_i^a \lambda_j^a
+V_{\eta}(\vec r_{ij}) \lambda_i^8 \lambda_j^8
+V_{\eta^{\prime}}(\vec r_{ij}) \lambda_i^0 \lambda_j^0\right\}
\vec\sigma_i\cdot\vec\sigma_j.
\label{RGMvchi}
\end{eqnarray}
\

The interaction (\ref{RGMvchi}) contains $\gamma=\pi,K,\eta,$ and $\eta'$ meson exchange terms and the radial part of the exchange potential is

\begin{equation}\label{RGMmodelII}
V_{\gamma}(r) = \frac{g_{\gamma}^2}{4\pi} \frac{1}{12m_im_j} \{ \mu_{\gamma}^2 \frac{e^{- \mu_{\gamma} r}}{r} - \Lambda_{\gamma}^2 \frac{e^{- \Lambda_{\gamma} r}}{r} \} ,
\end{equation}
\ 

\noindent where $\Lambda_{\gamma}=\Lambda_0+\kappa \mu_{\gamma}$ and the parameters are

$$\frac{g_{\pi q}^2}{4\pi} = \frac{g_{\eta q}^2}{4\pi} = 1.24,\ \frac{g_{\eta' q}^2}{4\pi} = 2.7652,$$
$$m_{u,d}=340\ {\rm MeV},\ \ C=0.77\ {\rm fm}^{-2},$$
$$\mu_{\pi}=139\ {\rm MeV},\ \mu_{\eta}=547\ {\rm MeV},\ \mu_{\eta'}=958\ {\rm MeV},$$
\begin{equation}\label{RGMparam2}
\ \ \Lambda_0=5.82\ {\rm fm}^{-1},\  \kappa = 1.34,\  V_0=-112\ {\rm MeV}.
\end{equation}
\

For a system of $u$ and $d$ quarks only, as in the NN interaction, we recall that the $K$ exchange does not contribute. 
\\

\section{The resonating group method}\label{rgmsection}
\

The resonating group method \cite{WHE37} is one of the well established methods used to study the interaction between two composite systems. It allows to calculate bound state energies and scattering phase shifts. It has been first applied to nuclear physics in the study of the nucleus-nucleus interaction \cite{KAM77,WIL77}. Its application to baryon-baryon systems was initiated by Oka and Yazaki \cite{OKA80} where the quark structure of the nucleons was assumed. In a baryon-baryon system, where each baryon is a 3q cluster, it takes explicitly into account the quark interchange between the two interacting baryons. This comes from the assumption that the total wave function can be written as

\begin{equation}\label{RGMpsy}
\psi=\mathcal{A}\left[ \Phi\ \chi (\vec{R}_{AB})\right]\ ,
\end{equation}
\ 

\noindent where ${\mathcal{A}}$ is an antisymmetrization operator defined below, $\Phi$ contains the product of the internal wave functions of the interacting baryons and $\chi(\vec{R}_{AB})$ is the wave function of the relative motion, depending on the relative coordinate $\vec{R}_{AB}$ between the clusters $A$ and $B$.
\\

The internal wave function of each cluster has orbital, flavor, spin and color parts. In $\Phi$ the flavor and spin are combined to give a definite total spin $S$ and isospin $I$ so that one has

\begin{equation}\label{PHI}
\Phi=\left[ \phi_A(\vec{\xi}_A) \phi_B(\vec{\xi}_B) \right]_{SI}\ ,
\end{equation}
\ 

\noindent where $\vec{\xi}_A = (\vec{\xi}_1,\vec{\xi}_2)$ and $\vec{\xi}_B = (\vec{\xi}_3,\vec{\xi}_4)$ are the internal coordinates of the clusters $A$ and $B$ :

\begin{eqnarray}\label{JACOBI}
\vec{\xi}_1 = \vec{r}_1 - \vec{r}_2\ ,\ \ &&\ \ \vec{\xi}_3 = \vec{r}_4 - \vec{r}_5\ ,\nonumber \\
\vec{\xi}_2 = \frac{\vec{r}_1 + \vec{r}_2 - 2 \vec{r}_3}{2}\ ,\ \ && \ \ \vec{\xi}_4 = \frac{\vec{r}_4 + \vec{r}_5 - 2 \vec{r}_6}{2}\ ,\nonumber \\
\vec{R}_A   = \frac{\vec{r}_1 + \vec{r}_2 + \vec{r}_3}{3}\ ,\ \ &&\ \ \vec{R}_B = \frac{\vec{r}_4 + \vec{r}_5 + \vec{r}_6}{3}\ .
\end{eqnarray}

The internal wave functions of the clusters $\phi_i(\xi_i),\ i=A,B$ are supposed to be known (see Section \ref{subsectionnucleon}). They are totally antisymmetric 3q states in orbital, spin, flavor and color space. The color part is a $[1^3]$ singlet when describing $N$ and $\Delta$. Usually the color part of a 3q state is not written explicitly. The same statement remains valid for the $6q$ state which is a $[222]_C$ singlet in every channel. The development below is based on the assumption that $\phi_i$ has a simple $s^3$ structure which is reasonable for the non-relativistic GBE models.
\\

The antisymmetrization operator ${\mathcal{A}}$ is defined by

\begin{equation}\label{RGMa}
{\mathcal{A}}=\frac{1}{10}\left( 1-\sum\limits_{i=1}^{3}\sum\limits_{j=4}^{6}P_{ij}\right) ,
\end{equation}
\ 

\noindent where $P_{ij}$ is the permutation operator of the quarks $i$ and $j$ belonging to clusters $A(1,2,3)$ and $B(4,5,6)$ respectively. Note that exchanges of more than one particle are redundant in the present case since the exchange of three particles such as $P_{14}P_{25}P_{36}$ are interpreted as the exchange of two baryons, i; e., $P_{AB}$ and the exchanges of two quarks can be expressed as the one particle exchanges times $P_{AB}$. The operator $P_{ij}$ acts in the orbital, flavor, spin and color space, so it can be written as $P_{ij}=P_{ij}^oP_{ij}^fP_{ij}^{\sigma}P_{ij}^c$ where

\begin{equation}\label{PIJ}
P_{ij}^f=\frac{1}{2}\lambda_i^f \cdot \lambda_j^f+\frac{1}{3}\ ,~\ \ \ \ 
P_{ij}^{\sigma}=\frac{1}{2}\vec{\sigma}_i \cdot \vec{\sigma}_j+\frac{1}{2}\ ,~\ \ \ \ 
P_{ij}^c=\frac{1}{2}\lambda_i^c \cdot \lambda_j^c+\frac{1}{3}\ ,
\end{equation}
\ 

\noindent with $\lambda_i^{f(c)}$ the Gell-Mann matrices of $SU_F(3)$ ($SU_C(3)$) and $\vec\sigma_i$ the Pauli matrices.
\\

When one knows the internal wave functions $\phi_A$ and $\phi_B$ from variational principle, the equation of motion for $\chi(\vec{R}_{AB})$ can be obtained in the form

\begin{equation}\label{RGM1}
\int \phi^+(\vec{\xi}_A)\phi^+(\vec{\xi}_B)(H-E)
{\cal A} [\phi(\vec{\xi}_A)\phi(\vec{\xi}_B)\chi(\vec{R}_{AB})] d^3\xi_A d^3\xi_B = 0\ ,
\end{equation}
\ 

\noindent where $H$ is the Hamiltonian of the six-quark system. We introduce the Hamiltonian kernel

\begin{eqnarray}\label{HKERNEL}
{\cal H}(\vec{R'},\vec{R})&=&\int \phi^+(\vec{\xi}_A)\phi^+(\vec{\xi}_B) 
\delta(\vec{R'}-\vec{R}_{AB}) H {\cal A} [\phi(\vec{\xi}_A)\phi(\vec{\xi}_B) 
\delta(\vec{R}-\vec{R}_{AB})] d^3\xi_A d^3\xi_B d^3R_{AB} \nonumber \\
&=& {\cal H}^{(d)}(\vec{R})\delta(\vec{R}-\vec{R'})-{\cal H}^{(ex)}(\vec{R'},\vec{R})\ ,
\end{eqnarray}
\ 

\noindent and the normalization kernel

\begin{eqnarray}\label{NKERNEL}
{\cal N}(\vec{R'},\vec{R})&=&\int \phi^+(\vec{\xi}_A)\phi^+(\vec{\xi}_B) 
\delta(\vec{R'}-\vec{R}_{AB}) {\cal A} [\phi(\vec{\xi}_A)\phi(\vec{\xi}_B) 
\delta(\vec{R}-\vec{R}_{AB})]d^3\xi_A d^3\xi_B d^3R_{AB} \nonumber \\
&=& {\cal N}^{(d)}(\vec{R})\delta(\vec{R}-\vec{R'})-{\cal N}^{(ex)}(\vec{R'},\vec{R})\ ,
\end{eqnarray}
\ 

\noindent with ${\cal H}^{(d)}(\vec{R})$, ${\cal H}^{(ex)}(\vec{R'},\vec{R})$ and ${\cal N}^{(ex)}(\vec{R'},\vec{R})$ defined below. The direct term of the normalization kernel is ${\cal N}^{(d)}(\vec{R})=1$. 
\\

On can rewrite (\ref{RGM1}) as

\begin{equation}\label{RGMrgm}
\int {\cal L}(\vec{R'},\vec{R}) \chi(\vec{R}) d^3R = 0\ ,
\end{equation}
where

\begin{equation}
{\cal L}(\vec{R'},\vec{R}) = {\cal H}(\vec{R'},\vec{R}) - E {\cal N}(\vec{R'},\vec{R}).
\end{equation}

This is the RGM equation. 
\\

The direct term of the Hamiltonian kernel, ${\cal H}^{(d)}(\vec{R})$, consists of the relative kinetic, the relative potential and the baryon internal Hamiltonians

\begin{equation}\label{HDIRECT}
{\cal H}^{(d)}(\vec{R})=-\frac{\nabla^2_R}{2\mu}+V_{rel}^{(d)}(\vec{R})+H_{int}\ ,
\end{equation}
\ 

\noindent where
\begin{equation}
\mu=\frac{3m}{2}\ ,
\end{equation}
\ 

\noindent is the reduced mass of the clusters $A$ and $B$. Then one can write

\begin{equation}
{\cal L}(\vec{R'},\vec{R}) = [-\frac{\nabla^2_R}{2\mu}+V_{rel}^{(d)}(\vec{R})-E_{rel}]\delta(\vec{R}-\vec{R'})-[{\cal H}^{(ex)}(\vec{R'},\vec{R})-E{\cal N}^{(ex)}(\vec{R'},\vec{R})]\ ,
\end{equation}
\ 

\noindent where 
\begin{equation}
E_{rel}=E-H_{int}
\end{equation}
\ 

\noindent is the energy of the relative motion and ${\cal H}^{(ex)}(\vec{R'},\vec{R})$ and ${\cal N}^{(ex)}(\vec{R'},\vec{R})$ are the contribution to ${\cal L}(\vec{R'},\vec{R})$ with $\vec{R'}\neq \vec{R}$..
\\

There are two important steps in solving this equation. One is to calculate the Hamiltonian kernel (\ref{HKERNEL}) by reducing the six-body matrix elements to two-body matrix elements. This is discussed in Section \ref{sectionRGMsixbody}. Another step is the discretization of the RGM equation. It is important both for bound and scattering states. The discretization has been performed by using the method of Ref. \cite{KAM77} and detailed later in this chapter. First we discuss the orbital part of the nucleon wave function.
\\

\subsection{The nucleon wave function}\label{subsectionnucleon}
\

In the RGM approach the wave function of the ground state nucleon must be known. As indicated above, its orbital part $\phi$ is supposed to have an $s^3$ structure.  This is a very good approximation to the exact wave function as seen in Chapter \ref{preliminarychapter}. The function $\phi$ is fully symmetric with respect to any permutation of $S_3$ and is chosen of the form
\begin{equation}\label{NUCLEONGROUND}
\phi=\prod_{i=1}^3 g(\vec{r}_i,\beta)\ ,
\end{equation}
with $g(\vec{r}_i,\beta)$ given by

\begin{equation}\label{QUARKGAUSS}
g(\vec{r},\beta)=(\frac{1}{\pi \beta^2})^{3/4} e^{-\frac{r^2}{2 \beta^2}}\ .
\end{equation}
\\

The size parameter $\beta$ appearing in (\ref{QUARKGAUSS}) is chosen such as the lowest configurations, $s^3[3]_O$ (Eq. \ref{NUCLEONGROUND}) and $sp^2[3]_O$, be minimally mixed in the nucleon wave function. This implies that the state $|N>$ is stable against the breathing mode oscillations. To this end we impose the additional condition $<N|H|N^*>=0$ which plays the same role as also does the so-called stability condition (see Ref.\cite{OKA80})

\begin{equation}\label{STABILITY}
\frac{\partial}{\partial \beta}\langle \phi|H|\phi\rangle =0\ ,
\end{equation}
\ 

\noindent where $H$ is the Hamiltonian (\ref{RGMhamiltonian}) written for a 3q system. This condition gives $\beta=0.437$ fm, which we shall use below. 
\\

\subsection{Bound states}\label{RGMsubsectionbound}
\

Here we describe the discretization procedure directly applicable to bound states. According to Ref. \cite{KAM77}, the relative wave function $\chi (\vec{R})$ has been expanded over a finite number of Gaussians $\chi_i$ centered at $\vec{R}_i$ ($i = 1,2,...,N$) where $R_i$ are points, equally spaced or not, between the origin and some value of $R$ depending on the range of the interaction. The expansion is

\begin{equation}\label{KAM1}
\chi(\vec{R})=\sum_{i=1}^N C_i \chi_i(\vec{R})\ ,
\end{equation}
\ 

\noindent with

\begin{equation}\label{KAM1b}
\chi_i(\vec{R})=g(\vec{R}-\vec{R_i},\sqrt{2/3 \beta})=(\frac{3}{2 \pi \beta^2})^{3/4} e^{-\frac{3}{4 \beta^2}(\vec{R}-\vec{R}_i)^2}\ . 
\end{equation}
\

If $g(\vec{r},\beta)$ is the quark normalized Gaussian wave function (\ref{QUARKGAUSS}), from the Jacobi transformations (\ref{JACOBI}) it follows that the relative wave function has to be expanded in terms of the Gaussians (\ref{KAM1b}) with the size parameter $\sqrt{2/3} \beta$.
\\

This method can be applied straightforwardly to the bound state problem. The modification necessary for treating the scattering problem will be explained later, in the next subsection. The binding energy $E$ and the expansion coefficients $C_i$ are given by the eigenvalues and eigenvectors of the following equation :

\begin{equation}\label{GCM}
\sum_{j=1}^N H_{ij}C_j = E \sum_{j=1}^N N_{ij}C_j\ ,
\end{equation}
\ 

\noindent where $N$ is the number of Gaussians considered in (\ref{KAM1}). The matrices

\begin{eqnarray}
H_{ij}&=&\int \phi^+(\vec{\xi}_A)\phi^+(\vec{\xi}_B) \chi(\vec{R}_{AB}-\vec{R}_i) H (1-{\cal A}') [\phi(\vec{\xi}_A)\phi(\vec{\xi}_B) \chi(\vec{R}_{AB}-\vec{R}_j)] d^3\xi_A d^3\xi_B d^3R_{AB}\ ,\nonumber \\
\end{eqnarray}
\ 

\noindent and

\begin{eqnarray}
N_{ij}&=&\int \phi^+(\vec{\xi}_A)\phi^+(\vec{\xi}_B) \chi(\vec{R}_{AB}-\vec{R}_i) (1-{\cal A}') [\phi(\vec{\xi}_A)\phi(\vec{\xi}_B)  \chi(\vec{R}_{AB}-\vec{R}_j)] d^3\xi_A d^3\xi_B d^3R_{AB}\ ,\nonumber \\
\end{eqnarray}
\ 

\noindent are obtained from (\ref{HKERNEL}) and (\ref{NKERNEL}) respectively. By including the center of mass coordinate $\vec{R}_G=(\vec{R}_A+\vec{R}_B)/2$ with the centre of mass wave function normalized as

\begin{equation}
\int  g(\vec{R}_G,\sqrt{1/6} \beta) g(\vec{R}_G,\sqrt{1/6} \beta) d^3R_G = 1,
\end{equation}
\ 

\noindent and transforming back from the Jacobi coordinates to $\vec{r}_i\ (i=1,...,6)$, we get the following formulas

\begin{eqnarray}
H_{ij}&=&\int \prod_{k=1}^3 \phi^+(\vec{r}_k-\frac{\vec{R}_i}{2}) \prod_{k'=4}^6 \phi^+(\vec{r}_{k'}+\frac{\vec{R}_i}{2}) H {\cal A} [\prod_{l=1}^3\phi(\vec{r}_l-\frac{\vec{R}_j}{2}) \prod_{l'=4}^6\phi(\vec{r}_{l'}+\frac{\vec{R}_j}{2})] d^3r_1 ...  d^3r_6\ ,\nonumber \\
\end{eqnarray}
\ 

\noindent and

\begin{eqnarray}
N_{ij}&=&\int \prod_{k=1}^3 \phi^+(\vec{r}_k-\frac{\vec{R}_i}{2}) \prod_{k'=4}^6 \phi^+(\vec{r}_{k'}+\frac{\vec{R}_i}{2}) {\cal A} [\prod_{l=1}^3\phi(\vec{r}_l-\frac{\vec{R}_j}{2}) \prod_{l'=4}^6\phi(\vec{r}_{l'}+\frac{\vec{R}_j}{2})] d^3r_1 ...  d^3r_6\ ,\nonumber  \\
\end{eqnarray}
\ 

\noindent with $\phi(\vec{r})\equiv g(\vec{r},\beta)$ given by (\ref{QUARKGAUSS}). These forms are much easier to handle in actual calculations. They allow to reduce the 6q matrix elements to two-body matrix elements. Moreover the distances $R_i$ play now the role of a generator coordinate \cite{SHI00} and lead to a better understanding of the relation between the resonating group method and the generator coordinate method \cite{CVE83}.
\\

\subsection{Scattering states}
\

For scattering states the expansion (\ref{KAM1}) holds up to a finite distance $R=R_c$, depending on the range of the interaction. Beyond $R_c$, $\chi(\vec{R})$ becomes the usual combination of Hankel functions containing the $S$-matrix. Because practical calculations of both bound state and scattering states are done in terms of partial waves, we first give the partial wave expansion of Eq. (\ref{KAM1}) in terms of locally peaked wave functions with a definite angular momentum $l$ and projection $m$ :

\begin{equation}\label{KAM3}
\chi_{lm}(\vec{R})=\sum_{i=1}^N C_i^{(l)} \chi_i^{(l)}(R)Y_{lm}(\hat{R})\ ,
\end{equation}
\ 

\noindent with the explicit form of $\chi_i^{(l)}$ given by

\begin{equation}\label{KAM4}
\chi_i^{(l)}(R)=4 \pi (\frac{3}{2 \pi b^2})^{3/4} e^{-\frac{3}{4 b^2}(R^2+R_i^2)}i_l(\frac{3}{2 b^2}R R_i)\ ,
\end{equation}
\ 

\noindent where $i_l$ is the modified spherical Bessel function \cite{ABR64}. When we treat the scattering problem, the form (\ref{KAM4}) holds up to $R\leq R_c$ only. In fact in this case the relative wave function is expanded in terms of ${\tilde{\chi}}^{(l)}$ as

\begin{equation}\label{KAM5}
\chi^{(l)}(R)=\sum_{i=1}^N C_i^{(l)}{\tilde{\chi}}_i^{(l)}(R)\ ,
\end{equation}
\ 

\noindent where

\begin{eqnarray}\label{RGMchiT}
{\tilde{\chi}}_i^{(l)}(R) = &\alpha_i^{(l)} \chi_i^{(l)} (R)\ ,\ \ \ \ \ \ \ \ \ \ \ \ \ \ \ \ \ \ \ \ &(R \leq R_c) \nonumber \\
{\tilde{\chi}}_i^{(l)}(R) = &h_l^{(-)}(kR) + S_i^{(l)}  h_l^{(+)} (kR)\ ,\ \ &(R \geq R_c)
\end{eqnarray}
\ 

\noindent with $\chi_i^{(l)}(R)$ defined by Eq. (\ref{KAM4}). Note that $\bar{\chi}_i^{(l)}(R)$ and $\chi_i^{(l)}(R)$ are proportional for $R\leq R_c$ only. Here $k$ is the wave number $k=\sqrt{2\mu E_{rel}/\hbar^2}$ and $h_l^{(-)}$ and $h_l^{(+)}$ are spherical Hankel functions \cite{ABR64}. The coefficients $\alpha_i^{(l)}$ and $S_i^{(l)}$ are determined from the continuity of ${\tilde{\chi}}_i^{(l)}$ and its derivative at $R=R_c$. The coefficients $C_i^{(l)}$ of (\ref{KAM3}) are normalized such that $\sum_{i=1}^N C_i^{(l)} = 1$. Then the $S$-matrix is given in terms of the coefficients $C_i^{(l)}$ as

\begin{equation}\label{KAM6}
S^{(l)}=\sum_{i=1}^N C_i^{(l)} S_i^{(l)}\ . 
\end{equation}
\

The method of determining the expansion coefficients is based on a functional approach of Oka and Yazaki \cite{OKA80}. One defines the functional $J$ by

\begin{equation}\label{RGMfunctional}
J[\chi^l]=S^{(l)}+i\frac{3m}{2k}\int \chi^{(l)}(R'){\cal L}^{(l)}(R',R) \chi^{(l)}(R) dR'dR\ ,
\end{equation}
\ 

\noindent where $m$ is the quark mass.
\\

Noting that $C_{N}$ is eliminated by the condition $\sum_{i=1}^N C_i=1$, we make $J$ stationary with respect to the variation of $C_i$'s ($i=1,2,...,N-1$), to obtain \cite{KAM77}

\begin{equation}\label{RGMfuncLT}
\sum_{j=1}^{N-1}\tilde{\cal L}^{(l)}_{ij}C^{(l)}_j={\cal M}^{(l)}_i
\end{equation}
\ 

\noindent with
\begin{equation}\label{RGMfuncL}
\tilde{\cal L}^{(l)}_{ij}={\cal K}^{(l)}_{ij} - {\cal K}^{(l)}_{iN} - {\cal K}^{(l)}_{Nj} + {\cal K}^{(l)}_{NN},
\end{equation}
\ 

\noindent and
\begin{equation}\label{RGMfuncM}
{\cal M}^{(l)}_i={\cal K}^{(l)}_{NN} - {\cal K}^{(l)}_{iN},
\end{equation}
\ 

\noindent where
\begin{equation}\label{RGMfuncK}
{\cal K}^{(l)}_{ij}=\int \tilde{\chi}_i^{(l)}(R'){\cal L}^{(l)}(R',R) \tilde{\chi}_j^{(l)}(R) dR'dR\ ,
\end{equation}
\ 

\noindent is calculated using Eq. (\ref{RGMchiT}). Note that this integral is quite tricky because of the difference between $\tilde{\chi}^{(l)}$ and $\chi^{(l)}$ in the outer part ($R>R_c$).
\\

Finally we give the convergence criterium of the scattering problem in the framework of the resonating group method. This criterium, which we use in practical calculation, has been given by Kamimura \cite{KAM77} in the following form

\begin{equation}\label{RGMcriteria}
\left\| \frac{3m}{2k}\sum_{i=1}^{N-1} {\cal K}^{(l)}_{Ni} C^{(l)}_i \right\| \ll 1.
\end{equation}
\ \\

This will be discussed in Section \ref{sectionnumerical}.
\\

\subsection{Coupled channels}\label{subsectionCC}
\

The ansatz (\ref{RGMpsy}) corresponds to the one channel approximation of RGM. Here we consider more than one channel in order to look to the effect of extra channels on the phase shift. We start from the observation that the function (\ref{RGMpsy}) can be written in terms of shell model configurations with a fixed number $N$ of excitation quanta. The most important shell model configurations for the NN problem in the framework of the GBE model have been introduced in the previous chapter in Eq. (\ref{PREbasis}). In terms of these shell model configurations, one has for example for $\left| s^6[6]_O[33]_{FS}\right>$ the identity

\begin{equation}\label{RGMs6}
\mathcal{A} \left[ \phi_N \phi_N \chi (\vec{R}_{NN})_{0s} \right]_{FS}=\sqrt{\frac{10}{9}} \left| s^6[6]_O[33]_{FS}\right>\ ,
\end{equation}
\ 

\noindent where $\chi_{0,s}(\vec{R})$ denotes the 0s-wave harmonic oscillator function. Remember that it is always assumed that the center of mass motion is removed from the shell-model wave function.
\\

However, the only contribution from the $s^4p^2$ shell comes from a fixed superposition of symmetry states which allows (\ref{RGMpsy}) to be rewritten as 

\begin{eqnarray}\label{RGMs4p2}
\mathcal{A} \left[ \phi_N \phi_N \chi (\vec{R}_{NN})_{2s} \right]_{FS}=&&\frac{3\sqrt{2}}{9} \left|(\sqrt{\frac{5}{6}}s^5 2s - \sqrt{\frac{1}{6}}s^4p^2)[6]_O[33]_{FS}\right> \nonumber \\
&&-\frac{4\sqrt{2}}{9}\left|s^4 p^2[42]_O[33]_{FS}\right> \nonumber \\
&&-\frac{4\sqrt{2}}{9}\left|s^4 p^2[42]_O[51]_{FS}\right>\ .
\end{eqnarray}
\ 

Note that the left hand side contains the 2s-relative motion wave function. This shows that each channel of the form (\ref{RGMpsy}) represents a given truncation of the shell-model space. Therefore a multi-channel treatment is desirable. Extending the ansatz (\ref{RGMpsy}) to include, besides $NN$, two new channels, the $\Delta\Delta$ and a particular ``hidden colour'' channel $CC$ defined explicitly later, Shimizu {\it et al.} \cite{SHI00b} demonstrated that these three channels become linearly independent. This provides a Hilbert space where the three compact shell-model configurations $\left|(\sqrt{\frac{5}{6}}s^5 2s - \sqrt{\frac{1}{6}}s^4p^2)[6]_O[33]_{FS}\right>$, $\left|s^4 p^2[42]_O[33]_{FS}\right>$ and $\left|s^4 p^2[42]_O[51]_{FS}\right>$ are relaxed as compared to (\ref{RGMs4p2}) and can participate as independent variational configurations when one applies the coupled channel RGM. Note that the other possible compact 6q configurations from the $s^4p^2$ shell, such as $[411]_{FS}$, $[321]_{FS}$ and  $[2211]_{FS}$ are not taken into account, but we have already seen in Sections \ref{sectionCasimir} and \ref{sectionCluster0} that they play only a minor role when one uses the GBE interaction.
\\

In the coupled channels RGM, the total wave function (\ref{RGMpsy}) becomes

\begin{equation}\label{RGMpsyCC}
\psi=\sum\limits_{\beta}^{}{\mathcal{A}}\left[\Phi_{\beta} \chi_{\beta}(\vec{R}_{AB})\right]\ ,
\end{equation}
\ 

\noindent where $\beta$ is a specific channel (here $\beta=NN$, $\Delta\Delta$ or $CC$), ${\mathcal{A}}$ is the antisymmetrization operator defined above, $\Phi_{\beta}$ contains the product of internal wave functions of the interacting baryons and $\chi_{\beta}(\vec{R}_{AB})$ is the wave function of the relative motion in the channel $\beta$.
\\

In $\Phi_{\beta}$ the flavor and spin are combined to give a definite total spin $S$ and isospin $I$ as indicated by (\ref{PHI}), carrying now an extra index $\beta$ on the left hand side to specify a given channel. Here the functions $\phi_i(\xi_i),\ i=A,B$ are still totally antisymmetric 3q states in orbital, spin, flavor and color space. The color part is a $[1^3]_C$ singlet for $N$ and $\Delta$ states and a $[21]_C$ octet for $C$ states.
\\

In this case, based on Eq. (\ref{RGMpsyCC}), the RGM equation becomes a system of coupled channel equations for $\chi_{\beta}$

\begin{eqnarray}
\sum_{\beta}\int{\cal L}_{\alpha\beta}(\vec{R'},\vec{R}) \chi_{\beta}(\vec{R}) d^3R &=& \sum_{\beta}\int [{\cal H}_{\alpha\beta}(\vec{R'},\vec{R})-E {\cal N}_{\alpha\beta}(\vec{R'},\vec{R})] \chi_{\beta}(\vec{R}) d^3R =0\ .\nonumber \\
\end{eqnarray}

Usually the normalization kernel ${\cal N}_{\alpha\beta}$ is not diagonal because of the antisymmetrization. For a given $SI$ sector one can establish which are the 6q states of (\ref{PHI}) allowed by the Pauli principle \cite{HAR81}. If we consider the $l = 0$ partial waves {\it i. e.} we study the $^3S_1$ and $^1S_0$ phase shifts, then according to \cite{HAR81}, the 6q allowed states are $NN, \Delta\Delta$ and $CC$. The $NN$ and $\Delta\Delta$ states are easy to define directly from Eq. (\ref{RGMpsyCC}). For $CC$ states we adopt the definition of Ref. \cite{FAE82} which is more appropriate for RGM calculations. This $CC$ state, composed of six quarks, allows some ``color polarization'' of the 6q system in the interaction region.
\\

First, because we assume that the orbital part of each nucleon has $[3]_O$ symmetry, the orbital part of the two-nucleon system in a relative s-state is either $[6]_O$ or $[42]_O$ as first discussed by Neudatchin {\it et al.} \cite{KRA85,NEU75,NEU77,NEU91}. Since the total wave function is antisymmetrized in the RGM calculation, the state which has $[6]_O$ symmetry in the orbital space should be antisymmetric in spin, isospin and color space. Thus the projection onto the orbital symmetry $[6]_O$ is equivalent to the projection onto an antisymmetric state of spin, isospin and color since the relative wave function contains an orbital part only. That is why we introduce the $CC$ state as the linear combination

\begin{equation}\label{RGMCCgeneral}
|CC\rangle = \alpha |NN\rangle +\beta |\Delta\Delta\rangle +\gamma {\cal A}_{\sigma f c}|\Delta\Delta\rangle\ ,
\end{equation}
\ 

\noindent with the operator ${\cal A}_{\sigma f c}$ defined as

\begin{equation}
{\cal A}_{\sigma f c}=\frac{1}{10}[1-\sum_{i=1}^3\sum_{j=4}^6P_{ij}^{\sigma}P_{ij}^{f}P_{ij}^{c}]\ ,
\end{equation}
\ 

\noindent where $P_{ij}^{\sigma}$,$P_{ij}^{f}$ and $P_{ij}^{c}$ are the exchange operators in the spin, flavor and color space respectively defined by (\ref{PIJ}). From the orthonormality conditions $\langle CC|CC\rangle =1$, $\langle CC|NN\rangle =0$ and $\langle CC|\Delta\Delta\rangle =0$ one can determine the coefficients $\alpha$, $\beta$ and $\gamma$ so that (see Appendix \ref{appendixCOUPLED})

\begin{equation}\label{HIDDEN}
|CC\rangle = -\frac{\sqrt{5}}{6} |NN\rangle +\frac{1}{3} |\Delta\Delta\rangle -\frac{15}{4} {\cal A}_{\sigma f c}|\Delta\Delta\rangle\ .
\end{equation}
\

The important feature in the definition (\ref{RGMCCgeneral}) of the $CC$-state is that the eigenvalue of the color $SU(3)$ Casimir operator is equal to 12 for each 3q cluster. This tells us that $C$ is a color octet state and thus explains why we call the $CC$-state a hidden color state. It must be stressed that $C$ does not have a definite spin and isospin but $CC$ does. Note that at zero separation between quarks (shell model basis) the $CC$ state above is the same as that introduced by Harvey. The two differ only at finite separation distances. To see the identity with Harvey's $CC$ state \cite{HAR81} at zero separation one can combine it with the $NN$ and $\Delta\Delta$ states as defined by Eq. (\ref{RGMpsyCC}) to get symmetry states of the form $| [f]_{FS} [222]_C ; {\tilde{g}}_{FSC}\rangle$ where ${\tilde{g}}$ is the representation resulting from the inner product of $[f]_{FS}$ and $ [222]_C$ which is conjugate with the representation $g$ of an orbital state such as to produce a totally antisymmetric 6q state. Comparing Table 3 of Ref. \cite{FAE82} with that of Harvey's \cite{HAR81} Table 1 one can see that the coefficients of this basis transformation are identical which proves the identity of the hidden color state (\ref{HIDDEN}) with that of Harvey at $R=0$. Note that Harvey's definition \cite{HAR81} of $CC$ is more appropriate for the generator coordinate method than for RGM calculations.
\\

\subsection{Six-body matrix elements}\label{sectionRGMsixbody}
\

The method to compute the necessary six-body matrix elements is explained here in some details, using the very simple example of $SI=10$ case. In this case the problem is decoupled in three independent parts, namely the orbital part, the spin-flavor part and the color part. First we shall present the method used to derive the spin-flavor contribution. Next the orbital part will be presented. The details on the color part are given in the Appendix \ref{appendixCOUPLED}.
\\

We know that for the nucleon, the spin-flavor wave function is given by

\begin{equation}
\psi_N=\frac{1}{\sqrt{2}}[\chi^{\rho}\phi^{\rho}+\chi^{\lambda}\phi^{\lambda}]\ ,
\end{equation}
\ 

\noindent where $\chi$ and $\phi$ are the spin and flavor parts respectively. For the spin parts we have

\begin{eqnarray}\label{PSINCHI}
\chi^{\rho}_{1/2}&=&\frac{1}{\sqrt{2}}(\uparrow \downarrow \uparrow - \downarrow \uparrow \uparrow)\ ,\nonumber \\
\chi^{\rho}_{-1/2}&=&\frac{1}{\sqrt{2}}(\uparrow \downarrow \downarrow - \downarrow \uparrow \downarrow)\ ,\nonumber \\
\chi^{\lambda}_{1/2}&=&\frac{1}{\sqrt{6}}(\uparrow \downarrow \uparrow + \downarrow \uparrow \uparrow - 2 \uparrow \uparrow \downarrow)\ ,\nonumber \\
\chi^{\lambda}_{-1/2}&=&\frac{-1}{\sqrt{6}}(\uparrow \downarrow \downarrow + \downarrow \uparrow \downarrow -2 \downarrow \downarrow \uparrow )\ ,
\end{eqnarray}
\ 

\noindent and similarly for the flavor parts with $\uparrow$ replaced by $u$ and $\downarrow$ replaced by $d$. Then for the channel $\beta=NN$, the definition (\ref{PHI}) becomes

\begin{equation}\label{PHISI}
\Phi_{NN}^{SI}=\frac{1}{2}\sum C^{\frac{1}{2} \frac{1}{2} S}_{s_1 s_2 s}C^{\frac{1}{2} \frac{1}{2} I}_{\tau_1 \tau_2 \tau} [\chi^{\rho}_{s_1}(1)\phi^{\rho}_{\tau_1}(1)+\chi^{\lambda}_{s_1}(1)\phi^{\lambda}_{\tau_1}(1)][\chi^{\rho}_{s_2}(2)\phi^{\rho}_{\tau_2}(2)+\chi^{\lambda}_{s_2}(2)\phi^{\lambda}_{\tau_2}(2)]\ ,
\end{equation}
\ 

\noindent where $S$ and $I$ are the spin and isospin of the NN system. $\chi(i)$ and $\phi(i)$ are the spin and flavor parts of the $i{\rm^{th}}$ nucleon. For $S=S_z=1$ and $I=I_z=0$, after inserting the values of the corresponding Clebsch-Gordan coefficients we have

\begin{eqnarray}\label{PHISI10}
\Phi_{NN}^{10}&=&\frac{1}{2\sqrt{2}} \{[\chi^{\rho}_{1/2}(1)\phi^{\rho}_{1/2}(1)+\chi^{\lambda}_{1/2}(1)\phi^{\lambda}_{1/2}(1)][\chi^{\rho}_{1/2}(2)\phi^{\rho}_{-1/2}(2)+\chi^{\lambda}_{1/2}(2)\phi^{\lambda}_{-1/2}(2)]\nonumber \\
& & -[\chi^{\rho}_{1/2}(1)\phi^{\rho}_{-1/2}(1)+\chi^{\lambda}_{1/2}(1)\phi^{\lambda}_{-1/2}(1)][\chi^{\rho}_{1/2}(2)\phi^{\rho}_{1/2}(2)+\chi^{\lambda}_{1/2}(2)\phi^{\lambda}_{1/2}(2)] \}\ .\nonumber \\
\end{eqnarray}
\

At this stage we use MATHEMATICA \cite{WOL96}. We introduce the expressions (\ref{PSINCHI}) in (\ref{PHISI10}) and the equivalents for the flavor parts. We get a huge expression with 338 terms depending now on the quantum numbers of the quarks. For an operator $O$ we then get a linear combination of $338^2=114244$ matrix elements of the form

\begin{equation}
\langle s_1s_2s_3s_4s_5s_6\tau_1\tau_2\tau_3\tau_4\tau_5\tau_6|O|s_1's_2's_3's_4's_5's_6'\tau_1'\tau_2'\tau_3'\tau_4'\tau_5'\tau_6'\rangle\ ,
\end{equation}
\ 

\noindent where $s_i$ and $\tau_i$ ($i=1,\ldots,6$) stand for the spin and isospin projection of the $i{\rm^{th}}$ quark. Note that the normal order of particles is implied. Now let us choose for example $O = \vec{\sigma}_1\cdot\vec{\sigma}_3~\vec{\lambda}_1^f\cdot\vec{\lambda}_3^f ~P_{36}^{\sigma f}$, which contains the permutation $P_{36}$. Then we have

\begin{eqnarray}
&&\langle s_1s_2s_3s_4s_5s_6\tau_1\tau_2\tau_3\tau_4\tau_5\tau_6|\vec{\sigma}_1\cdot\vec{\sigma}_3\vec{\lambda}_1^f\cdot\vec{\lambda}_3^f P_{36}^{\sigma f}|s_1's_2's_3's_4's_5's_6'\tau_1'\tau_2'\tau_3'\tau_4'\tau_5'\tau_6'\rangle \nonumber \\
&&= \langle s_1s_2s_3s_4s_5s_6\tau_1\tau_2\tau_3\tau_4\tau_5\tau_6|\vec{\sigma}_1\cdot\vec{\sigma}_3\vec{\lambda}_1^f\cdot\vec{\lambda}_3^f|s_1's_2's_6's_4's_5's_3'\tau_1'\tau_2'\tau_6'\tau_4'\tau_5'\tau_3'\rangle\nonumber \\
&&= \langle s_1s_3\tau_1\tau_3|\vec{\sigma}_1\cdot\vec{\sigma}_3\vec{\lambda}_1^f\cdot\vec{\lambda}_3^f|s_1's_6'\tau_1'\tau_6'\rangle \ \delta_{s_2}^{s_2'}\delta_{s_4}^{s_4'}\delta_{s_5}^{s_5'}\delta_{s_6}^{s_3'}\delta_{\tau_2}^{\tau_2'}\delta_{\tau_4}^{\tau_4'}\delta_{\tau_5}^{\tau_5'}\delta_{\tau_6}^{\tau_3'}\nonumber \\
&&= \langle s_1s_3|\vec{\sigma}_1\cdot\vec{\sigma}_3|s_1's_6' \rangle \langle \tau_1\tau_3|\vec{\lambda}_1^f\cdot\vec{\lambda}_3^f|\tau_1'\tau_6' \rangle \ \delta_{s_2}^{s_2'}\delta_{s_4}^{s_4'}\delta_{s_5}^{s_5'}\delta_{s_6}^{s_3'}\delta_{\tau_2}^{\tau_2'}\delta_{\tau_4}^{\tau_4'}\delta_{\tau_5}^{\tau_5'}\delta_{\tau_6}^{\tau_3'}\ .
\end{eqnarray}
\

This shows how a six-body matrix element can be reduced to the calculation of two-body matrix elements. The necessary nonzero two-body matrix elements are

\begin{eqnarray}
\langle \uparrow \uparrow|\vec{\sigma}_1\cdot\vec{\sigma}_2|\uparrow \uparrow \rangle =\langle \downarrow \downarrow|\vec{\sigma}_1\cdot\vec{\sigma}_2|\downarrow \downarrow \rangle &=&1\ ,\nonumber \\
\langle \uparrow \downarrow|\vec{\sigma}_1\cdot\vec{\sigma}_2|\uparrow \downarrow \rangle =\langle \downarrow \uparrow|\vec{\sigma}_1\cdot\vec{\sigma}_2|\downarrow \uparrow \rangle &=&-1\ ,\nonumber \\
\langle \uparrow \downarrow|\vec{\sigma}_1\cdot\vec{\sigma}_2|\downarrow \uparrow \rangle =\langle \downarrow \uparrow|\vec{\sigma}_1\cdot\vec{\sigma}_2|\uparrow \downarrow \rangle &=&2\ ,\nonumber \\
\langle uu|\vec{\lambda}_1^f\cdot\vec{\lambda}_2^f|uu \rangle =\langle dd|\vec{\lambda}_1^f\cdot\vec{\lambda}_2^f|dd \rangle &=&4/3\ ,\nonumber \\
\langle ud|\vec{\lambda}_1^f\cdot\vec{\lambda}_2^f|ud \rangle =\langle du|\vec{\lambda}_1^f\cdot\vec{\lambda}_2^f|du \rangle &=&-2/3\ ,\nonumber \\
\langle ud|\vec{\lambda}_1^f\cdot\vec{\lambda}_2^f|du \rangle =\langle du|\vec{\lambda}_1^f\cdot\vec{\lambda}_2^f|ud \rangle &=&2\ .
\end{eqnarray}

MATHEMATICA is then used to compute systematically and analytically the sum of the 114244 terms stemming from Eq. (\ref{PHISI10}).
\\

The same procedure is used for all the other matrix elements such as the kinetic energy and the confinement interaction, for the diagonal ($\Delta\Delta$, $CC$) and off-diagonal matrix elements of the coupled channels RGM as well. Details for the coupled channels matrix elements can be found in Appendix \ref{appendixCOUPLED}.
\\

In Tables \ref{RGMtable10} and \ref{RGMtable01} we give the results for the diagonal and off-diagonal matrix elements of the channels $NN$, $\Delta\Delta$ and $CC$ to be used in coupled or uncoupled channel calculations of the $^3S_1$ and $^1S_0$ phase shifts respectively. Although we apply the SU(3) version of the GBE model the matrix elements of $\sigma_{i} \cdot \sigma_{j}~\tau_{i} \cdot \tau_{j}$ and $\sigma_{i} \cdot \sigma_{j}~\tau_{i} \cdot \tau_{j} P^{f \sigma c}_{36}$ needed in SU(2) calculations are also indicated. In fact they are used in calculating the expectation value of $\sigma_{i} \cdot \sigma_{j}~ \lambda^8_i \cdot \lambda^8_j$ by subtracting them from $\sigma_{i} \cdot \sigma_{j}~\lambda^f_i \cdot \lambda^f_j$ because there is no $K$-meson exchange. Moreover, the values we found for $\sigma_{i} \cdot \sigma_{j}~\tau_{i} \cdot \tau_{j}$ and $\sigma_{i} \cdot \sigma_{j}~\tau_{i} \cdot \tau_{j} P^{f \sigma c}_{36}$ can be considered as a validity test of our method because they are in full agreement with Table 1 of Ref. \cite{SHI84}.
\\

\begin{figure}[H]
\begin{center}
\includegraphics[width=13cm]{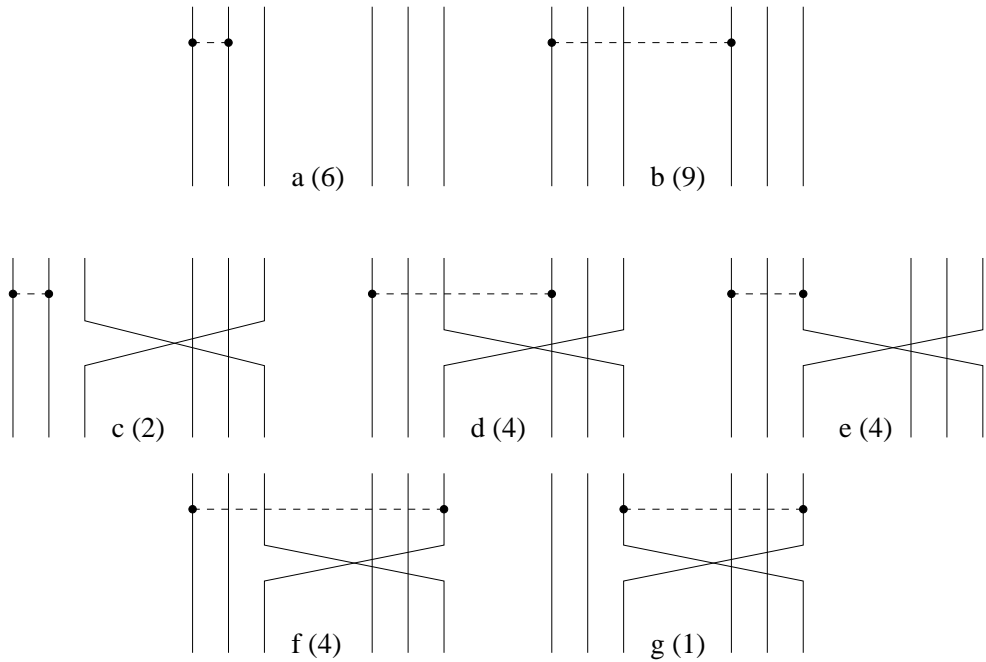}
\end{center}
\caption{\label{RGMinteractiondiagram} The different diagrams associated to two-quark interaction. The diagram $a$ and $b$ correspond to interaction without quark permutation, the other diagrams represent the contribution of the exchange terms. In brackets, next to the label $a$, ... $g$, we reproduce the multiplicity of each diagram.}
\end{figure}

Next, we have to compute the two-body matrix element of the orbital part. We decompose the antisymmetrization operator ${\mathcal{A}}$ by factorizing the permutation operator $P_{36}^{o \sigma f c}$ into its orbital $P_{36}^{o}$ and spin-flavor-color $P_{36}^{\sigma f c}$ part. We then obtain seven distinct ways of applying the permutation operator $P_{36}$, illustrated in Fig. \ref{RGMinteractiondiagram}. The first two diagrams come from the antisymmetrization operator ${\mathcal{A}}$ of (\ref{RGMa}) without permutation (no quark exchange), the others with the permutation of the quarks. By permutation of the quarks we mean that two quarks are interchanged between the two clusters during the interaction. In Fig. \ref{RGMinteractiondiagram} next to its label, the multiplicity of each diagram is indicated : 6 and 9 for diagrams without exchange and 2, 4, 4, 4, 1 for diagrams with exchange. Note that the total number of diagrams has to be $n(n-1)/2$ (namely 15 for the 6q problem) both with or without exchange.
\\

All the details of the calculation of theses diagrams are given in the appendices \ref{appendixNORM}-\ref{appendixTENSOR}. We summarize the results in Table \ref{RGMtableorbital}, where the expression of $D$, $E$ and $V$ are given by
\begin{eqnarray}\label{RGMDEV}
D(\vec{R}_i,\vec{R}_j)&=& e^{-\frac{(\vec{R}_i-\vec{R}_j)^2}{16\beta^2}},\nonumber \\
E(\vec{R}_i,\vec{R}_j)&=& D(\vec{R}_i,-\vec{R}_j)\ =\ e^{-\frac{(\vec{R}_i+\vec{R}_j)^2}{16\beta^2}},\nonumber \\
V(\vec{a})&=&\left( \frac{1}{\sqrt{2\pi} \beta}\right)^3 \int e^{-\frac{1}{2\beta^2}(\vec{r}-\frac{\vec{a}}{2})^2}V(\vec{r}) d^3r
\end{eqnarray}
\ 

\noindent and where $\vec{a}$ is defined in the last column of Table \ref{RGMtableorbital}
\\

\begin{table}[H]
\centering

\begin{tabular}{|c|l|c|}
\hline
\rule[-4mm]{0mm}{9.5mm}Diagram & Orbital matrix element & $\vec{a}$ \\

\hline
\hline

\rule[-3mm]{0mm}{8mm}$a$ & $D^6(\vec{R}_i,\vec{R}_j)\ V(\vec{a}_{12})$  & $0$\\
\rule[-3mm]{0mm}{5mm}$b$ & $D^6(\vec{R}_i,\vec{R}_j)\ V(\vec{a}_{14})$ & $\vec{R}_i+\vec{R}_j$\\
\rule[-3mm]{0mm}{5mm}$c$ & $D^4(\vec{R}_i,\vec{R}_j)\ E^2(\vec{R}_i,\vec{R}_j)\ V(\vec{a}_{12})$ & $0$\\
\rule[-3mm]{0mm}{5mm}$d$ & $D^4(\vec{R}_i,\vec{R}_j)\ E^2(\vec{R}_i,\vec{R}_j)\ V(\vec{a}_{14})$ & $\vec{R}_i+\vec{R}_j$ \\
\rule[-3mm]{0mm}{5mm}$e$ & $D^4(\vec{R}_i,\vec{R}_j)\ E^2(\vec{R}_i,\vec{R}_j)\ V(\vec{a}_{13})$ & $\vec{R}_j$\\
\rule[-3mm]{0mm}{5mm}$f$ & $D^4(\vec{R}_i,\vec{R}_j)\ E^2(\vec{R}_i,\vec{R}_j)\ V(\vec{a}_{16})$ & $\vec{R}_i$\\
\rule[-3mm]{0mm}{5mm}$g$ & $D^4(\vec{R}_i,\vec{R}_j)\ E^2(\vec{R}_i,\vec{R}_j)\ V(\vec{a}_{36})$ & $\vec{R}_i-\vec{R}_j$\\

\hline
\end{tabular}
\caption{Matrix elements of the different diagrams of the Fig. \ref{RGMinteractiondiagram}.}\label{RGMtableorbital}

\end{table}
\ 

\ \\

\begin{table}[H]
\centering

\begin{tabular}{|c|c|c|c|c|c|c|}
\hline
\rule[-4mm]{0mm}{5mm}$\alpha$ & $NN$ & $NN$  & $\Delta\Delta$ & $NN$ & $\Delta\Delta$ & $CC$ \\
$\beta$  & $NN$ & $\Delta\Delta$ & $\Delta\Delta$ & $CC$ & $CC$  & $CC$ \\
\hline
\hline

\rule[-3mm]{0mm}{5mm}$1$                                                                       
&    81 &    0 &    81 &    0 &    0 &    81 \\
\rule[-3mm]{0mm}{5mm}$P_{36}^{f \sigma c}$                                                     
&    -1 &    4 &     1 &  -12 &   24 &   -63 \\
\rule[-3mm]{0mm}{5mm}$\lambda_1^c.\lambda_2^c$                                                 
& -2592 &    0 & -2592 &    0 &    0 &  -648 \\
\rule[-3mm]{0mm}{5mm}$\lambda_3^c.\lambda_6^c$                                                 
&     0 &    0 &     0 &    0 &    0 & -1296 \\
\rule[-3mm]{0mm}{5mm}$\lambda_1^c.\lambda_2^c\ P_{36}^{f \sigma c}$                            
&    32 & -128 &   -32 &  384 & -768 &    72 \\
\rule[-3mm]{0mm}{5mm}$\lambda_3^c.\lambda_6^c\ P_{36}^{f \sigma c}$                            
&   -64 &  256 &    64 &   96 & -192 &  1152 \\
\rule[-3mm]{0mm}{5mm}$\lambda_1^c.\lambda_3^c\ P_{36}^{f \sigma c}$                            
&    32 & -128 &   -32 &  384 & -768 &   720 \\
\rule[-3mm]{0mm}{5mm}$\lambda_1^c.\lambda_6^c\ P_{36}^{f \sigma c}$                            
&    32 & -128 &   -32 &  -48 &   96 &   720 \\
\rule[-3mm]{0mm}{5mm}$\lambda_1^c.\lambda_4^c\ P_{36}^{f \sigma c}$                            
&   -16 &   64 &    16 &   24 &  -48 &  1260 \\
\rule[-3mm]{0mm}{5mm}$\sigma_1.\sigma_2\ \tau_1.\tau_2$                                        
&  4860 &    0 &   972 &    0 &    0 &   108 \\
\rule[-3mm]{0mm}{5mm}$\sigma_3.\sigma_6\ \tau_3.\tau_6$                                        
&  -900 &  576 &  1980 &    0 &    0 &  1116 \\
\rule[-3mm]{0mm}{5mm}$\sigma_1.\sigma_2\ \tau_1.\tau_2\ P_{36}^{f \sigma c}$                   
&  -444 &   48 &    12 & -720 &  288 &   588 \\
\rule[-3mm]{0mm}{5mm}$\sigma_3.\sigma_6\ \tau_3.\tau_6\ P_{36}^{f \sigma c}$                   
&   708 &   48 &  1596 &  240 &  672 & -1092 \\
\rule[-3mm]{0mm}{5mm}$\sigma_1.\sigma_3\ \tau_1.\tau_3\ P_{36}^{f \sigma c}$                   
&   132 &  336 &    12 & -720 &  288 &  -420 \\
\rule[-3mm]{0mm}{5mm}$\sigma_1.\sigma_6\ \tau_1.\tau_6\ P_{36}^{f \sigma c}$                   
&   132 &   48 &    12 &  336 &  -96 &  -420 \\
\rule[-3mm]{0mm}{5mm}$\sigma_1.\sigma_4\ \tau_1.\tau_4\ P_{36}^{f \sigma c}$                   
&    36 & -144 &   -36 &  228 &  288 & -1260 \\
\rule[-3mm]{0mm}{5mm}$\sigma_1.\sigma_2\ \lambda_1^f.\lambda_2^f$                              
&  4536 &    0 &  1296 &    0 &    0 &   -18 \\
\rule[-3mm]{0mm}{5mm}$\sigma_3.\sigma_6\ \lambda_3^f.\lambda_6^f$                              
&  -864 &  576 &  1584 &    0 &    0 &  1020 \\
\rule[-3mm]{0mm}{5mm}$\sigma_1.\sigma_2\ \lambda_1^f.\lambda_2^f\ P_{36}^{f \sigma c}$         
&  -376 &   64 &    16 & -672 &  384 &   706 \\
\rule[-3mm]{0mm}{5mm}$\sigma_3.\sigma_6\ \lambda_3^f.\lambda_6^f\ P_{36}^{f \sigma c}$         
&   784 &   32 &  1520 &  216 &  528 & -1024 \\
\rule[-3mm]{0mm}{5mm}$\sigma_1.\sigma_3\ \lambda_1^f.\lambda_3^f\ P_{36}^{f \sigma c}$         
&   104 &  304 &    16 & -672 &  384 &  -332 \\
\rule[-3mm]{0mm}{5mm}$\sigma_1.\sigma_6\ \lambda_1^f.\lambda_6^f\ P_{36}^{f \sigma c}$         
&   104 &   64 &    16 &  340 & -200 &  -332 \\
\rule[-3mm]{0mm}{5mm}$\sigma_1.\sigma_4\ \lambda_1^f.\lambda_4^f\ P_{36}^{f \sigma c}$         
&    44 & -152 &   -32 &  278 &  164 & -1197 \\
\rule[-3mm]{0mm}{5mm}$\sigma_1.\sigma_2\ \lambda_1^{f,0}.\lambda_2^{f,0}$                      
&  -648 &    0 &   648 &    0 &    0 &  -252 \\
\rule[-3mm]{0mm}{5mm}$\sigma_3.\sigma_6\ \lambda_3^{f,0}.\lambda_6^{f,0}$                      
&    72 &    0 &  -792 &    0 &    0 &  -192 \\
\rule[-3mm]{0mm}{5mm}$\sigma_1.\sigma_2\ \lambda_1^{f,0}.\lambda_2^{f,0}\ P_{36}^{f \sigma c}$ 
&   136 &   32 &     8 &   96 &  192 &   236 \\
\rule[-3mm]{0mm}{5mm}$\sigma_3.\sigma_6\ \lambda_3^{f,0}.\lambda_6^{f,0}\ P_{36}^{f \sigma c}$ 
&   152 &  -32 &  -152 &  -48 & -288 &   136 \\
\rule[-3mm]{0mm}{5mm}$\sigma_1.\sigma_3\ \lambda_1^{f,0}.\lambda_3^{f,0}\ P_{36}^{f \sigma c}$ 
&   -56 &  -64 &     8 &   96 &  192 &   176 \\
\rule[-3mm]{0mm}{5mm}$\sigma_1.\sigma_6\ \lambda_1^{f,0}.\lambda_6^{f,0}\ P_{36}^{f \sigma c}$ 
&   -56 &   32 &     8 &    8 & -208 &   176 \\
\rule[-3mm]{0mm}{5mm}$\sigma_1.\sigma_4\ \lambda_1^{f,0}.\lambda_4^{f,0}\ P_{36}^{f \sigma c}$ 
&    16 &  -16 &     8 &  -20 & -248 &   126 \\
\hline
\rule[-4mm]{0mm}{10mm}factor & $\frac{1}{972}$ & $\frac{\sqrt5}{972}$ & $\frac{1}{972}$ 
& $\frac{\sqrt5}{972}$ & $\frac{1}{972}$ & $\frac{1}{972}$ \\
\hline
\end{tabular}
\caption{Matrix elements $\langle \alpha|O|\beta \rangle$ of different operators $O$ for $SI = 10$.}\label{RGMtable10}
\end{table}

\begin{table}[H]
\centering

\begin{tabular}{|c|c|c|c|c|c|c|}
\hline
\rule[-4mm]{0mm}{5mm}$\alpha$ & $NN$ & $NN$ & $\Delta\Delta$ & $NN$ & $\Delta\Delta$ & $CC$ \\
$\beta$  & $NN$ & $\Delta\Delta$ & $\Delta\Delta$ & $CC$ & $CC$ & $CC$ \\
\hline
\hline

\rule[-3mm]{0mm}{5mm}$1$                                                                       
&    81 &    0 &    81 &    0 &    0 &    81 \\
\rule[-3mm]{0mm}{5mm}$P_{36}^{f \sigma c}$                                                     
&    -1 &    4 &     1 &  -12 &   24 &   -63 \\
\rule[-3mm]{0mm}{5mm}$\lambda_1^c.\lambda_2^c$                                                 
& -2592 &    0 & -2592 &    0 &    0 &  -648 \\
\rule[-3mm]{0mm}{5mm}$\lambda_3^c.\lambda_6^c$                                                 
&     0 &    0 &     0 &    0 &    0 & -1296 \\
\rule[-3mm]{0mm}{5mm}$\lambda_1^c.\lambda_2^c\ P_{36}^{f \sigma c}$                            
&    32 & -128 &   -32 &  384 & -768 &    72 \\
\rule[-3mm]{0mm}{5mm}$\lambda_3^c.\lambda_6^c\ P_{36}^{f \sigma c}$                            
&   -64 &  256 &    64 &   96 & -192 &  1152 \\
\rule[-3mm]{0mm}{5mm}$\lambda_1^c.\lambda_3^c\ P_{36}^{f \sigma c}$                            
&    32 & -128 &   -32 &  384 & -768 &   720 \\
\rule[-3mm]{0mm}{5mm}$\lambda_1^c.\lambda_6^c\ P_{36}^{f \sigma c}$                            
&    32 & -128 &   -32 &  -48 &   96 &   720 \\
\rule[-3mm]{0mm}{5mm}$\lambda_1^c.\lambda_4^c\ P_{36}^{f \sigma c}$                            
&   -16 &   64 &    16 &   24 &  -48 &  1260 \\
\rule[-3mm]{0mm}{5mm}$\sigma_1.\sigma_2\ \tau_1.\tau_2$                                        
&  4860 &    0 &   972 &    0 &    0 &   108 \\
\rule[-3mm]{0mm}{5mm}$\sigma_3.\sigma_6\ \tau_3.\tau_6$                                        
&  -900 &  576 &  1980 &    0 &    0 &  1116 \\
\rule[-3mm]{0mm}{5mm}$\sigma_1.\sigma_2\ \tau_1.\tau_2\ P_{36}^{f \sigma c}$                   
&  -444 &   48 &    12 & -720 &  288 &   588 \\
\rule[-3mm]{0mm}{5mm}$\sigma_3.\sigma_6\ \tau_3.\tau_6\ P_{36}^{f \sigma c}$                   
&   708 &   48 &  1596 &  240 &  672 & -1092 \\
\rule[-3mm]{0mm}{5mm}$\sigma_1.\sigma_3\ \tau_1.\tau_3\ P_{36}^{f \sigma c}$                   
&   132 &  336 &    12 & -720 &  288 &  -420 \\
\rule[-3mm]{0mm}{5mm}$\sigma_1.\sigma_6\ \tau_1.\tau_6\ P_{36}^{f \sigma c}$                   
&   132 &   48 &    12 &  336 &  -96 &  -420 \\
\rule[-3mm]{0mm}{5mm}$\sigma_1.\sigma_4\ \tau_1.\tau_4\ P_{36}^{f \sigma c}$                   
&    36 & -144 &   -36 &  228 &  288 & -1260 \\
\rule[-3mm]{0mm}{5mm}$\sigma_1.\sigma_2\ \lambda_1^f.\lambda_2^f$                              
&  4536 &    0 &  1296 &    0 &    0 &  -126 \\
\rule[-3mm]{0mm}{5mm}$\sigma_3.\sigma_6\ \lambda_3^f.\lambda_6^f$                              
& -1008 &  576 &  1440 &    0 &    0 &   948 \\
\rule[-3mm]{0mm}{5mm}$\sigma_1.\sigma_2\ \lambda_1^f.\lambda_2^f\ P_{36}^{f \sigma c}$         
&  -376 &   64 &    16 & -672 &  384 &   814 \\
\rule[-3mm]{0mm}{5mm}$\sigma_3.\sigma_6\ \lambda_3^f.\lambda_6^f\ P_{36}^{f \sigma c}$         
&   832 &   32 &  1568 &  232 &  496 &  -976 \\
\rule[-3mm]{0mm}{5mm}$\sigma_1.\sigma_3\ \lambda_1^f.\lambda_3^f\ P_{36}^{f \sigma c}$         
&   104 &  304 &    16 & -672 &  384 &  -260 \\
\rule[-3mm]{0mm}{5mm}$\sigma_1.\sigma_6\ \lambda_1^f.\lambda_6^f\ P_{36}^{f \sigma c}$         
&   104 &   64 &    16 &  364 & -248 &  -260 \\
\rule[-3mm]{0mm}{5mm}$\sigma_1.\sigma_4\ \lambda_1^f.\lambda_4^f\ P_{36}^{f \sigma c}$         
&    36 & -168 &   -48 &  298 &  124 & -1155 \\
\rule[-3mm]{0mm}{5mm}$\sigma_1.\sigma_2\ \lambda_1^{f,0}.\lambda_2^{f,0}$                      
&  -648 &    0 &   648 &    0 &    0 &  -468 \\
\rule[-3mm]{0mm}{5mm}$\sigma_3.\sigma_6\ \lambda_3^{f,0}.\lambda_6^{f,0}$                      
&  -216 &    0 & -1080 &    0 &    0 &  -336 \\
\rule[-3mm]{0mm}{5mm}$\sigma_1.\sigma_2\ \lambda_1^{f,0}.\lambda_2^{f,0}\ P_{36}^{f \sigma c}$ 
&   136 &   32 &     8 &   96 &  192 &   452 \\
\rule[-3mm]{0mm}{5mm}$\sigma_3.\sigma_6\ \lambda_3^{f,0}.\lambda_6^{f,0}\ P_{36}^{f \sigma c}$ 
&   248 &  -32 &   -56 &  -16 & -352 &   232 \\
\rule[-3mm]{0mm}{5mm}$\sigma_1.\sigma_3\ \lambda_1^{f,0}.\lambda_3^{f,0}\ P_{36}^{f \sigma c}$ 
&   -56 &  -64 &     8 &   96 &  192 &   320 \\
\rule[-3mm]{0mm}{5mm}$\sigma_1.\sigma_6\ \lambda_1^{f,0}.\lambda_6^{f,0}\ P_{36}^{f \sigma c}$ 
&   -56 &   32 &     8 &   56 & -304 &   320 \\
\rule[-3mm]{0mm}{5mm}$\sigma_1.\sigma_4\ \lambda_1^{f,0}.\lambda_4^{f,0}\ P_{36}^{f \sigma c}$ 
&     0 &  -48 &   -24 &   20 & -328 &   210 \\
\hline
\rule[-4mm]{0mm}{10mm}factor & $\frac{1}{972}$ & $\frac{\sqrt5}{972}$ & $\frac{1}{972}$ 
& $\frac{\sqrt5}{972}$ & $\frac{1}{972}$ & $\frac{1}{972}$ \\
\hline
\end{tabular}
\caption{Matrix elements $\langle \alpha|O|\beta \rangle$ of different operators $O$ for $SI = 01$.}\label{RGMtable01}

\end{table}

\section{Results for the phase shifts}
\

\subsection{The role of the coupled channels}
\ 

We shall now present our results for the $^3S_1$ and $^1S_0$ scattering phase shifts using the RGM approach for the GBE model (\ref{RGMhamiltonian}-\ref{RGMparam2}) and discuss the role of the coupled channels $NN$, $\Delta\Delta$ and $CC$ as introduced in Section \ref{subsectionCC}. According to Section \ref{RGMsubsectionbound}, the relative wave function $\chi (\vec{R})$ has been expanded over a finite number of equally spaced Gaussians where the peak of each Gaussian is given by $R_i=R_0+i*t$ with $R_0=0.3$ fm and $t=0.35$ fm ($i=1, .., N$). For scattering states this expansion holds up to a finite distance $R=R_c$, where  $R_c$ depends on the range of the interaction. Beyond $R_c$, $\chi(\vec{R})$ is written as the usual combination of Hankel functions containing the $S$-matrix (\ref{KAM6}). Then the phase shifts are determined by imposing the continuity of $\chi(\vec{R})$ and of its derivative with respect to $R$ at  $R=R_c$. Either if we take  one, two or three channels namely $NN$, $NN+\Delta\Delta$ or $NN+\Delta\Delta+CC$ we found that a number of $N=15$ Gaussians in the expansion (\ref{KAM1}) is large enough to obtain convergence. In all cases the result is stable at the matching radius $R_c = 4.5$ fm. The size parameter of the Gaussians is fixed at $\beta=0.437$ fm, as obtained from the stability condition (\ref{STABILITY}) in the GBE parametrization (\ref{RGMparam2}). All relevant two-body matrix elements in color and flavor-spin are extracted from Tables \ref{RGMtable10} and \ref{RGMtable01}.

\begin{figure}[H]
\begin{center}
\includegraphics[width=14cm]{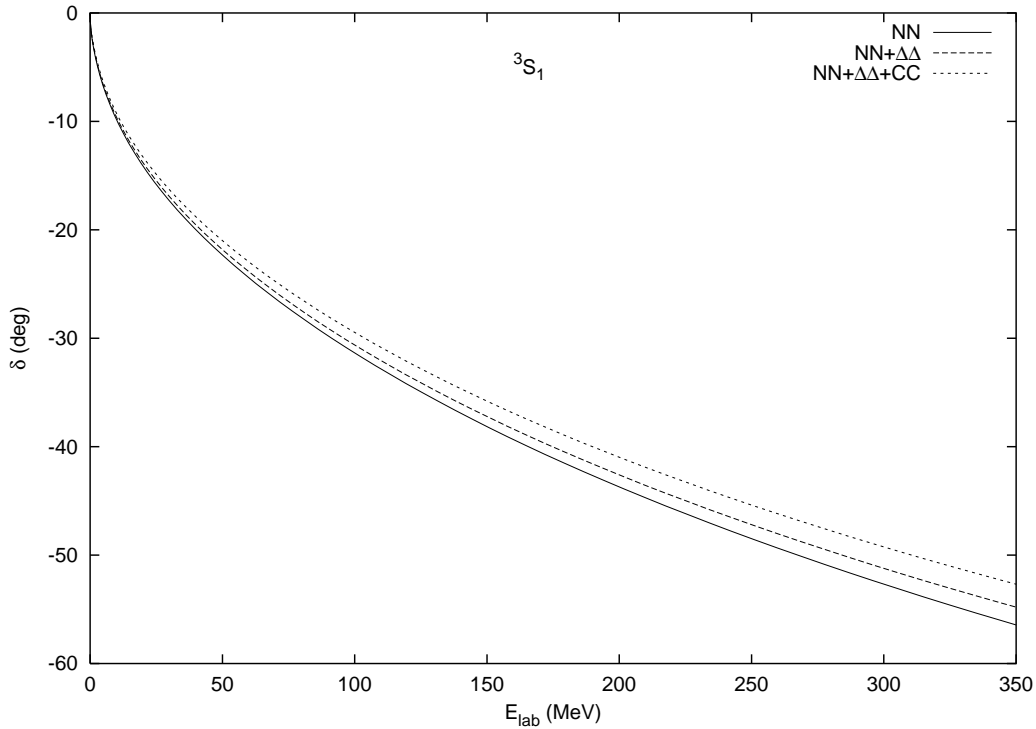}
\end{center}
\caption{\label{RGM133S1} $^3S_1$ NN scattering phase shift as a function of $E_{lab}$. The solid line shows the result for the $NN$ channel only, the dotted line for the $NN$+$\Delta\Delta$  and the dashed line for the $NN$+$\Delta\Delta$+$CC$ coupled channels.}
\end{figure}

\begin{figure}[H]
\begin{center}
\includegraphics[width=14cm]{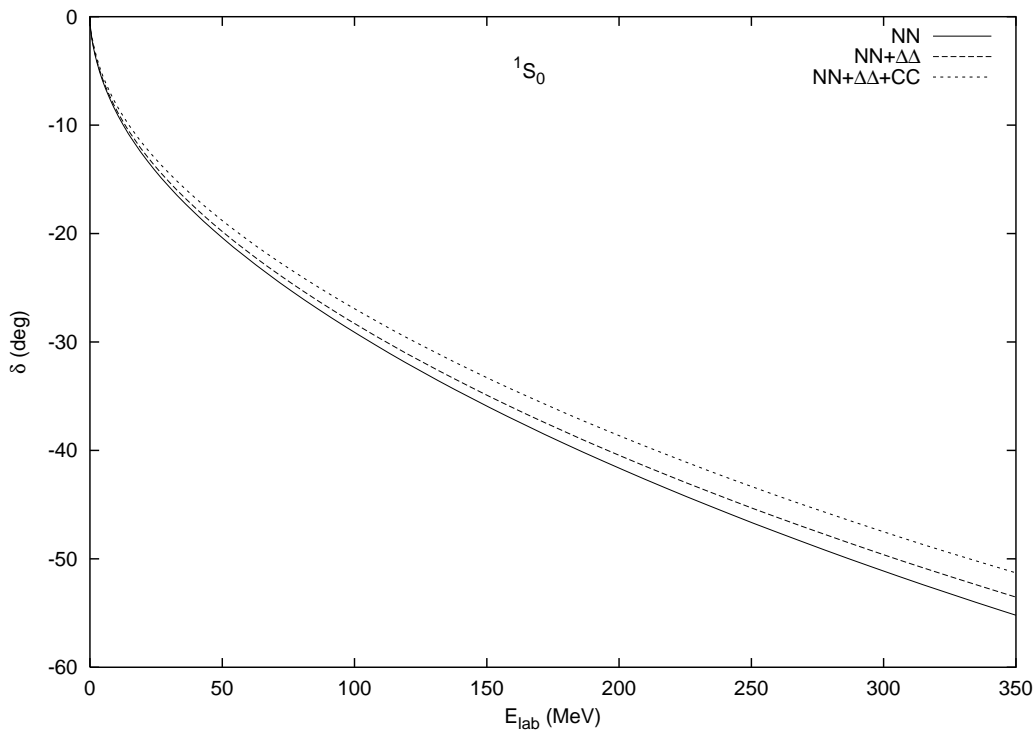}
\end{center}
\caption{\label{RGM131S0}  Same as Fig. \ref{RGM133S1} but for the $^1S_0$ partial wave.}
\end{figure}

In Figs. \ref{RGM133S1} and \ref{RGM131S0} we show the phase shifts obtained from one ($NN$), two ($NN+\Delta\Delta$) and three ($NN+\Delta\Delta+CC$) coupled channels as a function of the laboratory energy  $E_{lab}= 2 \hbar^2 k^2/3m$ with $m = m_{u,d}$ of (\ref{RGMparam2}). The negative value of the phase shifts of both $^3S_1$ and $^1S_0$ partial waves reveal the presence of a short-range repulsion in the NN interaction. This proves that the GBE interaction can explain the short-range repulsion, due to its flavour-spin quark-quark operator combined with the quark interchange between clusters in the framework of a dynamical approach. One can see that in the case of pure pseudoscalar exchange, the addition to $NN$ of the $\Delta\Delta$ channel alone or of both $\Delta\Delta$ and $CC$ channels brings a very small change in the $^3S_1$ and $^1S_0$ phase shifts below 300 MeV, making the repulsion slightly weaker. The $CC$ channel brings slightly more repulsion than the $\Delta\Delta$ channel. In fact the role of $CC$ channels is expected to increase for larger values of the relative momentum $k$, or alternatively smaller separation distances between nucleons, where they could bring an important contribution. Of course, the contribution of the $CC$ channels to the $NN$ phase shifts vanishes at larger separations because of their color structure. The conclusion regarding the minor contribution of $\Delta\Delta$ and $CC$ channels to the phase shifts below 300 MeV is similar for results  based on the OGE model (see for example \cite{FAE82}). Thus for $l=0$ waves it is good enough to perform one channel calculations in the lab energy interval 0-350 MeV.

We recall that the pseudoscalar exchange interaction (\ref{RGMmodelII}) contains both a short-range part, responsible for the repulsion, and a long-range Yukawa-type potential which brings attraction in the NN potential. In order to see the difference in the amount of repulsion induced by the GBE and that induced by the OGE interaction we repeated the one channel $(NN)$ calculations above by removing the Yukawa-type part, inexistent in OGE models. We compared  the resulting phase shifts with those of Fig. 2 of Ref. \cite{FAE82} obtained with an OGE interaction parametrized such as to satisfy the stability condition (\ref{STABILITY}). We found that in the GBE model the repulsion is much stronger and corresponds to a hard core radius $r^{GBE}_0 = 0.68$ fm (versus $r^{OGE}_0 = 0.30$ fm) in the $^3S_1$ and $r^{GBE}_0 = 0.81$ fm (versus $r^{OGE}_0 = 0.35$ fm) in the $^1S_0$ partial waves. The radius $r_0$ was extracted from the phase shifts at small $k$, which is approximately given by $\delta = -k\ r_0$.  This outcome is consistent with the findings of Ref. \cite{SHI00} where a simplified $SU(2)$ version of the GBE model has been used. One can also see that the repulsion induced by the GBE interaction in the $^3S_1$ partial wave is weaker than that induced in the $^1S_0$ partial wave. This is consistent with the previous chapter where we found that the height of the repulsive core is lower for $^3S_1$ than for $^1S_0$. Thus the OGE model gives less repulsion than the GBE model. In Ref. \cite{SHI00} the stronger repulsion induced by the GBE interaction is viewed as a welcome feature in correctly describing the phase shifts above $E_{lab}=350$ MeV. Our present view is somewhat different (see later).
\\

A note of caution is required regarding the removal of the long-range Yukawa part of the interaction (\ref{RGMmodelII}) with the parametrization (\ref{RGMparam2}) which contains a rather large coupling constant $g^2_{\eta^{'}q}/(4\pi)=2.7652$. The $\eta^{'}$-meson exchange is responsible for describing correctly the $\Delta- N$ splitting. If the long-range Yukawa part is removed, the model fails to describe this splitting because the contribution coming from the second term of (\ref{RGMmodelII}) for $\gamma = \eta^{'}$ becomes too large in a 3q system in the parametrization (\ref{RGMparam2}). We recall that the contribution to $N$ of the short-range $\eta^{'}$-meson exchange part is proportional to a factor of 2 and the contribution to $\Delta$ to a factor -2 \cite{GLO96a}, which brings $\Delta$ too low and $N$ too high if the Yukawa part is removed. In these circumstances two or three coupled channel calculations become meaningless.
\\

\begin{figure}[H]
\begin{center}
\includegraphics[width=14cm]{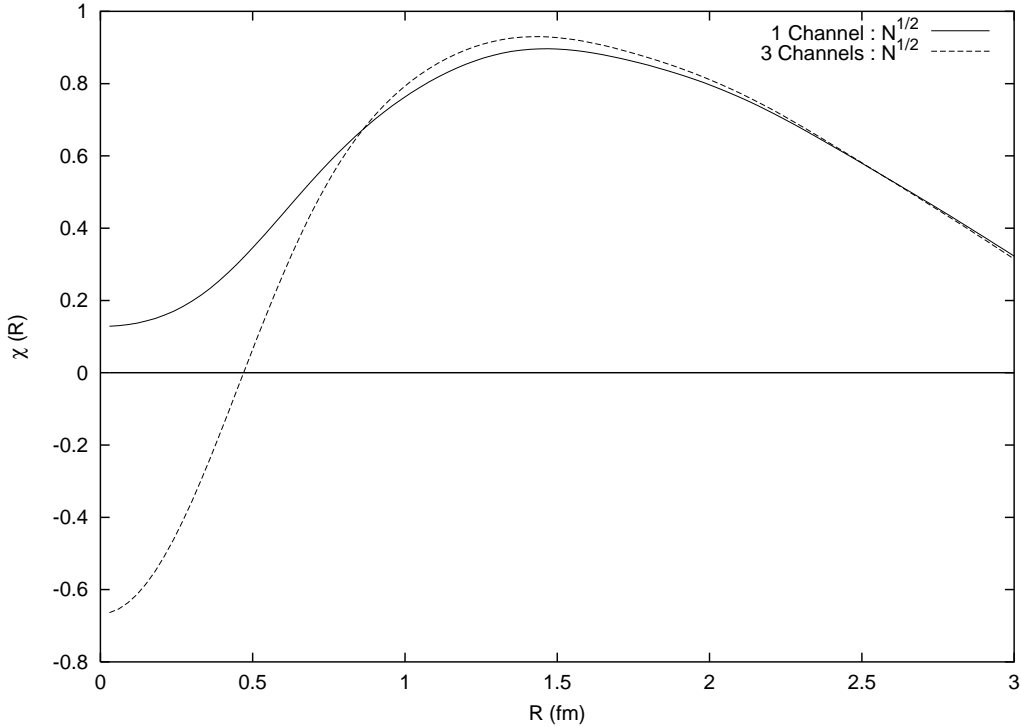}
\end{center}
\caption{\label{RGMkhiR3S1} The relative wave function of Eq. (\ref{RGMNORM}) for the $^3S_1$ partial wave for $k = 1$ fm$^{-1}$ obtained in one channel (solid line) and three channels (dashed line) calculations.}
\end{figure}

It is also interesting to see the behaviour of the relative wave function $\chi^{l=0}$ of Eq. (\ref{KAM5}) at short distances. Instead of $\chi^{l=0}$ it is more appropriate \cite{OKA80} to introduce a renormalized wave function as

\begin{equation}\label{RGMNORM}
\chi^{l=0}_{\alpha}(R) = \sum\limits_{\beta}\int{dR^{'}~[{N^{l=0}_{\beta\alpha}}(R,R')]^{1/2}
~\chi^{l=0}_{\beta}(R^{'})}\ ,
\end{equation}
\ 

\noindent where the quantity to be integrated contains the $l = 0$ component of the norm $N$.
\\

\begin{figure}[H]
\begin{center}
\includegraphics[width=14cm]{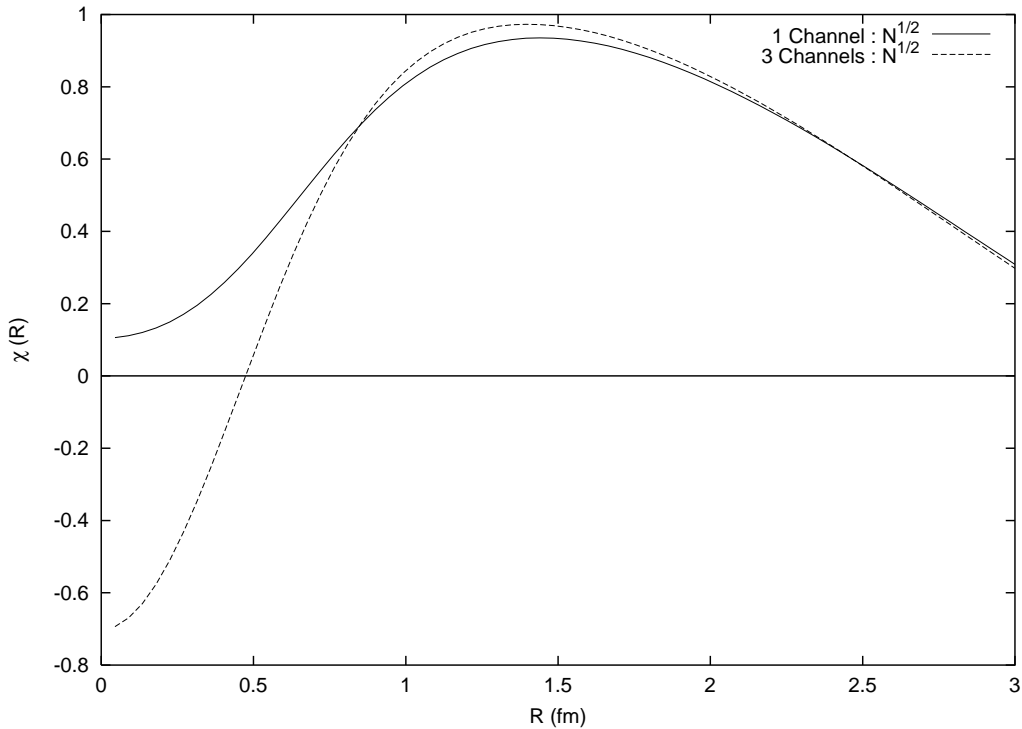}
\end{center}
\caption{\label{RGMkhiR1S0} Same as Fig. \ref{RGMkhiR3S1} but for the $^1S_0$ partial wave.}
\end{figure}

In Figs. \ref{RGMkhiR3S1} and \ref{RGMkhiR1S0} we show results for the function (\ref{RGMNORM}) for the $^3S_1$ and $^1S_0$ waves, respectively, at the relative momentum $k = 1$ fm$^{-1}$ both for the one and the three channel cases. One can see that for $R<1$ fm the one and three channel functions are entirely different, in the three channel case a node being present. If the renormalization was made with the norm $N$ instead of its square, as in Eq. (\ref{RGMNORM}), no node would have been present. The existence of a node is related to the presence of the $[42]_{O}$ configuration in the wave function (see e.g. \cite{STA97}). Here, whenever it appears, it is due to the cancellation of the positive and negative components of the wave function, but the lack of a node does not exclude a repulsive potential. In a renormalized wave function the amplitudes of positive and negative components change their values depending on the multiplicative factor $N$ or $N^{1/2}$ so the node could appear in one renormalization definition but not in the other. On the other hand, as discussed above, the phase shift changes insignificantly when one goes from one channel to three channels, and this can also be seen in the asymptotic form of the wave function beyond $R=1$ fm, although in the overlap region the two functions are entirely different. The above behaviour of the wave function is very similar to that found in Ref. \cite{SHI00} where no long-range part is present in the schematic quark-quark potential due to pion exchange. 
\\

\begin{figure}[H]
\begin{center}
\includegraphics[width=14cm]{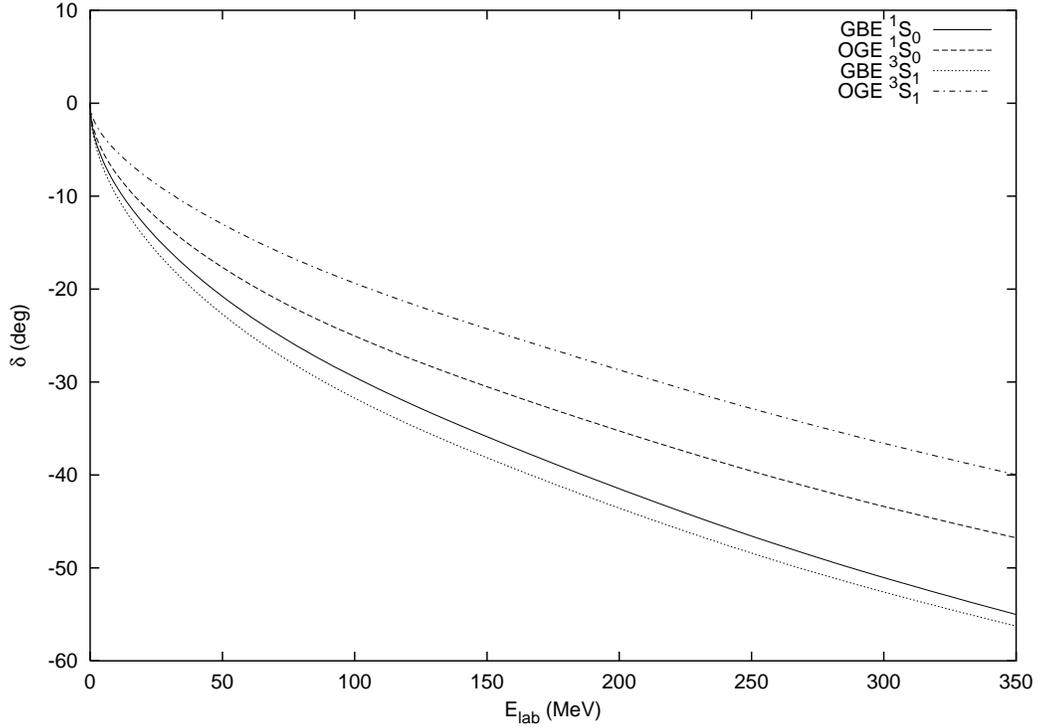}
\end{center}
\caption{\label{RGMfigGBEOGE} $^3S_1$ and $^1S_0$ $NN$ scattering phase shifts as a function of the laboratory energy $E_{lab}$. The solid and dotted lines show the result corresponding to the GBE model and the dashed and dot-dashed lines that of the OGE model (see Ref.\protect\cite{FAE82}). }
\end{figure}

In Figs. \ref{RGMfigGBEOGE} we represent the $^3S_1$ and $^1S_0$ phase shifts of Figs. \ref{RGM133S1} and \ref{RGM131S0} together, in the one channel case $(NN)$ again, with the Yukawa part included. This is to show that in the GBE model the two phase shifts are very near each other, with $\delta(^3S_1)$ slightly lower than $\delta(^1S_0)$. Contrary, in OGE calculations, as for example those of Fig. 2 of Ref. \cite{FAE82}, one obtains $\delta(^3S_1) > \delta(^1S_0)$. In calculations based on the OGE model the difference between the two phase shifts is reduced by the addition of a scalar potential acting at a nucleon level with a larger attractive strength in the $^1S_0$ channel than in the $^3S_1$ channel  \cite{OKA83}. A major difference between the GBE $\delta(^3S_1)$ and $\delta(^1S_0)$ is expected to appear after the inclusion of a quark-quark tensor force \cite{PLE99} because this will modify only the $^3S_1$ phase shift.
\\

At this stage, our conclusions are :

\begin{enumerate}

\item The phase shifts present a behaviour typical for strongly repulsive potentials. We find that this repulsion, which is induced by the pseudoscalar meson exchange is stronger than that produced by the OGE interaction.

\item In the $^1S_0$ partial wave the repulsion is stronger than in $^3S_1$ partial wave as our previous studies suggested.

\item Our results prove that in the laboratory energy interval 0-350 MeV the one channel approximation is entirely satisfactory for a GBE model with a hyperfine spin-spin interaction (no $\sigma$-meson exchange).

\end{enumerate}
\ 

\subsection{The role of the $\sigma$-meson exchange}
\ 

To describe the scattering data and the deuteron properties, intermediate- and long-range attraction potentials are necessary. In calculations based on OGE models they were phenomenologically simulated at the NN level by central and tensor potentials respectively \cite{OKA83} (see also Ref. \cite{FOR85}). However, for a consistent picture it is desirable to search for the origin of the attraction at the quark level. Here we incorporate a $\sigma$-meson exchange interaction at the quark level, missing in the original Models I and II, and we study its effect on the $^1S_0$ phase shift. Note that the $^1S_0$ phase shift is not influenced by a tensor potential, contrary to the $^3S_1$ phase shift. For the $\sigma$-meson exchange interaction we choose the following form

\begin{equation}\label{SIGMA}
V_{\sigma}=-\frac{g_{\sigma q}^2}{4\pi}~(\frac{e^{-\mu_{\sigma}r}}{r}-\frac{e^{-\Lambda_{\sigma}r}}{r})\ ,
\end{equation}
\ 

\noindent with parameters discussed below. The introduction of such an interaction is consistent with the spirit of the GBE model. It simulates the effect of two correlated pions \cite{RIS99}. The good quality of the baryon spectrum is not destroyed by the addition of this interaction which essentially leads to an overall shift of the spectrum \cite{STA00}.

\begin{figure}[H]
\begin{center}
\includegraphics[width=13.5cm]{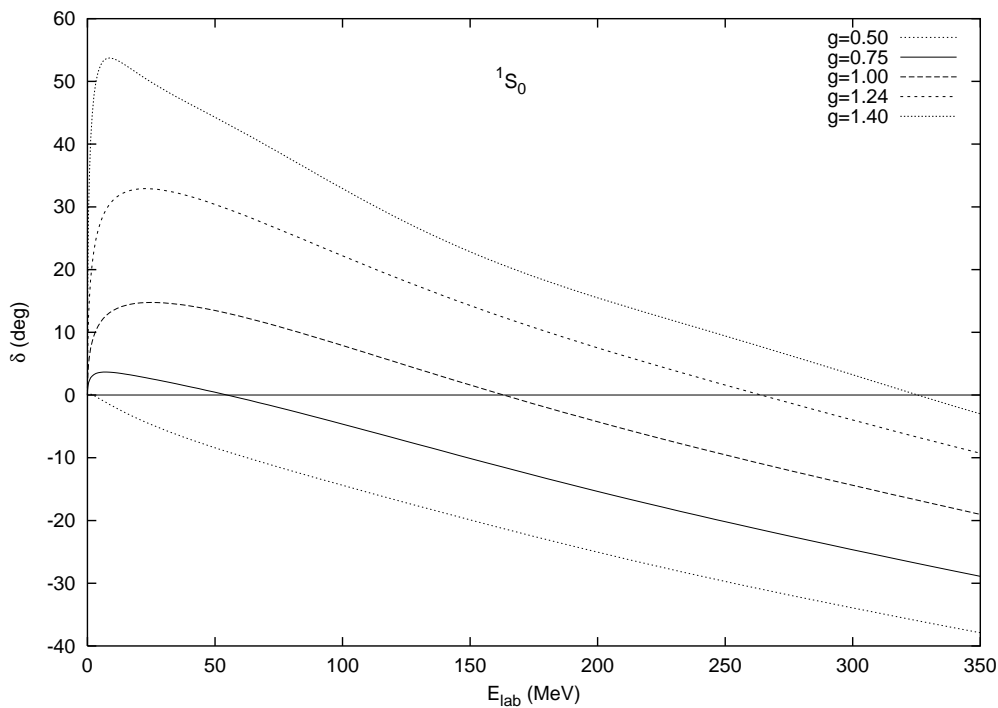}
\end{center}
\caption{\label{RGMfigscalarg} $^1S_0$ phase shift for various values of $g = g_{\sigma q}^2/4\pi$ at fixed $\mu_{\sigma} = 600$ MeV and $\Lambda_{\sigma} = 830$ MeV.}
\end{figure}

\begin{figure}[H]
\begin{center}
\includegraphics[width=13.5cm]{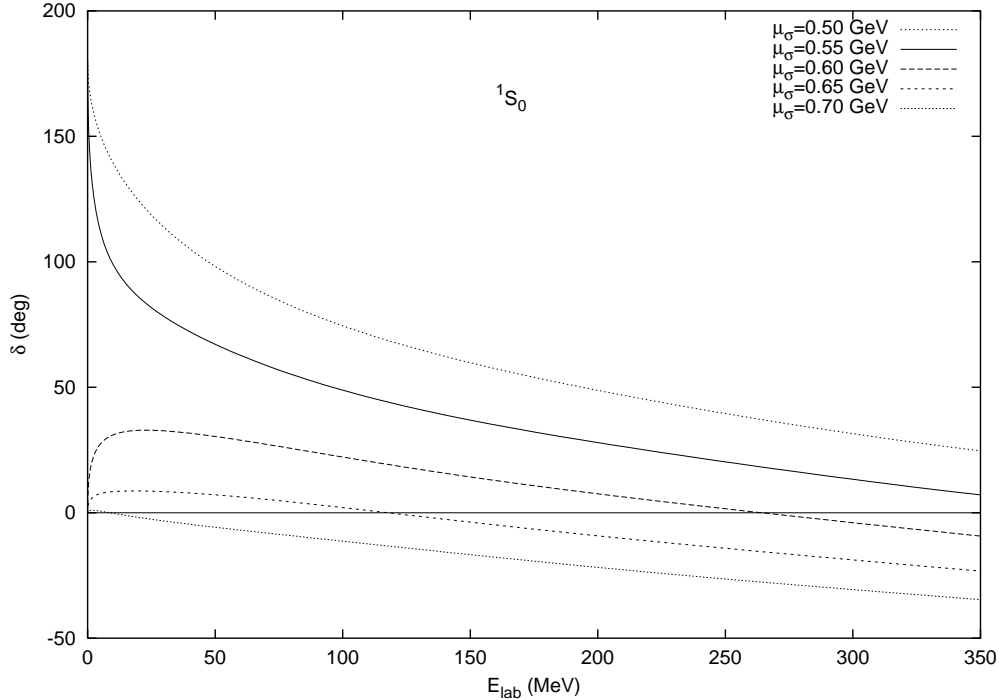}
\end{center}
\caption{\label{RGMfigscalarm} $^1S_0$ phase shift for various values of $\mu_{\sigma}$ at fixed  $\frac{g_{\sigma q}^2}{4\pi}=1.24$ and $\Lambda_{\sigma}=830$ MeV.}
\end{figure}

\begin{figure}[H]
\begin{center}
\includegraphics[width=12.8cm]{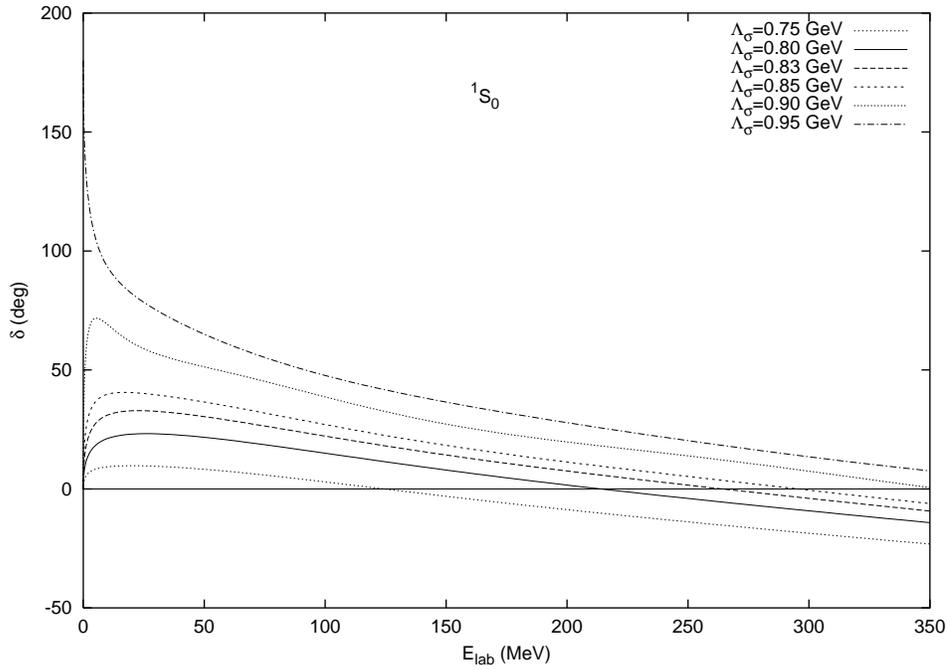}
\end{center}
\caption{\label{RGMfigscalarl} $^1S_0$ phase shift for various values of $\Lambda_{\sigma}$ at fixed  $\frac{g_{\sigma q}^2}{4\pi}=1.24$ and $\mu_{\sigma}= 600$ MeV.}
\end{figure}

The sensitivity of the $^1S_0$ phase shift with respect to the coupling constant $\frac{g_{\sigma q}^2}{4\pi}$, the mass $\mu_{\sigma}$ and the cut-off parameter $\Lambda_{\sigma}$ can be seen from Figs. \ref{RGMfigscalarg}-\ref{RGMfigscalarl}. As expected, the attraction in the NN potential increases with $\frac{g_{\sigma q}^2}{4\pi}$ and hence the value of $E_{lab}$ where the phase shift changes sign also increases with $\frac{g_{\sigma q}^2}{4\pi}$, as shown in Fig. \ref{RGMfigscalarg}. Note that the potential $V_{\sigma}$ of (\ref{SIGMA}) remains attractive as long as $\mu_{\sigma} < \Lambda_{\sigma}$. However $\mu_{\sigma}$ cannot be too close to $\Lambda_{\sigma}$. As suggested by Fig. \ref{RGMfigscalarm}, the attractive pocket in the NN potential becomes too small for $\mu_{\sigma} > $ 650 MeV, making the repulsion dominant and leading to negative phase shifts at all energies, when $\frac{g_{\sigma q}^2}{4\pi}$ = 1.24 and $\Lambda_{\sigma}$ = 830 MeV.
\\

A large difference between  $\mu_{\sigma}$ and $\Lambda_{\sigma}$ is not good either. From  Fig. \ref{RGMfigscalarl} one can see that when $\Lambda_{\sigma} > $ 950 MeV and  $\mu_{\sigma}$ = 600 MeV an undesired bound state in the $^1S_0$ phase shift is accommodated at $\frac{g_{\sigma q}^2}{4\pi} = 1.24$. This is due to the fact that the contribution of the second term in the right hand side of (\ref{SIGMA}) becomes negligible and $V_{\sigma}$ brings too much attraction in the NN potential. In this way we found  an optimal set of values

\begin{equation}\label{RGMsigamparamI}
\frac{g_{\sigma q}^2}{4\pi} = \frac{g_{\pi q}^2}{4\pi} = 1.24,~~~~~
\mu_{\sigma} = 600\ {\rm MeV}\ ,~~~~~\Lambda_{\sigma} = 830\ {\rm MeV}\ .
\end{equation}
\\

Applying the stability condition (\ref{STABILITY}) to this extended Hamiltonian, the size parameter of the Gaussians is now $\beta=0.351$ fm. The number of Gaussians as well as their peak positions are the same as in the GBE model without $V_\sigma$.
\\

\begin{figure}[H]
\begin{center}
\includegraphics[width=12.8cm]{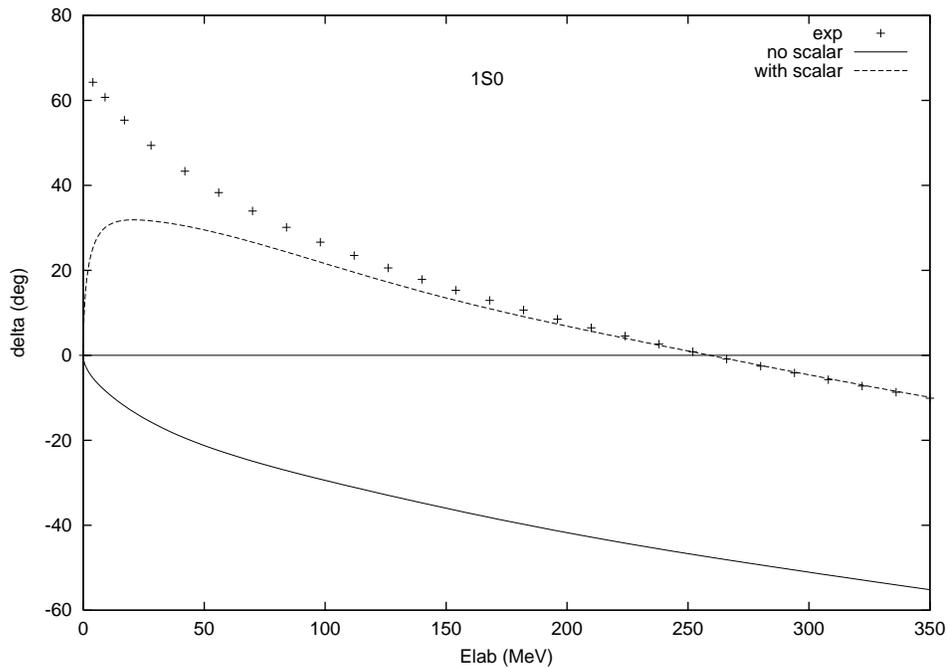}
\end{center}
\caption{\label{Fig.6} The $^1S_0$ NN scattering phase shift obtained in the GBE model as a function of $E_{lab}$. The solid line is the same that the one channel result of Fig. \ref{RGM131S0} and the dashed line includes the effect of the $\sigma$-meson exchange potential (\ref{SIGMA}) between quarks with $\mu_{\sigma}=600$ MeV and $\Lambda_{\sigma}=830$ MeV. Experimental data are from Ref. \cite{STO93}.}
\end{figure}

As one can see from Fig. \ref{Fig.6}, with these values the theoretical curve gets quite close to the experimental points without altering the good short-range behaviour, and in particular the change of sign of the phase shift at $E_{lab} \approx 260$ MeV. Thus the addition of a $\sigma$-meson exchange interaction alone leads to a good description of the phase shift in a large enough energy interval. One can argue that the still existing discrepancy at low energies could possibly be removed by the coupling of the $^5D_0$ N-$\Delta$ channel, suggested by Ref. \cite{VAL95} in the frame of a hybrid model, containing both gluon and meson exchange at the quark level. To achieve this coupling, as well as to describe the $^3S_1$ phase shift, the introduction of a tensor interaction is necessary.
\\

\begin{figure}[H]
\begin{center}
\includegraphics[width=13.5cm]{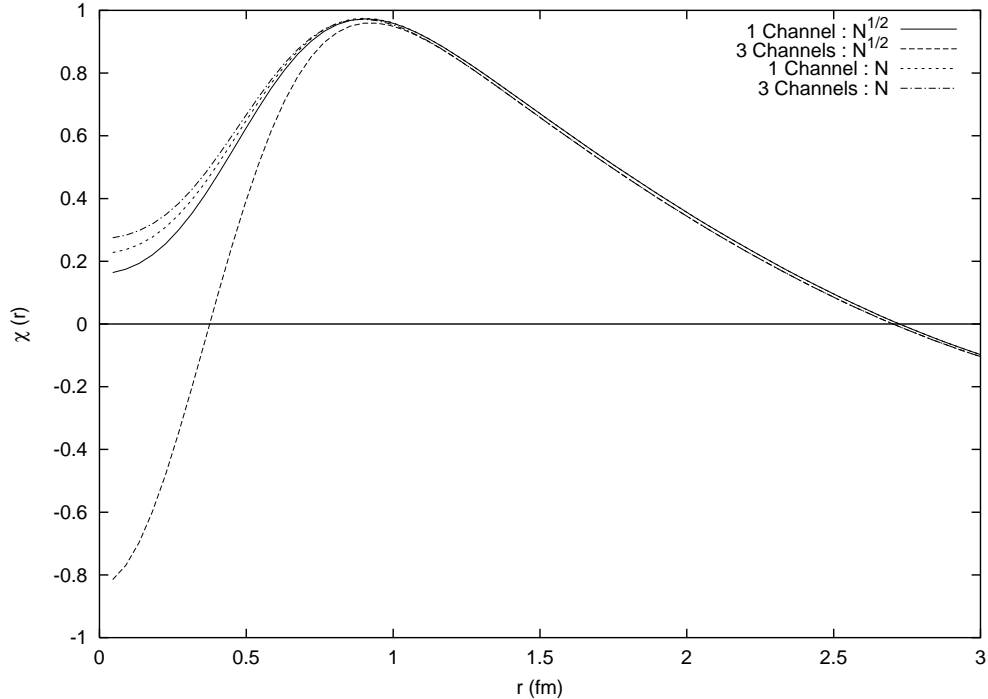}
\end{center}
\caption{\label{RGMkhiR1S0sigma} The relative wave function of Eq. (\ref{RGMNORM}) for the $^1S_0$ partial wave for $k = 1$ fm$^{-1}$ for the GBE model with the $\sigma$-meson exchange potential (\ref{SIGMA}) and the parameters (\ref{RGMsigamparamI}).}
\end{figure}

\begin{figure}[H]
\begin{center}
\includegraphics[width=13.5cm]{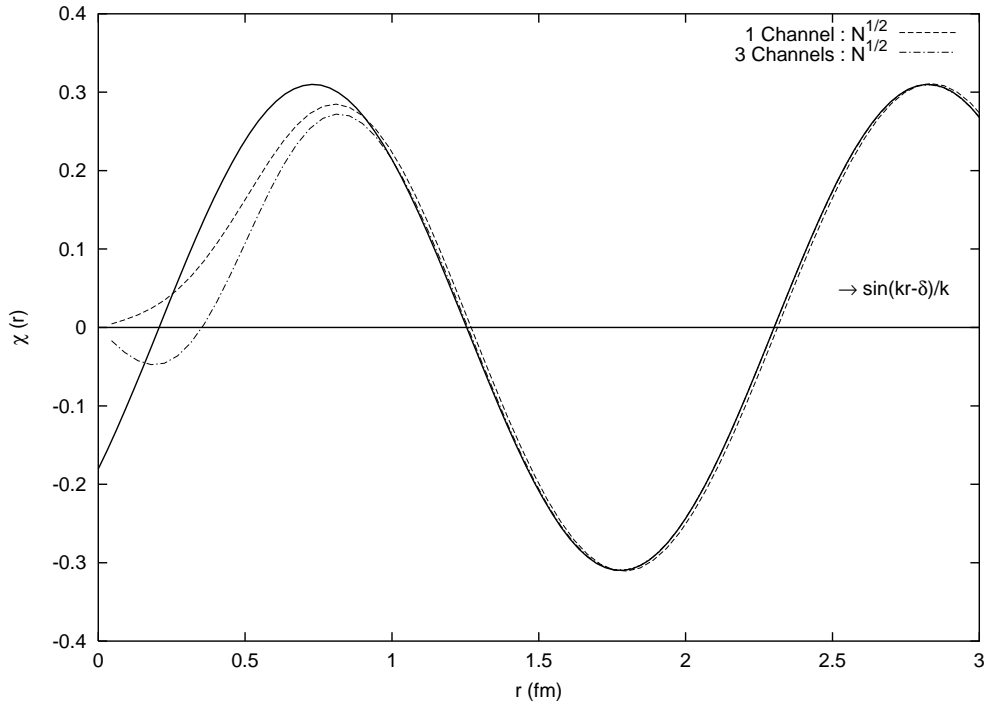}
\end{center}
\caption{\label{RGMkhiR1S0sigmaR} The relative wave function $\chi^{l=0}(R)*R$ for the $^1S_0$ partial wave for $k = 3$ fm$^{-1}$ for the GBE model with the $\sigma$-meson exchange potential (\ref{SIGMA}) and the parameters (\ref{RGMsigamparamI}).}
\end{figure}

In Fig. \ref{RGMkhiR1S0sigma} we show the renormalized wave function obtained from Eq. (\ref{RGMNORM}) in the GBE model where the $\sigma$-meson exchange potential has been added. The conclusion presented above are still valid. No node appear if we renormalize the wave function with the norm $N$ instead of $N^{1/2}$ and it is only in the three coupled channels that a node is present. However we observe that, whichever renormalization is chosen, the wave function reaches its maximum at approximatively 1 fm contrary to 1.5 fm without $\sigma$-meson exchange. This is an immediate consequence of the attraction introduced by the scalar potential (see later). For convenience in Fig. \ref{RGMkhiR1S0sigmaR} we also plot $\chi^{l=0}(R)*R$ for the $l=0$ wave function, proportional to $sin(kR-\delta)/k$ for large value of $R$.
\\

\ 

\begin{figure}[H]
\begin{center}
\includegraphics[width=12.6cm]{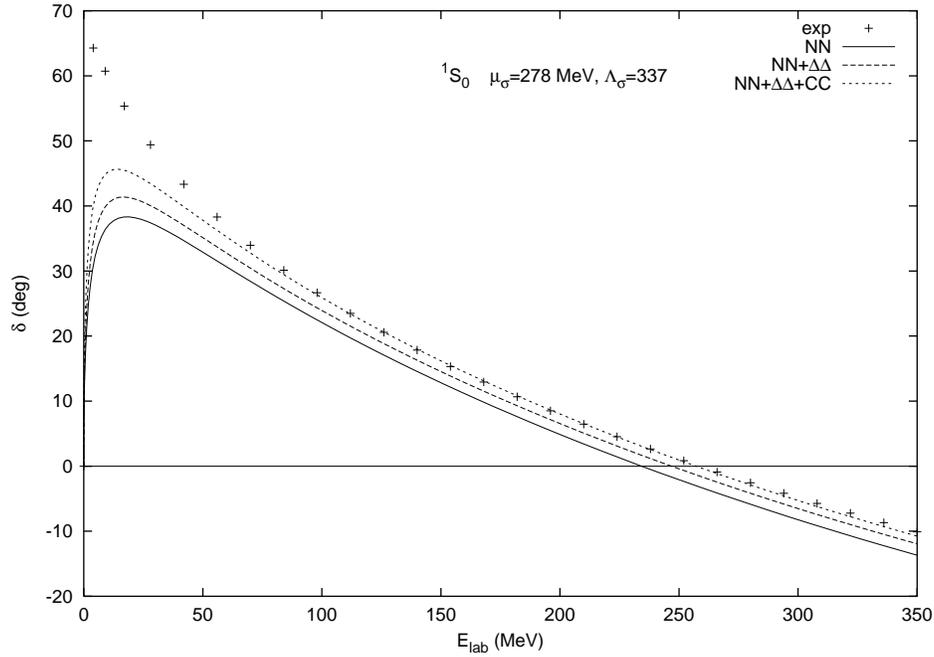}
\end{center}
\caption{\label{RGMfig1S0sigmaII} The $^1S_0$ NN scattering phase shift obtained in the GBE model with the $\sigma$-meson exchange potential (\ref{SIGMA}) between quarks with $\mu_{\sigma}=278$ MeV and $\Lambda_{\sigma}=337$ MeV as a function of $E_{lab}$. Experimental data are from Ref. \cite{STO93}.}
\end{figure}

Up to now we viewed the $\sigma$-meson exchange as coming from the spontaneous chiral symmetry breaking. This has led us to take the $\sigma$-meson mass as approximately given by $m_\sigma^2=m_\pi^2+4m_q^2$ consistent with the PCAC discussion of Section \ref{PCACsection}. However, if we consider that the $\sigma$-meson exchange simulates the exchange of two correlated pions, we can adopt another point of view, namely to take the $\sigma$-meson mass $m_\sigma=2*m_\pi=278$ MeV. With this choice, we decided to keep the coupling constant $\frac{g_{\pi q}^2}{4\pi} = 1.24$. We then adjusted $\Lambda_\sigma$ to reproduce the change of sign of the $^1S_0$ phase shift at $E_{lab} \approx 260$ MeV. We obtained  $\Lambda_\sigma=337$ MeV. As one can see in Fig. \ref{RGMfig1S0sigmaII} this choice improves the results a lot, still leaving room for a tensor type coupling of the $^5D_0$ N-$\Delta$ channel as discussed above. Interestingly, we find that the coupled channel $NN-\Delta\Delta-CC$ RGM calculation is much more effective now than in the GBE model without $\sigma$-meson exchange, and it substantially improves the low energy scattering region.
\\

\begin{figure}[H]
\begin{center}
\includegraphics[width=12.6cm]{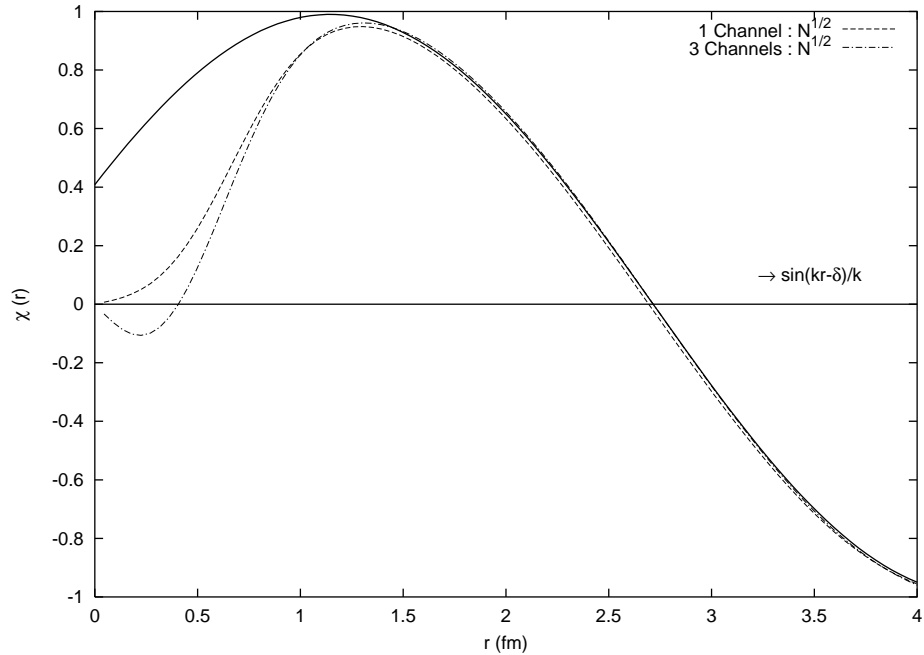}
\end{center}
\caption{\label{RGMkhiR1S0sigmaIIk1} The relative wave function of $\chi^{l=0}(R)*R$ for the $^1S_0$ partial wave for $k = 1$ fm$^{-1}$ for the GBE model with the $\sigma$-meson exchange potential (\ref{SIGMA}) and $\mu_{\sigma}=278$ MeV and $\Lambda_{\sigma}=337$ MeV.}
\end{figure}

\begin{figure}[H]
\begin{center}
\includegraphics[width=13cm]{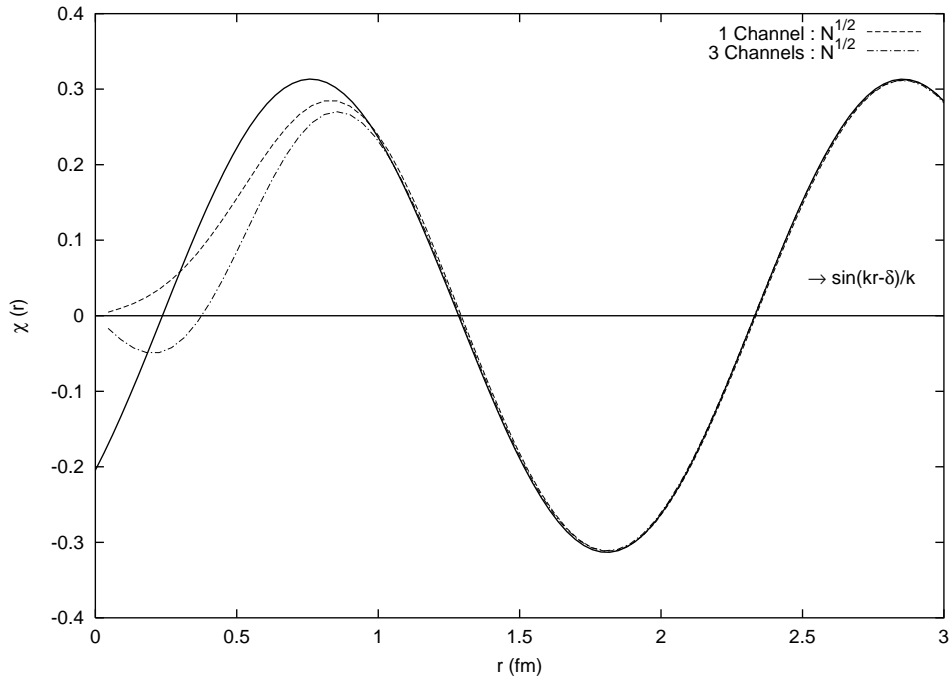}
\end{center}
\caption{\label{RGMkhiR1S0sigmaII} Same as Fig. \ref{RGMkhiR1S0sigmaIIk1} but for $k=3$ fm$^{-1}$.}
\end{figure}

In Fig. \ref{RGMkhiR1S0sigmaIIk1} we give the renormalized wave function for the relative momentum $k=1$ fm$^{-1}$ with the new parameters of the $\sigma$-meson exchange interaction. The corresponding phase shift is $\delta(k=1)=28^o$. We are then in an attractive region and the wave function is ``pulled in'' as compared with the sinusoidal wave in the free case. In Fig. \ref{RGMkhiR1S0sigmaII} we give the relative wave function of $\chi^{l=0}(R)*R$ for $k = 3$ fm$^{-1}$. The phase shift is now negative $\delta(k=3)=-38^o$, due to the short-range repulsion, which gives a wave function ``pushed out'' compared to the free case.
\\

\ 

Now if we compare Figs. \ref{RGMkhiR1S0sigmaR} and \ref{RGMkhiR1S0sigmaII} we see only a very small difference between the wave functions, where the position of the node is slightly shifted to a larger value of $R$ in the new parametrization of the $\sigma$-meson exchange interaction. The same considerations remain valid whichever value of the relative momentum $k$ is chosen. We also note that the position of the node is practically independent of $k$, as known from the $\alpha-\alpha$ system \cite{SAI69}.
\\

The existence of oscillations and nodes in the short distance part of the relative wave function of the NN system is a consequence of the composite structure of the nucleon. It happens that some of the 6q states, when antisymmetrized, vanish, {\it i. e.} they are Pauli forbidden. For the NN problem it is easier to see this effect in the context of the shell model description, developed in Chapter \ref{preliminarychapter}. There, the amplitude of some components of the NN wave function were practically negligible (see Tables \ref{PREBOdiag01}-\ref{PREBOstates10}) which means nearly forbidden.
\\

Taking forbidden states into account, Neudatchin {\it et al.} \cite{NEU75} constructed deep potentials which fit the deuteron properties. In these potentials the forbidden states are unphysical bound states. Various forms of these deep potentials are available (see {\it e. g.} \cite{KUK86}). On the other hand, phenomenological potentials, like Reid's soft core potential \cite{REI68} explain the scattering data through the introduction of a short-range repulsion for $l=0$ and other partial waves. The corresponding wave functions do not posses any node at short distance.
\\

It is a challenge to find out which behaviour is correct for the wave functions. It has been suggested by Khokhlov {\it et al.} \cite{KHO00} that the hard bremsstrahlung $pp\rightarrow pp\gamma$ processes, for example, could provide a test of the quality of the wave function.
\\
\ \\

So far we restricted our study of the role played by the $\sigma$-meson interaction potential to the $^1S_0$ phase shift. If the same $\sigma$-meson exchange interaction, with the same parametrization, is used for the $^3S_1$ case, very similar phase shift is obtained as compared to the $^1S_0$ case. In fact it is impossible to reproduce experimental data as we can see in Fig. \ref{RGMfig3S1sigmasame}. However, as already mentioned, we expect the tensor force to provide the necessary contribution in order to reproduce the experiment. Note that the tensor force does not alter the $^1S_0$ phase shift as long as we do not change the $\sigma$-meson exchange potential.
\\

\subsection{The role of the tensor force}
\ 

As we have seen in Chapter \ref{gbechapter}, the interaction potential (\ref{GBEsimplepotential}) between the constituent quarks contains both a spin-spin and a tensor part which, in the broken $SU_F(3)$, reads

\begin{equation}\label{RGMsimplepotential}
V_{\gamma}(r_{ij}) =  \vec{\lambda}^f_i \cdot  \vec{\lambda}^f_j \left\{ V^{SS}_{\gamma}(r_{ij}) \vec{\sigma}_i\cdot \vec{\sigma}_j + V^T_{\gamma}(r_{ij}) S^T_{ij}\right\}
\end{equation}
\ 

\noindent with contributions from the mesons $\gamma = \pi,\eta$ and $\eta '$ for the NN interaction and where $S_{ij}^T$ is given by

\begin{equation}\label{RGMSijT}
S_{ij}^T=\frac{3(\vec{r}_{ij}\cdot \vec{\sigma}_i)(\vec{r}_{ij}\cdot \vec{\sigma}_j)}{r^2}-\vec{\sigma}_i\cdot \vec{\sigma}_j\ .
\end{equation}

\begin{figure}[H]
\begin{center}
\includegraphics[width=13cm]{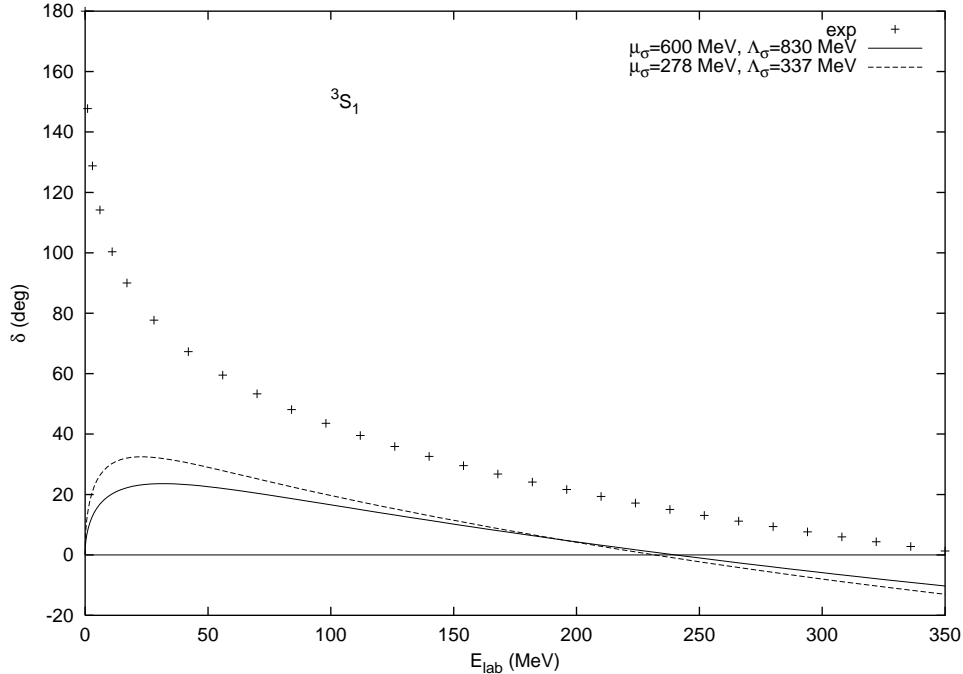}
\end{center}
\caption{\label{RGMfig3S1sigmasame} The $^3S_1$ scattering phase shift with $\sigma$-meson exchange potential. The solid line represent the parametrization $\mu_\sigma=600$ MeV, $\Lambda_\sigma=830$ MeV and the dashed line the parametrization $\mu_\sigma=278$ MeV, $\Lambda_\sigma=337$ MeV. Experimental data are from Ref. \cite{STO93}.}
\end{figure}

In Model II of the GBE interaction used in this chapter, the spin-spin part of the pseudoscalar exchange potential is given by 

\begin{equation}\label{RGMMODELIIspin}
V^{SS}_{\gamma}(r_{ij})= \frac{g_\gamma^2}{4\pi}\frac{1}{12 m_i m_j} \left\{ \mu_\gamma^2 \frac{e^{-\mu_\gamma r_{ij}}}{r_{ij}} -  \Lambda_\gamma^2 \frac{e^{-\Lambda_\gamma r_{ij}}}{r_{ij}} \right\},
\end{equation}
\ 

\noindent expression which is derived with the form factor

\begin{equation}\label{RGMformfactor}
F(q^2)=\sqrt{\frac{\Lambda_\gamma^2}{\Lambda_\gamma^2+\vec{q}^2}}.
\end{equation}
\\

If we use the same form factor (\ref{RGMformfactor}) in the derivation of the tensor potential, we obtain the following form for the tensor part of the pseudoscalar exchange potential

\begin{eqnarray}\label{RGMtensorpotential}
V^T_{\gamma}(r_{ij})&=&G_f\ \frac{g^2_{\gamma}}{4 \pi}\frac{1}{12 m_i m_j}\left\{ \mu^2_{\gamma}(1+\frac{3}{\mu_{\gamma}r}+\frac{3}{\mu^2_{\gamma}r^2})\frac{e^{-\mu_{\gamma}r}}{r}-\Lambda^2_{\gamma}(1+\frac{3}{\Lambda_{\gamma}r}+\frac{3}{\Lambda^2_{\gamma}r^2})\frac{e^{-\Lambda_{\gamma}r}}{r}\right\}\nonumber\\&&\nonumber\\
\end{eqnarray}
\ 

\noindent where the dimensionless global factor $G_f$ has been added in order to allow the adjustment of the strength of this interaction such as to be as close as possible to the experiment \cite{BAR02b}. It would be interesting to analyze the contribution of the ``regularized'' part of (\ref{RGMtensorpotential}). That is why in the following we shall split the tensor potential in the following way

\begin{equation}\label{RGMtensorpotentialsplit}
V^T_{\gamma}(r_{ij})=T_1(r_{ij})+T_2(r_{ij})\ ,
\end{equation}
\ 

\noindent where

\begin{equation}\label{RGMtensorpotentialT1}
T_1(r)=G_f\ \frac{g^2_{\gamma}}{4 \pi}\frac{1}{12 m_i m_j}\left\{ \mu^2_{\gamma}(1+\frac{3}{\mu_{\gamma}r}+\frac{3}{\mu^2_{\gamma}r^2})\frac{e^{-\mu_{\gamma}r}}{r}\right\}\ ,
\end{equation}
\\

\noindent and the ``regularized'' part
\ 

\begin{equation}\label{RGMtensorpotentialT2}
T_2(r)=G_f\ \frac{g^2_{\gamma}}{4 \pi}\frac{1}{12 m_i m_j}\left\{ -\Lambda^2_{\gamma}(1+\frac{3}{\Lambda_{\gamma}r}+\frac{3}{\Lambda^2_{\gamma}r^2})\frac{e^{-\Lambda_{\gamma}r}}{r}\right\}\ .
\end{equation}
\\

All the spin-flavor-color and spatial matrix elements used in the RGM calculations with the tensor force are detailed in Appendix \ref{appendixTENSOR}. The numerical parameters (see next section) are very similar to the previous cases, in particular the stability condition is not influenced by the tensor potential.
\\

\begin{figure}[H]
\begin{center}
\includegraphics[width=14.5cm]{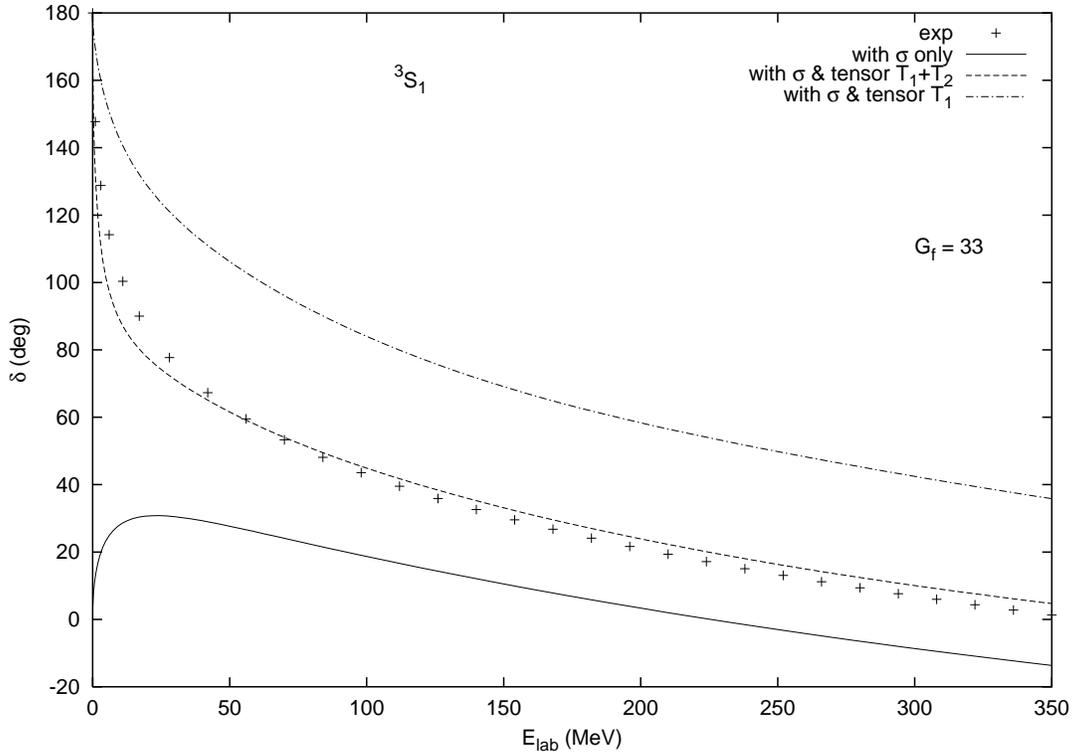}
\end{center}
\caption{\label{RGMfigtensor33} The $^3S_1$ scattering phase shift with the global factor of (\ref{RGMtensorpotential}) with $G_f=33$. The solid and dashed lines represent the $^3S_1$ phase shift without and with tensor force, respectively. The dot-dashed line represent the result of the tensor coupling if the ``regularized'' part ($T_2$, see description in the text) is taken zero.}
\end{figure}

In Fig. \ref{RGMfigtensor33} we show the result  for $^3S_1$ phase shift for the $^3S_1-^3D_1$ coupled RGM calculation. Note that the contribution to the coupled channel RGM calculation comes only from the tensor force (\ref{RGMtensorpotential}). The $\sigma$-meson mass and the cut-off parameter $\Lambda_\sigma$ are chosen the same as for the $^1S_0$ phase shift : $\mu_\sigma=278$ MeV and $\Lambda_\sigma=337$ MeV. If both $T_1$ and $T_2$ are taken into account an agreement is obtained provided the global factor is $G_f= 33$. However, we see in Fig. \ref{RGMfigtensor33} that if we drop the term $T_2$ from the potential (\ref{RGMtensorpotential}) there is more attraction induced by the tensor force. Note that in this case (no $T_2$) a global factor of $G_f=22$ only is required to reproduce the experimental data. We conclude that the ``regularized'' $T_2$ term decreases the attractive effect of the tensor force, as expected from (\ref{RGMtensorpotential}).
\\

\begin{figure}[H]
\begin{center}
\includegraphics[width=14.5cm]{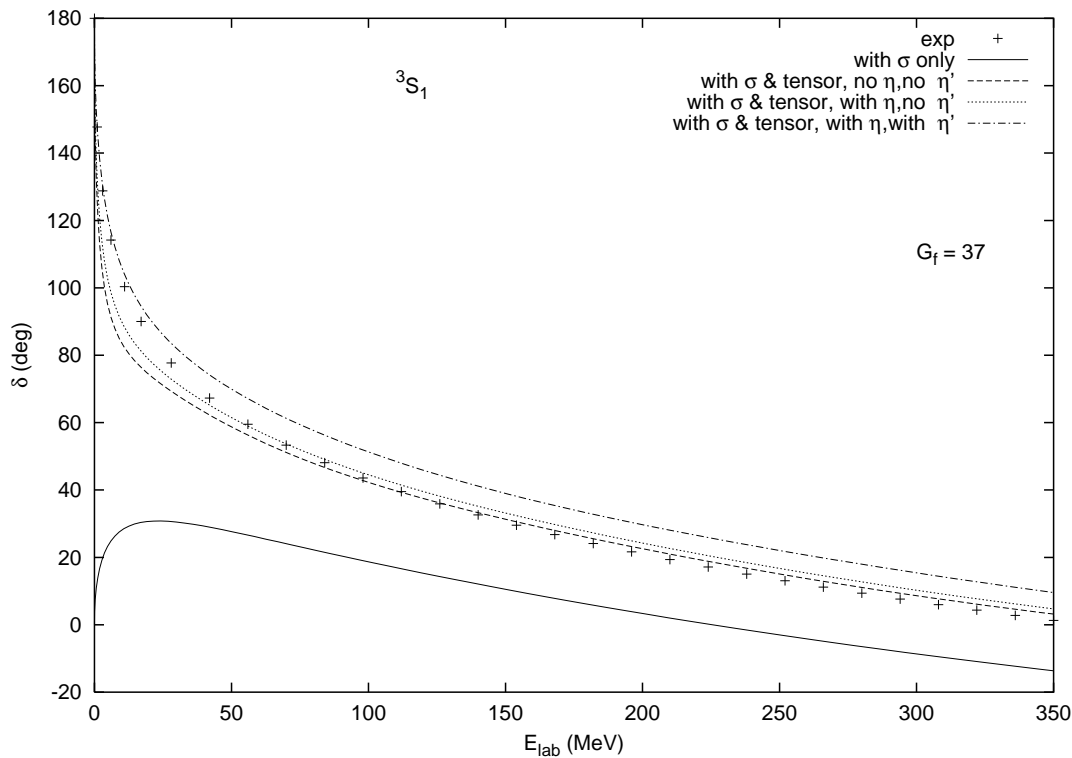}
\end{center}
\caption{\label{RGMfigtensor37} The $^3S_1$ scattering phase shift with the global factor $G_f$ equal to 37 and with the regularization term $T_2$ included. The solid line represents the $^3S_1$ phase shift without tensor force. The dashed, dotted and dot-dashed lines represent the tensor coupling results with the tensor part of the $\pi$-meson exchange only, $\pi$- and $\eta$-meson exchange and $\pi$-, $\eta$ and $\eta '$-meson exchange, respectively.}
\end{figure}

In order to see the influence of the breaking of the $SU_F(3)$ symmetry on the tensor force, in Fig. \ref{RGMfigtensor37} we show explicitely the contribution to the $^3S_1$ scattering phase shift of the $\pi$-, $\eta$- and $\eta$'-meson exchange tensor parts of the potential. In this figure, the global factor has been chosen to reproduce the experimental data when only the $\pi$-meson exchange part of the tensor potential is taken into account. This gives $G_f=37$ as compared to $G_f=33$ if $\eta$-meson and $\eta$'-meson exchange part of the tensor potential are included, as it is the case in Fig. \ref{RGMfigtensor33}. We see that it is the $\pi$-meson which give the largest contribution, but both the $\eta$ and $\eta$' tensor part increase the attraction. The same conclusion arise in the parametrization $\mu_\sigma=600$ MeV and $\Lambda_\sigma=830$ MeV, but where the global factor has to be taken $G_f=65$ \cite{BAR02}.
\\

\begin{figure}[H]
\begin{center}
\includegraphics[width=14cm]{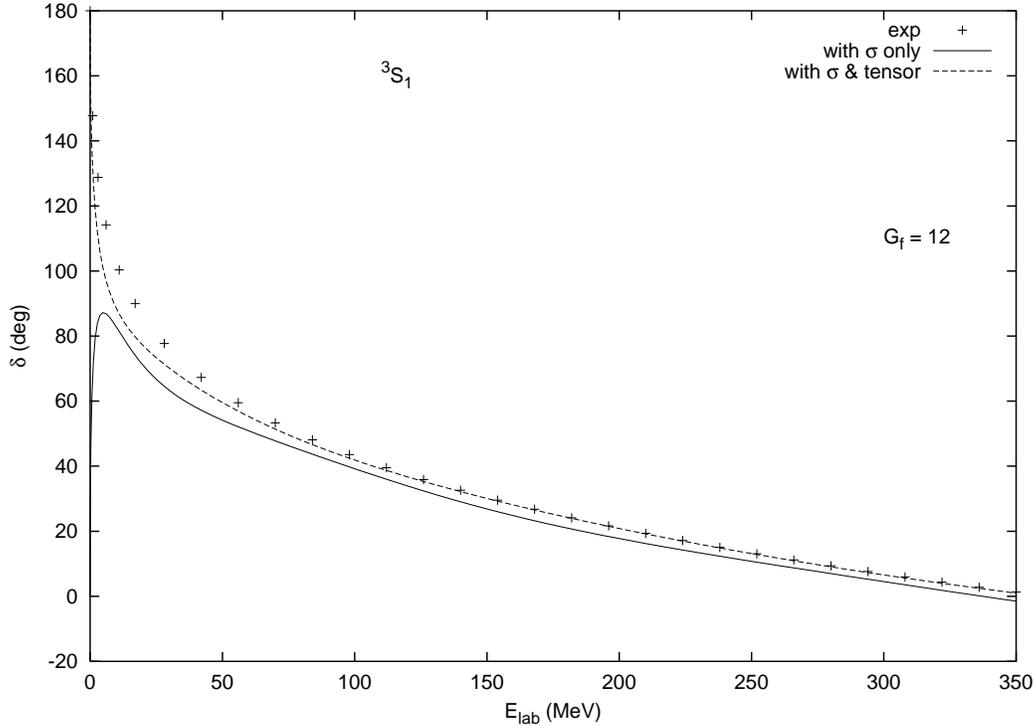}
\end{center}
\caption{\label{RGMfigtensor12} Same as Fig. \ref{RGMfigtensor33} but for $\Lambda_\sigma=357$ MeV and $G_f=12$.}
\end{figure}

Up to now we used an approach with the same $V_\sigma$ of Eq. (\ref{SIGMA}) both for $^1S_0$ and $^3S_1$. However, it is interesting to see how the introduction of a different choice of parameters in the $\sigma$-meson exchange potential for the $^3S_1$ case\footnote{A naive observation is that central part of the Reid soft core potential contains a supplementary term $\frac{e^{-2\mu_\pi r}}{r}$ in the $^3S_1$ potential case as compared to the $^1S_0$ potential, possibly responsible for a deeper middle-range attraction in the NN potential associated with the $^3S_1$ phase shift.} could change the value of the global factor $G_f$ of Eq. (\ref{RGMtensorpotential}). To do so, we have chosen the following parametrization : $\mu_{\sigma}=278$ MeV and $\Lambda_\sigma=357$ MeV which give more attraction than our previous choices. This is illustrated by the solid line in Fig. \ref{RGMfigtensor12}. In order to reproduce the experimental data, the global factor of the total ($T_1+T_2$) tensor potential has to be adjusted to the value $G_f=12$ as shown by the dashed line of Fig. \ref{RGMfigtensor12}. Note that because the $\sigma$-meson exchange potential has changed, the stability condition has to be checked. This leads to the nucleon size parameter $\beta=0.389$ fm which is very close to the previous parametrization, namely $\mu_{\sigma}=278$ MeV and $\Lambda_{\sigma}=337$ MeV, where $\beta$ was given by 0.388 fm.
\\

From Figs. \ref{RGMfig3S1sigmasame} and \ref{RGMfigtensor12} we conclude that the reproduction the $^3S_1$ scattering phase shift needs either a very strong tensor force ($G_f=33$), keeping the same $\sigma$-meson exchange potential as in the $^1S_0$ case, or a stronger $\sigma$-meson exchange potential which would destroy the $^1S_0$ phase shift, but would only require $G_f=12$. The origin of this difficulty could come from the dominance of the regularized delta-function of the spin-spin potential in the non-relativistic case. As seen in Chapter \ref{preliminarychapter}, the Yukawa-part of the potential is completely hidden by the regularized term as seen in Fig. \ref{PREmolNoYuk10}. This regularized term, essential for baryon spectra, leads to a small size for the nucleon, leaving very little room for middle- and long-range attraction.
\\

\section{General considerations about the numerical parameters}\label{sectionnumerical}
\ 

In this section we shall give some information about the stability of the results with the choice of the numerical parameters. By numerical parameters we understand all the parameters introduced in solving the RGM equation which are not part of the GBE model. Among these parameters there is the radius $R_c$ representing the range of the interaction, $N$ the number of locally peaked Gaussians used in the relative wave function expansion (\ref{KAM1}) and the points $R_i,\ (i=1,...,N)$ corresponding to the centers of the Gaussians.
\\

For the scattering problem, in order to see the stability of the results with the parameters, we use the following criterium, consistent with (\ref{RGMcriteria})

\begin{equation}\label{RGMcriteriaNUM}
\left|\ \left| S^{(l)}\right| -1 \right| \ll 1\ ,
\end{equation}

\begin{figure}[H]
\begin{center}
\includegraphics[width=14.5cm]{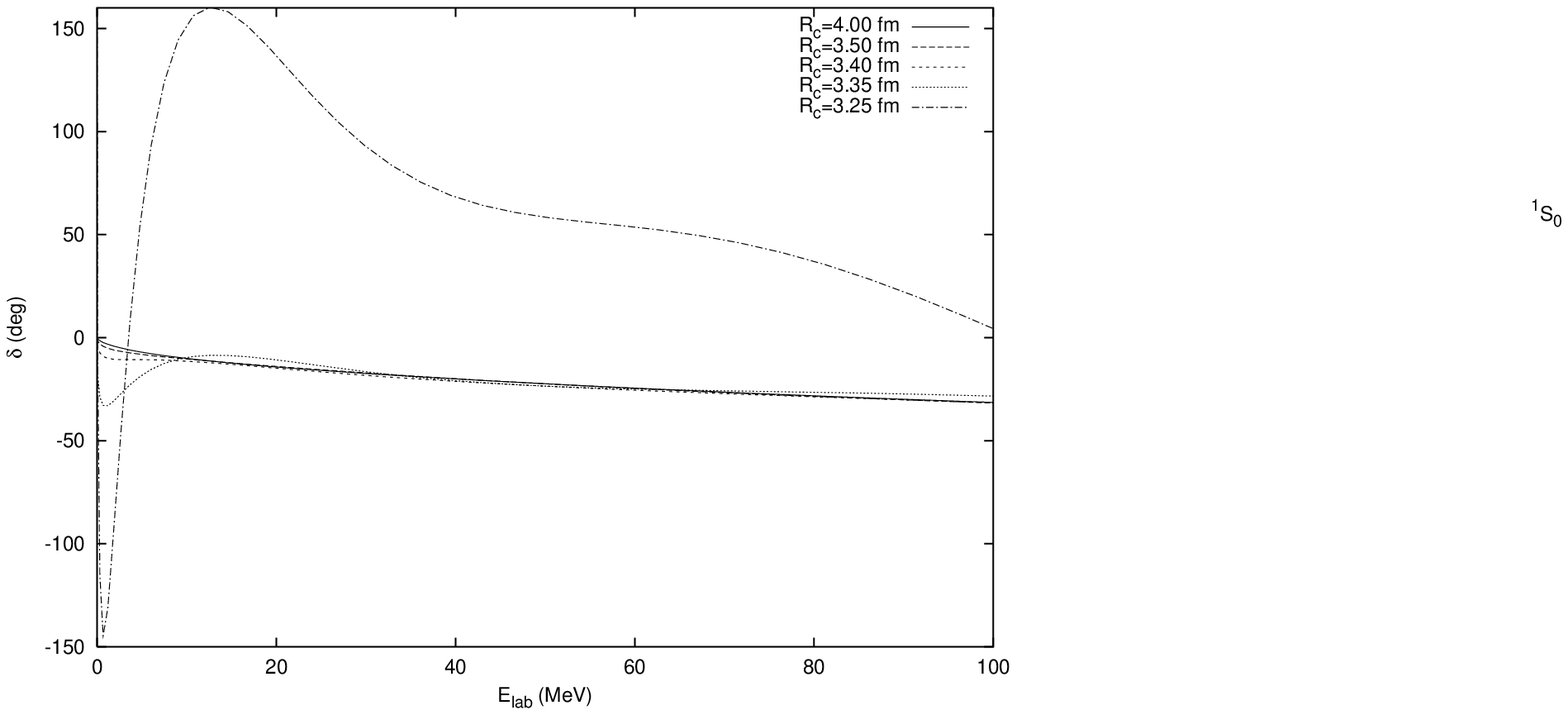}
\end{center}
\caption{\label{RGMfigRc} $^1S_0$ NN scattering phase shift as a function of $E_{lab}$ for different values of $R_c$.}
\end{figure}

Let us first analyze the influence of the parameter $R_c$ on the scattering results. In Table \ref{RGMtableRc} we give the value of the left hand side of (\ref{RGMcriteriaNUM}) for various values of $R_c$ in the particular case of $^1S_0$ wave when we choose the relative momentum $k=2$ fm$^{-1}$. In Fig. \ref{RGMfigRc} we give the $^1S_0$ phase shift for five values of $R_c$. Both the table and the figure correspond to the GBE model without $\sigma$-exchange. The other numerical parameters are

$$N=16;\ R_i=R_0+(i-1)\ t,\ (i=1,...,N); R_0= 0.3\ {\rm fm};\ t=0.35\ {\rm fm}$$

\begin{table}[H]
\centering

\begin{tabular}{|c|c|}
\hline
\rule[-4mm]{0mm}{9.5mm}$R_c$ (fm) & $\left| \left| S^{(0)}\right| -1 \right|$  \\

\hline
\hline

\rule[-3mm]{0mm}{8mm}3.25 & 0.08287 \\
\rule[-3mm]{0mm}{5mm}3.35 & 0.01560 \\
\rule[-3mm]{0mm}{5mm}3.40 & 0.01077 \\
\rule[-3mm]{0mm}{5mm}3.50 & 0.00031 \\
\rule[-3mm]{0mm}{5mm}4.00 & 0.00024 \\
\rule[-3mm]{0mm}{5mm}4.50 & 0.00001 \\
\rule[-3mm]{0mm}{5mm}5.00 & 0.00225 \\
\rule[-3mm]{0mm}{5mm}5.50 & 0.03634 \\

\hline
\end{tabular}
\caption{Value of the left hand side of (\ref{RGMcriteriaNUM}) for various values of the $R_c$ parameter.}\label{RGMtableRc}

\end{table}

From Table \ref{RGMtableRc} and Fig. \ref{RGMfigRc} we can see that for $R_c < 3.5$ fm the phase shift result does not converge. We conclude that the region where the interaction between the two nucleons is effective should go up to 3.5-4.0 fm. Note that for larger values of $R_c$ convergence also disappears. This is not due to the potential range but rather because of numerical considerations. Indeed in this case we maintain the same parameters $R_i$ and $N$. This means that for the inner region, where the Gaussian approximation is used, the relative wave function goes quickly to zero much bevore $R_c$. Then, if $R_c$ is chosen too large, numerical problems appear in the determination of the boundary condition.
\\

The same considerations have been applied to all the others cases ($^3S_1$, with or without $\sigma$-exchange, ...). It appears that the choice of $R_c=4.5$ fm is always reasonable.
\\

Next we shall examine the influence on the results of the number $N$ of Gaussians. However, this parameter has to be determined consistently with the points $R_i$. The reason for this is quite obvious : we have to cover the whole interaction region with sufficient precision. This means enough points, but also well distributed in the interaction region. For simplicity, in all our study, we adopted the equally spaced Gaussians choice for simplicity ($R_i=R_0+(i-1)\ t$). The only drawback of this choice is that we have to increase $N$, the number of Gaussians, in order to take properly into account all the ranges appearing in the GBE interaction. The remaining question is then to determinate the step $t$ and maybe the value of $R_0$.
\\

\begin{table}[H]
\centering

\begin{tabular}{|c|c|c|}
\hline
\rule[-4mm]{0mm}{9.5mm}$N$ & $t$ (fm) &$\left| \left| S^{(0)}\right| -1 \right|$   \\

\hline
\hline

\rule[-3mm]{0mm}{8mm}10 & 0.300 & 0.00001\\
\rule[-3mm]{0mm}{5mm}10 & 0.350 & 0.00001\\
\rule[-3mm]{0mm}{5mm}10 & 0.400 & 0.00789\\
\rule[-3mm]{0mm}{5mm}15 & 0.300 & 0.05812\\
\rule[-3mm]{0mm}{5mm}15 & 0.350 & 0.00086\\
\rule[-3mm]{0mm}{5mm}15 & 0.400 & 0.00011\\
\rule[-3mm]{0mm}{5mm}20 & 0.300 & 0.00004\\
\rule[-3mm]{0mm}{5mm}20 & 0.350 & 8.16185\\
\rule[-3mm]{0mm}{5mm}20 & 0.400 & 7.60720\\

\hline
\end{tabular}
\caption{Value of the convergence criterium (\ref{RGMcriteriaNUM}) in the scattering problem for the determination of the number of Gaussians and their positions $R_i$.}\label{RGMtableNt}

\end{table}
\ 

\begin{figure}[H]
\begin{center}
\includegraphics[width=14.5cm]{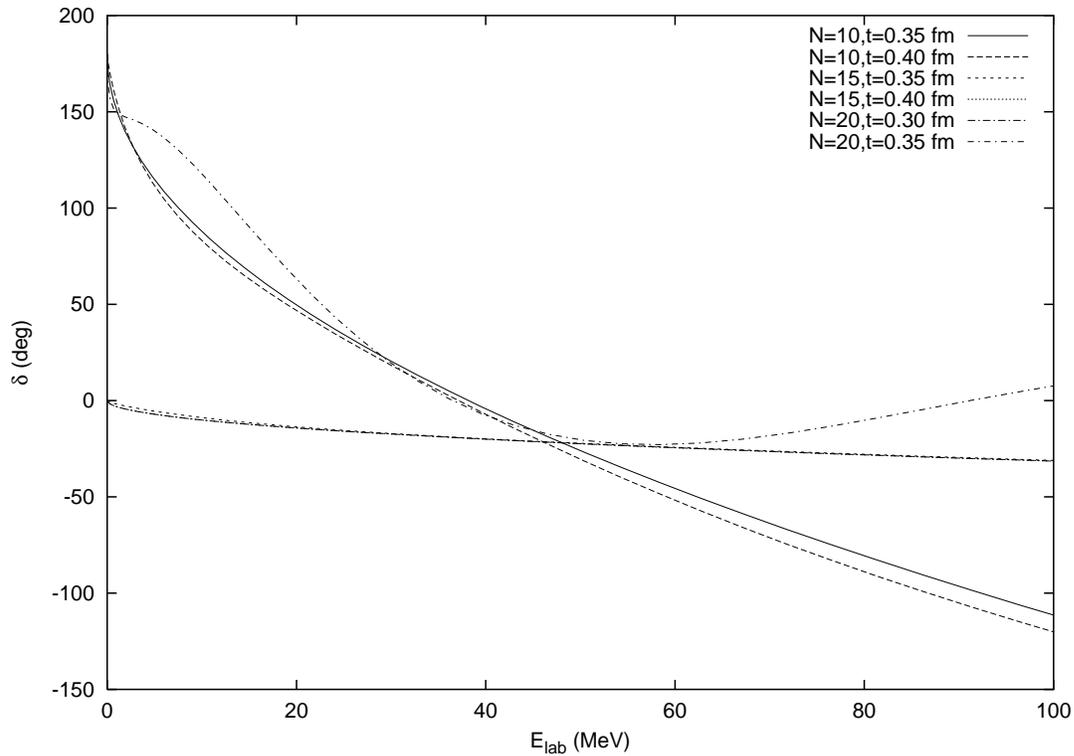}
\end{center}
\caption{\label{RGMfigNt} $^1S_0$ NN scattering phase shift as a function of $E_{lab}$ for different values of the parameters $N$ and $t$.}
\end{figure}

First we take $R_0=0$. Moreover we know that the range of the interaction is about 4.5 fm. Choosing different combinations of $N$ and $t$ covering this region, we obtain different values for the convergence criterium (\ref{RGMcriteriaNUM}). Note that here we are in the case of no bound state which corresponds to a minimum energy greater than $2m_N$. The $^1S_0$ results are presented in Table \ref{RGMtableNt} and in Fig. \ref{RGMfigNt} for $k=2$ fm$^{-1}$ in the GBE model version of this chapter (no $\sigma$-exchange).
\\

From Table \ref{RGMtableNt} and Fig. \ref{RGMfigNt} we conclude that 10 Gaussians is clearly an underestimate for reproducing the scattering phase shifts, because we get at least one bound state which is impossible here. For 20 or more Gaussians we obtain some numerical problems if the step parameter $t$ is taken too big. This indicates that an appropriate choice for the parameters could be $N=15$ Gaussians with a step of $t=0.35$ fm. But other near possibilities could have been chosen.
\\

\ 

Taking $R_0\neq0$ in order to decrease the number of Gaussians, we observe that the scattering results are identical except if $R_0$ is too big. This indicates that the choice of $R_0$ has to be operated in order to ensure the proper inclusion of the short-range part of the GBE Hamiltonian. For the particular case $^1S_0$, $R_0=0.3$ fm is a good choice.
\\

Finally to give an idea of the effectiveness of the RGM method from numerical point of view, we show in Table \ref{RGMcpu} the necessary time to calculate the phase shifts in different situations. These numbers correspond to the case $N=15$ Gaussians and the $^3S_1$ channel. The corresponding platform used here is a PC Athlon 800 MHz running under Linux 2.4 with glibc 2.1. Except the spin-flavor-color matrix elements, all components of the program has been written in C++.
\\

\begin{table}[H]
\centering

\begin{tabular}{|l|c|}
\hline
\rule[-4mm]{0mm}{9.5mm}Case & time (sec.) \\

\hline
\hline

\rule[-3mm]{0mm}{8mm}GBE no $\sigma$, no tensor & 14.7 \\
\rule[-3mm]{0mm}{5mm}GBE with $\sigma$, no tensor  & 17.9\\
\rule[-3mm]{0mm}{5mm}GBE no $\sigma$, no tensor, coupled channels & 29.6\\
\rule[-3mm]{0mm}{5mm}GBE with $\sigma$, no tensor, coupled channels & 18.9\\
\rule[-3mm]{0mm}{5mm}GBE no $\sigma$, with tensor & 32.3\\
\rule[-3mm]{0mm}{5mm}GBE with $\sigma$, with tensor & 31.4\\

\hline
\end{tabular}
\caption{Computational time of the $^3S_1$ phase shift for different cases. The hardware configuration is explained in the text.}\label{RGMcpu}

\end{table}
\ 

\section{Summary}
\ 

The non-relativistic GBE quark model which has been presented in details in Chapter \ref{gbechapter}, describes very well the static properties and excitation spectra of a single hadron. In this chapter we extended the study from a single hadron to two-nucleon systems, employing the resonating group method (RGM). We recalled and applied it to a six-quark system. In particular we derived the six-quark matrix elements as combinations of two-body matrix elements and calculated analytically the integrals in the spin-flavor or color space by use of Mathematica. The matrix elements contributing to the direct terms are simple but the ones corresponding to exchange of quarks between clusters are much more involved to calculate.
\\

The RGM is particularly well adapted in the region where two nucleons overlap each other. There, the internal quark structure of the nucleon is expected to play an important role. An important feature of the RGM is that the effect of the exchange of quarks between the two nucleons is treated exactly. In the overlap region the exchange terms are very important. They contribute to the short- and medium-range parts of the nucleon-nucleon interaction. We obtained results for the $^1S_0$ and the $^3S_1$ phase shifts which reflects the importance of the short-range repulsion and the middle- and long-range attraction.
\\

First, we studied the nucleon-nucleon interaction at short distances within the coupled channels RGM calculation, where the considered channels were $NN$, $\Delta\Delta$ and $CC$. The important ingredients of the problem considered here were the spin-flavor dependence of the Goldstone boson interaction and the quark exchange between the two nucleons. We obtained a short-range repulsion in the nucleon-nucleon system, as it was expected after the preliminary studies of Chapter \ref{preliminarychapter}. However, coupled channels improved only slightly the results, both for the $^1S_0$ and the $^3S_1$ phase shifts. This proved that in the laboratory energy interval 0-350 MeV one-channel approximation is entirely satisfactory for a GBE model, provided the stability condition was respected and one was far enough for the threshold.
\\

The coupled channel RGM calculations also revealed the role of the Pauli forbidden states implying a node in the short-distance part of the relative wave function, depending on the choice of the renormalization method.
\\

However, to reproduce the scattering data, short-, middle- and long-range interactions are all necessary. We then incorporated a $\sigma$-meson exchange interaction at the quark level, consistent with the spirit of the GBE model, and studied the effects of its parametrization on the phase shifts. For the $^1S_0$ case, the attraction induced by the $\sigma$-meson interaction lead to a good description of the experiment. A $\sigma$-meson mass chosen equal to $\mu_\sigma=600$ MeV has been considered as reasonable to reproduce the phase shifts in a large energy interval. But the result is improved if the $\sigma$-meson mass is reduced to 278 MeV. Only small discrepancy remains at small relative momentum. Surprisingly, we found that the coupled channels $NN$-$\Delta\Delta$-$CC$ RGM calculations improve the low energy scattering region when a $\sigma$-meson exchange interaction is added to the GBE hyperfine interaction.
\\

For the $^3S_1$ phase shift, in order to reproduce the experimental data, we added a tensor part to the pseudoscalar-meson exchange potential, consistent with the GBE model and performed the $^3S_1$-$^3D_1$ coupled channels RGM calculation . We showed how the introduction of this tensor force provided the necessary contribution to reproduce the $^3S_1$ phase shift. Note that good results of the $^1S_0$ phase shift are conserved because the tensor force does not play any role in that case. Subsequently we discussed how the strength of the tensor potential has to be significantly enlarged in order to reproduce the $^3S_1$ phase shift. Even a change in the parametrization of the $\sigma$-meson exchange potential which leads to more attraction still implies a strength for the tensor potential one order of magnitude larger than that for the spin-spin part.
\newpage


\chapter{Conclusions and Perspectives}\label{conclusionchapter}

\vspace{3.4cm}
\ 

The objective of this thesis was to derive the nucleon-nucleon (NN) interaction from a microscopic point of view where we suppose that the nucleon is a composite particle formed of three constituent quarks so that the NN system can be viewed as a six-quark system. The quark model we chose to describe the nucleon is the Goldstone boson exchange model, recently proposed and used in baryon spectroscopy. The GBE model is a constituent quark model based on a Hamiltonian where the most important characteristics of QCD, namely the colour confinement and the chiral symmetry are taken into account. This is the only model which can describe so far the correct order of levels of non-strange and strange light baryons, and this is due to the flavor-spin dependence of its hyperfine interaction. It is therefore interesting to find out if the GBE Hamiltonian can be as successful in describing the bound and scattering states of nucleons.
\\

The GBE model has been proposed in 1996 by Glozman and Riska \cite{GLO96a} in the frame of a harmonic oscillator confinement. In the GBE model the gluon exchange hyperfine interaction was completely dropped, leaving in the Hamiltonian only a pseudoscalar meson exchange. The main argument in doing so is that at the energy considered here, of the order of 1 GeV, the fundamental degrees of freedom of QCD, gluons and current quarks, have to be replaced by effective degrees of freedom such as the Goldstone bosons (pseudoscalar mesons) and constituent quarks, due to the spontaneous breaking of chiral symmetry. In this sense, there is no need of quark-gluon interaction anymore for describing light hadrons. The remarkable success of the GBE quark-quark interaction in reproducing the spectra has been explained as coming from the particular symmetry introduced through the spin-flavor operator $\vec{\sigma}_i\cdot\vec{\sigma}_j\ \vec{\lambda}_i^f \cdot\vec{\lambda}_j^f$ and by the short-range part of the hyperfine interaction carrying a proper sign. Several realistic parametrizations of the model have been successively produced by the Graz group \cite{GLO96b,GLO97a,GLO97c,GLO98,WAG00,WAG01} and they led to a closer and closer agreement with the experiment. This agreement concerns the spectra, the strong decays an the electro-weak form factors, as it has been described in Chapter \ref{gbechapter}.
\\

In the various GBE model parametrizations of the Graz group, either a non-relativistic or a relativistic kinematic has been used. If for some observables, such as the electroweak form factors, a relativistic kinematic description is of a crucial importance, the treatment of the relative motion of two baryons being essentially non-relativistic at the energies we are interested in, the study of nucleon-nucleon interaction in a non-relativistic model is a reasona\-ble approach and the first task to achieve. Moreover the study of a six-quark system in a relativistic approach is very difficult, particularly in a dynamical approach such as the resonating group method (RGM) used in Chapter \ref{rgmchapter} of this work, and it is not obvious it could change the quality of the results. A non-relativistic approach has therefore been adopted in this thesis, based on a Hamiltonian formed of a kinetic term, a linear color confinement potential and a hyperfine flavor-spin dependent interaction as presented in details in Chapter \ref{gbechapter}.
\\

In order to see if the GBE model is able to describe baryon-baryon system, we started by an exploratory step which consisted in calculating the NN interaction at zero separation between two nucleons, because the first arising question is whether or not the chiral constituent quark model is able to produce a short-range repulsion in the NN system. For this purpose, we diagonalized the corresponding Hamiltonian in a harmonic oscillator basis containing up to two excitations quanta. In Chapter \ref{preliminarychapter}, using the Born-Oppenheimer (adiabatic) approximation, we obtained an effective internucleon potential between nucleons from the difference between the lowest eigenvalue of the six-quark Hamiltonian and two times the nucleon (three-quark) mass calculated in the same model. We found a very strong effective repulsion of the order of 1 GeV in both $^3S_1$ and $^1S_0$ channels. This repulsion is the effect of the dominance of the $[51]_{FS}$ symmetry state in the system. As our work is the first one which applies the GBE model to the NN system, it was interesting to compare our results with those based on the one-gluon exchange model. We found that the GBE model induces more repulsion at short-range than the OGE model.
\\

Next we calculated the NN potential in the adiabatic approximation as a function of $Z$, the separation distance between the centers of the two three-quark clusters. The orbital part of the six-quark states was constructed either from cluster model or molecular orbital single particle states. The latter are more realistic, having the proper axially and reflectionally symmetries. Also technically they are more convenient. We explicitly showed that they are important at small values of $Z$. In particular we found that the NN potential obtained in the molecular orbital basis has a less repulsive core than the one obtained in the cluster model basis. The adiabatic approximation led then to the expected NN short-range repulsion. Indeed, the potential presents a hard core of about 1 GeV height and a radius of roughly 1 fm. However, the adiabatic potential does not present any attractive pocket neither in what we called Model I nor in Model II. Noting also that none of the bases led to an attractive pocket, we have introduced a $\sigma$-meson exchange interaction between quarks to account for this attraction. The obtained results confirmed our expectation that a scalar-meson exchange interaction can reproduce the necessary middle-range attraction.
\\

To have a better understanding of the two bases we have also calculated the quadrupole moment of the six-quark system as a function of $Z$.  The results show that in the molecular orbital basis the system acquires some small deformations at $Z=0$,  and as a function of the quadrupole moment the adiabatic potential looks more repulsive in the molecular orbital basis than in the cluster model basis.
\\

A important conclusion of Chapter \ref{preliminarychapter} is that in the two different versions of the GBE model the results are very similar. Only small differences appear in the shape of the potential. The parametrization of Model II, relying on more realistic grounds, has been used in the next step of our investigations, namely in the dynamical study based on RGM.
\\

The adiabatic approximation calculations gave us an idea about the size and shape of the hard core produced by the GBE interaction. But within this approximation, comparison with experiment is only qualitative. That is why, taking $Z$ as a generator coordinate, the following step has been to perform a dynamical study. That was the goal of the Chapter \ref{rgmchapter} where the resonating group method has been used in order to calculate the nucleon-nucleon scattering phase shifts. Restricting to quark degrees of freedom and non-relativistic kinematics, the resonating group method is particularly appropriate to treat the interaction between two composite systems and can straightforwardly be applied to the study of the baryon-baryon interaction in constituent quark models. A very important aspect of the resonating group method is the introduction of non-local effects in the potential. Moreover the RGM is particularly well adapted in the region where the two nucleons overlap and where the internal quark structure of the nucleon as well as the interchange of quarks between the two nucleons are expected to play an important role.
\\

The first step in applying the RGM calculation is to achieve the very difficult task of reducing the six-quark matrix elements to combinations of two-body matrix elements and in calculating the two-body matrix elements in the color and spin-flavor space as explained in Appendices \ref{appendixCOUPLED}-\ref{appendixTENSOR}. First, we studied the role of coupled channels on the $^1S_0$ and the $^3S_1$ phase shifts. We considered $NN$, $\Delta\Delta$ and $CC$ channels. We showed that the GBE interaction plus the interchange of quark between two cluster produced a short-range repulsion in the nucleon-nucleon system, confirming the preliminary studies of Chapter \ref{preliminarychapter}. The behaviour of the $^1S_0$ and the $^3S_1$ phase shifts is indeed typical for a short-range repulsion. However at this stage, coupled channels improved only slightly the results, both for the $^1S_0$ and the $^3S_1$ phase shifts. This proved that in the laboratory energy interval 0-350 MeV one-channel approximation is entirely satisfactory for a GBE model, provided the so-called stability condition is respected.
\\

However, to reproduce the scattering data, short-, middle- and long-range interactions are all necessary. Based on the encouraging results obtained in Chapter \ref{preliminarychapter} with the adia\-batic approximation, we then incorporated a $\sigma$-meson exchange interaction at the quark level, consistent with the spirit of the GBE model. We studied in details the effects of its parametrization on the phase shifts. For the $^1S_0$ case, the attraction induced by the $\sigma$-meson interaction led to a good description of the experiment. The region where the phase shift changes sign from positive to negative values is very well described. Only small discrepancies survive at small relative momentum. Interestingly, we found that the coupled channels $NN$-$\Delta\Delta$-$CC$ RGM calculation improves the low energy scattering region when a $\sigma$-meson exchange interaction is added to the pseudoscalar meson exchange.
\\

For the $^3S_1$ phase shift, we have seen that GBE model with the same $\sigma$-meson exchange interaction as for the $^1S_0$ phase shift, could not reproduce the experimental data. This is quite natural because the tensor force is still missing. We then added the tensor part stemming from the pseudoscalar meson exchange, consistent with the GBE model and performed the $^3S_1$-$^3D_1$ coupled channels RGM calculation . We showed how the introduction of this tensor force provided the necessary contribution to bind the NN system and to reproduce the $^3S_1$ phase shift. Note also that the good results for the $^1S_0$ phase shift are unchanged because the tensor force does not play any role in that case. We found that the strength of the tensor force has to be significantly larger than its spin-spin counterpart in order to reproduce the $^3S_1$ phase shift. Modifying the parametrization for the $\sigma$-meson exchange interaction does not change the conclusion. The smallest amplifying factor of the tensor has been found to be equal to 12.
\\

Along this thesis we therefore showed how the GBE model extended to a six-quark system can explain successfully the short-range repulsion for two interacting nucleons. Moreover, by incorporating a missing middle-range attraction through a $\sigma$-meson exchange interaction and adding a tensor force, both in a consistent manner with the GBE interaction, we have been able to reproduce the $^1S_0$ and $^3S_1$ phase shifts. Anyhow, the strong tensor force required to reproduce experimental data could be considered as a drawback. We expect this effect to come from the dominance of the short-range of the spin-spin interaction potential, in the non-relativistic case at least. As seen in Chapter \ref{preliminarychapter}, the regularized term completely hides the attraction due to the Yukawa-part of the potential, necessary for the long-range of the nucleon-nucleon interaction. This causes the regularized term, essential for baryon spectra, to lead to a small size for the nucleon, leaving very little room for middle- and long-range attraction.
\\

In this work we intentionally chose a parametrization of the GBE model consistent with both the strange and non-strange baryon spectra. In a sense this could be seen as a limitation. However our objective was to reproduce the  $^1S_0$ and the $^3S_1$ phase shifts keeping the quality of the spectra unchanged. Looking for new parametrizations of the GBE model, or extending the model to include other interactions such as coming from the vector-meson exchange for example, could be a subsequent task. This could lead to another set of parameters where the tensor force would be on the same level as the spin-spin interaction. Another extension for this work should concern the study of the other partial wave scattering phase shifts as well as the deuteron observables. These are the necessary following steps to conclude about the performances of the GBE model reproducing the bound and scattering states of the NN system and the baryon spectra within the same constituent quark model. Finally, with the present parametrization one can also consider the study of other types of baryon-baryon systems, including the strange ones, for which theoretical predictions on a microscopic ground are still needed.


\appendix

\textwidth 16cm

\chapter{Coupled Channels Matrix Elements}\label{appendixCOUPLED}
\ 

In this appendix we derive the matrix elements needed in the coupled channels RGM approach. In order to achieve this goal we shall introduce the definition of the $|\Delta\Delta>$ and $|CC>$ states consistently with the definition (\ref{PHISI10}) of $|NN>$. Then using these definitions we shall find all the operators appearing in the calculation of the matrix elements. To illustrate the procedure we use the example of the $SI=10$ case. The method can be directly extended to the other sectors. 
\\

\section{The definition of the $|\Delta\Delta>$ and $|CC>$ states}
\

In the same way as for the $|NN>$ channel (see Eq. (\ref{PHISI10})), the spin-flavor part of the $|\Delta\Delta>$ wave function reads

\begin{equation}\label{COUPLEDDD}
\Phi_{\Delta\Delta}^{SI}=\sum C^{\frac{3}{2} \frac{3}{2} S}_{s_1 s_2 s}C^{\frac{3}{2} \frac{3}{2} I}_{\tau_1 \tau_2 \tau} [\chi^{sym}_{s_1}(1)\phi^{sym}_{\tau_1}(1)][\chi^{sym}_{s_2}(2)\phi^{sym}_{\tau_2}(2)]\ ,
\end{equation}
\ 

\noindent with
\begin{eqnarray}\label{COUPLEDsym}
\chi^{sym}_{3/2}&=&\uparrow  \uparrow  \uparrow\ ,\nonumber \\
\chi^{sym}_{1/2}&=&\frac{1}{\sqrt{3}}(\uparrow \uparrow \downarrow + \uparrow \downarrow \uparrow + \downarrow \uparrow \uparrow)\ ,\nonumber \\
\chi^{sym}_{-1/2}&=&\frac{1}{\sqrt{3}}(\uparrow \downarrow \downarrow + \downarrow \uparrow \downarrow + \downarrow \downarrow \uparrow)\ ,\nonumber \\
\chi^{sym}_{-3/2}&=& \downarrow \downarrow \downarrow\ ,
\end{eqnarray}
\ 

\noindent and similarly for the flavor parts with $\uparrow$ replaced by $u$ and $\downarrow$ replaced by $d$. In (\ref{COUPLEDDD}) $S$ and $I$ are the spin and isospin of the  $\Delta\Delta$ system, $\chi(i)$ and $\phi(i)$ are the spin and flavor parts of the $i{\rm^{th}}$ $\Delta$. For example for $S=S_z=1$ and $I=I_z=0$, after inserting the values of the corresponding Clebsch-Gordan coefficients we get

\begin{eqnarray}\label{COUPLEDDD10}
\Phi_{\Delta\Delta}^{10} = \frac{1}{2 \sqrt{10}} &\{ &\sqrt{3}\chi^{sym}_{3/2}(1)\phi^{sym}_{3/2}(1)\chi^{sym}_{-1/2}(2)\phi^{sym}_{-3/2}(2)-\sqrt{4} \chi^{sym}_{1/2}(1)\phi^{sym}_{3/2}(1)\chi^{sym}_{1/2}(2)\phi^{sym}_{-3/2}(2)\nonumber \\
&+&\sqrt{3} \chi^{sym}_{-1/2}(1)\phi^{sym}_{3/2}(1)\chi^{sym}_{3/2}(2)\phi^{sym}_{-3/2}(2)- \sqrt{3} \chi^{sym}_{3/2}(1)\phi^{sym}_{1/2}(1)\chi^{sym}_{-1/2}(2)\phi^{sym}_{-1/2}(2)\nonumber \\
&+&\sqrt{4} \chi^{sym}_{1/2}(1)\phi^{sym}_{1/2}(1)\chi^{sym}_{1/2}(2)\phi^{sym}_{-1/2}(2)-\sqrt{3} \chi^{sym}_{-1/2}(1)\phi^{sym}_{1/2}(1)\chi^{sym}_{3/2}(2)\phi^{sym}_{-1/2}(2)\nonumber \\
&+&\sqrt{3}\chi^{sym}_{3/2}(1)\phi^{sym}_{-1/2}(1)\chi^{sym}_{-1/2}(2)\phi^{sym}_{1/2}(2)-\sqrt{4} \chi^{sym}_{1/2}(1)\phi^{sym}_{-1/2}(1)\chi^{sym}_{1/2}(2)\phi^{sym}_{1/2}(2)\nonumber \\
&+&\sqrt{3} \chi^{sym}_{-1/2}(1)\phi^{sym}_{-1/2}(1)\chi^{sym}_{3/2}(2)\phi^{sym}_{1/2}(2)- \sqrt{3} \chi^{sym}_{3/2}(1)\phi^{sym}_{-3/2}(1)\chi^{sym}_{-1/2}(2)\phi^{sym}_{3/2}(2)\nonumber \\
&+&\sqrt{4} \chi^{sym}_{1/2}(1)\phi^{sym}_{-3/2}(1)\chi^{sym}_{1/2}(2)\phi^{sym}_{3/2}(2)-\sqrt{3} \chi^{sym}_{-1/2}(1)\phi^{sym}_{-3/2}(1)\chi^{sym}_{3/2}(2)\phi^{sym}_{3/2}(2)\} \nonumber
\end{eqnarray}
\

For the $|CC>$ state we introduce the following definition 

\begin{equation}
|CC\rangle = \alpha |NN\rangle +\beta |\Delta\Delta\rangle +\gamma {\cal A}_{\sigma f c}|\Delta\Delta\rangle\ ,
\end{equation}
\ 

\noindent with

\begin{equation}
{\cal A}_{\sigma f c}=\frac{1}{10}[1-\sum_{i=1}^3\sum_{j=4}^6P_{ij}^{\sigma}P_{ij}^{f}P_{ij}^{c}]\ ,
\end{equation}
\ 

\noindent where $P_{ij}^{\sigma}$,$P_{ij}^{f}$ and $P_{ij}^{c}$ are the exchange operators in the spin, isospin and color space definded by Eqs. (\ref{PIJ}), respectively. From the orthonormality conditions $\langle CC|CC\rangle =1$, $\langle CC|NN\rangle =0$ and $\langle CC|\Delta\Delta\rangle =0$ one can determine the coefficients $\alpha$, $\beta$ and $\gamma$ from the equations

\begin{eqnarray}
\langle CC|CC\rangle &=& \alpha^2+ \beta^2 +\gamma(\gamma+2\beta) \langle \Delta\Delta|{\cal A}_{\sigma f c}|\Delta\Delta\rangle \nonumber \\
&&+ 2 \alpha \gamma \langle NN|{\cal A}_{\sigma f c}|\Delta\Delta\rangle =1 \label{COUPLEDalpha}\\
\langle NN|CC\rangle &=& \alpha + \gamma \langle NN|{\cal A}_{\sigma f c}|\Delta\Delta\rangle =0 \label{COUPLEDbeta}\\
\langle \Delta\Delta|CC\rangle &=& \beta + \gamma \langle \Delta\Delta|{\cal A}_{\sigma f c}|\Delta\Delta\rangle =0 \label{COUPLEDgamma}
\end{eqnarray}
\

Introducing (\ref{COUPLEDbeta}) and (\ref{COUPLEDgamma}) in (\ref{COUPLEDalpha}) we obtain

\begin{equation}\label{COUPLEDabg}
\alpha^2+\beta^2+\gamma \beta+1=0
\end{equation}\

Now, using the easy to obtain following results

\begin{eqnarray}
\langle NN|{\cal A}_{\sigma f c}|\Delta\Delta\rangle &=&- \frac{2\sqrt{5}}{45}\nonumber \\
\langle \Delta\Delta|{\cal A}_{\sigma f c}|\Delta\Delta\rangle &=& \frac{4}{45}\nonumber \\
\end{eqnarray}
\ 

\noindent we get $\alpha=\frac{2\sqrt{5}}{45}\gamma$  from (\ref{COUPLEDbeta}) and $\beta=-\frac{4}{45}\gamma$ from (\ref{COUPLEDgamma}), which introduced in (\ref{COUPLEDabg}) give $\gamma=\frac{15}{4}$.
\\

Thus the $|CC>$ state is

\begin{equation}\label{COUPLEDCC}
|CC\rangle = -\frac{\sqrt{5}}{6} |NN\rangle +\frac{1}{3} |\Delta\Delta\rangle -\frac{15}{4} {\cal A}_{\sigma f c}|\Delta\Delta\rangle\ .
\end{equation}
\ 

This is the definition of $|CC>$ used in Section \ref{subsectionCC}.

\section{The coupled channels matrix elements}
\

For $|NN>$ and $|\Delta\Delta>$ it is easy to see that the antisymmetrization operator introduced in Eq. (\ref{RGMpsy}) can be reduced to $\frac{1}{10}\left( 1-9 P_{36}^{o \sigma f c}\right)$. Here we shall first show that we can still write

\begin{equation}\label{COUPLEDP36}
{\cal A}|CC\rangle = \frac{1}{10}\left( 1-9 P_{36}^{o \sigma f c}\right) |CC\rangle  .
\end{equation}
\ 

\noindent which is less trivial. For the two first terms of (\ref{COUPLEDCC}) it is obvious. For the last term, because of the presence of the operator ${\cal A}_{\sigma f c}$, it is a bit more tricky. However, for this last term we have

\begin{equation}
{\cal A}\ {\cal A}_{\sigma f c}|\Delta \Delta \rangle = \frac{1}{10}\left(1+9 P_{36}^{o}\right)  {\cal A}_{\sigma f c}|\Delta \Delta\rangle
\end{equation}
\ 

\noindent because ${\cal A}_{\sigma f c}|\Delta \Delta \rangle$ is $FSC$-antisymmetric, so that the role of ${\cal A}$ reduces to acting with $P_{36}^o$ on the orbital part because the orbital part of each $\Delta$ is already symmetric.
\\

On the other hand, we have

\begin{equation}
(1-9 P_{36}^{o \sigma f c}){\cal A}_{\sigma f c}|\Delta \Delta \rangle = {\cal A}_{\sigma f c}|\Delta \Delta \rangle- 9 P_{36}^{o \sigma f c} {\cal A}_{\sigma f c}|\Delta \Delta\rangle\ ,
\end{equation}
\ 

\noindent which means that we have to show that

\begin{equation}\label{COUPLEDOOFS}
P_{36}^{o}{\cal A}_{\sigma f c}|\Delta \Delta \rangle = - P_{36}^{o \sigma f c} {\cal A}_{\sigma f c}|\Delta \Delta\rangle\ .
\end{equation}
\

As the orbital and the $FSC$ parts are uncoupled, it follows that

\begin{equation}
P_{36}^{o \sigma f c}{\cal A}_{\sigma f c}|\Delta \Delta \rangle =  P_{36}^{o} P_{36}^{\sigma f c} {\cal A}_{\sigma f c}|\Delta \Delta\rangle = P_{36}^{o} \left( - {\cal A}_{\sigma f c}|\Delta \Delta \rangle\right)
\end{equation}
\ 

\noindent because ${\cal A}_{\sigma f c}|\Delta \Delta\rangle$ is antisymmetric in $FSC$, so that (\ref{COUPLEDOOFS}) is proved from which (\ref{COUPLEDP36}) follows.
\\

To solve the RGM, in addition to the $NN$ and $\Delta\Delta$ diagonal and off-diagonal matrix elements, we need matrix elements of the type $\langle XX|O|CC\rangle$ where $XX$ stands for $NN$, $\Delta\Delta$ and $CC$ and $O$ stands for $V_{12},\ V_{36},\ V_{12}P_{36}^{o \sigma f c},\ V_{36}P_{36}^{o \sigma f c},\ V_{13}P_{36}^{o \sigma f c},\ V_{16}P_{36}^{o \sigma f c}$ and $V_{14}P_{36}^{o \sigma f c}$.
\\

We then have to calculate (in the $SI=10$ case)

\begin{equation}
\langle XX|O|CC\rangle=-\frac{\sqrt{5}}{6}\langle XX|O|NN\rangle-\frac{1}{24}\langle XX|O|\Delta \Delta\rangle+\frac{3}{8}\langle XX|O \sum_{i,j} P_{ij}^{\sigma f c}|\Delta \Delta\rangle
\end{equation}

Let us considered the three possible cases
\\

\begin{itemize}
\item $|XX\rangle=|NN\rangle$
\ \\

We get

\begin{equation}
-\frac{\sqrt{5}}{6}\langle NN|O|NN\rangle-\frac{1}{24}\langle NN|O|\Delta \Delta\rangle+\frac{3}{8}\langle NN|O \sum_{i,j}P_{ij}^{\sigma f c}|\Delta \Delta\rangle
\end{equation}
where the only unknown term is

\begin{equation}\label{COUPLEDitem1}
\langle NN|O \sum_{i,j}P_{ij}^{\sigma f c}|\Delta \Delta\rangle
\end{equation}

\item $|XX\rangle=|\Delta \Delta\rangle$
\ \\

In the same way, the only unknown term is

\begin{equation}\label{COUPLEDitem2}
\langle \Delta\Delta|O \sum_{i,j} P_{ij}^{\sigma f c}|\Delta \Delta\rangle
\end{equation}

\item $|XX\rangle=|CC\rangle$
\ \\

This time we have

\begin{eqnarray}\label{COUPLEDdure}
&&-\frac{\sqrt{5}}{6}\langle CC|O|NN\rangle-\frac{1}{24}\langle CC|O|\Delta \Delta\rangle+\frac{3}{8}\sum_{i,j}\langle CC|O P_{ij}^{\sigma f c}|\Delta \Delta\rangle \nonumber \\
&=&-\frac{\sqrt{5}}{6}\langle CC|O|NN\rangle-\frac{1}{24}\langle CC|O|\Delta \Delta\rangle+\frac{3}{8} \left\{ -\frac{\sqrt{5}}{6}\langle NN|O \sum_{i,j}P_{ij}^{\sigma f c}|\Delta \Delta\rangle  \right. \nonumber \\
&&\left. +\frac{1}{3}\langle \Delta \Delta|O \sum_{i,j}P_{ij}^{\sigma f c}|\Delta \Delta\rangle - \frac{15}{4}\langle \Delta \Delta| \sum_{i,j}P_{ij}^{\sigma f c} O^+  {\cal A}_{\sigma f c}|\Delta \Delta\rangle \right\}
\end{eqnarray}
where $O^+$ is the Hermitian conjugate of $O$.
\\

The four first terms of the last equation of (\ref{COUPLEDdure}) can be derived from the previous cases. It turns out that the only unknown term is

\begin{equation}\label{COUPLEDitem3}
\langle \Delta \Delta| \sum_{i,j}P_{ij}^{\sigma f c} O^+  {\cal A}_{\sigma f c}|\Delta \Delta\rangle
\end{equation}

\end{itemize}

The matrix elements of the Eqs. (\ref{COUPLEDitem1}) and (\ref{COUPLEDitem2}) are computed using a program based on MATHEMATICA, with similar techniques as those described in Chapter \ref{rgmchapter}. For the Eq. (\ref{COUPLEDitem3}) we need to proceed further before using MATHEMATICA. For this we need the relations

\begin{equation}\label{COUPLEDrelationdetails}
P_{ij}^{\sigma f c}O^+_{kl}{\cal A}_{\sigma f c}|\Psi\rangle = \left\{ \begin{array}{ll}
 - O^+_{jl}{\cal A}_{\sigma f c}|\Psi\rangle & {\rm if}\ k=i,l\neq j \nonumber \\
 - O^+_{il}{\cal A}_{\sigma f c}|\Psi\rangle & {\rm if}\ k=j,l\neq i \nonumber \\
 - O^+_{kj}{\cal A}_{\sigma f c}|\Psi\rangle & {\rm if}\ k\neq j,l=i \nonumber \\
 - O^+_{ki}{\cal A}_{\sigma f c}|\Psi\rangle &  {\rm if}\ k\neq i,l=j \nonumber \\
 - O^+_{ji}{\cal A}_{\sigma f c}|\Psi\rangle & {\rm if}\ k=i,l= j \nonumber \\
 - O^+_{ij}{\cal A}_{\sigma f c}|\Psi\rangle & {\rm if}\ k=j,l= i \nonumber \\
 - O^+_{kl}{\cal A}_{\sigma f c}|\Psi\rangle & {\rm if}\ (k,l)\neq  (i,j)\ \forall i,j,k,l \nonumber \\
\end{array} \right.
\end{equation}
\ 

\noindent where the property $P_{ij}^{\sigma f c}{\cal A}_{\sigma f c}|\Psi\rangle=-{\cal A}_{\sigma f c}|\Psi\rangle$ has been used. For example if we consider the operator $O=O_{12}P_{36}^{\sigma f c}$, we have $O^+=P_{36}^{\sigma f c}O_{12}=O_{12}P_{36}^{\sigma f c}$, so in this case

\begin{eqnarray}\label{COUPLEDrelationexample}
\sum_{i\in A,j\in B}P_{ij}^{\sigma f c} O^+ {\cal A}_{\sigma f c}|X_AX_B\rangle &=& - \sum_{i\in A,j\in B}P_{ij}^{\sigma f c} O_{12} {\cal A}_{\sigma f c}|X_AX_B\rangle \nonumber \\
&=&(O_{42} + O_{51} + O_{62} + O_{14} + O_{15} + O_{16} \nonumber \\
&&+ O_{12} + O_{12} + O_{12})  {\cal A}_{\sigma f c}|X_AX_B\rangle \nonumber \\ \nonumber\\
&=&(3\ O_{12}+6\ O_{36}){\cal A}_{\sigma f c}|X_AX_B\rangle
\end{eqnarray}
\

In the same way we get the other operators. The results are gathered in Table \ref{COUPLEDtable}. The last task is then to use the relation ${\cal A}_{\sigma f c}=\frac{1}{10}\left(1-9\ P_{36}^{\sigma f c}\right)$ with the usual techniques in order to obtain the spin-flavor-color matrix elements.
\\

\begin{table}[H]
\centering

\begin{tabular}{|c|c|}
\hline
\rule[-3mm]{0mm}{7mm}$O$ & $\sum_{i\in A,j\in B}P_{ij}^{\sigma f c} O^+ {\cal A}_{\sigma f c}|X_AX_B\rangle$ \\

\hline
\hline

\rule[-3mm]{0mm}{8mm}$O_{12}$ & $-(3\ O_{12}+6\ O_{36}){\cal A}_{\sigma f c}|X_AX_B\rangle$ \\
\rule[-3mm]{0mm}{5mm}$O_{36}$ & $-(4\ O_{12}+5\ O_{36}){\cal A}_{\sigma f c}|X_AX_B\rangle$ \\
\rule[-3mm]{0mm}{5mm}$O_{12}P_{36}^{\sigma f c}$ & $(3\ O_{12}+6\ O_{36}){\cal A}_{\sigma f c}|X_AX_B\rangle$ \\
\rule[-3mm]{0mm}{5mm}$O_{36}P_{36}^{\sigma f c}$ & $(4\ O_{12}+5\ O_{36}){\cal A}_{\sigma f c}|X_AX_B\rangle$ \\
\rule[-3mm]{0mm}{5mm}$O_{13}P_{36}^{\sigma f c}$ & $(3\ O_{12}+6\ O_{36}){\cal A}_{\sigma f c}|X_AX_B\rangle$ \\
\rule[-3mm]{0mm}{5mm}$O_{16}P_{36}^{\sigma f c}$ & $(4\ O_{12}+5\ O_{36}){\cal A}_{\sigma f c}|X_AX_B\rangle$ \\
\rule[-3mm]{0mm}{5mm}$O_{14}P_{36}^{\sigma f c}$ & $(4\ O_{12}+5\ O_{36}){\cal A}_{\sigma f c}|X_AX_B\rangle$ \\

\hline

\end{tabular}
\caption{Expression of $\sum_{i\in A,j\in B}P_{ij}^{\sigma f c} O^+ {\cal A}_{\sigma f c}|X_AX_B\rangle$ for different operators $O$.}\label{COUPLEDtable}

\end{table}

\newpage
\thispagestyle{empty} 
\ 
\newpage
\renewcommand{\chaptermark}[1]%
{\markboth{#1}{#1}}
\renewcommand{\sectionmark}[1]%
{\markboth{ \thesection\ \  #1}}
\rhead[\fancyplain{}{\bfseries\leftmark}]{\fancyplain{}{\bfseries\thepage}}
\lhead[\fancyplain{}{\bfseries\thepage}]{\fancyplain{}{\bfseries\rightmark}}
\cfoot{}
\chapter{The Norm}\label{appendixNORM}
\ 

In this appendix we calculate the orbital matrix elements of the norm presented in the Table \ref{RGMtableorbital} of Chapter \ref{rgmchapter}. In coupled channels RGM, the norm is given by

\begin{equation}
N_{\alpha,\beta}=\frac{1}{10}\left( <\alpha|{\bf 1}^{\sigma fc}|\beta> <\Psi_{6q}|\Psi_{6q}> - 9 <\alpha|P_{36}^{\sigma fc}|\beta> <\Psi_{6q}|P^o_{36}|\Psi_{6q}> \right)
\end{equation}
\\

We then have to compute $<\Psi_{6q}|\Psi_{6q}>$ and $<\Psi_{6q}|P^{o}_{36}|\Psi_{6q}>$ with

\begin{equation}
|\Psi_{6q}>=\prod_{k=1}^3 \phi(\vec{r}_k-\frac{\vec{R}_i}{2},b) \prod_{k'=4}^6 \phi(\vec{r}_{k'}+\frac{\vec{R}_i}{2},b).
\end{equation}
\ 

\noindent where $\phi$ is the wave function of the individual quark. We introduce

\begin{eqnarray}
N_d=<\Psi_{6q}|\Psi_{6q}>&=&\int dr^3_1...dr^3_6 \Psi^*_{6q}(R_i) \Psi_{6q}(R_j)\nonumber\\
&=&\prod_{k=1}^3 \prod_{k'=4}^6 \int dr^3_1...dr^3_6 \phi^*(\vec{r}_k-\frac{\vec{R}_i}{2},b) \phi^*(\vec{r}_{k'}+\frac{\vec{R}_i}{2},b)\nonumber\\
&&  \phi(\vec{r}_k-\frac{\vec{R}_j}{2},b) \phi(\vec{r}_{k'}+\frac{\vec{R}_j}{2},b)\nonumber\\
&=&(\int dr^3_1 \phi^*(\vec{r}_1-\frac{\vec{R}_i}{2},b) \phi(\vec{r}_1-\frac{\vec{R}_j}{2},b))^6 \nonumber\\
&=&((\frac{1}{\sqrt{\pi}b})^3 \int dr^3 e^{-\frac{(\vec{r}-\frac{\vec{R}_i}{2})^2}{2b^2}} e^{-\frac{(\vec{r}-\frac{\vec{R}_j}{2})^2}{2b^2}})^6,
\end{eqnarray}

\noindent and using

\begin{equation}
-\frac{(\vec{r}-\frac{\vec{R}_i}{2})^2}{2b^2}-\frac{(\vec{r}-\frac{\vec{R}_j}{2})^2}{2b^2}=\frac{-1}{b^2}((\frac{\vec{R}_i-\vec{R}_j}{4})^2+(\vec{r}-\frac{\vec{R}_i+\vec{R}_j}{4})^2)
\end{equation}

\noindent we obtain

\begin{eqnarray}\label{NORMdirect}
N_d&=&( (\frac{1}{\sqrt{\pi}b})^3 e^{-\frac{1}{b^2}(\frac{\vec{R}_i-\vec{R}_j}{4})^2} 4\pi \int_0^{\infty} dr\ r^2 e^{-\frac{r^2}{b^2}})^6 \nonumber\\
&=&((\frac{1}{\sqrt{\pi}b})^3 e^{-\frac{1}{b^2}(\frac{\vec{R}_i-\vec{R}_j}{4})^2} 4\pi \frac{b^3 \sqrt{\pi}}{4})^6 \nonumber\\
&=& ( e^{-\frac{1}{b^2}(\frac{\vec{R}_i-\vec{R}_j}{4})^2})^6 \nonumber\\
&=& \Delta^6(\vec{R}_i,\vec{R}_j)
\end{eqnarray}

\noindent where

\begin{equation}
\Delta(\vec{R}_i,\vec{R}_j)= e^{-\frac{1}{b^2}(\frac{\vec{R}_i-\vec{R}_j}{4})^2}.
\end{equation}
\\

In the same way we have

\begin{eqnarray}
<\Psi_{6q}|P^o_{36}|\Psi_{6q}>&=&\Delta^4(\vec{R}_i,\vec{R}_j)\cdot \nonumber\\
&&\int dr^3_3 \phi^*(\vec{r}_3-\frac{\vec{R}_i}{2},b) \phi(\vec{r}_3+\frac{\vec{R}_j}{2},b) \cdot \nonumber\\
&& \int dr^3_6 \phi^*(\vec{r}_6+\frac{\vec{R}_i}{2},b) \phi(\vec{r}_6-\frac{\vec{R}_j}{2},b).
\end{eqnarray}

Using

\begin{eqnarray}
\int dr^3_3 \phi^*(\vec{r}_3-\frac{\vec{R}_i}{2},b) \phi(\vec{r}_3+\frac{\vec{R}_j}{2},b)&=&(\frac{1}{\sqrt{\pi}b})^3 \int dr^3 e^{-\frac{(\vec{r}-\frac{\vec{R}_i}{2})^2}{2b^2}} e^{-\frac{(\vec{r}+\frac{\vec{R}_j}{2})^2}{2b^2}}\nonumber\\
&=&(\frac{1}{\sqrt{\pi}b})^3 e^{-\frac{1}{b^2}(\frac{\vec{R}_i+\vec{R}_j}{4})^2} \int dr^3 e^{-\frac{r^2}{b^2}} \nonumber\\
&=&e^{-\frac{1}{b^2}(\frac{\vec{R}_i+\vec{R}_j}{4})^2}\nonumber\\
&=&\Delta(\vec{R}_i,-\vec{R}_j),
\end{eqnarray}

\noindent we get

\begin{equation}
<\Psi_{6q}|P^o_{36}|\Psi_{6q}>=\Delta^4(\vec{R}_i,\vec{R}_j)\Delta^2(\vec{R}_i,-\vec{R}_j)=N^e(\vec{R}_i,\vec{R}_j)
\end{equation}
\ 

\noindent with 

\begin{equation}\label{NORMexchange}
N^e(\vec{R}_i,\vec{R}_j)=\Delta^4(\vec{R}_i,\vec{R}_j)\Delta^2(\vec{R}_i,-\vec{R}_j)
\end{equation}
\\

In Tables \ref{NORM00}, \ref{NORM01} and \ref{NORM11} we give all the flavor-spin-color matrix elements associated to the norm calculation.
\\

\begin{table}[H]
\centering

\begin{tabular}{|c|c|c|c|c|c|c|}
\hline
\rule[-2mm]{0mm}{7mm}$\alpha$ & $NN$ & $NN$  & $\Delta\Delta$ & $NN$ & $\Delta\Delta$ & $CC$ \\
\rule[-4mm]{0mm}{5mm}$\beta$  & $NN$ & $\Delta\Delta$ & $\Delta\Delta$ & $CC$ & $CC$  & $CC$ \\

\hline
\hline

\rule[-3mm]{0mm}{8mm}$1$                      &   135 &    0 &   135 &    0 &    0 &   135 \\
\rule[-3mm]{0mm}{5mm}$P_{36}^{f \sigma c}$    &    35 &   20 &     5 &    8 &  -16 &  -154 \\

\hline

\rule[-4mm]{0mm}{10mm}factor & $\frac{1}{135}$ & $\frac{1}{135}$ & $\frac{1}{135}$ 
& $\frac{\sqrt{10}}{135}$ & $\frac{\sqrt{10}}{135}$ & $\frac{1}{135}$ \\
\hline

\end{tabular}
\caption{Matrix elements $\langle \alpha|O|\beta \rangle$ of different operators for ($S$,$I$) = (0,0).}\label{NORM00}

\end{table}
\ 

\begin{table}[H]
\centering

\begin{tabular}{|c|c|c|c|c|c|c|}
\hline
\rule[-2mm]{0mm}{7mm}$\alpha$ & $NN$ & $NN$  & $\Delta\Delta$ & $NN$ & $\Delta\Delta$ & $CC$ \\
\rule[-4mm]{0mm}{5mm}$\beta$  & $NN$ & $\Delta\Delta$ & $\Delta\Delta$ & $CC$ & $CC$  & $CC$ \\

\hline
\hline

\rule[-3mm]{0mm}{8mm}$1$                    &    81 &    0 &    81 &    0 &    0 &    81 \\
\rule[-3mm]{0mm}{5mm}$P_{36}^{f \sigma c}$  &    -1 &    4 &     1 &  -12 &   24 &   -63 \\

\hline

\rule[-4mm]{0mm}{10mm}factor & $\frac{1}{81}$ & $\frac{\sqrt5}{81}$ & $\frac{1}{81}$ 
& $\frac{\sqrt5}{81}$ & $\frac{1}{81}$ & $\frac{1}{81}$ \\

\hline

\end{tabular}
\caption{Matrix elements $\langle \alpha|O|\beta \rangle$ of different operators for ($S$,$I$) = (1,0) and (0,1).}\label{NORM01}

\end{table}
\ 

\begin{table}[H]
\centering

\begin{tabular}{|c|c|c|c|c|c|c|}
\hline
\rule[-2mm]{0mm}{7mm}$\alpha$ & $NN$ & $NN$  & $\Delta\Delta$ & $NN$ & $\Delta\Delta$ & $CC$ \\
\rule[-4mm]{0mm}{5mm}$\beta$  & $NN$ & $\Delta\Delta$ & $\Delta\Delta$ & $CC$ & $CC$  & $CC$ \\

\hline
\hline
\rule[-3mm]{0mm}{8mm}$1$                    &   243 &    0 &   243 &    0 &     0 &     243 \\
\rule[-3mm]{0mm}{5mm}$P_{36}^{f \sigma c}$  &   31  &   20 &     1 &  320 & -3080 & -169756 \\

\hline

\rule[-4mm]{0mm}{10mm}factor & $\frac{1}{243}$ & $\frac{1}{243}$ & $\frac{1}{243}$ 
& $\frac{1}{243 \cdot \sqrt{1486}}$ & $\frac{1}{243 \cdot \sqrt{1486}}$ & $\frac{1}{243 \cdot 743}$ \\

\hline

\end{tabular}
\caption{Matrix elements $\langle \alpha|O|\beta \rangle$ of different operators for ($S$,$I$) = (1,1).}\label{NORM11}

\end{table}

\newpage
\thispagestyle{empty} 
\ 
\newpage
\chapter{The Kinetic Energy}\label{appendixKE}
\ 

Here, we calcultate the kinetic energy matrix element used in RGM. There are two different ways to achieve this task.
\\
\ 

{\bf Method I} : The first method is to sum up the kinetic energy of all six quarks, extracting the center of mass movement only after. This method requires the calculation of single particle operators only. We first introduce

\begin{equation}\label{KET1}
T_{sp}= \sum_i \frac{p_i^2}{2m} = \sum_i t_i
\end{equation}
\ 

\noindent where all the quarks are supposed to have the same mass. From (\ref{KET1}) we still have to extract $T_R=\frac{1}{2M}(\sum_i p_i)^2$, the center of mass motion, with $M=\sum_i m_i=6m$ to get $T=T_{sp}-T_R$.
\\
\ \\

{\bf Method II} : This method takes into account the center of mass movement from the very beginning, so we have

\begin{eqnarray}
T&=&\sum_i \frac{p_i^2}{2m}-T_R = \sum_i \frac{p_i^2}{2m}-\frac{1}{2M}(\sum_i p_i)^2\nonumber\\
&=&\frac{5}{12m}\sum_i p_i^2-\frac{1}{6m}\sum_{i<j}\vec{p}_i\cdot\vec{p}_j\nonumber\\
&=&\sum_{i<j} t_{ij} = \sum_{i<j} \frac{1}{12m} (\vec{p}_i-\vec{p}_j)^2 \ .
\end{eqnarray}
\ 

We then see the presence of the two-body operators

\begin{equation}
t_{ij}= \frac{1}{12m} (p_i^2 + p_j^2)-\frac{1}{6m} \vec{p}_i\cdot\vec{p}_j
\end{equation}
\\

The matrix elements to compute are then $<\Psi_{6q}|t_i|\Psi_{6q}>$,$<\Psi_{6q}|t_i P_{36}|\Psi_{6q}>$,$<\Psi_{6q}|t_{ij}|\Psi_{6q}>$ and $<\Psi_{6q}|t_{ij}P_{36}|\Psi_{6q}>$ where $P_{36}$ stands for $P^{o\sigma fc}_{36}$.
\\

We have

\begin{eqnarray}
<\Psi_{6q}|t_k|\Psi_{6q}>&=&\Delta^5(\vec{R}_i,\vec{R}_j)\frac{1}{2m}\int dr^3_k \phi^*(\vec{r}_k-\frac{\vec{R}_i}{2},b) p_k^2 \phi(\vec{r}_k-\frac{\vec{R}_j}{2},b)\nonumber\\
&=&\Delta^5(\vec{R}_i,\vec{R}_j)\frac{1}{2m}\int dr^3_k \vec{\nabla} \phi^*(\vec{r}_k-\frac{\vec{R}_i}{2},b) \cdot \vec{\nabla} \phi(\vec{r}_k-\frac{\vec{R}_j}{2},b)\nonumber\\
&=&\Delta^5(\vec{R}_i,\vec{R}_j)\frac{1}{2m}(\frac{1}{\sqrt{\pi}b})^3\int dr^3_k \frac{1}{b^4} (\vec{r}_k-\frac{\vec{R}_i}{2})\cdot (\vec{r}_k-\frac{\vec{R}_j}{2}) \cdot \nonumber\\
&& e^{\frac{-1}{2b^2}((\vec{r}_k-\frac{\vec{R}_i}{2})^2+ (\vec{r}_k-\frac{\vec{R}_j}{2})^2)}\nonumber\\
&=&\Delta^5(\vec{R}_i,\vec{R}_j)\frac{1}{2mb^4}(\frac{1}{\sqrt{\pi}b})^3\int dr^3_k ((\vec{r}_k-\frac{\vec{R}_i+\vec{R}_j}{4})^2-(\frac{\vec{R}_i-\vec{R}_j}{4})^2) \cdot \nonumber\\
&& e^{\frac{-1}{b^2}((\vec{r}_k-\frac{\vec{R}_i+\vec{R}_j}{4})^2 + (\frac{\vec{R}_j+\vec{R}_j}{4})^2)}\nonumber\\
&=&\Delta^6(\vec{R}_i,\vec{R}_j)\frac{1}{2mb^4} (\frac{1}{\sqrt{\pi}b})^3 \int dr^3 (r^2-(\frac{\vec{R}_i-\vec{R}_j}{4})^2) e^{\frac{-r^2}{b^2}}\nonumber\\
&=&\Delta^6(\vec{R}_i,\vec{R}_j)\frac{4\pi}{2mb^4} (\frac{1}{\sqrt{\pi}b})^3 \int_0^{\infty} dr (r^4e^{\frac{-r^2}{b^2}}-r^2(\frac{\vec{R}_i-\vec{R}_j}{4})^2) e^{\frac{-r^2}{b^2}})\nonumber\\
&=&\Delta^6(\vec{R}_i,\vec{R}_j)\frac{4\pi}{2mb^4} (\frac{1}{\sqrt{\pi}b})^3 (\frac{3\sqrt{\pi}b^5}{8}-(\frac{\vec{R}_i-\vec{R}_j}{4})^2)\frac{\sqrt{\pi}b^3}{4})\nonumber\\
&=&\Delta^6(\vec{R}_i,\vec{R}_j)\frac{3}{4mb^2} (1 -\frac{1}{6b^2} (\frac{\vec{R}_i-\vec{R}_j}{2})^2)\nonumber\\
&=&N^d(\vec{R}_i,\vec{R}_j)\frac{3}{4mb^2} (1 -\frac{1}{6b^2} (\frac{\vec{R}_i-\vec{R}_j}{2})^2)
\end{eqnarray}
\ 

In the same way, we have

\begin{eqnarray}
<\Psi_{6q}|t_k P_{36}|\Psi_{6q}>&=&<P^{\sigma fc}_{36}>N^e(\vec{R}_i,\vec{R}_j) \frac{3}{4mb^2} (1 -\frac{1}{6b^2} (\frac{\vec{R}_i-\vec{R}_j}{2})^2)
\end{eqnarray}
\ 

\noindent for $k=1,2,4,5$, and

\begin{eqnarray}
<\Psi_{6q}|t_k P_{36}|\Psi_{6q}>&=&<P^{\sigma fc}_{36}>N^e(\vec{R}_i,\vec{R}_j) \frac{3}{4mb^2} (1 -\frac{1}{6b^2} (\frac{\vec{R}_i+\vec{R}_j}{2})^2)
\end{eqnarray}
\ 

\noindent for $k=3,6$.
\\

Next, we calculate

\begin{eqnarray}
<\Psi_{6q}|\frac{-\vec{p}_1\cdot\vec{p}_2}{6m}|\Psi_{6q}>&=&\Delta^4(\vec{R}_i,\vec{R}_j) \frac{1}{6m} \int dr^3_1 dr^3_2 \phi^*(\vec{r}_1-\frac{\vec{R}_i}{2},b)\phi^*(\vec{r}_2-\frac{\vec{R}_i}{2},b) \cdot \nonumber\\
&&  \vec{\nabla}\phi(\vec{r}_1-\frac{\vec{R}_j}{2},b) \cdot \vec{\nabla} \phi(\vec{r}_2-\frac{\vec{R}_j}{2},b)\nonumber\\
&=&\Delta^4(\vec{R}_i,\vec{R}_j) \frac{1}{6mb^4} (\frac{1}{\sqrt{\pi}b})^6 \int dr^3_1 dr^3_2 (\vec{r}_1-\frac{\vec{R}_j}{2},b) \cdot (\vec{r}_2-\frac{\vec{R}_j}{2},b) \cdot \nonumber\\
&& e^{\frac{-1}{2b^2}((\vec{r}_1-\frac{\vec{R}_i}{2})^2+(\vec{r}_1-\frac{\vec{R}_j}{2})^2+ (\vec{r}_2-\frac{\vec{R}_i}{2})^2+(\vec{r}_2-\frac{\vec{R}_j}{2})^2)}
\end{eqnarray}
\\

Let us assume $\vec{r}=\vec{r}_1-\frac{\vec{R}_j}{2}$ and $\vec{t}=\vec{r}_2-\frac{\vec{R}_j}{2}$, we get

\begin{eqnarray}
<\Psi_{6q}|\frac{-\vec{p}_1\cdot\vec{p}_2}{6m}|\Psi_{6q}>&=&\Delta^4(\vec{R}_i,\vec{R}_j) \frac{1}{6mb^4} (\frac{1}{\sqrt{\pi}b})^6 \nonumber\\
&&\int dr^3 dt^3 \vec{r}\cdot \vec{t} e^{\frac{-1}{2b^2}((\vec{r}-\frac{\vec{R}_i-\vec{R}_j}{2})^2+r^2+t^2+\vec{t}-\frac{\vec{R}_i-\vec{R}_j}{2})^2)}\nonumber\\
&=&\Delta^4(\vec{R}_i,\vec{R}_j) \frac{1}{6mb^4} (\frac{1}{\sqrt{\pi}b})^6  e^{\frac{-(\vec{R}_i-\vec{R}_j)^2}{4b^2}} \cdot \nonumber\\
&&  \int dr^3 dt^3 \vec{r}\cdot \vec{t} e^{\frac{-1}{b^2}(r^2+t^2-(\vec{r}+\vec{t})\cdot(\frac{\vec{R}_i-\vec{R}_j}{2}))}\nonumber\\
&=&\Delta^4(\vec{R}_i,\vec{R}_j) \frac{1}{6mb^4} (\frac{1}{\sqrt{\pi}b})^6  e^{\frac{-(\vec{R}_i-\vec{R}_j)^2}{4b^2}} \cdot \nonumber\\
&&  \int dr^3 dt^3 \vec{r}\cdot \vec{t} e^{\frac{-1}{b^2}((\vec{r}-\frac{\vec{R}_i-\vec{R}_j}{4})^2+(\vec{t}-\frac{\vec{R}_i-\vec{R}_j}{2})^2-\frac{(\vec{R}_i-\vec{R}_j)^2}{8})}\nonumber\\
\end{eqnarray}
\\

If we take $\vec{u}=\vec{r}-\frac{\vec{R}_i-\vec{R}_j}{4}$ and $\vec{v}=\vec{t}-\frac{\vec{R}_i-\vec{R}_j}{4}$, we obtain

\begin{eqnarray}
<\Psi_{6q}|\frac{-\vec{p}_1\cdot\vec{p}_2}{6m}|\Psi_{6q}>&=&\Delta^6(\vec{R}_i,\vec{R}_j) \frac{1}{6mb^4} (\frac{1}{\sqrt{\pi}b})^6 \cdot \nonumber\\
&&  \int du^3 dv^3 (\vec{u}+\frac{\vec{R}_i-\vec{R}_j}{4})\cdot (\vec{v}+\frac{\vec{R}_i-\vec{R}_j}{4}) e^{\frac{-(u^2+v^2)}{b^2}}\nonumber\\
&=&\Delta^6(\vec{R}_i,\vec{R}_j) \frac{1}{6mb^4} (\frac{1}{\sqrt{\pi}b})^6  (\int du^3 dv^3 \vec{u} \cdot \vec{v} e^{\frac{-(u^2+v^2)}{b^2}} \nonumber\\
&&  +  \int du^3 dv^3 \frac{\vec{R}_i-\vec{R}_j}{4}\cdot(\vec{u}+\vec{v})e^{\frac{-(u^2+v^2)}{b^2}} \nonumber\\
&& + \int du^3 dv^3 (\frac{\vec{R}_i-\vec{R}_j}{4})^2 e^{\frac{-(u^2+v^2)}{b^2}})\nonumber\\
&=&\Delta^6(\vec{R}_i,\vec{R}_j) \frac{1}{6mb^4} (\frac{1}{\sqrt{\pi}b})^6 \int du^3 dv^3 (\frac{\vec{R}_i-\vec{R}_j}{4})^2 e^{\frac{-(u^2+v^2)}{b^2}}\nonumber\\
&=&\Delta^6(\vec{R}_i,\vec{R}_j) \frac{1}{24mb^4} (\frac{\vec{R}_i-\vec{R}_j}{2})^2\nonumber\\
&=&N^d(\vec{R}_i,\vec{R}_j) \frac{1}{24mb^4} (\frac{\vec{R}_i-\vec{R}_j}{2})^2
\end{eqnarray}
\\

In the same way, we get

\begin{eqnarray}
<\Psi_{6q}|\frac{-\vec{p}_3\cdot\vec{p}_6}{6m}|\Psi_{6q}>&=&-N^d(\vec{R}_i,\vec{R}_j) \frac{1}{24mb^4} (\frac{\vec{R}_i-\vec{R}_j}{2})^2\nonumber\\
<\Psi_{6q}|\frac{-\vec{p}_1\cdot\vec{p}_2 P_{36}}{6m}|\Psi_{6q}>&=&<P^{\sigma fc}_{36}>N^e(\vec{R}_i,\vec{R}_j) \frac{1}{24mb^4} (\frac{\vec{R}_i-\vec{R}_j}{2})^2\nonumber\\
<\Psi_{6q}|\frac{-\vec{p}_3\cdot\vec{p}_6 P_{36}}{6m}|\Psi_{6q}>&=&-<P^{\sigma fc}_{36}>N^e(\vec{R}_i,\vec{R}_j) \frac{1}{24mb^4} (\frac{\vec{R}_i+\vec{R}_j}{2})^2\nonumber\\
<\Psi_{6q}|\frac{-\vec{p}_1\cdot\vec{p}_3 P_{36}}{6m}|\Psi_{6q}>&=&<P^{\sigma fc}_{36}>N^e(\vec{R}_i,\vec{R}_j) \frac{1}{24mb^4} \frac{R_i^2 -R_j^2}{4}\nonumber\\
<\Psi_{6q}|\frac{-\vec{p}_1\cdot\vec{p}_6 P_{36}}{6m}|\Psi_{6q}>&=&-<P^{\sigma fc}_{36}>N^e(\vec{R}_i,\vec{R}_j) \frac{1}{24mb^4} \frac{R_i^2 -R_j^2}{4}\nonumber\\
<\Psi_{6q}|\frac{-\vec{p}_1\cdot\vec{p}_4 P_{36}}{6m}|\Psi_{6q}>&=&-<P^{\sigma fc}_{36}>N^e(\vec{R}_i,\vec{R}_j) \frac{1}{24mb^4} (\frac{\vec{R}_i-\vec{R}_j}{2})^2
\end{eqnarray}
\\

In summary, for the single particle kinetic energy operators, we have three different types of matrix elements
\begin{eqnarray}\label{KEsummary1}
<\Psi_{6q}|t_1|\Psi_{6q}>&=&N^d(\vec{R}_i,\vec{R}_j)\frac{K_0}{3}(1-\frac{(\vec{R}_i-\vec{R}_j)^2}{24b^2})\nonumber\\
<\Psi_{6q}|t_1P_{36}|\Psi_{6q}>&=&<P^{\sigma fc}_{36}>N^e(\vec{R}_i,\vec{R}_j)\frac{K_0}{3}(1-\frac{(\vec{R}_i-\vec{R}_j)^2}{24b^2})\nonumber\\
<\Psi_{6q}|t_3P_{36}|\Psi_{6q}>&=&<P^{\sigma fc}_{36}>N^e(\vec{R}_i,\vec{R}_j)\frac{K_0}{3}(1-\frac{(\vec{R}_i+\vec{R}_j)^2}{24b^2})\nonumber\\
\end{eqnarray}

\noindent where $K_0=\frac{3}{4mb^2}$, and six distinct types of matrix elements for the two-body operators. These are

\begin{eqnarray}\label{KEsummary2}
<\Psi_{6q}|t_{12}|\Psi_{6q}>&=&N^d(\vec{R}_i,\vec{R}_j)\frac{K_0}{3}\nonumber\\
<\Psi_{6q}|t_{36}|\Psi_{6q}>&=&N^d(\vec{R}_i,\vec{R}_j)\frac{K_0}{3}(1-\frac{(\vec{R}_i-\vec{R}_j)^2}{12b^2})\nonumber\\
<\Psi_{6q}|t_{12}P_{36}|\Psi_{6q}>&=&<P^{\sigma fc}_{36}>N^e(\vec{R}_i,\vec{R}_j)\frac{K_0}{3}\nonumber\\
<\Psi_{6q}|t_{36}P_{36}|\Psi_{6q}>&=&<P^{\sigma fc}_{36}>N^e(\vec{R}_i,\vec{R}_j)\frac{K_0}{3}(1-\frac{(\vec{R}_i+\vec{R}_j)^2}{12b^2})\nonumber\\
<\Psi_{6q}|t_{13}P_{36}|\Psi_{6q}>&=&<P^{\sigma fc}_{36}>N^e(\vec{R}_i,\vec{R}_j)\frac{K_0}{3}(1-\frac{R_i^2}{12b^2})\nonumber\\
<\Psi_{6q}|t_{16}P_{36}|\Psi_{6q}>&=&<P^{\sigma fc}_{36}>N^e(\vec{R}_i,\vec{R}_j)\frac{K_0}{3}(1-\frac{R_j^2}{12b^2})\nonumber\\
<\Psi_{6q}|t_{14}P_{36}|\Psi_{6q}>&=&<P^{\sigma fc}_{36}>N^e(\vec{R}_i,\vec{R}_j)\frac{K_0}{3}(1-\frac{(\vec{R}_i-\vec{R}_j)^2}{12b^2})
\end{eqnarray}
\\

Now, let us project these matrix element on a fixed orbital momentum in the bra and in the ket. We obtain

\begin{eqnarray}
<Y_L^{M}(R_j) t_1 Y_{L'}^{M'*}(R_i)>&=&K_0 e^{-\alpha_d(R_i^2+r_j^2)}\int Y_L^{M}(R_j)(e^{\gamma_d \vec{R}_i\cdot \vec{R}_j}(1-\frac{R_i^2+R_j^2}{24b^2}) \nonumber\\
&& +e^{\gamma_d \vec{R}_i\cdot \vec{R}_j}\frac{\vec{R}_i\cdot \vec{R}_j}{12b^2}) Y_{L'}^{M'*}(R_i)dR^3_i dR^3_j\nonumber\\
&=&4\pi K_0(1-\frac{R_i^2+R_j^2}{24b^2})e^{-\alpha_d(R_i^2+r_j^2)} (i_L(\gamma_d R_iR_j) \delta_{L,L'}^{M,M'}  \nonumber\\
&& + \frac{\pi R_iR_j}{9b^2} \sum_{lmm'} \int Y_{L'}^{M'*}(R_i) Y_L^{M}(R_j) Y_l^{m*}(R_i) Y_l^{m*}(R_j) \cdot \nonumber\\
&& Y_1^{m'*}(R_j) Y_1^{m}(R_i) i_l(\gamma_d R_iR_j) dR^3_i dR^3_j)\nonumber\\
&=&4\pi K_0(1-\frac{R_i^2+R_j^2}{24b^2})e^{-\alpha_d(R_i^2+r_j^2)}\cdot \nonumber\\
&&  (i_L(\gamma_d R_iR_j) \delta_{L,L'}^{M,M'}+ \frac{\pi R_iR_j}{9b^2} X(R_i,R_j))
\end{eqnarray}

\noindent with
\begin{eqnarray}\label{KEX}
X(R_i,R_j)&=&\sum_{lmm'}\int Y_{L'}^{M'*}(R_i) Y_L^{M}(R_j) Y_l^{m*}(R_i) Y_l^{m*}(R_j)\nonumber\\
&& Y_1^{m'*}(R_i) Y_1^{m'*}(R_j) i_l(\gamma_d R_iR_j) dR^3_i dR^3_j
\end{eqnarray}

\noindent and where the relation

\begin{eqnarray}
\vec{R}_i\cdot \vec{R}_j=\frac{4\pi}{3}R_iR_j\sum_m Y_1^{m*}(R_i) Y_1^{m}(R_j)
\end{eqnarray}

\noindent has been used. Now, based on the relation

\begin{eqnarray}
Y_l^{m}(\theta,\phi) Y_1^{m'}(\theta,\phi) &=& \sum_{kq}(\theta,\phi)\sqrt{\frac{(2l+1)2}{4\pi(2k+1)}}Y_k^{q}<l1mm'|k,q><l100|k,0>\nonumber\\
&=&\sum_q \left( \sqrt{\frac{(2l+1)3}{4\pi(2l+3)}}Y_{l+1}^{q}(\theta,\phi)<l1mm'|l+1,q> \right.\nonumber\\
&& \left.-\sqrt{\frac{3l}{4\pi(2l-1)}}Y_{l-1}^{q}(\theta,\phi)<l1mm'|l-1,q>\right),
\end{eqnarray}
\ 

\noindent the Eq. (\ref{KEX}) becomes

\begin{eqnarray}
X(R_i,R_j) &=& \sum_{lmm'qq'} \int Y_{L'}^{M'*}(R_i) (\sqrt{\frac{(2l+1)3}{4\pi(2l+3)}}Y_{l+1}^{q}(R_i)<l1mm'|l+1,q>\nonumber\\
&&-\sqrt{\frac{3l}{4\pi(2l-1)}}Y_{l-1}^{q}(R_i)<l1mm'|l-1,q>)dR^3_i \nonumber\\
&&\int Y_L^{M}(R_j) (\sqrt{\frac{(2l+1)3}{4\pi(2l+3)}}Y_{l+1}^{q'*}(R_j)<l1mm'|l+1,q'>\nonumber\\
&&-\sqrt{\frac{3l}{4\pi(2l-1)}}Y_{l-1}^{q'*}(R_j)<l1mm'|l-1,q'>)i_l(\gamma_d R_iR_j)dR^3_j\nonumber\\&&\nonumber\\
&=& \sum_{lmm'q} \int Y_{L'}^{M'*}(R_i) (\sqrt{\frac{(2l+1)3}{4\pi(2l+3)}}Y_{l+1}^{q}(R_i)<l1mm'|l+1,q>\nonumber\\
&&-\sqrt{\frac{3l}{4\pi(2l-1)}}Y_{l-1}^{q}(R_i)<l1mm'|l-1,q>)dR^3_i \nonumber\\
&&(\sqrt{\frac{3L}{4\pi(2L+1)}}<L-1,1,m,m'|L,M>i_{L-1}(\gamma_d R_iR_j) \delta_{l,L-1}\nonumber\\
&&-\sqrt{\frac{3(L+1)}{4\pi(2L+1)}}<L+1,1,m,m'|L,M>i_{L+1}(\gamma_d R_iR_j) \delta_{l,L+1})\nonumber\\&&\nonumber\\
&=& \sum_{mm'} \frac{3L}{4\pi(2L+1)} <L-1,1,m,m'|L,M>\cdot \nonumber\\
&& <L-1,1,m,m'|L,M'> i_{L-1}(\gamma_d R_iR_j)\delta_{L,L'}\nonumber\\
&&+\frac{3(L+1)}{4\pi(2L+1)}<L+1,1,m,m'|L,M>\cdot \nonumber\\
&& <L+1,1,m,m'|L,M'>i_{L+1}(\gamma_d R_iR_j)\delta_{L,L'}\nonumber\\&&\nonumber\\
&=& \left( \frac{3L}{4\pi(2L+1)}  i_{L-1}(\gamma_d R_iR_j)+\frac{3(L+1)}{4\pi(2L+1)}i_{L+1}(\gamma_d R_iR_j) \right) \delta_{L,L'}^{M,M'}\nonumber\\
\end{eqnarray}

\noindent which finally leads to

\begin{eqnarray}
<Y_L^M(R_j) t_1 Y_{L'}^{M*'}(R_i)>&=&4\pi K_0 e^{-\alpha_d(R_i^2+r_j^2)}\left\{(1-\frac{R_i^2+R_j^2}{24b^2})i_L(\gamma_d R_iR_j)\right. \nonumber\\
&&+\frac{R_iR_j}{12b^2}\left(\frac{L}{2L+1}  i_{L-1}(\gamma_d R_iR_j) \right.\nonumber\\
&&\left.\left. +\frac{L+1}{2L+1}i_{L+1}(\gamma_d R_iR_j)\right)\right\}\delta_{L,L'}^{M,M'}
\end{eqnarray}
\\

In the same way we get

\begin{eqnarray}
<Y_L^M(R_j) t_1 P_{36} Y_{L'}^{M*'}(R_i)>&=&<P^{\sigma fc}_{36}>4\pi K_0 e^{-\alpha_d(R_i^2+r_j^2)}\left\{(1-\frac{R_i^2+R_j^2}{24b^2})i_L(\gamma_e R_iR_j)\right.\nonumber\\
&&+\frac{R_iR_j}{12b^2}\left(\frac{L}{2L+1}  i_{L-1}(\gamma_e R_iR_j)\right. \nonumber\\
&& \left.\left. +\frac{L+1}{2L+1}i_{L+1}(\gamma_e R_iR_j)\right)\right\}\delta_{L,L'}^{M,M'}\nonumber\\&&\nonumber\\
<Y_L^M(R_j) t_3 P_{36} Y_{L'}^{M*'}(R_i)>&=&<P^{\sigma fc}_{36}>4\pi K_0 e^{-\alpha_d(R_i^2+r_j^2)}\left\{(1-\frac{R_i^2+R_j^2}{24b^2})i_L(\gamma_e R_iR_j)\right.\nonumber\\
&&-\frac{R_iR_j}{12b^2}\left(\frac{L}{2L+1}  i_{L-1}(\gamma_e R_iR_j) \right.\nonumber\\
&&\left.\left. +\frac{L+1}{2L+1}i_{L+1}(\gamma_e R_iR_j)\right)\right\}\delta_{L,L'}^{M,M'}\nonumber\\&&\nonumber\\
<Y_L^M(R_j) t_{12} Y_{L'}^{M*'}(R_i)>&=&4\pi \frac{K_0}{3} e^{-\alpha_d(R_i^2+r_j^2)}i_L(\gamma_d R_iR_j)\delta_{L,L'}^{M,M'}\nonumber\\&&\nonumber\\
<Y_L^M(R_j) t_{36} Y_{L'}^{M*'}(R_i)>&=&4\pi \frac{K_0}{3} e^{-\alpha_d(R_i^2+r_j^2)}\left\{(1-\frac{R_i^2+R_j^2}{12b^2})i_L(\gamma_d R_iR_j)\right.\nonumber\\
&&+\frac{R_iR_j}{6b^2}\left(\frac{L}{2L+1}  i_{L-1}(\gamma_d R_iR_j) \right.\nonumber\\
&& \left.\left.+\frac{L+1}{2L+1}i_{L+1}(\gamma_d R_iR_j)\right)\right\} \delta_{L,L'}^{M,M'}\nonumber\\
<Y_L^M(R_j) t_{12} P_{36}Y_{L'}^{M*'}(R_i)>&=&4<P^{\sigma fc}_{36}>\pi \frac{K_0}{3} e^{-\alpha_d(R_i^2+r_j^2)}i_L(\gamma_e R_iR_j)\delta_{L,L'}^{M,M'}\nonumber\\
<Y_L^M(R_j) t_{36} P_{36}Y_{L'}^{M*'}(R_i)>&=&4<P^{\sigma fc}_{36}>\pi \frac{K_0}{3} e^{-\alpha_d(R_i^2+r_j^2)}\left\{(1-\frac{R_i^2+R_j^2}{12b^2})i_L(\gamma_e R_iR_j)\right.\nonumber\\
&&-\frac{R_iR_j}{6b^2}\left(\frac{L}{2L+1}  i_{L-1}(\gamma_e R_iR_j) \right. \nonumber\\
&&\left.\left. +\frac{L+1}{2L+1}i_{L+1}(\gamma_e R_iR_j)\right)\right\} \delta_{L,L'}^{M,M'}\nonumber\\&&\nonumber\\
<Y_L^M(R_j) t_{13} P_{36}Y_{L'}^{M*'}(R_i)>&=&4<P^{\sigma fc}_{36}>\pi \frac{K_0}{3} e^{-\alpha_d(R_i^2+r_j^2)}i_L(\gamma_e R_iR_j)(1-\frac{R_i^2}{12b^2})\delta_{L,L'}^{M,M'}\nonumber\\&&\nonumber\\
<Y_L^M(R_j) t_{16} P_{36}Y_{L'}^{M*'}(R_i)>&=&4<P^{\sigma fc}_{36}>\pi \frac{K_0}{3} e^{-\alpha_d(R_i^2+r_j^2)}i_L(\gamma_e R_iR_j)(1-\frac{R_j^2}{12b^2})\delta_{L,L'}^{M,M'}\nonumber\\&&\nonumber\\
<Y_L^M(R_j) t_{14} P_{36}Y_{L'}^{M*'}(R_i)>&=&4<P^{\sigma fc}_{36}>\pi \frac{K_0}{3} e^{-\alpha_d(R_i^2+r_j^2)}\left\{(1-\frac{R_i^2+R_j^2}{12b^2})i_L(\gamma_e R_iR_j)\right.\nonumber\\
&&+\frac{R_iR_j}{6b^2}\left(\frac{L}{2L+1}  i_{L-1}(\gamma_e R_iR_j)\right. \nonumber\\
&& \left.\left.+\frac{L+1}{2L+1}i_{L+1}(\gamma_e R_iR_j)\right)\right\} \delta_{L,L'}^{M,M'}
\end{eqnarray}
\\
to be used in conjuction with the matrix elements (\ref{KEsummary1}-\ref{KEsummary2})
\newpage
\thispagestyle{empty} 
\ 
\newpage
\chapter{The Confinement}\label{appendixCONF}
\ 
In this appendix we derive the matrix elements of the confinement potential given by
\begin{equation}
V_{Conf}(r_{ij})=(\frac{-3}{8}\lambda^{c}_i\cdot \lambda^{c}_j) V_{conf}(r_{ij})\ ,
\end{equation}

with
\begin{equation}
V_{conf}(r)=C\ r^n\ ,
\end{equation}
\ 

\noindent where $n=1$ for a linear confinement, $n=2$ for a quadratic confinement and $C$ is the strength of the confinement.
\\

For that, we use the two-body potential formula (\ref{RGMDEV}) given in Chapter \ref{rgmchapter}

\begin{equation}
<\Psi_{6q}|V_{ij}|\Psi_{6q}>=V(\vec{a})N(\vec{R}_i,\vec{R}_j)
\end{equation}

\noindent where

\begin{equation}\label{CONFVa}
V(\vec{a}) = (\frac{1}{\sqrt{2 \pi}b})^3 \int e^{\frac{-(\vec{r}-\vec{a}/2)^2}{2 b^2}}V(\vec{r})d\vec{r}
\end{equation}

\noindent with
\begin{eqnarray}\label{CONFVadetail}
& v_{12} &:\  \vec{a} = 0 \nonumber\\
& v_{36} &:\  \vec{a} = \vec{R}_i + \vec{R}_j \nonumber\\
& v_{12}P_{36} &:\  \vec{a} = 0 \nonumber\\
V_{ij} =& v_{36}P_{36} &:\  \vec{a} = \vec{R}_i - \vec{R}_j \nonumber\\
& v_{13}P_{36} &:\  \vec{a} = \vec{R}_j \nonumber\\
& v_{16}P_{36} &:\  \vec{a} = \vec{R}_i \nonumber\\
& v_{14}P_{36} &:\  \vec{a} = \vec{R}_i + \vec{R}_j
\end{eqnarray}
\ 

\noindent and where $N(\vec{R}_i,\vec{R}_j)$ is either the direct norm (\ref{NORMdirect})

\begin{equation}
N^d(\vec{R}_i,\vec{R}_j) = <\Psi_{6q}|\Psi_{6q}> = \Delta(\vec{R}_i,\vec{R}_j)^6
\end{equation}
\ 

\noindent given in Appendix \ref{appendixNORM} or the exchange part (\ref{NORMexchange}) of the norm

\begin{equation}
N^e(\vec{R}_i,\vec{R}_j) = <\Psi_{6q}|P^o_{36}|\Psi_{6q}> = \Delta(\vec{R}_i,\vec{R}_j)^4 \Delta(\vec{R}_i,-\vec{R}_j)^2
\end{equation}
\ 

Now, for a linear confinement, $V(\vec{r})=r$ Eq. (\ref{CONFVa}) becomes

\begin{eqnarray}
V(\vec{a}) &=& 2\pi (\frac{1}{\sqrt{2 \pi}b})^3 e^{\frac{-a^2}{8b^2}} \int_0^{\infty} r^3 e^{\frac{-r^2}{2 b^2}} dr \int_{-1}^1 e^{\frac{-arx}{2b^2}} dx\nonumber\\
&=& \frac{2 e^{\frac{-a^2}{8b^2}}}{\sqrt{2 \pi}ab}  (\int_0^{\infty} r^2 e^{\frac{-r^2}{2 b^2}+\frac{ar}{2b^2}} dr  - (\int_0^{\infty} r^2 e^{\frac{-r^2}{2 b^2}-\frac{ar}{2b^2}} dr )
\end{eqnarray}

Let make the change of variables

\begin{equation}
t=r\pm\frac{a}{2}\ \Rightarrow \ dr= dt,
\end{equation}

\noindent which leads to

\begin{eqnarray}
V(\vec{a}) &=&  \frac{2}{\sqrt{2 \pi}ab}  (\int_{\frac{-a}{2}}^{\infty} (t+\frac{a}{2})^2 e^{\frac{-t^2}{2 b^2}} dt  - \int_{\frac{a}{2}}^{\infty} (t-\frac{a}{2})^2 e^{\frac{-t^2}{2 b^2}} dt )\nonumber\\
&=& \frac{2}{\sqrt{2 \pi}ab}  (\int_{\frac{-a}{2}}^{\frac{a}{2}} (t+\frac{a}{2})^2 e^{\frac{-t^2}{2 b^2}} dt  +2a \int_{\frac{a}{2}}^{\infty} t e^{\frac{-t^2}{2 b^2}} dt )\nonumber\\
&=&\frac{2}{\sqrt{2 \pi}ab} (b^3\sqrt{2\pi}erf(\frac{a}{\sqrt{8}b})-ab^2e^{\frac{-a^2}{8b^2}}+\frac{\sqrt{2\pi}a^2b}{4}erf(\frac{a}{\sqrt{8}b})  + 2ab^2e^{\frac{-a^2}{8b^2}})\nonumber\\
&=&\frac{2}{ab\sqrt{2\pi}}((b^3+\frac{a^2b}{4})\sqrt{2\pi}erf(\frac{a}{\sqrt{8}b})+ab^2e^{\frac{-a^2}{8b^2}})\nonumber\\
&=&\sqrt{\frac{2}{\pi}}be^{\frac{-a^2}{8b^2}}+\frac{a^2+4b^2}{2a}erf(\frac{a}{\sqrt{8}b})
\end{eqnarray}
\\

Note the particular case $a\rightarrow 0$ where one obtains the limiting value

\begin{equation}
V(0)=\sqrt{\frac{8}{\pi}}b.
\end{equation}
\\

In the case of a quadratic confinement, $V(\vec{r})=r^2$ Eq. (\ref{CONFVa}) becomes

\begin{eqnarray}
V(\vec{a}) &=& 2\pi (\frac{1}{\sqrt{2 \pi}b})^3 e^{\frac{-a^2}{8b^2}} \int_0^{\infty} r^2 e^{\frac{-r^2}{2 b^2}} dr \int_{-1}^1 e^{\frac{-arx}{2b^2}} dx\nonumber\\
&=& 3b^2+\frac{a^2}{4}
\end{eqnarray}
\ 

If $a\rightarrow 0$, we get

\begin{equation}
V(0)=3b^2
\end{equation}

Let us now consider the different cases of Eq. (\ref{CONFVadetail}) for the linear confinement potential. First we calculate the $<V_{Conf}(12)>$ matrix element. In this case $\vec{a}=0$, such we obtain

\begin{eqnarray}
<V_{Conf}(12)>&=&<\Psi_{6q}|v_{12}\left( -\frac{3}{8} \lambda^{c}_1\cdot \lambda^{c}_2 \right)|\Psi_{6q}>\nonumber\\
&=&<\frac{-3}{8}\lambda^{c}_1\cdot \lambda^{c}_2 N^d(\vec{R}_i,\vec{R}_j) (\frac{1}{\sqrt{2 \pi}b})^3 \int e^{\frac{-r^2}{2 b^2}}V_{conf}(r) d\vec{r}> \nonumber\\
&=& <\frac{-3}{8}\lambda^{c}_1\cdot \lambda^{c}_2> N^d(\vec{R}_i,\vec{R}_j) C \sqrt{\frac{8}{\pi}}b
\end{eqnarray}
\\

This correspond to the diagram $a$ of Fig. \ref{RGMinteractiondiagram}. Projecting the relative wave function on a partial wave $L$, we have

\begin{eqnarray}
<Y_L^M(R_j) V_{Conf}(12) Y_{L'}^{M'}(R_i)>&=& <\frac{-3}{8} \lambda^{c}_1\cdot \lambda^{c}_2> 4\pi e^{-\alpha_d (R_i^2+R_j^2)} i_L(\gamma_d R_i R_j)\cdot \nonumber\\
&&  C \sqrt{\frac{8}{\pi}}b \delta_{L,L'}^{M,M'}.
\end{eqnarray}

The diagram $b$ of Fig. \ref{RGMinteractiondiagram} corresponds to $<V_{Conf}(36)>$, defined as above

\begin{equation}
<V_{Conf}(36)>= <\frac{-3}{8}\lambda^{c}_3\cdot\lambda^{c}_6> N^d(\vec{R}_i,\vec{R}_j) (\frac{1}{\sqrt{2 \pi}b})^3 \int e^{\frac{-(\vec{r}-\frac{\vec{R}_i+\vec{R}_j}{2})^2}{2 b^2}}V_{conf}(r)d\vec{r}
\end{equation}
\ 

Here, we need the error function $erf(x)$ defined by

\begin{equation}
erf(x) = \frac{2x}{\sqrt{\pi}} \int^1_0 e^{-y^2x^2} dy
\end{equation}

Then

\begin{eqnarray}
<V_{conf}(36)>&=& <\frac{-3}{8}\lambda^{c}_3\cdot\lambda^{c}_6> N^d(\vec{R}_i,\vec{R}_j) C \sqrt{\frac{2}{\pi}}b(e^{-\frac{(\vec{R}_i+\vec{R}_j)^2}{8 b^2}} \nonumber\\
&& +\frac{(\vec{R}_i+\vec{R}_j)^2+4b^2}{4 b^2} \int^1_0 e^{-y^2\frac{(\vec{R}_i+\vec{R}_j)^2}{8 b^2}} dy)\nonumber\\
&=& <\frac{-3}{8}\lambda^{c}_3\cdot\lambda^{c}_6>  4\pi e^{-\alpha_d (R_i^2+R_j^2)} C\sqrt{\frac{2}{\pi}}b  \sum_{lm} Y_l^m(R_i) Y_l^{m*}(R_j) \nonumber\\
&&( e^{\frac{-(R_i^2+R_j^2)}{8b^2}} i_l((\gamma_d-\frac{1}{4b^2})R_iR_j) + (1+\frac{R_i^2+R_j^2}{4b^2}) \nonumber\\
&&  \int_0^1 e^{\frac{-y^2(R_i^2+R_j^2)}{8b^2}} i_l((\gamma_d-\frac{y^2}{4b^2})R_iR_j)dy \nonumber\\
&&+ \frac{4\pi}{6b^2} \sum_{m'} Y_1^{m'}(R_i) Y_1^{m'*}(R_j) \int_0^1 e^{\frac{-y^2(R_i^2+R_j^2)}{8b^2}} i_l((\gamma_d-\frac{y^2}{4b^2})R_iR_j)dy )\nonumber\\
\end{eqnarray}

Projecting, we obtain

\begin{eqnarray}
<Y_L^M(R_j) V_{Conf}(36) Y_{L'}^{M'}(R_i)>&=&  <\frac{-3}{8}\lambda^{c}_3\cdot\lambda^{c}_6>  4\pi e^{-\alpha_d (R_i^2+R_j^2)} C \sqrt{\frac{2}{\pi}}b \cdot \nonumber\\
&&   ( e^{\frac{-(R_i^2+R_j^2)}{8b^2}} i_L((\gamma_d-\frac{1}{4b^2})R_iR_j)  \nonumber\\
&&+ (1+\frac{R_i^2+R_j^2}{4b^2}) \int_0^1 e^{\frac{-y^2(R_i^2+R_j^2)}{8b^2}} i_L((\gamma_d-\frac{y^2}{4b^2})R_iR_j)dy \nonumber\\
&&+ \frac{R_iR_j}{2b^2} \int_0^1 e^{\frac{-y^2(R_i^2+R_j^2)}{8b^2}} (\frac{L}{2L+1} i_{L-1}((\gamma_d-\frac{y^2}{4b^2})R_iR_j)\nonumber\\
&&+\frac{L+1}{2L+1} i_{L+1}((\gamma_d-\frac{y^2}{4b^2})R_iR_j))dy ) \delta_{L,L'}^{M,M'}
\end{eqnarray}
\\

For $<V_{Conf}(12)P_{36}>$, the diagram $c$ of Fig. \ref{RGMinteractiondiagram}, we have

\begin{eqnarray}
<V_{Conf}(12)P_{36}>&=& <\frac{-3}{8}\lambda^{c}_1\cdot \lambda^{c}_2 P^{\sigma fc}_{36}> N^e(\vec{R}_i,\vec{R}_j) C \sqrt{\frac{8}{\pi}}b
\end{eqnarray}
\ 

After the projection we get

\begin{eqnarray}
<Y_L^M(R_j) V_{Conf}(12)P_{36} Y_{L'}^{M'}(R_i)>&=& <\frac{-3}{8} \lambda^{c}_1\cdot \lambda^{c}_2 P^{\sigma fc}_{36}> 4\pi e^{-\alpha_d (R_i^2+R_j^2)} \nonumber\\
&&  i_L(\gamma_e R_i R_j) C \sqrt{\frac{8}{\pi}}b \delta_{L,L'}^{M,M'}
\end{eqnarray}

For the diagram $g$ of Fig. \ref{RGMinteractiondiagram} $<V_{Conf}(36)P_{36}>$ we have

\begin{eqnarray}
<V_{Conf}(36)P_{36}>&=& <\frac{-3}{8}\lambda^{c}_3\cdot\lambda^{c}_6 P^{\sigma fc}_{36}>  4\pi e^{-\alpha_d (R_i^2+R_j^2)} C \sqrt{\frac{2}{\pi}}b  \sum_{lm} Y_l^m(R_i) Y_l^{m*}(R_j) \nonumber\\
&&( e^{\frac{-(R_i^2+R_j^2)}{8b^2}} i_l((\gamma_e+\frac{1}{4b^2})R_iR_j) + (1+\frac{R_i^2+R_j^2}{4b^2})\nonumber\\
&&  \int_0^1 e^{\frac{-y^2(R_i^2+R_j^2)}{8b^2}} i_l((\gamma_e+\frac{y^2}{4b^2})R_iR_j)dy \nonumber\\
&&- \frac{4\pi}{6b^2} \sum_{m'} Y_1^{m'}(R_i) Y_1^{m'*}(R_j) \nonumber\\
&& \int_0^1 e^{\frac{-y^2(R_i^2+R_j^2)}{8b^2}} i_l((\gamma_e+\frac{y^2}{4b^2})R_iR_j)dy )
\end{eqnarray}
\

and for the projection

\begin{eqnarray}
<Y_L^M(R_j) V_{Conf}(36) P_{36} Y_{L'}^{M'}(R_i)>&=&  <\frac{-3}{8}\lambda^{c}_3\cdot\lambda^{c}_6 P^{\sigma fc}_{36}>  4\pi e^{-\alpha_d (R_i^2+R_j^2)} C \sqrt{\frac{2}{\pi}}b  \nonumber\\
&&  ( e^{\frac{-(R_i^2+R_j^2)}{8b^2}} i_L((\gamma_e+\frac{1}{4b^2})R_iR_j)  \nonumber\\
&&+ (1+\frac{R_i^2+R_j^2}{4b^2}) \int_0^1 e^{\frac{-y^2(R_i^2+R_j^2)}{8b^2}} i_L((\gamma_e+\frac{y^2}{4b^2})R_iR_j)dy \nonumber\\
&&- \frac{R_iR_j}{2b^2} \int_0^1 e^{\frac{-y^2(R_i^2+R_j^2)}{8b^2}} (\frac{L}{2L+1} i_{L-1}((\gamma_e+\frac{y^2}{4b^2})R_iR_j)\nonumber\\
&&+\frac{L+1}{2L+1} i_{L+1}((\gamma_e+\frac{y^2}{4b^2})R_iR_j))dy ) \delta_{L,L'}^{M,M'}\nonumber
\end{eqnarray}

Next, for the diagram $e$ of Fig. \ref{RGMinteractiondiagram},

\begin{eqnarray}
<V_{Conf}(13)P_{36}>&=& <\frac{-3}{8}\lambda^{c}_1\cdot \lambda^{c}_3 P^{\sigma fc}_{36}> N^e(\vec{R}_i,\vec{R}_j) C\cdot \nonumber\\
&&  (\sqrt{\frac{2}{\pi}}be^{\frac{-R_j^2}{8b^2}}+\frac{R_j^2+4b^2}{2R_j}erf(\frac{R_j}{\sqrt{8}b}))
\end{eqnarray}
\ 

and for the projection, we obtain

\begin{eqnarray}
<Y_L^M(R_j) V_{Conf}(13)P_{36} Y_{L'}^{M'}(R_i)>&=&<\frac{-3}{8} \lambda^{c}_1\cdot \lambda^{c}_3 P^{\sigma fc}_{36}> 4\pi e^{-\alpha_d (R_i^2+R_j^2)} i_L(\gamma_e R_i R_j) \nonumber\\
&&C (\sqrt{\frac{2}{\pi}}be^{\frac{-R_j^2}{8b^2}}+\frac{R_j^2+4b^2}{2R_j}erf(\frac{R_j}{\sqrt{8}b})) \delta_{L,L'}^{M,M'}.\nonumber\\
\end{eqnarray}
\

In the same way, for the diagram $f$ of Fig. \ref{RGMinteractiondiagram} we have

\begin{eqnarray}
<V_{Conf}(16)P_{36}>&=&<\frac{-3}{8}\lambda^{c}_1\cdot \lambda^{c}_6 P^{\sigma fc}_{36}> N^e(\vec{R}_i,\vec{R}_j) C\nonumber\\
&&  (\sqrt{\frac{2}{\pi}}be^{\frac{-R_i^2}{8b^2}}+\frac{R_i^2+4b^2}{2R_i}erf(\frac{R_i}{\sqrt{8}b}))
\end{eqnarray}

with its projection

\begin{eqnarray}
<Y_L^M(R_j) V_{Conf}(16)P_{36} Y_{L'}^{M'}(R_i)>&=&<\frac{-3}{8} \lambda^{c}_1\cdot \lambda^{c}_6 P^{\sigma fc}_{36}> 4\pi e^{-\alpha_d (R_i^2+R_j^2)} i_L(\gamma_e R_i R_j)  \nonumber\\
&&C (\sqrt{\frac{2}{\pi}}be^{\frac{-R_i^2}{8b^2}}+\frac{R_i^2+4b^2}{2R_i}erf(\frac{R_i}{\sqrt{8}b})) \delta_{L,L'}^{M,M'}\nonumber\\
\end{eqnarray}
\ 

Finally, for the diagram $d$ of Fig. \ref{RGMinteractiondiagram}, we have

\begin{eqnarray}
<V_{Conf}(14)P_{36}>&=&<\frac{-3}{8}\lambda^{c}_1\cdot\lambda^{c}_4 P^{\sigma fc}_{36}>  4\pi e^{-\alpha_d (R_i^2+R_j^2)} C \sqrt{\frac{2}{\pi}}b  \sum_{lm} Y_l^m(R_i) Y_l^{m*}(R_j) \nonumber\\
&&( e^{\frac{-(R_i^2+R_j^2)}{8b^2}} i_l((\gamma_e-\frac{1}{4b^2})R_iR_j) + (1+\frac{R_i^2+R_j^2}{4b^2}) \nonumber\\
&&  \int_0^1 e^{\frac{-y^2(R_i^2+R_j^2)}{8b^2}} i_l((\gamma_e-\frac{y^2}{4b^2})R_iR_j)dy \nonumber\\
&&+ \frac{4\pi}{6b^2} \sum_{m'} Y_1^{m'}(R_i) Y_1^{m'*}(R_j) \nonumber\\
&& \int_0^1 e^{\frac{-y^2(R_i^2+R_j^2)}{8b^2}} i_l((\gamma_e-\frac{y^2}{4b^2})R_iR_j)dy )
\end{eqnarray}

with

\begin{eqnarray}
<Y_L^M(R_j) V_{Conf}(14)P_{36} Y_{L'}^{M'}(R_i)>&=& <\frac{-3}{8}\lambda^{c}_1\cdot\lambda^{c}_4 P^{\sigma fc}_{36}>  4\pi e^{-\alpha_d (R_i^2+R_j^2)} C \sqrt{\frac{2}{\pi}}b  \nonumber\\
&&  ( e^{\frac{-(R_i^2+R_j^2)}{8b^2}} i_L((\gamma_e-\frac{1}{4b^2})R_iR_j)  \nonumber\\
&&+ (1+\frac{R_i^2+R_j^2}{4b^2}) \int_0^1 e^{\frac{-y^2(R_i^2+R_j^2)}{8b^2}} i_L((\gamma_e-\frac{y^2}{4b^2})R_iR_j)dy \nonumber\\
&&+ \frac{R_iR_j}{2b^2} \int_0^1 e^{\frac{-y^2(R_i^2+R_j^2)}{8b^2}} (\frac{L}{2L+1} i_{L-1}((\gamma_e-\frac{y^2}{4b^2})R_iR_j)\nonumber\\
&&+\frac{L+1}{2L+1} i_{L+1}((\gamma_e-\frac{y^2}{4b^2})R_iR_j))dy ) \delta_{L,L'}^{M,M'}
\end{eqnarray}
\\

In Tables \ref{CONF00}, \ref{CONF01} and \ref{CONF11} we give all the flavor-spin-color matrix elements associated to the confinement potential.
\ \\

\begin{table}[H]
\centering

\begin{tabular}{|c|c|c|c|c|c|c|}
\hline
\rule[-2mm]{0mm}{7mm}$\alpha$ & $NN$ & $NN$  & $\Delta\Delta$ & $NN$ & $\Delta\Delta$ & $CC$ \\
\rule[-4mm]{0mm}{5mm}$\beta$  & $NN$ & $\Delta\Delta$ & $\Delta\Delta$ & $CC$ & $CC$  & $CC$ \\

\hline
\hline

\rule[-3mm]{0mm}{8mm}$\lambda_1^c.\lambda_2^c$                                                 
& -1080 &    0 & -1080 &    0 &    0 &  -108 \\
\rule[-3mm]{0mm}{5mm}$\lambda_3^c.\lambda_6^c$                                                 
&     0 &    0 &     0 &    0 &    0 &  -648 \\
\rule[-3mm]{0mm}{5mm}$\lambda_1^c.\lambda_2^c\ P_{36}^{f \sigma c}$                            
&  -280 & -160 &   -40 &  -64 &  128 &   260 \\
\rule[-3mm]{0mm}{5mm}$\lambda_3^c.\lambda_6^c\ P_{36}^{f \sigma c}$                            
&   560 &  320 &    80 &  -16 &   32 &   380 \\
\rule[-3mm]{0mm}{5mm}$\lambda_1^c.\lambda_3^c\ P_{36}^{f \sigma c}$                            
&  -280 & -160 &   -40 &  -64 &  128 &   494 \\
\rule[-3mm]{0mm}{5mm}$\lambda_1^c.\lambda_6^c\ P_{36}^{f \sigma c}$                            
&  -280 & -160 &   -40 &    8 &  -16 &   548 \\
\rule[-3mm]{0mm}{5mm}$\lambda_1^c.\lambda_4^c\ P_{36}^{f \sigma c}$                            
&   140 &   80 &    20 &   -4 &    8 &   581 \\

\hline

\rule[-4mm]{0mm}{10mm}factor & $\frac{1}{405}$ & $\frac{1}{405}$ & $\frac{1}{405}$ 
& $\frac{\sqrt10}{405}$ & $\frac{\sqrt10}{405}$ & $\frac{1}{405}$ \\

\hline

\end{tabular}
\caption{Matrix elements $\langle \alpha|O|\beta \rangle$ of different operators $O$ for ($S$,$I$) = (0,0).}\label{CONF00}

\end{table}

\begin{table}[H]
\centering

\begin{tabular}{|c|c|c|c|c|c|c|}
\hline
\rule[-2mm]{0mm}{7mm}$\alpha$ & $NN$ & $NN$  & $\Delta\Delta$ & $NN$ & $\Delta\Delta$ & $CC$ \\
\rule[-4mm]{0mm}{5mm}$\beta$  & $NN$ & $\Delta\Delta$ & $\Delta\Delta$ & $CC$ & $CC$  & $CC$ \\

\hline
\hline

\rule[-3mm]{0mm}{8mm}$\lambda_1^c.\lambda_2^c$                                                 
& -2592 &    0 & -2592 &    0 &    0 &  -648 \\
\rule[-3mm]{0mm}{5mm}$\lambda_3^c.\lambda_6^c$                                                 
&     0 &    0 &     0 &    0 &    0 & -1296 \\
\rule[-3mm]{0mm}{5mm}$\lambda_1^c.\lambda_2^c\ P_{36}^{f \sigma c}$                            
&    32 & -128 &   -32 &  384 & -768 &    72 \\
\rule[-3mm]{0mm}{5mm}$\lambda_3^c.\lambda_6^c\ P_{36}^{f \sigma c}$                            
&   -64 &  256 &    64 &   96 & -192 &  1152 \\
\rule[-3mm]{0mm}{5mm}$\lambda_1^c.\lambda_3^c\ P_{36}^{f \sigma c}$                            
&    32 & -128 &   -32 &  384 & -768 &   720 \\
\rule[-3mm]{0mm}{5mm}$\lambda_1^c.\lambda_6^c\ P_{36}^{f \sigma c}$                            
&    32 & -128 &   -32 &  -48 &   96 &   720 \\
\rule[-3mm]{0mm}{5mm}$\lambda_1^c.\lambda_4^c\ P_{36}^{f \sigma c}$                            
&   -16 &   64 &    16 &   24 &  -48 &  1260 \\

\hline

\rule[-4mm]{0mm}{10mm}factor & $\frac{1}{972}$ & $\frac{\sqrt5}{972}$ & $\frac{1}{972}$ 
& $\frac{\sqrt5}{972}$ & $\frac{1}{972}$ & $\frac{1}{972}$ \\

\hline

\end{tabular}
\caption{Matrix elements $\langle \alpha|O|\beta \rangle$ of different operators $O$ for ($S$,$I$) =  (1,0) and (0,1).}\label{CONF01}
\end{table}
\ \\

\begin{table}[H]
\centering

\begin{tabular}{|c|c|c|c|c|c|c|}
\hline
\rule[-2mm]{0mm}{7mm}$\alpha$ & $NN$ & $NN$  & $\Delta\Delta$ & $NN$ & $\Delta\Delta$ & $CC$ \\
\rule[-4mm]{0mm}{5mm}$\beta$  & $NN$ & $\Delta\Delta$ & $\Delta\Delta$ & $CC$ & $CC$  & $CC$ \\

\hline
\hline

\rule[-3mm]{0mm}{8mm}$\lambda_1^c.\lambda_2^c$                                                 
& -1944 &    0 & -1944 &     0 &     0 &  -420876 \\
\rule[-3mm]{0mm}{5mm}$\lambda_3^c.\lambda_6^c$                                                 
&     0 &    0 &     0 &     0 &     0 &  -682344 \\
\rule[-3mm]{0mm}{5mm}$\lambda_1^c.\lambda_2^c\ P_{36}^{f \sigma c}$                            
&  -248 & -160 &    -8 & -2560 & 24640 &   334532 \\
\rule[-3mm]{0mm}{5mm}$\lambda_3^c.\lambda_6^c\ P_{36}^{f \sigma c}$                            
&   496 &  320 &    16 & -7840 &  2560 &   583196 \\
\rule[-3mm]{0mm}{5mm}$\lambda_1^c.\lambda_3^c\ P_{36}^{f \sigma c}$                            
&  -248 & -160 &    -8 & -2560 & 24640 &   531038 \\
\rule[-3mm]{0mm}{5mm}$\lambda_1^c.\lambda_6^c\ P_{36}^{f \sigma c}$                            
&  -248 & -160 &    -8 &  3920 & -1280 &   535412 \\
\rule[-3mm]{0mm}{5mm}$\lambda_1^c.\lambda_4^c\ P_{36}^{f \sigma c}$                            
&   124 &   80 &     4 & -1960 &   640 &   657557 \\

\hline

\rule[-4mm]{0mm}{10mm}factor & $\frac{1}{729}$ & $\frac{1}{729}$ & $\frac{1}{729}$ 
& $\frac{1}{729 \cdot \sqrt{1486}}$ & $\frac{1}{729 \cdot \sqrt{1486}}$ & $\frac{1}{729 \cdot 743}$ \\

\hline

\end{tabular}
\caption{Matrix elements $\langle \alpha|O|\beta \rangle$ of different operators $O$ for ($S$,$I$) = (1,1).}\label{CONF11}

\end{table}

\newpage
\thispagestyle{empty} 
\ 
\newpage
\chapter{The Hyperfine Interaction}\label{appendixHYP}
\ 

The derivation of the matrix elements of the hyperfine interaction is very similar to that of the confinement potential but there are however important differences which come from the more complicated spatial dependence and the presence of the the spin-flavor operator instead of a simple color operator. The hyperfine interaction has the form

\begin{equation}
V_{\chi}(r_{ij})=\left\{\sum^{3}_{a=1}V_{\pi}(r_{ij})\lambda^a_i\lambda^a_j+\sum^{7}_{a=4}V_K(r_{ij})\lambda^a_i\lambda^a_j+V_{\eta}(r_{ij})\lambda^8_i\lambda^8_j+\frac{2}{3}V_{\eta'}(r_{ij})\right\} \vec{\sigma}_i\cdot \vec{\sigma}_j
\end{equation}
\ 

\noindent where $\lambda^a$ are the Gell-Mann flavour matrices, and

\begin{equation}\label{HYPVg}
V_{\gamma}(r)=\frac{g^2_{\gamma}}{4 \pi}\frac{1}{12 m_i m_j}\left\{ \mu^2_{\gamma}\frac{e^{-\mu_{\gamma}r}}{r}-\Lambda^2_{\gamma}\frac{e^{-\Lambda_{\gamma}r}}{r}\right\}
\end{equation}
\ 

\noindent for $\gamma=\pi,K,\eta$ and $\eta$'. In Eq. (\ref{HYPVg}), $g_\gamma, m_i, m_j, \mu_\gamma$ and $\Lambda_\gamma$ are the parameters of the model.
\\\ 
\\

Now, if we take $V(\vec{r})=\frac{e^{-\mu r}}{r}$ in Eq. (\ref{CONFVa}) of Appendix \ref{appendixCONF} we obtain

\begin{eqnarray}
V(\vec{a}) &=& 2\pi (\frac{1}{\sqrt{2 \pi}b})^3 e^{\frac{-a^2}{8b^2}} \int_0^{\infty} r e^{\frac{-r^2}{2 b^2}-\mu r}\frac{2b^2}{ar}(e^{\frac{ar}{2b^2}}-e^{\frac{-ar}{2b^2}})dr\nonumber\\
&=&\frac{2 e^{\frac{-a^2}{8b^2}}}{\sqrt{2 \pi}ab} \int_0^{\infty} e^{\frac{-r^2}{2 b^2}-(\mu-\frac{a}{2b^2} )r}-e^{\frac{-r^2}{2 b^2}-(\mu+\frac{a}{2b^2} )r}dr\nonumber\\
&=&\frac{2 e^{\frac{-a^2}{8b^2}}}{\sqrt{2 \pi}ab} (e^{\frac{b^2}{2}(\mu-\frac{a}{2b^2})^2} \int_0^{\infty} e^{\frac{-1}{2}(\frac{r}{b}+b(\mu-\frac{a}{2b^2}))^2}  - e^{\frac{b^2}{2}(\mu+\frac{a}{2b^2})^2}\cdot \nonumber\\
&&  \int_0^{\infty} e^{\frac{-1}{2}(\frac{r}{b}+b(\mu+\frac{a}{2b^2}))^2})
\end{eqnarray}

and if we make the following change of variable

\begin{equation}
t=\frac{1}{\sqrt{2}}(\frac{r}{b}+b(\mu\pm \frac{a}{2b^2}))\ \Rightarrow \ dr=\sqrt{2}b dt
\end{equation}

we get

\begin{eqnarray}\label{HYPyukawa}
V(\vec{a}) &=& \frac{2}{\sqrt{\pi}a} e^{\frac{b^2\mu^2}{2}}(e^{\frac{-a\mu}{2}} \int_{\frac{b(\mu-\frac{a}{2b^2})}{\sqrt{2}}}^{\infty} e^{-t^2} dt - e^{\frac{a\mu}{2}} \int_{\frac{b(\mu+\frac{a}{2b^2})}{\sqrt{2}}}^{\infty} e^{-t^2} dt)\nonumber\\
&=& \frac{e^{\frac{b^2\mu^2}{2}}}{a}(e^{\frac{-a\mu}{2}} erfc(\frac{b(\mu-\frac{a}{2b^2})}{\sqrt{2}})-e^{\frac{a\mu}{2}} erfc(\frac{b(\mu+\frac{a}{2b^2})}{\sqrt{2}}))
\end{eqnarray}
\\

Note that if $a\rightarrow 0$, we get

\begin{equation}
V(0)=\sqrt{\frac{2}{\pi}}\frac{1}{b}-\mu e^{\frac{\mu^2 b^2}{2}}erfc(\frac{\mu b}{\sqrt{2}}).
\end{equation}
\\

Similarly to the confinement potential we have to work out the matrix elements of the seven types of two-body operators associated with diagrams of Fig. \ref{RGMinteractiondiagram}.
\\

For the diagram $a$ of Fig. \ref{RGMinteractiondiagram} we have

\begin{eqnarray}
<V_{\chi}(12)>&=&<\sum_{\gamma}\lambda^{\gamma}_1\lambda^{\gamma}_2 \vec{\sigma}_1\cdot \vec{\sigma}_2 N^d(\vec{R}_i,\vec{R}_j) (\frac{1}{\sqrt{2 \pi}b})^3 \int e^{\frac{-r^2}{2 b^2}}V_{\gamma}(r) d\vec{r}> \nonumber\\
&=&\sum_{\gamma=0}^8 <\lambda^{\gamma}_1\lambda^{\gamma}_2  \vec{\sigma}_1\cdot \vec{\sigma}_2> N^d(\vec{R}_i,\vec{R}_j) (\frac{1}{\sqrt{2 \pi}b})^3 \int e^{\frac{-r^2}{2 b^2}}V_{\gamma}(r)d\vec{r} \nonumber\\
&=&\sum_{\gamma=0}^8 <\lambda^{\gamma}_1\lambda^{\gamma}_2  \vec{\sigma}_1\cdot \vec{\sigma}_2> N^d(\vec{R}_i,\vec{R}_j) \cdot \nonumber\\
&&((\mu_{\gamma}^2-\Lambda_{\gamma}^2)\sqrt{\frac{2}{\pi}}\frac{1}{b}-\mu_{\gamma}^3 e^{\frac{\mu_{\gamma}^2 b^2}{2}}erfc(\frac{\mu_{\gamma} b}{\sqrt{2}})+\Lambda_{\gamma}^3 e^{\frac{\Lambda_{\gamma}^2 b^2}{2}}erfc(\frac{\Lambda_{\gamma} b}{\sqrt{2}})) \nonumber\\
\end{eqnarray}
\ 

\noindent where Eq. (\ref{HYPyukawa}) has been used. Now if we project on the $L^{th}$ partial wave, we obtain

\begin{eqnarray}
<Y_L^M(R_j) V_{\chi}(12) Y_{L'}^{M'}(R_i)>&=&\sum_{\gamma=0}^8 <\lambda^{\gamma}_1\lambda^{\gamma}_2  \vec{\sigma}_1\cdot \vec{\sigma}_2> 4\pi e^{-\alpha_d (R_i^2+R_j^2)} i_L(\gamma_d R_i R_j)\cdot  \nonumber\\
&&((\mu_{\gamma}^2-\Lambda_{\gamma}^2)\sqrt{\frac{2}{\pi}}\frac{1}{b}-\mu_{\gamma}^3 e^{\frac{\mu_{\gamma}^2 b^2}{2}}erfc(\frac{\mu_{\gamma} b}{\sqrt{2}}) \nonumber\\
&& +\Lambda_{\gamma}^3 e^{\frac{\Lambda_{\gamma}^2 b^2}{2}}erfc(\frac{\Lambda_{\gamma} b}{\sqrt{2}})) \delta_{L,L'}^{M,M'}
\end{eqnarray}
\\

For the case $<V_{\chi}(36)>$, equivalent to the diagram $b$ of Fig. \ref{RGMinteractiondiagram},

\begin{equation}
<V_{\chi}(36)>=\sum_{\gamma=0}^8 <\lambda^{\gamma}_3\lambda^{\gamma}_6  \vec{\sigma}_3\cdot \vec{\sigma}_6> N^d(\vec{R}_i,\vec{R}_j) (\frac{1}{\sqrt{2 \pi}b})^3 \int e^{\frac{-(\vec{r}-\frac{\vec{R}_i+\vec{R}_j}{2})^2}{2 b^2}}V_{\gamma}(r)d\vec{r}
\end{equation}
\ 

\noindent Here we need to do the following approximation of the $V_{\gamma}(r)$ potential

\begin{equation}
V_{\gamma}(r) \approx \sum^{p}_{k=1}h_{\gamma}^k e^{\frac{-r^2}{{c_{\gamma}^k}^2}}.
\end{equation}
\ 

\noindent This is equivalent to calculate $V(\vec{a})$ with the potential $V(\vec{r})=e^{\frac{-r^2}{c^2}}$

\begin{eqnarray}
V(\vec{a}) &=& 2\pi (\frac{1}{\sqrt{2 \pi}b})^3 e^{\frac{-a^2}{8b^2}} \int_0^{\infty} r^2 e^{-(\frac{1}{2 b^2}+\frac{1}{c^2})r^2}\frac{2b^2}{ar}(e^{\frac{ar}{2b^2}}-e^{\frac{-ar}{2b^2}})dr\nonumber\\
&=&(\frac{c}{\sqrt{c^2+2b^2}})^3 e^{\frac{-a^2}{4(c^2+2b^2)}}
\end{eqnarray}
\\

Doing so, we obtain

\begin{eqnarray}
<V_{\chi}(36)>&=&\sum_{\gamma=0}^8 <\lambda^{\gamma}_3\lambda^{\gamma}_6  \vec{\sigma}_3\cdot \vec{\sigma}_6> N^d(\vec{R}_i,\vec{R}_j) \sum^{p}_{k=1}h_{\gamma}^k (\frac{c_{\gamma}^k}{\sqrt{{c_{\gamma}^k}^2+2b^2}})^3 e^{\frac{-(\vec{R}_i+\vec{R}_j)^2}{2({c_{\gamma}^k}^2+2b^2)}}\nonumber\\
&=&\sum_{\gamma=0}^8 <\lambda^{\gamma}_3\lambda^{\gamma}_6  \vec{\sigma}_3\cdot \vec{\sigma}_6>  e^{-\alpha_d (R_i^2+R_j^2)} e^{\gamma_d \vec{R}_i \cdot \vec{R}_j} \cdot\nonumber\\
&& \sum^{p}_{k=1}h_{\gamma}^k (\frac{c_{\gamma}^k}{\sqrt{{c_{\gamma}^k}^2+2b^2}})^3 e^{\frac{-(R_i^2+R_j^2)}{4({c_{\gamma}^k}^2+2b^2)}} e^{\frac{-\vec{R}_i\cdot \vec{R}_j}{2({c_{\gamma}^k}^2+2b^2)}}
\end{eqnarray}
\\

The projection leads to

\begin{eqnarray}
<Y_L^M(R_j) V_{\chi}(36) Y_{L'}^{M'}(R_i)>&=&\sum_{\gamma=0}^8 <\lambda^{\gamma}_3\lambda^{\gamma}_6  \vec{\sigma}_3\cdot \vec{\sigma}_6>4\pi e^{-\alpha_d (R_i^2+R_j^2)} \sum^{p}_{k=1}h_{\gamma}^k (\frac{c_{\gamma}^k}{\sqrt{{c_{\gamma}^k}^2+2b^2}})^3  \cdot\nonumber\\
&&e^{\frac{-(R_i^2+R_j^2)}{4({c_{\gamma}^k}^2+2b^2)}} i_L((\gamma_d-\frac{1}{2({c_{\gamma}^k}^2+2b^2)}) R_i R_j) \delta_{L,L'}^{M,M'}
\end{eqnarray}
\\

For the diagram $c$ of Fig. \ref{RGMinteractiondiagram} we have

\begin{eqnarray}
<V_{\chi}(12)P_{36}>&=&<\sum_{\gamma}\lambda^{\gamma}_1\lambda^{\gamma}_2 \vec{\sigma}_1\cdot \vec{\sigma}_2 N^e(\vec{R}_i,\vec{R}_j) (\frac{1}{\sqrt{2 \pi}b})^3 \int e^{\frac{-r^2}{2 b^2}}V_{\gamma}(r) d\vec{r} P^{\sigma fc}_{36}> \nonumber\\
&=&\sum_{\gamma=0}^8 <\lambda^{\gamma}_1\lambda^{\gamma}_2  \vec{\sigma}_1\cdot \vec{\sigma}_2 P^{\sigma fc}_{36}> N^e(\vec{R}_i,\vec{R}_j) (\frac{1}{\sqrt{2 \pi}b})^3 \int e^{\frac{-r^2}{2 b^2}}V_{\gamma}(r)d\vec{r} \nonumber\\
&=&\sum_{\gamma=0}^8 <\lambda^{\gamma}_1\lambda^{\gamma}_2  \vec{\sigma}_1\cdot \vec{\sigma}_2 P^{\sigma fc}_{36}> N^e(\vec{R}_i,\vec{R}_j)\cdot \nonumber\\
&&((\mu_{\gamma}^2-\Lambda_{\gamma}^2)\sqrt{\frac{2}{\pi}}\frac{1}{b}-\mu_{\gamma}^3 e^{\frac{\mu_{\gamma}^2 b^2}{2}}erfc(\frac{\mu_{\gamma} b}{\sqrt{2}})+\Lambda_{\gamma}^3 e^{\frac{\Lambda_{\gamma}^2 b^2}{2}}erfc(\frac{\Lambda_{\gamma} b}{\sqrt{2}}))\nonumber\\
\end{eqnarray}
\ 

And its projection

\begin{eqnarray}
<Y_L^M(R_j) V_{\chi}(12)P_{36} Y_{L'}^{M'}(R_i)>&=&\sum_{\gamma=0}^8 <\lambda^{\gamma}_1\lambda^{\gamma}_2  \vec{\sigma}_1\cdot \vec{\sigma}_2 P^{\sigma fc}_{36}> 4\pi e^{-\alpha_e (R_i^2+R_j^2)} i_L(\gamma_e R_i R_j)\cdot  \nonumber\\
&&((\mu_{\gamma}^2-\Lambda_{\gamma}^2)\sqrt{\frac{2}{\pi}}\frac{1}{b}-\mu_{\gamma}^3 e^{\frac{\mu_{\gamma}^2 b^2}{2}}erfc(\frac{\mu_{\gamma} b}{\sqrt{2}}) \nonumber\\
&& +\Lambda_{\gamma}^3 e^{\frac{\Lambda_{\gamma}^2 b^2}{2}}erfc(\frac{\Lambda_{\gamma} b}{\sqrt{2}})) \delta_{L,L'}^{M,M'}
\end{eqnarray}
\\

For the diagram $g$ of Fig. \ref{RGMinteractiondiagram}

\begin{eqnarray}
<V_{\chi}(36)P_{36}>&=&\sum_{\gamma=0}^8 <\lambda^{\gamma}_3\lambda^{\gamma}_6  \vec{\sigma}_3\cdot \vec{\sigma}_6 P^{\sigma fc}_{36}> N^e(\vec{R}_i,\vec{R}_j) (\frac{1}{\sqrt{2 \pi}b})^3 \int e^{\frac{-(\vec{r}-\frac{\vec{R}_i-\vec{R}_j}{2})^2}{2 b^2}}V_{\gamma}(r)d\vec{r}\nonumber\\
&=&\sum_{\gamma=0}^8 <\lambda^{\gamma}_3\lambda^{\gamma}_6  \vec{\sigma}_3\cdot \vec{\sigma}_6 P^{\sigma fc}_{36}> N^e(\vec{R}_i,\vec{R}_j) \sum^{p}_{k=1}h_{\gamma}^k (\frac{c_{\gamma}^k}{\sqrt{{c_{\gamma}^k}^2+2b^2}})^3 e^{\frac{-(\vec{R}_i-\vec{R}_j)^2}{2({c_{\gamma}^k}^2+2b^2)}}\nonumber\\
&=&\sum_{\gamma=0}^8 <\lambda^{\gamma}_3\lambda^{\gamma}_6  \vec{\sigma}_3\cdot \vec{\sigma}_6 P^{\sigma fc}_{36}>  e^{-\alpha_e (R_i^2+R_j^2)} e^{\gamma_e \vec{R}_i \cdot \vec{R}_j} \cdot\nonumber\\
&& \sum^{p}_{k=1}h_{\gamma}^k (\frac{c_{\gamma}^k}{\sqrt{{c_{\gamma}^k}^2+2b^2}})^3 e^{\frac{-(R_i^2+R_j^2)}{4({c_{\gamma}^k}^2+2b^2)}} e^{\frac{\vec{R}_i\cdot \vec{R}_j}{2({c_{\gamma}^k}^2+2b^2)}},
\end{eqnarray}
\ 

\noindent with its projection given by

\begin{eqnarray}
<Y_L^M(R_j) V_{\chi}(36)P_{36} Y_{L'}^{M'}(R_i)>&=&\sum_{\gamma=0}^8 <\lambda^{\gamma}_3\lambda^{\gamma}_6  \vec{\sigma}_3\cdot \vec{\sigma}_6 P^{\sigma fc}_{36}>4\pi e^{-\alpha_e (R_i^2+R_j^2)}\cdot \nonumber\\
&&  \sum^{p}_{k=1}h_{\gamma}^k (\frac{c_{\gamma}^k}{\sqrt{{c_{\gamma}^k}^2+2b^2}})^3 e^{\frac{-(R_i^2+R_j^2)}{4({c_{\gamma}^k}^2+2b^2)}}\cdot \nonumber\\
&&  i_L((\gamma_e+\frac{1}{2({c_{\gamma}^k}^2+2b^2)}) R_i R_j) \delta_{L,L'}^{M,M'}
\end{eqnarray}

For $<V_{\chi}(13)P_{36}>$, the diagram $e$ of Fig. \ref{RGMinteractiondiagram},

\begin{eqnarray}
<V_{\chi}(13)P_{36}>&=&\sum_{\gamma=0}^8 <\lambda^{\gamma}_1\lambda^{\gamma}_3  \vec{\sigma}_1\cdot \vec{\sigma}_3 P^{\sigma fc}_{36}> N^e(\vec{R}_i,\vec{R}_j) (\frac{1}{\sqrt{2 \pi}b})^3 \int e^{\frac{-(\vec{r}-\frac{\vec{R}_j}{2})^2}{2 b^2}}V_{\gamma}(r)d\vec{r}\nonumber\\
&=&\sum_{\gamma=0}^8 <\lambda^{\gamma}_1\lambda^{\gamma}_3  \vec{\sigma}_1\cdot \vec{\sigma}_3 P^{\sigma fc}_{36}> N^e(\vec{R}_i,\vec{R}_j)\cdot\nonumber\\
&& (\frac{\mu_{\gamma}^2 e^{\frac{b^2\mu_{\gamma}^2}{2}}}{R_j}(e^{\frac{-R_j\mu_{\gamma}}{2}} erfc(\frac{b(\mu_{\gamma}-\frac{R_j}{2b^2})}{\sqrt{2}})-e^{\frac{R_j\mu_{\gamma}}{2}} erfc(\frac{b(\mu_{\gamma}+\frac{R_j}{2b^2})}{\sqrt{2}}))\nonumber\\
&&  -  \frac{\Lambda_{\gamma}^2 e^{\frac{b^2\Lambda_{\gamma}^2}{2}}}{R_j}(e^{\frac{-R_j\Lambda_{\gamma}}{2}} erfc(\frac{b(\Lambda_{\gamma}-\frac{R_j}{2b^2})}{\sqrt{2}})) \nonumber\\
&& -e^{\frac{R_j\Lambda_{\gamma}}{2}} erfc(\frac{b(\Lambda_{\gamma}+\frac{R_j}{2b^2})}{\sqrt{2}})))
\end{eqnarray}
\ 

\noindent and

\begin{eqnarray}
<Y_L^M(R_j) V_{\chi}(13)P_{36} Y_{L'}^{M'}(R_i)>&=&\sum_{\gamma=0}^8 <\lambda^{\gamma}_1\lambda^{\gamma}_3  \vec{\sigma}_1\cdot \vec{\sigma}_3 P^{\sigma fc}_{36}> 4 \pi e^{-\alpha_e (R_i^2+R_j^2)} i_L(\gamma_e R_i R_j) \cdot\nonumber\\
&& (\frac{\mu_{\gamma}^2 e^{\frac{b^2\mu_{\gamma}^2}{2}}}{R_j}(e^{\frac{-R_j\mu_{\gamma}}{2}} erfc(\frac{b(\mu_{\gamma}-\frac{R_j}{2b^2})}{\sqrt{2}})\cdot \nonumber\\
&& -e^{\frac{R_j\mu_{\gamma}}{2}} erfc(\frac{b(\mu_{\gamma}+\frac{R_j}{2b^2})}{\sqrt{2}}))\nonumber\\
&&  -  \frac{\Lambda_{\gamma}^2 e^{\frac{b^2\Lambda_{\gamma}^2}{2}}}{R_j}(e^{\frac{-R_j\Lambda_{\gamma}}{2}} erfc(\frac{b(\Lambda_{\gamma}-\frac{R_j}{2b^2})}{\sqrt{2}})) \nonumber\\
&& -e^{\frac{R_j\Lambda_{\gamma}}{2}} erfc(\frac{b(\Lambda_{\gamma}+\frac{R_j}{2b^2})}{\sqrt{2}})))\delta_{L,L'}^{M,M'}
\end{eqnarray}
\\

In the same way, for the diagram $f$ of Fig. \ref{RGMinteractiondiagram}

\begin{eqnarray}
<V_{\chi}(16)P_{36}>&=&\sum_{\gamma=0}^8 <\lambda^{\gamma}_1\lambda^{\gamma}_6  \vec{\sigma}_1\cdot \vec{\sigma}_6 P^{\sigma fc}_{36}> N^e(\vec{R}_i,\vec{R}_j)\cdot\nonumber\\
&& (\frac{\mu_{\gamma}^2 e^{\frac{b^2\mu_{\gamma}^2}{2}}}{R_i}(e^{\frac{-R_i\mu_{\gamma}}{2}} erfc(\frac{b(\mu_{\gamma}-\frac{R_i}{2b^2})}{\sqrt{2}})-e^{\frac{R_i\mu_{\gamma}}{2}} erfc(\frac{b(\mu_{\gamma}+\frac{R_i}{2b^2})}{\sqrt{2}}))\nonumber\\
&&  -  \frac{\Lambda_{\gamma}^2 e^{\frac{b^2\Lambda_{\gamma}^2}{2}}}{R_i}(e^{\frac{-R_i\Lambda_{\gamma}}{2}} erfc(\frac{b(\Lambda_{\gamma}-\frac{R_i}{2b^2})}{\sqrt{2}})) \nonumber\\
&& -e^{\frac{R_i\Lambda_{\gamma}}{2}} erfc(\frac{b(\Lambda_{\gamma}+\frac{R_i}{2b^2})}{\sqrt{2}})))
\end{eqnarray}

\noindent with

\begin{eqnarray}
<Y_L^M(R_j) V_{\chi}(16)P_{36} Y_{L'}^{M'}(R_i)>&=&\sum_{\gamma=0}^8 <\lambda^{\gamma}_1\lambda^{\gamma}_6 \vec{\sigma}_1\cdot \vec{\sigma}_6 P^{\sigma fc}_{36}> 4 \pi e^{-\alpha_e (R_i^2+R_j^2)} i_L(\gamma_e R_i R_j) \cdot\nonumber\\
&& (\frac{\mu_{\gamma}^2 e^{\frac{b^2\mu_{\gamma}^2}{2}}}{R_i}(e^{\frac{-R_i\mu_{\gamma}}{2}} erfc(\frac{b(\mu_{\gamma}-\frac{R_i}{2b^2})}{\sqrt{2}})\cdot \nonumber\\
&& -e^{\frac{R_i\mu_{\gamma}}{2}} erfc(\frac{b(\mu_{\gamma}+\frac{R_i}{2b^2})}{\sqrt{2}}))\nonumber\\
&&  -  \frac{\Lambda_{\gamma}^2 e^{\frac{b^2\Lambda_{\gamma}^2}{2}}}{R_i}(e^{\frac{-R_i\Lambda_{\gamma}}{2}} erfc(\frac{b(\Lambda_{\gamma}-\frac{R_i}{2b^2})}{\sqrt{2}})) \nonumber\\
&& -e^{\frac{R_i\Lambda_{\gamma}}{2}} erfc(\frac{b(\Lambda_{\gamma}+\frac{R_i}{2b^2})}{\sqrt{2}})))\delta_{L,L'}^{M,M'}
\end{eqnarray}
\\

Finally, for the diagram $d$ of Fig. \ref{RGMinteractiondiagram}

\begin{eqnarray}
<V_{\chi}(14)P_{36}>&=&\sum_{\gamma=0}^8 <\lambda^{\gamma}_1\lambda^{\gamma}_4  \vec{\sigma}_1\cdot \vec{\sigma}_4 P^{\sigma fc}_{36}> N^e(\vec{R}_i,\vec{R}_j) \sum^{p}_{k=1}h_{\gamma}^k (\frac{c_{\gamma}^k}{\sqrt{{c_{\gamma}^k}^2+2b^2}})^3 e^{\frac{-(\vec{R}_i+\vec{R}_j)^2}{2({c_{\gamma}^k}^2+2b^2)}}\nonumber\\
&=&\sum_{\gamma=0}^8 <\lambda^{\gamma}_1\lambda^{\gamma}_4 \vec{\sigma}_1\cdot \vec{\sigma}_4 P^{\sigma fc}_{36}>  e^{-\alpha_e (R_i^2+R_j^2)} e^{\gamma_e \vec{R}_i \cdot \vec{R}_j} \cdot\nonumber\\
&& \sum^{p}_{k=1}h_{\gamma}^k (\frac{c_{\gamma}^k}{\sqrt{{c_{\gamma}^k}^2+2b^2}})^3 e^{\frac{-(R_i^2+R_j^2)}{4({c_{\gamma}^k}^2+2b^2)}} e^{\frac{-\vec{R}_i\cdot \vec{R}_j}{2({c_{\gamma}^k}^2+2b^2)}}
\end{eqnarray}

\noindent and

\begin{eqnarray}
<Y_L^M(R_j) V_{\chi}(14)P_{36} Y_{L'}^{M'}(R_i)>&=&\sum_{\gamma=0}^8 <\lambda^{\gamma}_1\lambda^{\gamma}_4 \vec{\sigma}_1\cdot \vec{\sigma}_4 P^{\sigma fc}_{36}>  4\pi e^{-\alpha_e (R_i^2+R_j^2)}\cdot \nonumber\\
&&  \sum^{p}_{k=1}h_{\gamma}^k (\frac{c_{\gamma}^k}{\sqrt{{c_{\gamma}^k}^2+2b^2}})^3  e^{\frac{-(R_i^2+R_j^2)}{4({c_{\gamma}^k}^2+2b^2)}}\cdot \nonumber\\
&&  i_L((\gamma_e-\frac{1}{2({c_{\gamma}^k}^2+2b^2)}) R_i R_j) \delta_{L,L'}^{M,M'}
\end{eqnarray}
\\\ \\

The Tables \ref{HYP00}-\ref{HYPdelta} give all the flavor-spin-color matrix elements needed for the hyperfine interaction.
\ \\

\begin{table}[H]
\centering

\begin{tabular}{|c|c|c|c|c|c|c|}
\hline
\rule[-2mm]{0mm}{7mm}$\alpha$ & $NN$ & $NN$  & $\Delta\Delta$ & $NN$ & $\Delta\Delta$ & $CC$ \\
\rule[-4mm]{0mm}{5mm}$\beta$  & $NN$ & $\Delta\Delta$ & $\Delta\Delta$ & $CC$ & $CC$  & $CC$ \\
\hline
\hline

\rule[-3mm]{0mm}{8mm}$\sigma_1.\sigma_2\ \tau_1.\tau_2$                                        
&  2025 &    0 &   405 &    0 &    0 &  -891 \\
\rule[-3mm]{0mm}{5mm}$\sigma_3.\sigma_6\ \tau_3.\tau_6$                                        
&  1125 &  720 &  1125 &    0 &    0 &  -315 \\
\rule[-3mm]{0mm}{5mm}$\sigma_1.\sigma_2\ \tau_1.\tau_2\ P_{36}^{f \sigma c}$                   
&   645 &   60 &    15 &   24 &  -48 &   -30 \\
\rule[-3mm]{0mm}{5mm}$\sigma_3.\sigma_6\ \tau_3.\tau_6\ P_{36}^{f \sigma c}$                   
&   465 &   60 &   735 & -192 & -192 &  -822 \\
\rule[-3mm]{0mm}{5mm}$\sigma_1.\sigma_3\ \tau_1.\tau_3\ P_{36}^{f \sigma c}$                   
&   465 &  420 &    15 &  168 &   24 &   114 \\
\rule[-3mm]{0mm}{5mm}$\sigma_1.\sigma_6\ \tau_1.\tau_6\ P_{36}^{f \sigma c}$                   
&   465 &   60 &    15 &  -12 &  -48 &   294 \\
\rule[-3mm]{0mm}{5mm}$\sigma_1.\sigma_4\ \tau_1.\tau_4\ P_{36}^{f \sigma c}$                   
&   315 &  360 &   135 &   18 & -180 &    72 \\
\rule[-3mm]{0mm}{5mm}$\sigma_1.\sigma_2\ \lambda_1^f.\lambda_2^f$                              
&  1890 &    0 &   540 &    0 &    0 &  -972 \\
\rule[-3mm]{0mm}{5mm}$\sigma_3.\sigma_6\ \lambda_3^f.\lambda_6^f$                              
&  1080 &  720 &   900 &    0 &    0 &  -396 \\
\rule[-3mm]{0mm}{5mm}$\sigma_1.\sigma_2\ \lambda_1^f.\lambda_2^f\ P_{36}^{f \sigma c}$         
&   590 &   80 &    20 &   32 &  -64 &   176 \\
\rule[-3mm]{0mm}{5mm}$\sigma_3.\sigma_6\ \lambda_3^f.\lambda_6^f\ P_{36}^{f \sigma c}$         
&   460 &   40 &   700 & -164 & -152 &  -644 \\
\rule[-3mm]{0mm}{5mm}$\sigma_1.\sigma_3\ \lambda_1^f.\lambda_3^f\ P_{36}^{f \sigma c}$         
&   440 &  380 &    20 &  152 &   -4 &   200 \\
\rule[-3mm]{0mm}{5mm}$\sigma_1.\sigma_6\ \lambda_1^f.\lambda_6^f\ P_{36}^{f \sigma c}$         
&   440 &   80 &    20 &  -22 &  -16 &   350 \\
\rule[-3mm]{0mm}{5mm}$\sigma_1.\sigma_4\ \lambda_1^f.\lambda_4^f\ P_{36}^{f \sigma c}$         
&   315 &  330 &   120 &   15 & -150 &   129 \\
\rule[-3mm]{0mm}{5mm}$\sigma_1.\sigma_2\ \lambda_1^{f,0}.\lambda_2^{f,0}$                      
&  -270 &    0 &   270 &    0 &    0 &  -162 \\
\rule[-3mm]{0mm}{5mm}$\sigma_3.\sigma_6\ \lambda_3^{f,0}.\lambda_6^{f,0}$                      
&   -90 &    0 &  -450 &    0 &    0 &  -162 \\
\rule[-3mm]{0mm}{5mm}$\sigma_1.\sigma_2\ \lambda_1^{f,0}.\lambda_2^{f,0}\ P_{36}^{f \sigma c}$ 
&  -110 &   40 &    10 &   16 &  -32 &   412 \\
\rule[-3mm]{0mm}{5mm}$\sigma_3.\sigma_6\ \lambda_3^{f,0}.\lambda_6^{f,0}\ P_{36}^{f \sigma c}$ 
&   -10 &  -40 &   -70 &   56 &   80 &   356 \\
\rule[-3mm]{0mm}{5mm}$\sigma_1.\sigma_3\ \lambda_1^{f,0}.\lambda_3^{f,0}\ P_{36}^{f \sigma c}$ 
&   -50 &  -80 &    10 &  -32 &  -56 &   172 \\
\rule[-3mm]{0mm}{5mm}$\sigma_1.\sigma_6\ \lambda_1^{f,0}.\lambda_6^{f,0}\ P_{36}^{f \sigma c}$ 
&   -50 &   40 &    10 &  -20 &   64 &   112 \\
\rule[-3mm]{0mm}{5mm}$\sigma_1.\sigma_4\ \lambda_1^{f,0}.\lambda_4^{f,0}\ P_{36}^{f \sigma c}$ 
&     0 &  -60 &   -30 &   -6 &   60 &   114 \\
\hline
\rule[-4mm]{0mm}{10mm}factor & $\frac{1}{405}$ & $\frac{1}{405}$ & $\frac{1}{405}$ 
& $\frac{\sqrt10}{405}$ & $\frac{\sqrt10}{405}$ & $\frac{1}{405}$ \\
\hline
\end{tabular}
\caption{Matrix elements $\langle \alpha|O|\beta \rangle$ of different operators $O$ for (S,I) = (0,0).}\label{HYP00}

\end{table}

\begin{table}[H]
\centering

\begin{tabular}{|c|c|c|c|c|c|c|}
\hline
\rule[-2mm]{0mm}{7mm}$\alpha$ & $NN$ & $NN$  & $\Delta\Delta$ & $NN$ & $\Delta\Delta$ & $CC$ \\
\rule[-4mm]{0mm}{5mm}$\beta$  & $NN$ & $\Delta\Delta$ & $\Delta\Delta$ & $CC$ & $CC$  & $CC$ \\
\hline
\hline

\rule[-3mm]{0mm}{8mm}$\sigma_1.\sigma_2\ \tau_1.\tau_2$                                        
&  4860 &    0 &   972 &    0 &    0 &   108 \\
\rule[-3mm]{0mm}{5mm}$\sigma_3.\sigma_6\ \tau_3.\tau_6$                                        
&  -900 &  576 &  1980 &    0 &    0 &  1116 \\
\rule[-3mm]{0mm}{5mm}$\sigma_1.\sigma_2\ \tau_1.\tau_2\ P_{36}^{f \sigma c}$                   
&  -444 &   48 &    12 & -720 &  288 &   588 \\
\rule[-3mm]{0mm}{5mm}$\sigma_3.\sigma_6\ \tau_3.\tau_6\ P_{36}^{f \sigma c}$                   
&   708 &   48 &  1596 &  240 &  672 & -1092 \\
\rule[-3mm]{0mm}{5mm}$\sigma_1.\sigma_3\ \tau_1.\tau_3\ P_{36}^{f \sigma c}$                   
&   132 &  336 &    12 & -720 &  288 &  -420 \\
\rule[-3mm]{0mm}{5mm}$\sigma_1.\sigma_6\ \tau_1.\tau_6\ P_{36}^{f \sigma c}$                   
&   132 &   48 &    12 &  336 &  -96 &  -420 \\
\rule[-3mm]{0mm}{5mm}$\sigma_1.\sigma_4\ \tau_1.\tau_4\ P_{36}^{f \sigma c}$                   
&    36 & -144 &   -36 &  228 &  288 & -1260 \\
\rule[-3mm]{0mm}{5mm}$\sigma_1.\sigma_2\ \lambda_1^f.\lambda_2^f$                              
&  4536 &    0 &  1296 &    0 &    0 &   -18 \\
\rule[-3mm]{0mm}{5mm}$\sigma_3.\sigma_6\ \lambda_3^f.\lambda_6^f$                              
&  -864 &  576 &  1584 &    0 &    0 &  1020 \\
\rule[-3mm]{0mm}{5mm}$\sigma_1.\sigma_2\ \lambda_1^f.\lambda_2^f\ P_{36}^{f \sigma c}$         
&  -376 &   64 &    16 & -672 &  384 &   706 \\
\rule[-3mm]{0mm}{5mm}$\sigma_3.\sigma_6\ \lambda_3^f.\lambda_6^f\ P_{36}^{f \sigma c}$         
&   784 &   32 &  1520 &  216 &  528 & -1024 \\
\rule[-3mm]{0mm}{5mm}$\sigma_1.\sigma_3\ \lambda_1^f.\lambda_3^f\ P_{36}^{f \sigma c}$         
&   104 &  304 &    16 & -672 &  384 &  -332 \\
\rule[-3mm]{0mm}{5mm}$\sigma_1.\sigma_6\ \lambda_1^f.\lambda_6^f\ P_{36}^{f \sigma c}$         
&   104 &   64 &    16 &  340 & -200 &  -332 \\
\rule[-3mm]{0mm}{5mm}$\sigma_1.\sigma_4\ \lambda_1^f.\lambda_4^f\ P_{36}^{f \sigma c}$         
&    44 & -152 &   -32 &  278 &  164 & -1197 \\
\rule[-3mm]{0mm}{5mm}$\sigma_1.\sigma_2\ \lambda_1^{f,0}.\lambda_2^{f,0}$                      
&  -648 &    0 &   648 &    0 &    0 &  -252 \\
\rule[-3mm]{0mm}{5mm}$\sigma_3.\sigma_6\ \lambda_3^{f,0}.\lambda_6^{f,0}$                      
&    72 &    0 &  -792 &    0 &    0 &  -192 \\
\rule[-3mm]{0mm}{5mm}$\sigma_1.\sigma_2\ \lambda_1^{f,0}.\lambda_2^{f,0}\ P_{36}^{f \sigma c}$ 
&   136 &   32 &     8 &   96 &  192 &   236 \\
\rule[-3mm]{0mm}{5mm}$\sigma_3.\sigma_6\ \lambda_3^{f,0}.\lambda_6^{f,0}\ P_{36}^{f \sigma c}$ 
&   152 &  -32 &  -152 &  -48 & -288 &   136 \\
\rule[-3mm]{0mm}{5mm}$\sigma_1.\sigma_3\ \lambda_1^{f,0}.\lambda_3^{f,0}\ P_{36}^{f \sigma c}$ 
&   -56 &  -64 &     8 &   96 &  192 &   176 \\
\rule[-3mm]{0mm}{5mm}$\sigma_1.\sigma_6\ \lambda_1^{f,0}.\lambda_6^{f,0}\ P_{36}^{f \sigma c}$ 
&   -56 &   32 &     8 &    8 & -208 &   176 \\
\rule[-3mm]{0mm}{5mm}$\sigma_1.\sigma_4\ \lambda_1^{f,0}.\lambda_4^{f,0}\ P_{36}^{f \sigma c}$ 
&    16 &  -16 &     8 &  -20 & -248 &   126 \\
\hline
\rule[-4mm]{0mm}{10mm}factor & $\frac{1}{972}$ & $\frac{\sqrt5}{972}$ & $\frac{1}{972}$ 
& $\frac{\sqrt5}{972}$ & $\frac{1}{972}$ & $\frac{1}{972}$ \\
\hline
\end{tabular}
\caption{Matrix elements $\langle \alpha|O|\beta \rangle$ of different operators $O$ for (S,I) = (1,0).}\label{HYP10}
\end{table}

\begin{table}[H]
\centering

\begin{tabular}{|c|c|c|c|c|c|c|}
\hline
\rule[-2mm]{0mm}{7mm}$\alpha$ & $NN$ & $NN$ & $\Delta\Delta$ & $NN$ & $\Delta\Delta$ & $CC$ \\
\rule[-4mm]{0mm}{5mm}$\beta$  & $NN$ & $\Delta\Delta$ & $\Delta\Delta$ & $CC$ & $CC$ & $CC$ \\
\hline
\hline

\rule[-3mm]{0mm}{8mm}$\sigma_1.\sigma_2\ \tau_1.\tau_2$                                        
&  4860 &    0 &   972 &    0 &    0 &   108 \\
\rule[-3mm]{0mm}{5mm}$\sigma_3.\sigma_6\ \tau_3.\tau_6$                                        
&  -900 &  576 &  1980 &    0 &    0 &  1116 \\
\rule[-3mm]{0mm}{5mm}$\sigma_1.\sigma_2\ \tau_1.\tau_2\ P_{36}^{f \sigma c}$                   
&  -444 &   48 &    12 & -720 &  288 &   588 \\
\rule[-3mm]{0mm}{5mm}$\sigma_3.\sigma_6\ \tau_3.\tau_6\ P_{36}^{f \sigma c}$                   
&   708 &   48 &  1596 &  240 &  672 & -1092 \\
\rule[-3mm]{0mm}{5mm}$\sigma_1.\sigma_3\ \tau_1.\tau_3\ P_{36}^{f \sigma c}$                   
&   132 &  336 &    12 & -720 &  288 &  -420 \\
\rule[-3mm]{0mm}{5mm}$\sigma_1.\sigma_6\ \tau_1.\tau_6\ P_{36}^{f \sigma c}$                   
&   132 &   48 &    12 &  336 &  -96 &  -420 \\
\rule[-3mm]{0mm}{5mm}$\sigma_1.\sigma_4\ \tau_1.\tau_4\ P_{36}^{f \sigma c}$                   
&    36 & -144 &   -36 &  228 &  288 & -1260 \\
\rule[-3mm]{0mm}{5mm}$\sigma_1.\sigma_2\ \lambda_1^f.\lambda_2^f$                              
&  4536 &    0 &  1296 &    0 &    0 &  -126 \\
\rule[-3mm]{0mm}{5mm}$\sigma_3.\sigma_6\ \lambda_3^f.\lambda_6^f$                              
& -1008 &  576 &  1440 &    0 &    0 &   948 \\
\rule[-3mm]{0mm}{5mm}$\sigma_1.\sigma_2\ \lambda_1^f.\lambda_2^f\ P_{36}^{f \sigma c}$         
&  -376 &   64 &    16 & -672 &  384 &   814 \\
\rule[-3mm]{0mm}{5mm}$\sigma_3.\sigma_6\ \lambda_3^f.\lambda_6^f\ P_{36}^{f \sigma c}$         
&   832 &   32 &  1568 &  232 &  496 &  -976 \\
\rule[-3mm]{0mm}{5mm}$\sigma_1.\sigma_3\ \lambda_1^f.\lambda_3^f\ P_{36}^{f \sigma c}$         
&   104 &  304 &    16 & -672 &  384 &  -260 \\
\rule[-3mm]{0mm}{5mm}$\sigma_1.\sigma_6\ \lambda_1^f.\lambda_6^f\ P_{36}^{f \sigma c}$         
&   104 &   64 &    16 &  364 & -248 &  -260 \\
\rule[-3mm]{0mm}{5mm}$\sigma_1.\sigma_4\ \lambda_1^f.\lambda_4^f\ P_{36}^{f \sigma c}$         
&    36 & -168 &   -48 &  298 &  124 & -1155 \\
\rule[-3mm]{0mm}{5mm}$\sigma_1.\sigma_2\ \lambda_1^{f,0}.\lambda_2^{f,0}$                      
&  -648 &    0 &   648 &    0 &    0 &  -468 \\
\rule[-3mm]{0mm}{5mm}$\sigma_3.\sigma_6\ \lambda_3^{f,0}.\lambda_6^{f,0}$                      
&  -216 &    0 & -1080 &    0 &    0 &  -336 \\
\rule[-3mm]{0mm}{5mm}$\sigma_1.\sigma_2\ \lambda_1^{f,0}.\lambda_2^{f,0}\ P_{36}^{f \sigma c}$ 
&   136 &   32 &     8 &   96 &  192 &   452 \\
\rule[-3mm]{0mm}{5mm}$\sigma_3.\sigma_6\ \lambda_3^{f,0}.\lambda_6^{f,0}\ P_{36}^{f \sigma c}$ 
&   248 &  -32 &   -56 &  -16 & -352 &   232 \\
\rule[-3mm]{0mm}{5mm}$\sigma_1.\sigma_3\ \lambda_1^{f,0}.\lambda_3^{f,0}\ P_{36}^{f \sigma c}$ 
&   -56 &  -64 &     8 &   96 &  192 &   320 \\
\rule[-3mm]{0mm}{5mm}$\sigma_1.\sigma_6\ \lambda_1^{f,0}.\lambda_6^{f,0}\ P_{36}^{f \sigma c}$ 
&   -56 &   32 &     8 &   56 & -304 &   320 \\
\rule[-3mm]{0mm}{5mm}$\sigma_1.\sigma_4\ \lambda_1^{f,0}.\lambda_4^{f,0}\ P_{36}^{f \sigma c}$ 
&     0 &  -48 &   -24 &   20 & -328 &   210 \\
\hline
\rule[-4mm]{0mm}{10mm}factor & $\frac{1}{972}$ & $\frac{\sqrt5}{972}$ & $\frac{1}{972}$ 
& $\frac{\sqrt5}{972}$ & $\frac{1}{972}$ & $\frac{1}{972}$ \\
\hline
\end{tabular}
\caption{Matrix elements $\langle \alpha|O|\beta \rangle$ of different operators $O$ for (S,I) = (0,1).}\label{HYP01}

\end{table}

\begin{table}[H]
\centering

\begin{tabular}{|c|c|c|c|c|c|c|}
\hline
\rule[-2mm]{0mm}{7mm}$\alpha$ & $NN$ & $NN$  & $\Delta\Delta$ & $NN$ & $\Delta\Delta$ & $CC$ \\
\rule[-4mm]{0mm}{5mm}$\beta$  & $NN$ & $\Delta\Delta$ & $\Delta\Delta$ & $CC$ & $CC$  & $CC$ \\
\hline
\hline

\rule[-3mm]{0mm}{8mm}$\sigma_1.\sigma_2\ \tau_1.\tau_2$                                        
&  3645 &    0 &   729 &      0 &      0 &    2187 \\
\rule[-3mm]{0mm}{5mm}$\sigma_3.\sigma_6\ \tau_3.\tau_6$                                        
&   225 &  720 &  1089 & -14400 &  -7200 &  185967 \\
\rule[-3mm]{0mm}{5mm}$\sigma_1.\sigma_2\ \tau_1.\tau_2\ P_{36}^{f \sigma c}$                   
&   633 &   60 &     3 &  -3360 &  -9240 & -185808 \\
\rule[-3mm]{0mm}{5mm}$\sigma_3.\sigma_6\ \tau_3.\tau_6\ P_{36}^{f \sigma c}$                   
&   525 &   60 &  1083 & -14160 & -13560 & -323508 \\
\rule[-3mm]{0mm}{5mm}$\sigma_1.\sigma_3\ \tau_1.\tau_3\ P_{36}^{f \sigma c}$                   
&   381 &  420 &     3 &   8880 &  -5640 & -188688 \\
\rule[-3mm]{0mm}{5mm}$\sigma_1.\sigma_6\ \tau_1.\tau_6\ P_{36}^{f \sigma c}$                   
&   381 &   60 &     3 &  -5880 &    480 & -186888 \\
\rule[-3mm]{0mm}{5mm}$\sigma_1.\sigma_4\ \tau_1.\tau_4\ P_{36}^{f \sigma c}$                   
&   219 &  240 &     3 &  -4980 & -26880 & -125418 \\
\rule[-3mm]{0mm}{5mm}$\sigma_1.\sigma_2\ \lambda_1^f.\lambda_2^f$                              
&  3402 &    0 &   972 &      0 &      0 &  -77274 \\
\rule[-3mm]{0mm}{5mm}$\sigma_3.\sigma_6\ \lambda_3^f.\lambda_6^f$                              
&   252 &  720 &   729 & -14400 &  -7200 &  138636 \\
\rule[-3mm]{0mm}{5mm}$\sigma_1.\sigma_2\ \lambda_1^f.\lambda_2^f\ P_{36}^{f \sigma c}$         
&   574 &   80 &     4 &  -2320 & -12320 &  -59554 \\
\rule[-3mm]{0mm}{5mm}$\sigma_3.\sigma_6\ \lambda_3^f.\lambda_6^f\ P_{36}^{f \sigma c}$         
&   584 &   40 &  1064 & -11960 &  -9760 & -284504 \\
\rule[-3mm]{0mm}{5mm}$\sigma_1.\sigma_3\ \lambda_1^f.\lambda_3^f\ P_{36}^{f \sigma c}$         
&   364 &  380 &     4 &   7880 &  -9320 & -128914 \\
\rule[-3mm]{0mm}{5mm}$\sigma_1.\sigma_6\ \lambda_1^f.\lambda_6^f\ P_{36}^{f \sigma c}$         
&   364 &   80 &     4 &  -6580 &   4420 & -127414 \\
\rule[-3mm]{0mm}{5mm}$\sigma_1.\sigma_4\ \lambda_1^f.\lambda_4^f\ P_{36}^{f \sigma c}$         
&   229 &  230 &     4 &  -4750 & -22700 & -105349 \\
\rule[-3mm]{0mm}{5mm}$\sigma_1.\sigma_2\ \lambda_1^{f,0}.\lambda_2^{f,0}$                      
&  -486 &    0 &   486 &      0 &      0 & -158922 \\
\rule[-3mm]{0mm}{5mm}$\sigma_3.\sigma_6\ \lambda_3^{f,0}.\lambda_6^{f,0}$                      
&    54 &    0 &  -594 &      0 &      0 &  -94662 \\
\rule[-3mm]{0mm}{5mm}$\sigma_1.\sigma_2\ \lambda_1^{f,0}..\lambda_2^{f,0}\ P_{36}^{f \sigma c}$ 
&  -118 &   40 &     2 &  -7721 &  -6160 &  154498 \\
\rule[-3mm]{0mm}{5mm}$\sigma_3.\sigma_6\ \lambda_3^{f,0}.\lambda_6^{f,0}\ P_{36}^{f \sigma c}$ 
&   118 &  -40 &   -38 &   9179 &   7600 &  125798 \\
\rule[-3mm]{0mm}{5mm}$\sigma_1.\sigma_3\ \lambda_1^{f,0}.\lambda_3^{f,0}\ P_{36}^{f \sigma c}$ 
&   -34 &  -80 &     2 &  -6941 &  -7360 &  137638 \\
\rule[-3mm]{0mm}{5mm}$\sigma_1.\sigma_6\ \lambda_1^{f,0}.\lambda_6^{f,0}\ P_{36}^{f \sigma c}$ 
&   -34 &   40 &     2 &   7159 &   7880 &  137038 \\
\rule[-3mm]{0mm}{5mm}$\sigma_1.\sigma_4\ \lambda_1^{f,0}.\lambda_4^{f,0}\ P_{36}^{f \sigma c}$ 
&    20 &  -20 &     2 &   8749 &   8360 &  123028 \\
\hline
\rule[-4mm]{0mm}{10mm}factor & $\frac{1}{729}$ & $\frac{1}{729}$ & $\frac{1}{729}$ 
& $\frac{1}{729 \cdot \sqrt{1486}}$ & $\frac{1}{729 \cdot \sqrt{1486}}$ & $\frac{1}{729 \cdot 743}$ \\
\hline
\end{tabular}
\caption{Matrix elements $\langle \alpha|O|\beta \rangle$ of different operators $O$ for (S,I) = (1,1).}\label{HYP11}

\end{table}

\begin{table}[H]
\centering

\begin{turn}{90}
\begin{tabular}{|c|c|c|c|c|c|c|c|c|c|c|c|c|c|c|c|c|}

\hline

\rule[-3mm]{0mm}{8mm}$(S,I)$ & (3,3) & (3,2)  & (2,3) & (3,1) & (1,3) & (3,0) & (0,3) & (2,2) & (2,1) & (1,2) & (2,0) & (0,2) & (1,1) & (1,0) & (0,1) & (0,0)\\
\hline
\hline

\rule[-2mm]{0mm}{7mm}$ P_{36}^{f \sigma c}$                                        
       &   243  &     81 &    81 &   -27 &   -27 &   -81 &   -81 &    27 &    -9 &    -9 &   -27 &   -27 &     3 &     9 &     9 &    27 \\
\rule[-2mm]{0mm}{4mm}$\sigma_1.\sigma_2\ \tau_1.\tau_2$                                        
       &   729  &    729 &   729 &   729 &   729 &   729 &   729 &   729 &   729 &   729 &   729 &   729 &   729 &   729 &   729 &   729 \\
\rule[-2mm]{0mm}{4mm}$\sigma_3.\sigma_6\ \tau_3.\tau_6$                                        
       &   729  &   -243 &  -243 &  -891 &  -891 & -1215 & -1215 &    81 &   297 &   297 &   155 &   155 &  1089 &  1485 &  1485 &  2025 \\
\rule[-2mm]{0mm}{4mm}$\sigma_1.\sigma_2\ \tau_1.\tau_2\ P_{36}^{f \sigma c}$                   
       &   243  &     81 &    81 &   -27 &   -27 &   -81 &   -81 &    27 &    -9 &    -9 &   -27 &   -27 &     3 &     9 &     9 &    27 \\
\rule[-2mm]{0mm}{4mm}$\sigma_3.\sigma_6\ \tau_3.\tau_6\ P_{36}^{f \sigma c}$                   
       &   243  &    405 &   405 &   513 &   513 &   567 &   567 &   675 &   855 &   855 &   945 &   945 &  1083 &  1197 &  1197 &  1323 \\
\rule[-2mm]{0mm}{4mm}$\sigma_1.\sigma_3\ \tau_1.\tau_3\ P_{36}^{f \sigma c}$                   
       &   243  &     81 &    81 &   -27 &   -27 &   -81 &   -81 &    27 &    -9 &    -9 &   -27 &   -27 &     3 &     9 &     9 &    27 \\
\rule[-2mm]{0mm}{4mm}$\sigma_1.\sigma_6\ \tau_1.\tau_6\ P_{36}^{f \sigma c}$                   
       &   243  &     81 &    81 &   -27 &   -27 &   -81 &   -81 &    27 &    -9 &    -9 &   -27 &   -27 &     3 &     9 &     9 &    27 \\
\rule[-2mm]{0mm}{4mm}$\sigma_1.\sigma_4\ \tau_1.\tau_4\ P_{36}^{f \sigma c}$                   
       &   243  &   -243 &  -243 &   -27 &   -27 &   243 &   243 &   243 &    27 &    27 &  -243 &  -243 &     3 &   -27 &   -27 &   243 \\
\rule[-2mm]{0mm}{4mm}$\sigma_1.\sigma_2\ \lambda_1^f.\lambda_2^f$                              
       &   972  &    972 &   972 &   972 &   972 &   972 &   972 &   972 &   972 &   972 &   972 &   972 &   972 &   972 &   972 &   972 \\
\rule[-2mm]{0mm}{4mm}$\sigma_3.\sigma_6\ \lambda_3^f.\lambda_6^f$                              
       &   972  &      0 &  -324 &  -648 & -1188 &  -972 & -1620 &     0 &   216 &     0 &   324 &     0 &   792 &  1188 &  1080 &  1620 \\
\rule[-2mm]{0mm}{4mm}$\sigma_1.\sigma_2\ \lambda_1^f.\lambda_2^f\ P_{36}^{f \sigma c}$         
       &   324  &    108 &   108 &   -36 &   -36 &  -108 &  -108 &    36 &   -12 &   -12 &   -36 &   -36 &     4 &    12 &    12 &    36 \\
\rule[-2mm]{0mm}{4mm}$\sigma_3.\sigma_6\ \lambda_3^f.\lambda_6^f\ P_{36}^{f \sigma c}$         
       &   324  &    432 &   540 &   504 &   684 &   432 &   756 &   720 &   840 &   912 &   900 &  1008 &  1064 &  1140 &  1176 &  1260 \\
\rule[-2mm]{0mm}{4mm}$\sigma_1.\sigma_3\ \lambda_1^f.\lambda_3^f\ P_{36}^{f \sigma c}$         
       &   324  &    108 &   108 &   -36 &   -36 &  -108 &  -108 &    36 &   -12 &   -12 &   -36 &   -36 &     4 &    12 &    12 &    36 \\
\rule[-2mm]{0mm}{4mm}$\sigma_1.\sigma_6\ \lambda_1^f.\lambda_6^f\ P_{36}^{f \sigma c}$         
       &   324  &    108 &   108 &   -36 &   -36 &  -108 &  -108 &    36 &   -12 &   -12 &   -36 &   -36 &     4 &    12 &    12 &    36 \\
\rule[-2mm]{0mm}{4mm}$\sigma_1.\sigma_4\ \lambda_1^f.\lambda_4^f\ P_{36}^{f \sigma c}$         
       &   324  &   -216 &  -324 &   -36 &   -36 &   216 &   324 &   216 &    36 &    24 &  -216 &  -216 &     4 &   -24 &   -36 &   216 \\
\rule[-2mm]{0mm}{4mm}$\sigma_1.\sigma_2\ \lambda_1^{f,0}.\lambda_2^{f,0}$                      
       &   486  &    486 &   486 &   486 &   486 &   486 &   486 &   486 &   486 &   486 &   486 &   486 &   486 &   486 &   486 &   486 \\
\rule[-2mm]{0mm}{4mm}$\sigma_3.\sigma_6\ \lambda_3^{f,0}.\lambda_6^{f,0}$                      
       &   486  &    486 &  -162 &   486 &  -594 &   486 &  -810 &  -162 &  -162 &  -594 &  -162 &  -810 &  -594 &  -594 &  -810 &  -810 \\
\rule[-2mm]{0mm}{4mm}$\sigma_1.\sigma_2\ \lambda_1^{f,0}..\lambda_2^{f,0}\ P_{36}^{f \sigma c}$ 
       &   162  &     54 &    54 &   -18 &   -18 &   -54 &   -54 &    18 &    -6 &    -6 &   -18 &   -18 &     2 &     6 &     6 &    18 \\
\rule[-2mm]{0mm}{4mm}$\sigma_3.\sigma_6\ \lambda_3^{f,0}.\lambda_6^{f,0}\ P_{36}^{f \sigma c}$ 
       &   162  &     54 &   270 &   -18 &   342 &   -54 &   378 &    90 &   -30 &   114 &   -30 &   126 &   -38 &  -114 &   -42 &  -126 \\
\rule[-2mm]{0mm}{4mm}$\sigma_1.\sigma_3\ \lambda_1^{f,0}.\lambda_3^{f,0}\ P_{36}^{f \sigma c}$ 
       &   162  &     54 &    54 &   -18 &   -18 &   -54 &   -54 &    18 &    -6 &    -6 &   -18 &   -18 &     2 &     6 &     6 &    18 \\
\rule[-2mm]{0mm}{4mm}$\sigma_1.\sigma_6\ \lambda_1^{f,0}.\lambda_6^{f,0}\ P_{36}^{f \sigma c}$ 
       &   162  &     54 &    54 &   -18 &   -18 &   -54 &   -54 &    18 &    -6 &    -6 &   -18 &   -18 &     2 &     6 &     6 &    18 \\
\rule[-2mm]{0mm}{6mm}$\sigma_1.\sigma_4\ \lambda_1^{f,0}.\lambda_4^{f,0}\ P_{36}^{f \sigma c}$ 
       &   162  &     54 &  -162 &   -18 &   -18 &   -54 &   162 &   -54 &    18 &    -6 &    54 &    54 &     2 &     6 &   -18 &   -54 \\
\hline
\rule[-4mm]{0mm}{10mm}factor & $\frac{1}{729}$ & $\frac{1}{729}$ & $\frac{1}{729}$ & $\frac{1}{729}$ & $\frac{1}{729}$ & $\frac{1}{729}$ & $\frac{1}{729}$ & $\frac{1}{729}$ & $\frac{1}{729}$ & $\frac{1}{729}$ & $\frac{1}{729}$ & $\frac{1}{729}$ & $\frac{1}{729}$ & $\frac{1}{729}$ & $\frac{1}{729}$ & $\frac{1}{729}$ \\
\hline
\end{tabular}
\end{turn}
\caption{Matrix elements $\langle \Delta\Delta|O|\Delta \Delta\rangle$ of different operators $O$ and (S,I) configurations.}\label{HYPdelta}

\end{table}

\newpage
\thispagestyle{empty} 
\ 
\newpage
\chapter{The Tensor Force}\label{appendixTENSOR}
\ 
In this appendix we present the calculation of the matrix elements needed in the $^3S_1- {^3D_1}$ tensor coupling in the RGM approach. The main difficulties arise from the coupling of three spaces, namely the spin, flavor and orbital spaces. In order to keep this appendix self-consistent, we reproduce here the tensor potential

\begin{equation}
V^T_{\chi}(r_{ij})=\left\{\sum^{3}_{a=1}V_{\pi}(r_{ij})\lambda^a_i\lambda^a_j+\sum^{7}_{a=4}V_K(r_{ij})\lambda^a_i\lambda^a_j+V_{\eta}(r_{ij})\lambda^8_i\lambda^8_j+\frac{2}{3}V_{\eta'}(r_{ij})\right\} S^T_{ij}
\end{equation}
\ 

\noindent where $\lambda^a_i$ are the Gell-Mann flavour matrices, $S^T_{ij}$ is given by

\begin{equation}\label{TENSORSTij}
S^T_{ij}=\frac{3(\vec{r}_{ij}\cdot \vec{\sigma}_i)(\vec{r}_{ij}\cdot \vec{\sigma}_j)}{r^2}-\vec{\sigma}_i\cdot \vec{\sigma}_j
\end{equation}

\noindent and

\begin{equation}\label{TENSORVg}
V^T_{\gamma}(r)=G_f\ \frac{g^2_{\gamma}}{4 \pi}\frac{1}{12 m_i m_j}[\mu^2_{\gamma}(1+\frac{3}{\mu_{\gamma}r}+\frac{3}{\mu^2_{\gamma}r^2})\frac{e^{-\mu_{\gamma}r}}{r}-\Lambda^2_{\gamma}(1+\frac{3}{\Lambda_{\gamma}r}+\frac{3}{\Lambda^2_{\gamma}r^2})\frac{e^{-\Lambda_{\gamma}r}}{r}]
\end{equation}
\ 

\noindent for $\gamma=\pi,K,\eta$ and $\eta$'. In Eq. (\ref{TENSORVg}), $G_f, g_\gamma, m_i, m_j, \mu_\gamma$ and $\Lambda_\gamma$ are the parameters of the model.
\\

For $^3S_1- {^3D_1}$ coupling, we need the states $|^3S_1>$ and $|^3D_1>$ ($I=0$, $S=1$ and $J=1$). Because the tensor force conserves $J$ (and $j_z$), we choose the particular projection $j_z=1$ for $|^3S_1>$ and $|^3D_1>$. Then the wave functions have the following forms 

\begin{equation}
|^3S_1>=f(R)\ Y_0^0\ \chi|_{s_z=1}\ \phi|_{i_z=0}
\end{equation}
and
\begin{eqnarray}
|^3D_1> & = & C^{S=1\;\;\,L=2\;\,\,J=1}_{s_z=1\ l_z=0\ j_z=1}\ f(R)\ Y_2^0(\hat{R}_i)\ \chi|_{s_z=1}\ \phi|_{i_z=0} \nonumber \\
& + & C^{S=1\;\;\,L=2\;\,\,J=1}_{s_z=0\ l_z=1\ j_z=1}\ f(R)\ Y_2^1(\hat{R}_i)\ \chi|_{s_z=0}\ \phi|_{i_z=0} \nonumber \\
& + & C^{\ S=1\;\;\;\:L=2\;\,\,J=1}_{s_z=-1\ l_z=2\ j_z=1}\ f(R)\ Y_2^2(\hat{R}_i)\ \chi|_{s_z=-1}\ \phi|_{i_z=0}
\end{eqnarray}
\ 

\noindent which leads to

\begin{eqnarray}\label{TENSOR3D1}
|^3D_1>& = f(R) & (\sqrt{\frac{1}{10}}\ Y_2^0(\hat{R}_i)\ \chi|_{s_z=1}\ \phi|_{i_z=0}\nonumber \\
& & - \sqrt{\frac{3}{10}}\ Y_2^1(\hat{R}_i)\ \chi|_{s_z=0}\ \phi|_{i_z=0}\nonumber \\
& & +\sqrt{\frac{6}{10}}\ Y_2^2(\hat{R}_i)\ \chi|_{s_z=-1}\ \phi|_{i_z=0})
\end{eqnarray}
and
\begin{eqnarray}
<^3D_1|& = f(R) & (\sqrt{\frac{1}{10}}\ Y_2^{0*}(\hat{R}_i)\ \chi|_{s_z=1}\ \phi|_{i_z=0}\nonumber \\
& & - \sqrt{\frac{3}{10}}\ Y_2^{1*}(\hat{R}_i)\ \chi|_{s_z=0}\ \phi|_{i_z=0}\nonumber \\
& & +\sqrt{\frac{6}{10}}\ Y_2^{2*}(\hat{R}_i)\ \chi|_{s_z=-1}\ \phi|_{i_z=0})
\end{eqnarray}
\\

Let us rewrite the operator $S^T_{ij}$ (\ref{TENSORSTij}) in the following form

\begin{eqnarray}
S^T_{ij}&=&\sqrt{\frac{\pi}{5}} Y_2^0(\hat{r}_{ij})(\sigma_{iz}\sigma_{jz}) \nonumber\\
&-&\sqrt{\frac{\pi}{5}}\frac{1}{4} Y_2^0(\hat{r}_{ij})(\sigma_{i+}\sigma_{j-}+\sigma_{i-}\sigma_{j+})\nonumber\\
&+&\sqrt{\frac{6 \pi}{5}}\frac{1}{4} Y_2^{-1}(\hat{r}_{ij})(\sigma_{iz}\sigma_{j+}+\sigma_{i+}\sigma_{jz})\nonumber\\
&-&\sqrt{\frac{6 \pi}{5}}\frac{1}{4} Y_2^1(\hat{r}_{ij})(\sigma_{iz}\sigma_{j-}+\sigma_{i-}\sigma_{jz})\nonumber\\
&+&\sqrt{\frac{6 \pi}{5}}\frac{1}{4} Y_2^{-2}(\hat{r}_{ij})(\sigma_{i+}\sigma_{j+})\nonumber\\
&+&\sqrt{\frac{6 \pi}{5}}\frac{1}{4} Y_2^2(\hat{r}_{ij})(\sigma_{i-}\sigma_{j-})
\end{eqnarray}

\noindent or

\begin{eqnarray}
S^T_{ij}&=&Y_2^0(\hat{r}_{ij})\ O_{ij}^0 + Y_2^{-1}(\hat{r}_{ij})\ O_{ij}^{-1} + Y_2^1(\hat{r}_{ij})\ O_{ij}^1 +Y_2^{-2}(\hat{r}_{ij})\ O_{ij}^{-2} + Y_2^2(\hat{r}_{ij})\ O_{ij}^2\nonumber\\
\end{eqnarray}

\noindent where
\begin{eqnarray}
O_{ij}^0 &=& \sqrt{\frac{\pi}{5}}\frac{1}{4}(4\ \sigma_{iz}\sigma_{jz}-\sigma_{i+}\sigma_{j-}-\sigma_{i-}\sigma_{j+}) \nonumber\\
O_{ij}^{-1} &=& \sqrt{\frac{6 \pi}{5}}\frac{1}{4} (\sigma_{iz}\sigma_{j+}+\sigma_{i+}\sigma_{jz})\nonumber\\
O_{ij}^1 &=& - \sqrt{\frac{6 \pi}{5}}\frac{1}{4} (\sigma_{iz}\sigma_{j-}+\sigma_{i-}\sigma_{jz})\nonumber\\
O_{ij}^{-2} &=& \sqrt{\frac{6 \pi}{5}}\frac{1}{4} (\sigma_{i+}\sigma_{j+})\nonumber\\
O_{ij}^2 &=& \sqrt{\frac{6 \pi}{5}}\frac{1}{4} (\sigma_{i-}\sigma_{j-})
\end{eqnarray}

We use the formula (\ref{CONFVa}) of two-body potential matrix element given in Appendix \ref{appendixCONF}

\begin{equation}
<\Psi_{6q}|V_{ij}|\Psi_{6q}>=V(\vec{a})N(\vec{R}_i,\vec{R}_j)
\end{equation}

\noindent where

\begin{equation}
V(\vec{a}) = (\frac{1}{\sqrt{2 \pi}b})^3 \int e^{\frac{-(\vec{r}-\vec{a}/2)^2}{2 b^2}}V(\vec{r})d\vec{r}
\end{equation}

\noindent with for

\begin{eqnarray}
& v_{12} &:\ \vec{a} = 0 \nonumber\\
& v_{36} &:\ \vec{a} = \vec{R}_i + \vec{R}_j \nonumber\\
& v_{12}P_{36} &:\ \vec{a} = 0 \nonumber\\
V_{ij} =& v_{36}P_{36} &:\ \vec{a} = \vec{R}_i - \vec{R}_j \nonumber\\
& v_{13}P_{36} &:\ \vec{a} = \vec{R}_j \nonumber\\
& v_{16}P_{36} &:\ \vec{a} = \vec{R}_i \nonumber\\
& v_{14}P_{36} &:\ \vec{a} = \vec{R}_i + \vec{R}_j
\end{eqnarray}
\ 

\noindent and where $N(\vec{R}_i,\vec{R}_j)$ is given by Eqs. (\ref{NORMdirect}) for the direct term and (\ref{NORMexchange}) for the exchange term.
\\\ \\

For the $<V^T(12)>$ matrix element associated to the diagram $a$ of Fig. \ref{RGMinteractiondiagram}, we then get

\begin{eqnarray}
<V^T(12)>&=&<\sum_{\gamma}\lambda^{\gamma}_1\lambda^{\gamma}_2 N^d(\vec{R}_i,\vec{R}_j) (\frac{1}{\sqrt{2 \pi}b})^3 \int e^{\frac{-r^2}{2 b^2}}V^T_{\gamma}S^T_{12}(\vec{r})d\vec{r}> \nonumber\\
&=&\sum_{\gamma=0}^8\sum_{k=-2}^2 <\lambda^{\gamma}_1\lambda^{\gamma}_2 O^k_{12}> \nonumber\\
&&N^d(\vec{R}_i,\vec{R}_j) (\frac{1}{\sqrt{2 \pi}b})^3 \int e^{\frac{-r^2}{2 b^2}}V^T_{\gamma}(r)Y_2^k(r)d\vec{r} \nonumber\\
&=&0\nonumber
\end{eqnarray}
\\

For the diagram $b$ of Fig. \ref{RGMinteractiondiagram} we obtain

\begin{eqnarray}
<V^T(36)>&=&<\sum_{\gamma}\lambda^{\gamma}_3\lambda^{\gamma}_6 N^d(\vec{R}_i,\vec{R}_j) (\frac{1}{\sqrt{2 \pi}b})^3 \int e^{\frac{-(\vec{r}-\frac{\vec{R}_i+\vec{R}_j}{2})^2}{2 b^2}}V^T_{\gamma}S^T_{36}(\vec{r})d\vec{r} > \nonumber\\
&=&\sum_{\gamma=0}^8\sum_{k=-2}^2<\lambda^{\gamma}_3\lambda^{\gamma}_6 O^k_{36}> N^d(\vec{R}_i,\vec{R}_j) (\frac{1}{\sqrt{2 \pi}b})^3 \int e^{\frac{-(\vec{r}-\frac{\vec{R}_i+\vec{R}_j}{2})^2}{2 b^2}}V^T_{\gamma}(r)Y_2^k(r)d\vec{r} \nonumber\\
&=&\sum_{\gamma=0}^8\sum_{k=-2}^2<\lambda^{\gamma}_3\lambda^{\gamma}_6 O^k_{36}> \times e^{-\alpha_d(R^2_i+R^2_j)} e^{3 \frac{\vec{R}_i\cdot \vec{R}_j}{4b^2}}\nonumber\\
(\alpha_d=\frac{3}{8b^2})&& (\frac{1}{\sqrt{2 \pi}b})^3 e^{- \frac{\vec{R}_i\cdot \vec{R}_j}{4b^2}} e^{\frac{-(R^2_i+R^2_j)}{8 b^2}} \int e^{\frac{-r^2}{2 b^2}} e^{\frac{\vec{r}\cdot \vec{R}_i}{2b^2}} e^{\frac{\vec{r}\cdot \vec{R}_j}{2b^2}} V^T_{\gamma}(r)Y_2^k(r)d\vec{r} \nonumber\\
&=&\sum_{\gamma=0}^8\sum_{k=-2}^2<\lambda^{\gamma}_3\lambda^{\gamma}_6 O^k_{36}> \times e^{-\alpha_d(R^2_i+R^2_j)} \sqrt{\frac{2}{\pi}}\frac{1}{b^3} e^{\frac{-(R^2_i+R^2_j)}{8 b^2}}\nonumber\\
&&\sum_{pq} i_p(\frac{R_i R_j}{2b^2}) Y_p^{q*}(R_i) Y_p^{q}(R_j) \int e^{\frac{-r^2}{2 b^2}} e^{\frac{\vec{r}\cdot \vec{R}_i}{2b^2}} e^{\frac{\vec{r}\cdot \vec{R}_j}{2b^2}} V^T_{\gamma}(r)Y_2^k(r)d\vec{r} \nonumber\\
&=&\sum_{\gamma=0}^8\sum_{k=-2}^2<\lambda^{\gamma}_3\lambda^{\gamma}_6 O^k_{36}> \times e^{-\alpha_d(R^2_i+R^2_j)} \sqrt{\frac{2}{\pi}}\frac{1}{b^3}  e^{\frac{-(R^2_i+R^2_j)}{8 b^2}} \nonumber\\
&& \sum_{pq} i_p(\frac{R_i R_j}{2b^2}) Y_p^{q*}(R_i) Y_p^{q}(R_j) \times (4\pi)^2\sum_{LM,lm} Y_L^M(R_i) Y_l^{m*}(R_j) \nonumber\\
&& \int i_L(\frac{r R_i}{2b^2}) i_l(\frac{r R_j}{2b^2}) Y_L^{M*}(r) Y_l^m(r) e^{\frac{-r^2}{2 b^2}} V^T_{\gamma}(r)Y_2^k(r) r^2 dr d\Omega_r \nonumber
\end{eqnarray}
\ 

\noindent Using the following relation

$$Y_l^{m} Y_2^{k}=\sum_{xy} \sqrt{\frac{(2l+1)5}{4\pi(2x+1)}}<l,2,m,k|x,y><l,2,0,0|x,0>Y_x^{y},$$
\ 

\noindent we get $x=L$ and $y=M$ in the previous equation. We then have

\begin{eqnarray}
<V^T(36)>&=&\sum_{\gamma=0}^8\sum_{k=-2}^2<\lambda^{\gamma}_3\lambda^{\gamma}_6 O^k_{36}> e^{-\alpha_d(R^2_i+R^2_j)} \sqrt{\frac{2}{\pi}}\frac{1}{b^3} e^{\frac{-(R^2_i+R^2_j)}{8 b^2}} \nonumber\\
&& \sum_{pq} i_p(\frac{R_i R_j}{2b^2}) Y_p^{q*}(R_i) Y_p^{q}(R_j) \ (4\pi)^2\sum_{LM,lm} Y_L^M(R_i) Y_l^{m*}(R_j)  \nonumber\\
&& \sqrt{\frac{(2l+1)5}{4\pi(2L+1)}} <l,2,m,k|L,M><l,2,0,0|L,0>  \nonumber\\
&& \int i_L(\frac{r R_i}{2b^2}) i_l(\frac{r R_j}{2b^2}) e^{\frac{-r^2}{2 b^2}} V^T_{\gamma}(r) r^2 dr \nonumber\\
&=&\sum_{\gamma=0}^8\sum_{k=-2}^2<\lambda^{\gamma}_3\lambda^{\gamma}_6 O^k_{36}> e^{-\alpha_d(R^2_i+R^2_j)} \sqrt{\frac{2}{\pi}}\frac{1}{b^3}  e^{\frac{-(R^2_i+R^2_j)}{8 b^2}}\nonumber\\
&& \sum_{pq} i_p(\frac{R_i R_j}{2b^2}) Y_p^{q*}(R_i) Y_p^{q}(R_j) (4\pi)^2\sum_{LM,lm} Y_L^M(R_i) Y_l^{m*}(R_j) \nonumber\\
&& \sqrt{\frac{(2l+1)5}{4\pi(2L+1)}} <l,2,m,k|L,M><l,2,0,0|L,0> K^{t,\gamma}_{L,l}(R_i,R_j) \nonumber
\end{eqnarray}
\\

\noindent where $K^{t,\gamma}_{L,l}(R_i,R_j)=\int i_L(\frac{r R_i}{2b^2}) i_l(\frac{r R_j}{2b^2}) e^{\frac{-r^2}{2 b^2}} V^T_{\gamma}(r) r^2 dr$.
\\\ \\

The non-zero matrix elements for the flavor-spin operator are given in the Table \ref{TENSORtable}
\\\ \\

Let us now consider the following three cases

\begin{itemize}
\item $^3S_1 - {^3S_1}$
\\

\begin{eqnarray}
<Y_0^0(R_j) V^T(36) Y_0^0(R_i)>&=&\sum_{\gamma=0}^8\sum_{k=-2}^2<\lambda^{\gamma}_3\lambda^{\gamma}_6 O^k_{36}>  e^{-\alpha_d(R^2_i+R^2_j)} \sqrt{\frac{2}{\pi}}\frac{1}{b^3}\nonumber\\
&&4\pi e^{\frac{-(R^2_i+R^2_j)}{8 b^2}} \sum_{pq} i_p(\frac{R_i R_j}{2b^2}) Y_p^{q*}(R_i) Y_p^{q}(R_j) \nonumber\\
&& \sum_{LM,lm} Y_L^M(R_i) Y_l^{m*}(R_j) K^{t,\gamma}_{L,l}(R_i,R_j)  \nonumber\\
&& \sqrt{\frac{(2l+1)5}{4\pi(2L+1)}} <l,2,m,k|L,M><l,2,0,0|L,0> \nonumber
\end{eqnarray}

The only non-zero values are obtained if we take $p=l,q=m$ and  $l=L,m=M$. This leads to

\begin{eqnarray}
<Y_0^0(R_j) V^T(36) Y_0^0(R_i)>&=&\sum_{\gamma=0}^8\sum_{k=-2}^2<\lambda^{\gamma}_3\lambda^{\gamma}_6 O^k_{36}> e^{-\alpha_d(R^2_i+R^2_j)} \sqrt{\frac{2}{\pi}}\frac{1}{b^3}\nonumber\\
&&4\pi e^{\frac{-(R^2_i+R^2_j)}{8 b^2}} \sum_{LM}  i_L(\frac{R_i R_j}{2b^2}) K^{t,\gamma}_{L,L}(R_i,R_j) \nonumber\\
&& \sqrt{\frac{5}{4\pi}} <L,2,M,k|L,M><L,2,0,0|L,0> \nonumber
\end{eqnarray}

From which it follows that $k$ has to be zero. If we note that $<L,2,M,0|L,M>=\frac{\sqrt{2}(L+L^2-3M^2)}{\sqrt{L(2L-1)(2L+2)(2L+3)}}$, and $<L,2,0,0|L,0>$ does not depend on $M$, we get

\begin{eqnarray}
\sum_{M=-L}^L <L,2,M,0|L,M>&\propto& \sum_{M=-L}^L (L+L^2-3M^2) \nonumber\\
&=& (2L+1) (L+L^2) - L(L+1)(2L+1)\nonumber\\
&=& 0\nonumber
\end{eqnarray}

\item $^3D_1- {^3S_1}$
\\

With the $|^3D_1>$ state defined in Eq. (\ref{TENSOR3D1}), for each $k'$ we have to calculate the matrix elements proportional to $<Y_2^{k'*}(R_j) V^T_{36} Y_0^0(R_i)>$ . We obtain

\begin{eqnarray}
<Y_2^{k'*}(R_j) V^T(36) Y_0^0(R_i)>&=&\sum_{\gamma=0}^8\sum_{k=-2}^2<\lambda^{\gamma}_3\lambda^{\gamma}_6 O^k_{36}> e^{-\alpha_d(R^2_i+R^2_j)} \sqrt{\frac{2}{\pi}}\frac{1}{b^3} \nonumber\\
&& e^{\frac{-(R^2_i+R^2_j)}{8 b^2}} (4\pi)^{\frac{3}{2}}\sum_{pq} i_p(\frac{R_i R_j}{2b^2}) Y_p^{q*}(R_i) Y_p^{q}(R_j) \nonumber\\
&&\sum_{LM,lm}  K^{t,\gamma}_{L,l}(R_i,R_j) Y_2^{k'*}(R_j) Y_L^M(R_i) Y_l^{m*}(R_j) \nonumber\\
&& \sqrt{\frac{(2l+1)5}{4\pi(2L+1)}} <l,2,m,k|L,M><l,2,0,0|L,0> \nonumber
\end{eqnarray}
\ 

It follow that $p=L$ and $q=M$. We then get

\begin{eqnarray}
<Y_2^{k'*}(R_j) V^T(36) Y_0^0(R_i)>&=&\sum_{\gamma=0}^8\sum_{k=-2}^2<\lambda^{\gamma}_3\lambda^{\gamma}_6 O^k_{36}>  e^{-\alpha_d(R^2_i+R^2_j)} \sqrt{\frac{2}{\pi}}\frac{1}{b^3}  \nonumber\\
&& e^{\frac{-(R^2_i+R^2_j)}{8 b^2}} (4\pi)^{\frac{3}{2}} \sum_{LM,lm} i_L(\frac{R_i R_j}{2b^2})  K^{t,\gamma}_{L,l}(R_i,R_j) \nonumber\\
&& Y_L^{M}(R_j) Y_2^{k'*}(R_j) Y_l^{m*}(R_j) \nonumber\\
&& \sqrt{\frac{(2l+1)5}{4\pi(2L+1)}} <l,2,m,k|L,M><l,2,0,0|L,0> \nonumber
\end{eqnarray}
\ 

\noindent The relation

$$Y_2^{k'*} Y_l^{m*}=\sum_{xy} \sqrt{\frac{5(2l+1)}{4\pi(2x+1)}}<2,l,-k',-m|x,y><2,l,0,0|x,0>Y_x^{-y*},$$

leads to $x=L$ and $y=-M$ in the previous equation

\begin{eqnarray}
<Y_2^{k'*}(R_j) V^T(36) Y_0^0(R_i)>&=&\sum_{\gamma=0}^8\sum_{k=-2}^2<\lambda^{\gamma}_3\lambda^{\gamma}_6 O^k_{36}> e^{-\alpha_d(R^2_i+R^2_j)} \sqrt{\frac{2}{\pi}}\frac{1}{b^3} \nonumber\\
&& e^{\frac{-(R^2_i+R^2_j)}{8 b^2}} 5\sqrt{4\pi} \sum_{LM,lm} i_L(\frac{R_i R_j}{2b^2})  K^{t,\gamma}_{L,l}(R_i,R_j) \frac{2l+1}{2L+1} \nonumber\\
&& <2,l,-k',-m|L,-M><2,l,0,0|L,0>  \nonumber\\
&&  <l,2,m,k|L,M><l,2,0,0|L,0> \nonumber
\end{eqnarray}
\ 

We see that the only non-zero terms come from $k=k'$ and $m=M-k'$. Moreover, with the CG propriety

$$<2,l,-k',-m|L,-M>=(-)^{2+l-L}<2,l,k',m|L,M>=<l,2,m,k'|L,M>$$

\noindent and

$$<l,2,0,0|L,0>=(-)^{l+2-L}<2,l,0,0|L,0>$$

\noindent we have

\begin{eqnarray}
<Y_2^{k'*}(R_j) V^T(36) Y_0^0(R_i)>&=&\sum_{\gamma=0}^8<\lambda^{\gamma}_3\lambda^{\gamma}_6 O^{k'}_{36}> e^{-\alpha_d(R^2_i+R^2_j)} \sqrt{\frac{2}{\pi}}\frac{1}{b^3} \nonumber\\
&& e^{\frac{-(R^2_i+R^2_j)}{8 b^2}} 5\sqrt{4\pi} \sum_{LM,l} i_L(\frac{R_i R_j}{2b^2})  K^{t,\gamma}_{L,l}(R_i,R_j)  \frac{2l+1}{2L+1} (-)^{l-L}\nonumber\\
&&  <l,2,M-k',k'|L,M>^2<2,l,0,0|L,0>^2. \nonumber
\end{eqnarray}
\\

Now if we look at Table \ref{TENSORtable}, we see that only $k'=0$ gives a non-zero contribution. Moreover, from the properties of the CG, only the values $l=L-2$, $l=L$ and $l=L+2$ are allowed. We can then write

\begin{eqnarray}
<Y_2^{k'*}(R_j) V^T(36) Y_0^0(R_i)>&=& \delta(k',0) \sum_{\gamma=0}^8<\lambda^{\gamma}_3\lambda^{\gamma}_6 O^0_{36}> e^{-\alpha_d(R^2_i+R^2_j)} \sqrt{\frac{2}{\pi}}\frac{1}{b^3} \nonumber\\
&& e^{\frac{-(R^2_i+R^2_j)}{8 b^2}} 5\sqrt{4\pi} \sum_{LM,l} i_L(\frac{R_i R_j}{2b^2})  K^{t,\gamma}_{L,l}(R_i,R_j)  \frac{2l+1}{2L+1}  (-)^{l-L}\nonumber\\
&&  <l,2,M,0|L,M>^2<2,l,0,0|L,0>^2\nonumber\\
&=& \delta(k',0) \sum_{\gamma=0}^8<\lambda^{\gamma}_3\lambda^{\gamma}_6 O^0_{36}> \times e^{-\alpha_d(R^2_i+R^2_j)} \sqrt{\frac{2}{\pi}}\frac{1}{b^3} \nonumber\\
&& e^{\frac{-(R^2_i+R^2_j)}{8 b^2}} \sum_L i_L(\frac{R_i R_j}{2b^2})  [C_1(L) K^{t,\gamma}_{L,L-2}(R_i,R_j) + \nonumber\\
&& C_2(L) K^{t,\gamma}_{L,L}(R_i,R_j)+ C_3(L) K^{t,\gamma}_{L,L+2}(R_i,R_j)]\nonumber
\end{eqnarray}

\noindent where

\begin{eqnarray}
C_1(L)&=&5\sqrt{4\pi}\frac{2L-3}{2L+1} \sum_M <L-2,2,M,0|L,M>^2<2,L-2,0,0|L,0>^2 \nonumber\\
&=&3\sqrt{\pi}\frac{L(L-1)}{2L-1} \nonumber\\
C_2(L)&=&5\sqrt{4\pi} \sum_M <L,2,M,0|L,M>^2<2,L,0,0|L,0>^2 \nonumber\\
&=&2\sqrt{\pi}\frac{L(L+1)(2L+1)}{(2L-1)(2L+3)} \nonumber\\
C_3(L)&=&5\sqrt{4\pi}\frac{2L+5}{2L+1} \sum_M <L+2,2,M,0|L,M>^2<2,L+2,0,0|L,0>^2 \nonumber\\
&=&3\sqrt{\pi}\frac{(L+1)(L+2)}{2L+3} \nonumber
\end{eqnarray}

\item $^3D_1- {^3D_1}$
\\

For each $k'$ and $z$, we need

\begin{eqnarray}
<Y_2^{k'*}(R_j) V^T(36) Y_2^z(R_i)>&=&\sum_{\gamma=0}^8\sum_{k=-2}^2<\lambda^{\gamma}_3\lambda^{\gamma}_6 O^k_{36}> e^{-\alpha_d(R^2_i+R^2_j)} \sqrt{\frac{2}{\pi}}\frac{1}{b^3} \nonumber\\
&& e^{\frac{-(R^2_i+R^2_j)}{8 b^2}} (4\pi)^2 \sum_{pq} i_p(\frac{R_i R_j}{2b^2}) Y_p^{q*}(R_i) Y_p^{q}(R_j)\nonumber\\
&&\sum_{LM,lm}  K^{t,\gamma}_{L,l}(R_i,R_j) Y_2^{k'*}(R_j) Y_2^z(R_i) Y_L^M(R_i) Y_l^{m*}(R_j) \nonumber\\
&& \sqrt{\frac{(2l+1)5}{4\pi(2L+1)}} <l,2,m,k|L,M><l,2,0,0|L,0> \nonumber
\end{eqnarray}
\ 

Here we use the relations

$$Y_2^{k'*} Y_l^{m*}=\sum_{xy} \sqrt{\frac{5(2l+1)}{4\pi(2x+1)}}<2,l,-k',-m|x,y><2,l,0,0|x,0>Y_x^{-y*}$$

\noindent and

$$Y_2^z Y_L^M=\sum_{x'y'} \sqrt{\frac{5(2L+1)}{4\pi(2x'+1)}}<2,L,z,M|x',y'><2,L,0,0|x',0>Y_{x'}^{y'}.$$
\\

This leads to $x=p$,$y=-q$,$x'=p$ and $y'=q$ :

\begin{eqnarray}
<Y_2^{k'*}(R_j) V^T(36) Y_2^z(R_i)>&=&\sum_{\gamma=0}^8\sum_{k=-2}^2<\lambda^{\gamma}_3\lambda^{\gamma}_6 O^k_{36}> \ e^{-\alpha_d(R^2_i+R^2_j)} \sqrt{\frac{2}{\pi}}\frac{1}{b^3}  \nonumber\\
&& e^{\frac{-(R^2_i+R^2_j)}{8 b^2}} (4\pi)^2 \sum_{pq} i_p(\frac{R_i R_j}{2b^2}) \sum_{LM,lm}  K^{t,\gamma}_{L,l}(R_i,R_j)  \nonumber\\
&&\sqrt{\frac{5(2l+1)}{4\pi(2p+1)}}<2,l,-k',-m|p,-q><2,l,0,0|p,0> \nonumber\\
&& \sqrt{\frac{5(2L+1)}{4\pi(2p+1)}}<2,L,z,M|p,q><2,L,0,0|p,0> \nonumber\\
&& \sqrt{\frac{(2l+1)5}{4\pi(2L+1)}} <l,2,m,k|L,M><l,2,0,0|L,0> \nonumber
\end{eqnarray}
\ 

Then, $q=k'+m$, $k=k'-z$, and using the proprieties

$$<2,l,0,0|p,0>=\sqrt{\frac{2p+1}{2l+1}}<2,p,0,0|l,0>,$$
$$<2,L,0,0|p,0>=\sqrt{\frac{2p+1}{2L+1}}<2,p,0,0|L,0>,$$
$$<l,2,0,0|L,0>=(-)^{l-L}\sqrt{\frac{2L+1}{2l+1}}<2,L,0,0|l,0>,$$

we obtain

\begin{eqnarray}
<Y_2^{k'*}(R_j) V^T(36) Y_2^z(R_i)>&=&\sum_{\gamma=0}^8<\lambda^{\gamma}_3\lambda^{\gamma}_6 O^{k'-z}_{36}> \ e^{-\alpha_d(R^2_i+R^2_j)} \sqrt{\frac{2}{\pi}}\frac{1}{b^3} 5 \sqrt{20\pi} \nonumber\\
&& e^{\frac{-(R^2_i+R^2_j)}{8 b^2}}  \sum_p  \sum_{LM,lm}  i_p(\frac{R_i R_j}{2b^2}) K^{t,\gamma}_{L,l}(R_i,R_j) \nonumber\\
&&<2,l,-k',-m|p,-k'-m><2,p,0,0|l,0> \nonumber\\
&& <2,L,z,M|p,k'+m><2,p,0,0|L,0> \nonumber\\
&& <l,2,m,k'-z|L,M><2,L,0,0|l,0> (-)^{l-L}\nonumber\\
&=&\sum_{\gamma=0}^8<\lambda^{\gamma}_3\lambda^{\gamma}_6 O^{k'-z}_{36}> e^{-\alpha_d(R^2_i+R^2_j)} \sqrt{\frac{2}{\pi}}\frac{1}{b^3} 5 \sqrt{20\pi} \nonumber\\
&& e^{\frac{-(R^2_i+R^2_j)}{8 b^2}}  \sum_p  \sum_{LM,l}  i_p(\frac{R_i R_j}{2b^2}) K^{t,\gamma}_{L,l}(R_i,R_j)  \nonumber\\
&&<2,l,-k',-z-M+k'|p,-z-M><2,p,0,0|l,0> \nonumber\\
&& <2,L,z,M|p,z+M><2,p,0,0|L,0> \nonumber\\
&& <l,2,z+M-k',k'-z|L,M> \nonumber\\
&& <2,L,0,0|l,0>  (-)^{l-L}\nonumber
\end{eqnarray}

\noindent where we used the relation $m=z+M-k'$. Now, noting that $p=l-2$, $p=l$ or $p=l+2$, we have

\begin{eqnarray}
<Y_2^{k'*}(R_j) V^T(36) Y_2^z(R_i)>&=&\sum_{\gamma=0}^8<\lambda^{\gamma}_3\lambda^{\gamma}_6 O^{k'-z}_{36}> e^{-\alpha_d(R^2_i+R^2_j)} \sqrt{\frac{2}{\pi}}\frac{1}{b^3} 5 \sqrt{20\pi}  \nonumber\\
&& e^{\frac{-(R^2_i+R^2_j)}{8 b^2}} \sum_{LM,l} K^{t,\gamma}_{L,l}(R_i,R_j) (-)^{l-L}\ \nonumber\\
&& <l,2,z+M-k',k'-z|L,M><2,L,0,0|l,0>\ \nonumber\\
&& [i_{l-2}(\frac{R_i R_j}{2b^2}) <2,l,-k',-z-M+k'|l-2,-z-M> \nonumber\\
&&<2,l-2,0,0|l,0> <2,L,z,M|l-2,z+M> \nonumber\\
&&<2,l-2,0,0|L,0> \nonumber\\
&& +  i_l(\frac{R_i R_j}{2b^2}) <2,l,-k',-z-M+k'|l,-z-M>\nonumber\\
&& <2,l,0,0|l,0><2,L,z,M|l,z+M><2,l,0,0|L,0>  \nonumber\\
&& + i_{l+2}(\frac{R_i R_j}{2b^2}) <2,l,-k',-z-M+k'|l+2,-z-M> \nonumber\\
&& <2,l+2,0,0|l,0><2,L,z,M|l+2,z+M> \nonumber\\
&&<2,l+2,0,0|L,0>]\nonumber
\end{eqnarray}
\ 

In the same way, because $l=L-2$, $l=L$ or $l=L+2$, we get

\begin{eqnarray}
<Y_2^{k'*}(R_j) V^T(36) Y_2^z(R_i)>&=&\sum_{\gamma=0}^8<\lambda^{\gamma}_3\lambda^{\gamma}_6 O^{k'-z}_{36}>  e^{-\alpha_d(R^2_i+R^2_j)} \sqrt{\frac{2}{\pi}}\frac{1}{b^3} 5 \sqrt{20\pi}  \nonumber\\
&& e^{\frac{-(R^2_i+R^2_j)}{8 b^2}} \sum_{LM}[ K^{t,\gamma}_{L,L-2}(R_i,R_j) \nonumber\\
&& <L-2,2,z+M-k',k'-z|L,M> \nonumber\\
&&<2,L,0,0|L-2,0> [i_{L-2}(\frac{R_i R_j}{2b^2})  \nonumber\\
&& <2,L-2,-k',-z-M+k'|L-2,-z-M>\nonumber\\
&& <2,L-2,0,0|L-2,0><2,L,z,M|L-2,z+M> \nonumber\\
&&<2,L-2,0,0|L,0>  \nonumber\\
&& + i_L(\frac{R_i R_j}{2b^2}) <2,L-2,-k',-z-M+k'|L,-z-M> \nonumber\\
&& <2,L,0,0|L-2,0><2,L,z,M|L,z+M> \nonumber\\
&&<2,L,0,0|L,0> ] \nonumber\\
&&+K^{t,\gamma}_{L,L}(R_i,R_j) \nonumber\\
&& <L,2,z+M-k',k'-z|L,M><2,L,0,0|L,0> \nonumber\\
&& [ i_{L-2}(\frac{R_i R_j}{2b^2}) <2,L,-k',-z-M+k'|L-2,-z-M> \nonumber\\
&&<2,L-2,0,0|L,0> <2,L,z,M|L-2,z+M> \nonumber\\
&& <2,L-2,0,0|L,0> \nonumber\\
&& +  i_L(\frac{R_i R_j}{2b^2}) <2,L,-k',-z-M+k'|L,-z-M>\nonumber\\
&& <2,L,0,0|L,0><2,L,z,M|L,z+M> \nonumber\\
&&<2,L,0,0|L,0>  + i_{L+2}(\frac{R_i R_j}{2b^2}) \nonumber\\
&& <2,L,-k',-z-M+k'|L+2,-z-M> \nonumber\\
&& <2,L+2,0,0|L,0><2,L,z,M|L+2,z+M> \nonumber\\
&&<2,L+2,0,0|L,0> ] \nonumber\\
&&+K^{t,\gamma}_{L,L+2}(R_i,R_j) <L+2,2,z+M-k',k'-z|L,M> \nonumber\\
&&<2,L,0,0|L+2,0> \nonumber\\
&& [ i_L(\frac{R_i R_j}{2b^2}) <2,L+2,-k',-z-M+k'|L,-z-M> \nonumber\\
&&<2,L,0,0|L+2,0> <2,L,z,M|L,z+M> \nonumber\\
&&<2,L,0,0|L,0> +  i_{L+2}(\frac{R_i R_j}{2b^2})  \nonumber\\
&&<2,L+2,-k',-z-M+k'|L+2,-z-M>\nonumber\\
&& <2,L+2,0,0|L+2,0><2,L,z,M|L+2,z+M> \nonumber\\
&&<2,L+2,0,0|L,0>] ]\nonumber
\end{eqnarray}
\ 

\noindent because the factor in front of the Bessel function $i_{L-4}$ and $i_{L+4}$ is zero. Using some CG symmetry properties, we have

\begin{eqnarray}
<Y_2^{k'*}(R_j) V^T(36) Y_2^z(R_i)>&=&\sum_{\gamma=0}^8<\lambda^{\gamma}_3\lambda^{\gamma}_6 O^{k'-z}_{36}> \times e^{-\alpha_d(R^2_i+R^2_j)} \sqrt{\frac{2}{\pi}}\frac{1}{b^3} \nonumber\\
&& e^{\frac{-(R^2_i+R^2_j)}{8 b^2}} <2,2,k'-z,z|2,k'>\nonumber\\
&&  \sum_L [K^{t,\gamma}_{L,L-2}(R_i,R_j) [D_1(L) i_{L-2}(\frac{R_i R_j}{2b^2}) + D_2(L) i_L(\frac{R_i R_j}{2b^2})] \nonumber\\
&&+K^{t,\gamma}_{L,L}(R_i,R_j) [D_3(L) i_{L-2}(\frac{R_i R_j}{2b^2}) +D_4(L) i_L(\frac{R_i R_j}{2b^2}) \nonumber\\
&& + D_5(L) i_{L+2}(\frac{R_i R_j}{2b^2})] \nonumber\\
&&+K^{t,\gamma}_{L,L+2}(R_i,R_j) [D_6(L) i_L(\frac{R_i R_j}{2b^2}) + D_7(L) i_{L+2}(\frac{R_i R_j}{2b^2})]] \nonumber
\end{eqnarray}

\noindent where

\begin{eqnarray}
D_1(L)&=&-3\sqrt{\frac{5\pi}{14}} \frac{L(L-1)(L-2)}{(2L-1)^2} \nonumber\\
D_2(L)=D_3(L)&=&-3\sqrt{\frac{5\pi}{14}} \frac{L(L+1)(L-1)}{(2L-1)^2}\nonumber\\
D_4(L)&=&\sqrt{\frac{5\pi}{14}} \frac{L(L+1)(2L+1)(2L+5)(2L-3)}{(2L-1)^2(2L+3)^2} \nonumber\\
D_5(L)=D_6(L)&=&-3\sqrt{\frac{5\pi}{14}} \frac{L(L+1)(L+2)}{(2L+3)^2} \nonumber\\
D_7(L)&=&-3\sqrt{\frac{5\pi}{14}} \frac{(L+1)(L+2)(L+3)}{(2L+3)^2} \nonumber
\end{eqnarray}
\\

\end{itemize}

Let us now look at the case $<V^T(12)P_{36}>$ corresponding to the diagram $c$ of Fig. \ref{RGMinteractiondiagram}

\begin{eqnarray}
<V^T(12)P_{36}>&=&<\sum_{\gamma}\lambda^{\gamma}_1\lambda^{\gamma}_2 N^e(\vec{R}_i,\vec{R}_j) (\frac{1}{\sqrt{2 \pi}b})^3 \int e^{\frac{-r^2}{2 b^2}}V^T_{\gamma}S^T_{12}(\vec{r})d\vec{r} P^{\sigma fc}_{36}> \nonumber\\
&=&\sum_{\gamma=0}^8\sum_{k=-2}^2<\lambda^{\gamma}_1\lambda^{\gamma}_2 O^k_{12} P^{\sigma fc}_{36}> N^e(\vec{R}_i,\vec{R}_j) (\frac{1}{\sqrt{2 \pi}b})^3 \int e^{\frac{-r^2}{2 b^2}}V^T_{\gamma}(r)Y_2^k(r)d\vec{r} \nonumber\\
&=&0\nonumber
\end{eqnarray}

For $<V^T(36)P_{36}>$, associated to the diagram $g$ of Fig. \ref{RGMinteractiondiagram}, we obtain

\begin{eqnarray}
<V^T(36)P_{36}>&=&<\sum_{\gamma}\lambda^{\gamma}_3\lambda^{\gamma}_6 N^e(\vec{R}_i,\vec{R}_j) (\frac{1}{\sqrt{2 \pi}b})^3 \int e^{\frac{-(\vec{r}-\frac{\vec{R}_i-\vec{R}_j}{2})^2}{2 b^2}}V^T_{\gamma}S^T_{36}(\vec{r})d\vec{r} P^{\sigma fc}_{36}> \nonumber\\
&=&\sum_{\gamma=0}^8\sum_{k=-2}^2<\lambda^{\gamma}_3\lambda^{\gamma}_6 O^k_{36} P^{\sigma fc}_{36}>  \nonumber\\
&&N^e(\vec{R}_i,\vec{R}_j) (\frac{1}{\sqrt{2 \pi}b})^3 \int e^{\frac{-(\vec{r}-\frac{\vec{R}_i-\vec{R}_j}{2})^2}{2 b^2}}V^T_{\gamma}(r)Y_2^k(r)d\vec{r} \nonumber\\
&=&\sum_{\gamma=0}^8\sum_{k=-2}^2<\lambda^{\gamma}_3\lambda^{\gamma}_6 O^k_{36} P^{\sigma fc}_{36}>  e^{-\alpha_e(R^2_i+R^2_j)} e^{\frac{\vec{R}_i\cdot \vec{R}_j}{4b^2}}\nonumber\\
(\alpha_e=\alpha_d)&& (\frac{1}{\sqrt{2 \pi}b})^3 e^{\frac{\vec{R}_i\cdot \vec{R}_j}{4b^2}} e^{\frac{-(R^2_i+R^2_j)}{8 b^2}} \int e^{\frac{-r^2}{2 b^2}} e^{\frac{\vec{r}\cdot \vec{R}_i}{2b^2}} e^{-\frac{\vec{r}\cdot \vec{R}_j}{2b^2}} V^T_{\gamma}(r)Y_2^k(r)d\vec{r} \nonumber\\
&=&\sum_{\gamma=0}^8\sum_{k=-2}^2<\lambda^{\gamma}_3\lambda^{\gamma}_6 O^k_{36} P^{\sigma fc}_{36}>  e^{-\alpha_e(R^2_i+R^2_j)} \sqrt{\frac{2}{\pi}}\frac{1}{b^3}  \nonumber\\
&& e^{\frac{-(R^2_i+R^2_j)}{8 b^2}} \sum_{pq} i_p(\frac{R_i R_j}{2b^2}) Y_p^{q*}(R_i) Y_p^{q}(R_j)  \nonumber\\
&&\int e^{\frac{-r^2}{2 b^2}} e^{\frac{\vec{r}\cdot \vec{R}_i}{2b^2}} e^{-\frac{\vec{r}\cdot \vec{R}_j}{2b^2}} V^T_{\gamma}(r)Y_2^k(r)d\vec{r} \nonumber\\
&=&\sum_{\gamma=0}^8\sum_{k=-2}^2<\lambda^{\gamma}_3\lambda^{\gamma}_6 O^k_{36} P^{\sigma fc}_{36}>  e^{-\alpha_e(R^2_i+R^2_j)} \sqrt{\frac{2}{\pi}}\frac{1}{b^3}  e^{\frac{-(R^2_i+R^2_j)}{8 b^2}} \nonumber\\
&& \sum_{pq} i_p(\frac{R_i R_j}{2b^2}) Y_p^{q*}(R_i) Y_p^{q}(R_j)  (4\pi)^2\sum_{LM,lm} Y_L^M(R_i) Y_l^{m*}(R_j)  \nonumber\\
&& (-)^l \int i_L(\frac{r R_i}{2b^2}) i_l(\frac{r R_j}{2b^2}) Y_L^{M*}(r) Y_l^m(r) e^{\frac{-r^2}{2 b^2}} V^T_{\gamma}(r)Y_2^k(r) r^2 dr d\Omega_r \nonumber
\end{eqnarray}

Where we used the property  $i_n(-y)=(-)^n i_n(y)$.
\\

\begin{itemize}

\item $^3S_1- {^3S_1}$
\\

$$<V^T(36)P_{36}>=0$$

\item $^3D_1- {^3S_1}$
\\

\begin{eqnarray}
<Y_2^{k'}(R_j) V^T(36)P_{36} Y_0^0(R_i)>&=& \delta(k',0) \sum_{\gamma=0}^8<\lambda^{\gamma}_3\lambda^{\gamma}_6 O^0_{36} P^{\sigma fc}_{36}>  e^{-\alpha_e(R^2_i+R^2_j)} \sqrt{\frac{2}{\pi}}\frac{1}{b^3}  \nonumber\\
&& e^{\frac{-(R^2_i+R^2_j)}{8 b^2}} \sum_L  (-)^L i_L(\frac{R_i R_j}{2b^2})  [C_1(L) K^{t,\gamma}_{L,L-2}(R_i,R_j) + \nonumber\\
&& C_2(L) K^{t,\gamma}_{L,L}(R_i,R_j)+ C_3(L) K^{t,\gamma}_{L,L+2}(R_i,R_j)]\nonumber
\end{eqnarray}

\item $^3D_1- {^3D_1}$
\\

\begin{eqnarray}
<Y_2^{k'*}(R_j) V^T(36)P_{36} Y_2^z(R_i)>&=&\sum_{\gamma=0}^8<\lambda^{\gamma}_3\lambda^{\gamma}_6 O^{k'-z}_{36} P^{\sigma fc}_{36}>  e^{-\alpha_e(R^2_i+R^2_j)} \sqrt{\frac{2}{\pi}}\frac{1}{b^3} \nonumber\\
&& e^{\frac{-(R^2_i+R^2_j)}{8 b^2}}  <2,2,k'-z,z|2,k'> \nonumber\\
&&  \sum_L (-)^L [K^{t,\gamma}_{L,L-2}(R_i,R_j) [D_1(L) i_{L-2}(\frac{R_i R_j}{2b^2}) \nonumber\\
&&  + D_2(L) i_L(\frac{R_i R_j}{2b^2})] \nonumber\\
&&+K^{t,\gamma}_{L,L}(R_i,R_j) [D_3(L) i_{L-2}(\frac{R_i R_j}{2b^2}) +D_4(L) i_L(\frac{R_i R_j}{2b^2}) \nonumber\\
&&  + D_5(L) i_{L+2}(\frac{R_i R_j}{2b^2})] \nonumber\\
&&+K^{t,\gamma}_{L,L+2}(R_i,R_j) [D_6(L) i_L(\frac{R_i R_j}{2b^2}) \nonumber\\
&&  + D_7(L) i_{L+2}(\frac{R_i R_j}{2b^2})]]\nonumber
\end{eqnarray}
\\

\end{itemize}

For the case $<V^T(13)P_{36}>$ corresponding to the diagram $e$ of Fig. \ref{RGMinteractiondiagram}, we get

\begin{eqnarray}
<V^T(13)P_{36}>&=&<\sum_{\gamma}\lambda^{\gamma}_1\lambda^{\gamma}_3 N^e(\vec{R}_i,\vec{R}_j) (\frac{1}{\sqrt{2 \pi}b})^3 \int e^{\frac{-(\vec{r}-\frac{\vec{R}_j}{2})^2}{2 b^2}}V^T_{\gamma}S^T_{13}(\vec{r})d\vec{r} P^{\sigma fc}_{36}> \nonumber\\
&=&\sum_{\gamma=0}^8\sum_{k=-2}^2<\lambda^{\gamma}_1\lambda^{\gamma}_3 O^k_{13} P^{\sigma fc}_{36}> \nonumber\\
&&N^e(\vec{R}_i,\vec{R}_j) (\frac{1}{\sqrt{2 \pi}b})^3 \int e^{\frac{-(\vec{r}-\frac{\vec{R}_j}{2})^2}{2 b^2}}V^T_{\gamma}(r)Y_2^k(r)d\vec{r} \nonumber\\
&=&\sum_{\gamma=0}^8\sum_{k=-2}^2<\lambda^{\gamma}_1\lambda^{\gamma}_3 O^k_{13} P^{\sigma fc}_{36}>  \nonumber\\
(\alpha_e=\alpha_d)&&e^{-\alpha_e(R^2_i+R^2_j)}\sqrt{\frac{2}{\pi}}\frac{1}{b^3} \sum_{pq} i_p(\gamma_e R_i R_j) Y_p^{q*}(R_i) Y_p^{q}(R_j)   \nonumber\\
(\gamma_e=\frac{2}{8b^2})&& \int e^{\frac{-(\vec{r}-\frac{\vec{R}_j}{2})^2}{2 b^2}}V^T_{\gamma}(r)Y_2^k(r)d\vec{r} \nonumber\\
&=&\sum_{\gamma=0}^8\sum_{k=-2}^2<\lambda^{\gamma}_1\lambda^{\gamma}_3 O^k_{13} P^{\sigma fc}_{36}>   \nonumber\\
&&e^{-\alpha_e(R^2_i+R^2_j)}\sqrt{\frac{2}{\pi}}\frac{1}{b^3} \sum_{pq} i_p(\gamma_e R_i R_j) Y_p^{q*}(R_i) Y_p^{q}(R_j)  \nonumber\\
&& 4 \pi e^{\frac{-R^2_j}{8 b^2}} \sum_{lm} Y_l^{m}(R_j) \int i_l(\frac{r R_j}{2b^2}) Y_l^{m*}(r) e^{\frac{-r^2}{2 b^2}}V^T_{\gamma}(r)Y_2^k(r) r^2 dr d\Omega_r \nonumber\\
(\Rightarrow l=2,m=k)&=&\sum_{\gamma=0}^8\sum_{k=-2}^2<\lambda^{\gamma}_1\lambda^{\gamma}_3 O^k_{13} P^{\sigma fc}_{36}> e^{-\alpha_e(R^2_i+R^2_j)} \frac{4\sqrt{2 \pi}}{b^3} \nonumber\\
&& e^{\frac{-R^2_j}{8 b^2}} \sum_{pq} i_p(\gamma_e R_i R_j) Y_p^{q*}(R_i) Y_p^{q}(R_j) Y_2^{k}(R_j) \nonumber\\
&& \int i_2(\frac{r R_j}{2b^2})  e^{\frac{-r^2}{2 b^2}}V^T_{\gamma}(r) r^2 dr  \nonumber\\
&=&\sum_{\gamma=0}^8\sum_{k=-2}^2<\lambda^{\gamma}_1\lambda^{\gamma}_3 O^k_{13} P^{\sigma fc}_{36}> e^{-\alpha_e(R^2_i+R^2_j)} \frac{4\sqrt{2 \pi}}{b^3}  \nonumber\\
&& e^{\frac{-R^2_j}{8 b^2}} \sum_{pq} i_p(\gamma_e R_i R_j) Y_p^{q*}(R_i) Y_p^{q}(R_j) Y_2^{k}(R_j) L^T_{\gamma}(R_j) \nonumber
\end{eqnarray}
\ 

\noindent where

\begin{equation}\label{TENSORLT}
L^T_{\gamma}(R_j)= \int i_2(\frac{r R_j}{2b^2})  e^{\frac{-r^2}{2 b^2}}V^T_{\gamma}(r) r^2 dr.
\end{equation}
\\

Let us now consider the three cases
\\

\begin{itemize}
\item $^3S_1- {^3S_1}$
\\

Obviously equals to zero.

\item $^3D_1- {^3S_1}$
\\

With our definition of the $|^3D_1>$ state, for each $k'$ we have to calculate the matrix elements proportional to $<Y_2^{k'*}(R_j) V^T_{13}P_{36} Y_0^0(R_i)>$. We obtain

\begin{eqnarray}
<Y_2^{k'*}(R_j) V^T(13)P_{36} Y_0^0(R_i)>&=&\sum_{\gamma=0}^8\sum_{k=-2}^2<\lambda^{\gamma}_1\lambda^{\gamma}_3 O^k_{13} P^{\sigma fc}_{36}>  e^{-\alpha_e(R^2_i+R^2_j)} \frac{4\sqrt{2 \pi}}{b^3}  \nonumber\\
&& e^{\frac{-R^2_j}{8 b^2}} L^T_{\gamma}(R_j) \sum_{pq} i_p(\gamma_e R_i R_j)  \nonumber\\
&&Y_2^{k'*}(R_j) Y_p^{q*}(R_i) Y_p^{q}(R_j) Y_2^{k}(R_j) Y_0^0(R_i) \nonumber\\
(\Rightarrow p=0,q=0)&=&\sum_{\gamma=0}^8\sum_{k=-2}^2<\lambda^{\gamma}_1\lambda^{\gamma}_3 O^k_{13} P^{\sigma fc}_{36}>  e^{-\alpha_e(R^2_i+R^2_j)} \frac{4\sqrt{2 \pi}}{b^3}  \nonumber\\
&& e^{\frac{-R^2_j}{8 b^2}} L^T_{\gamma}(R_j) i_0(\gamma_e R_i R_j) Y_2^{k'*}(R_j) Y_0^{0}(R_j) Y_2^{k}(R_j) \nonumber\\
(\Rightarrow k=k')&=&\sum_{\gamma=0}^8<\lambda^{\gamma}_1\lambda^{\gamma}_3 O^{k'}_{13} P^{\sigma fc}_{36}>  e^{-\alpha_e(R^2_i+R^2_j)} \frac{4\sqrt{2 \pi}}{b^3} \nonumber\\
&& \frac{1}{\sqrt{4 \pi}} e^{\frac{-R^2_j}{8 b^2}} L^T_{\gamma}(R_j) i_0(\gamma_e R_i R_j) \nonumber
\end{eqnarray}

\item $^3D_1- {^3D_1}$
\\

This time, we have a matrix element like $<Y_2^{k'*}(R_j) V^T_{13}P_{36} Y_2^z(R_i)>$
\begin{eqnarray}
<Y_2^{k'*}(R_j) V^T(13)P_{36} Y_2^z(R_i)>&=&\sum_{\gamma=0}^8\sum_{k=-2}^2<\lambda^{\gamma}_1\lambda^{\gamma}_3 O^k_{13} P^{\sigma fc}_{36}> e^{-\alpha_e(R^2_i+R^2_j)} \frac{4\sqrt{2 \pi}}{b^3} \nonumber\\
&& e^{\frac{-R^2_j}{8 b^2}} L^T_{\gamma}(R_j) \sum_{pq} i_p(\gamma_e R_i R_j) \nonumber\\
&&Y_2^{k'*}(R_j) Y_p^{q*}(R_i) Y_p^{q}(R_j) Y_2^{k}(R_j) Y_2^z(R_i) \nonumber\\
(\Rightarrow p=2,q=z)&=&\sum_{\gamma=0}^8\sum_{k=-2}^2<\lambda^{\gamma}_1\lambda^{\gamma}_3 O^k_{13} P^{\sigma fc}_{36}> e^{-\alpha_e(R^2_i+R^2_j)} \frac{4\sqrt{2 \pi}}{b^3} \nonumber\\
&& e^{\frac{-R^2_j}{8 b^2}} L^T_{\gamma}(R_j) i_2(\gamma_e R_i R_j) Y_2^{k'*}(R_j) Y_2^{k}(R_j) Y_2^{z}(R_j)\nonumber
\end{eqnarray}
\ 

Here we use the relation

$$Y_2^{k} Y_2^{z}=\sum_{xy}\frac{5}{\sqrt{4\pi(2x+1)}}<2,2,k,z|x,y><2,2,0,0|x,0>Y_x^{y}$$
\\

It follows that in the previous equation we need $x=2$ and $y=k'$. Using $<2,2,0,0|2,0>=-\sqrt{\frac{2}{7}}$ and $k=k'-z$, we have

\begin{eqnarray}
<Y_2^{k'*}(R_j) V^T(13)P_{36} Y_2^z(R_i)>&=&-\sum_{\gamma=0}^8<\lambda^{\gamma}_1\lambda^{\gamma}_3 O^{k'-z}_{13} P^{\sigma fc}_{36}> e^{-\alpha_e(R^2_i+R^2_j)} \frac{4\sqrt{2 \pi}}{b^3}  \nonumber\\
&& \sqrt{\frac{5}{14 \pi}} e^{\frac{-R^2_j}{8 b^2}} L^T_{\gamma}(R_j) i_2(\gamma_e R_i R_j)  <2,2,k'-z,z|2,k'> \nonumber
\end{eqnarray}
\\

\end{itemize}

For $<V^T(16)P_{36}>$, associated to the diagram $f$ of Fig. \ref{RGMinteractiondiagram},

\begin{eqnarray}
<V^T(16)P_{36}>&=&<\sum_{\gamma}\lambda^{\gamma}_1\lambda^{\gamma}_6 N^e(\vec{R}_i,\vec{R}_j) (\frac{1}{\sqrt{2 \pi}b})^3 \int e^{\frac{-(\vec{r}-\frac{\vec{R}_i}{2})^2}{2 b^2}}V^T_{\gamma}S^T_{16}(\vec{r})d\vec{r} P^{\sigma fc}_{36}> \nonumber\\
&=&\sum_{\gamma=0}^8\sum_{k=-2}^2<\lambda^{\gamma}_1\lambda^{\gamma}_6 O^k_{16} P^{\sigma fc}_{36}>  \nonumber\\
&&N^e(\vec{R}_i,\vec{R}_j) (\frac{1}{\sqrt{2 \pi}b})^3 \int e^{\frac{-(\vec{r}-\frac{\vec{R}_i}{2})^2}{2 b^2}}V^T_{\gamma}(r)Y_2^k(r)d\vec{r} \nonumber\\
&=&\sum_{\gamma=0}^8\sum_{k=-2}^2<\lambda^{\gamma}_1\lambda^{\gamma}_6 O^k_{16} P^{\sigma fc}_{36}>   \nonumber\\
(\alpha_e=\frac{3}{8b^2})&&e^{-\alpha_e(R^2_i+R^2_j)}\sqrt{\frac{2}{\pi}}\frac{1}{b^3} \sum_{pq} i_p(\gamma_e R_i R_j) Y_p^{q*}(R_i) Y_p^{q}(R_j)  \nonumber\\
(\gamma_e=\frac{2}{8b^2})&& \int e^{\frac{-(\vec{r}-\frac{\vec{R}_i}{2})^2}{2 b^2}}V^T_{\gamma}(r)Y_2^k(r)d\vec{r} \nonumber\\
&=&\sum_{\gamma=0}^8\sum_{k=-2}^2<\lambda^{\gamma}_1\lambda^{\gamma}_6 O^k_{16} P^{\sigma fc}_{36}>  \nonumber\\
&&e^{-\alpha_e(R^2_i+R^2_j)}\sqrt{\frac{2}{\pi}}\frac{1}{b^3} \sum_{pq} i_p(\gamma_e R_i R_j) Y_p^{q*}(R_i) Y_p^{q}(R_j)  \nonumber\\
&& 4 \pi e^{\frac{-R^2_i}{8 b^2}} \sum_{lm} Y_l^{m}(R_i) \int i_l(\frac{r R_i}{2b^2}) Y_l^{m*}(r) e^{\frac{-r^2}{2 b^2}}V^T_{\gamma}(r)Y_2^k(r) r^2 dr d\Omega_r \nonumber\\
(\Rightarrow l=2,m=k)&=&\sum_{\gamma=0}^8\sum_{k=-2}^2<\lambda^{\gamma}_1\lambda^{\gamma}_6 O^k_{16} P^{\sigma fc}_{36}> e^{-\alpha_e(R^2_i+R^2_j)} \frac{4\sqrt{2 \pi}}{b^3} \nonumber\\
&& e^{\frac{-R^2_i}{8 b^2}} \sum_{pq} i_p(\gamma_e R_i R_j) Y_p^{q*}(R_i) Y_p^{q}(R_j) Y_2^{k}(R_i) \nonumber\\
&& \int i_2(\frac{r R_i}{2b^2})  e^{\frac{-r^2}{2 b^2}}V^T_{\gamma}(r) r^2 dr  \nonumber\\
&=&\sum_{\gamma=0}^8\sum_{k=-2}^2<\lambda^{\gamma}_1\lambda^{\gamma}_6 O^k_{16} P^{\sigma fc}_{36}> e^{-\alpha_e(R^2_i+R^2_j)} \frac{4\sqrt{2 \pi}}{b^3} \nonumber\\
&& e^{\frac{-R^2_i}{8 b^2}} \sum_{pq} i_p(\gamma_e R_i R_j) Y_p^{q*}(R_i) Y_p^{q}(R_j) Y_2^{k}(R_i) L^T_{\gamma}(R_i)\nonumber
\end{eqnarray}

where $L^T_{\gamma}(R_i)$ is define in Eq. (\ref{TENSORLT}).
\\\ \\

For the projections, we have

\begin{itemize}
\item $^3S_1- {^3S_1}$
\\

Obviously equals to zero.

\item $^3D_1- {^3S_1}$
\\

For each $k'$ we get

\begin{eqnarray}
<Y_2^{k'*}(R_j) V^T(16)P_{36} Y_0^0(R_i)>&=&\sum_{\gamma=0}^8\sum_{k=-2}^2<\lambda^{\gamma}_1\lambda^{\gamma}_6 O^k_{16} P^{\sigma fc}_{36}> e^{-\alpha_e(R^2_i+R^2_j)} \frac{4\sqrt{2 \pi}}{b^3}  \nonumber\\
&& e^{\frac{-R^2_i}{8 b^2}} L^T_{\gamma}(R_i) \sum_{pq} i_p(\gamma_e R_i R_j) \nonumber\\
&&Y_2^{k'*}(R_j) Y_p^{q*}(R_i) Y_p^{q}(R_j) Y_2^{k}(R_i) Y_0^0(R_i) \nonumber\\
(\Rightarrow p=2,q=k)&=&\sum_{\gamma=0}^8\sum_{k=-2}^2<\lambda^{\gamma}_1\lambda^{\gamma}_6 O^k_{16} P^{\sigma fc}_{36}> e^{-\alpha_e(R^2_i+R^2_j)} \frac{4\sqrt{2 \pi}}{b^3} \nonumber\\
&&  \frac{1}{\sqrt{4 \pi}} e^{\frac{-R^2_i}{8 b^2}} L^T_{\gamma}(R_i) i_2(\gamma_e R_i R_j) Y_2^{k'*}(R_j) Y_2^{k}(R_j) \nonumber\\
(\Rightarrow k=k')&=&\sum_{\gamma=0}^8<\lambda^{\gamma}_1\lambda^{\gamma}_6 O^{k'}_{16} P^{\sigma fc}_{36}>  e^{-\alpha_e(R^2_i+R^2_j)} \frac{4\sqrt{2 \pi}}{b^3} \nonumber\\
&& \frac{1}{\sqrt{4 \pi}} e^{\frac{-R^2_i}{8 b^2}} L^T_{\gamma}(R_i) i_2(\gamma_e R_i R_j) \nonumber
\end{eqnarray}

\item $^3D_1- {^3D_1}$
\\

This time, we have elements like $<Y_2^{k'*}(R_j) V^T_{16}P_{36} Y_2^z(R_i)>$ to calculate

\begin{eqnarray}
<Y_2^{k'*}(R_j) V^T(16)P_{36} Y_2^z(R_i)>&=&\sum_{\gamma=0}^8\sum_{k=-2}^2<\lambda^{\gamma}_1\lambda^{\gamma}_6 O^k_{16} P^{\sigma fc}_{36}> e^{-\alpha_e(R^2_i+R^2_j)} \frac{4\sqrt{2 \pi}}{b^3} \nonumber\\
&& e^{\frac{-R^2_i}{8 b^2}} L^T_{\gamma}(R_i) \sum_{pq} i_p(\gamma_e R_i R_j) \nonumber\\
&&Y_2^{k'*}(R_j) Y_p^{q*}(R_i) Y_p^{q}(R_j) Y_2^{k}(R_i) Y_2^z(R_i) \nonumber\\
(\Rightarrow p=2,q=k')&=&\sum_{\gamma=0}^8\sum_{k=-2}^2<\lambda^{\gamma}_1\lambda^{\gamma}_6 O^k_{16} P^{\sigma fc}_{36}> e^{-\alpha_e(R^2_i+R^2_j)} \frac{4\sqrt{2 \pi}}{b^3} \nonumber\\
&& e^{\frac{-R^2_i}{8 b^2}} L^T_{\gamma}(R_i) i_2(\gamma_e R_i R_j) Y_2^{k'*}(R_i) Y_2^{k}(R_i) Y_2^{z}(R_i)\nonumber
\end{eqnarray}
\ 

\noindent Here we use the relation

$$Y_2^{k} Y_2^{z}=\sum_{xy}\frac{5}{\sqrt{4\pi(2x+1)}}<2,2,k,z|x,y><2,2,0,0|x,0>Y_x^{y}$$
\ 

This leads to take $x=2$ and $y=k'$ in the previous equation. Using  $<2,2,0,0|2,0>=-\sqrt{\frac{2}{7}}$ and $k=k'-z$, we have obtain
\begin{eqnarray}
<Y_2^{k'*}(R_j) V^T(16)P_{36} Y_2^z(R_i)>&=&-\sum_{\gamma=0}^8<\lambda^{\gamma}_1\lambda^{\gamma}_6 O^{k'-z}_{16} P^{\sigma fc}_{36}>  e^{-\alpha_e(R^2_i+R^2_j)} \frac{4\sqrt{2 \pi}}{b^3}  \nonumber\\
&& \sqrt{\frac{5}{14 \pi}} e^{\frac{-R^2_i}{8 b^2}} L^T_{\gamma}(R_i) i_2(\gamma_e R_i R_j)  <2,2,k'-z,z|2,k'> \nonumber
\end{eqnarray}
\\

\end{itemize}

Finally, for the $<V^T(14)P_{36}>$ case, corresponding to the diagram $d$ of Fig. \ref{RGMinteractiondiagram}

\begin{eqnarray}
<V^T(14)P_{36}>&=&<\sum_{\gamma}\lambda^{\gamma}_1\lambda^{\gamma}_4 N^e(\vec{R}_i,\vec{R}_j) (\frac{1}{\sqrt{2 \pi}b})^3 \int e^{\frac{-(\vec{r}-\frac{\vec{R}_i+\vec{R}_j}{2})^2}{2 b^2}}V^T_{\gamma}S^T_{14}(\vec{r})d\vec{r} P^{\sigma fc}_{36}> \nonumber\\
&=&\sum_{\gamma=0}^8\sum_{k=-2}^2<\lambda^{\gamma}_1\lambda^{\gamma}_4 O^k_{14} P^{\sigma fc}_{36}> \nonumber\\
&&N^e(\vec{R}_i,\vec{R}_j) (\frac{1}{\sqrt{2 \pi}b})^3 \int e^{\frac{-(\vec{r}-\frac{\vec{R}_i+\vec{R}_j}{2})^2}{2 b^2}}V^T_{\gamma}(r)Y_2^k(r)d\vec{r} \nonumber\\
&=&\sum_{\gamma=0}^8\sum_{k=-2}^2<\lambda^{\gamma}_1\lambda^{\gamma}_4 O^k_{14} P^{\sigma fc}_{36}> e^{-\alpha_e(R^2_i+R^2_j)} e^{\frac{\vec{R}_i\cdot \vec{R}_j}{4b^2}}\nonumber\\
&& (\frac{1}{\sqrt{2 \pi}b})^3 e^{-\frac{\vec{R}_i\cdot \vec{R}_j}{4b^2}} e^{\frac{-(R^2_i+R^2_j)}{8 b^2}} \int e^{\frac{-r^2}{2 b^2}} e^{\frac{\vec{r}\cdot \vec{R}_i}{2b^2}} e^{\frac{\vec{r}\cdot \vec{R}_j}{2b^2}} V^T_{\gamma}(r)Y_2^k(r)d\vec{r} \nonumber\\
&=&\sum_{\gamma=0}^8\sum_{k=-2}^2<\lambda^{\gamma}_1\lambda^{\gamma}_4 O^k_{14} P^{\sigma fc}_{36}>  e^{-\alpha_e(R^2_i+R^2_j)}  (\frac{1}{\sqrt{2 \pi}b})^3 \nonumber\\
&& e^{\frac{-(R^2_i+R^2_j)}{8 b^2}} \int e^{\frac{-r^2}{2 b^2}} e^{\frac{\vec{r}\cdot \vec{R}_i}{2b^2}} e^{\frac{\vec{r}\cdot \vec{R}_j}{2b^2}} V^T_{\gamma}(r)Y_2^k(r)d\vec{r} \nonumber\\
&=&\sum_{\gamma=0}^8\sum_{k=-2}^2<\lambda^{\gamma}_1\lambda^{\gamma}_4 O^k_{14} P^{\sigma fc}_{36}> e^{-\alpha_e(R^2_i+R^2_j)} \sqrt{\frac{2}{\pi}}\frac{1}{b^3} \nonumber\\
&& e^{\frac{-(R^2_i+R^2_j)}{8 b^2}} 4\pi \sum_{LM,lm} Y_L^M(R_i) Y_l^{m*}(R_j) \nonumber\\
&& \int i_L(\frac{r R_i}{2b^2}) i_l(\frac{r R_j}{2b^2}) Y_L^{M*}(r) Y_l^m(r) e^{\frac{-r^2}{2 b^2}} V^T_{\gamma}(r)Y_2^k(r) r^2 dr d\Omega_r \nonumber
\end{eqnarray}
\\

\begin{itemize}

\item $^3S_1- {^3S_1}$
\\

We see immediately that $l=0$, $m=0$, $L=0$ and $M=0$, then

$$<V^T(14)P_{36}>=0$$

\item $^3D_1- {^3S_1}$
\\

For this case $L=0$ and $M=0$

\begin{eqnarray}
<Y_2^{k'*}(R_j) V^T(14)P_{36} Y_0^0(R_i)>&=& \sum_{\gamma=0}^8\sum_{k=-2}^2<\lambda^{\gamma}_1\lambda^{\gamma}_4 O^k_{14} P^{\sigma fc}_{36}>  e^{-\alpha_e(R^2_i+R^2_j)} \sqrt{\frac{2}{\pi}}\frac{1}{b^3}  \nonumber\\
&& e^{\frac{-(R^2_i+R^2_j)}{8 b^2}} 4\pi \sum_{lm} Y_2^{k'*}(R_j) Y_l^{m*}(R_j)  \nonumber\\
&& \int i_0(\frac{r R_i}{2b^2}) i_l(\frac{r R_j}{2b^2}) Y_l^m(r) e^{\frac{-r^2}{2 b^2}} V^T_{\gamma}(r)Y_2^k(r) r^2 dr d\Omega_r \nonumber
\end{eqnarray}
\ 

\noindent then $l=2$ and $m=-k$ and this leads to take $k=k'$

\begin{eqnarray}
<Y_2^{k'*}(R_j) V^T(14)P_{36} Y_0^0(R_i)>&=& \sum_{\gamma=0}^8 <\lambda^{\gamma}_1\lambda^{\gamma}_4 O^{k'}_{14} P^{\sigma fc}_{36}> e^{-\alpha_e(R^2_i+R^2_j)} \sqrt{\frac{2}{\pi}}\frac{1}{b^3} \nonumber\\
&& e^{\frac{-(R^2_i+R^2_j)}{8 b^2}} 4\pi \int i_0(\frac{r R_i}{2b^2}) i_2(\frac{r R_j}{2b^2}) e^{\frac{-r^2}{2 b^2}} V^T_{\gamma}(r) r^2 dr \nonumber\\
&=&\sum_{\gamma=0}^8 <\lambda^{\gamma}_1\lambda^{\gamma}_4 O^{k'}_{14} P^{\sigma fc}_{36}>  e^{-\alpha_e(R^2_i+R^2_j)} \sqrt{\frac{2}{\pi}}\frac{1}{b^3} \nonumber\\
&& e^{\frac{-(R^2_i+R^2_j)}{8 b^2}} 4\pi  K^{t,\gamma}_{0,2}(R_i,R_j) \nonumber
\end{eqnarray}
\ 

\item $^3D_1- {^3D_1}$
\\

Here, we have  $L=2$, $M=-z$, $l=2$ $m=-k'$

\begin{eqnarray}
<Y_2^{k'*}(R_j) V^T(14)P_{36} Y_2^z(R_i)>&=&\sum_{\gamma=0}^8\sum_{k=-2}^2<\lambda^{\gamma}_1\lambda^{\gamma}_4 O^k_{14} P^{\sigma fc}_{36}> e^{-\alpha_e(R^2_i+R^2_j)} \sqrt{\frac{2}{\pi}}\frac{1}{b^3} \nonumber\\
&& e^{\frac{-(R^2_i+R^2_j)}{8 b^2}} 4\pi \int i_2(\frac{r R_i}{2b^2}) i_2(\frac{r R_j}{2b^2})  \nonumber\\
&& Y_2^{-z*}(r) Y_2^{-k'}(r) e^{\frac{-r^2}{2 b^2}} V^T_{\gamma}(r)Y_2^k(r) r^2 dr d\Omega_r \nonumber
\end{eqnarray}

Using the relation

$$Y_2^{-k'} Y_2^k=\sum_{xy} \sqrt{\frac{25}{4\pi(2x+1)}}<2,2,-k',k|x,y><2,2,0,0|x,0>Y_x^y$$
\ 

\noindent with $x=2$ and $y=-z$, we get

\begin{eqnarray}
<Y_2^{k'*}(R_j) V^T(14)P_{36} Y_2^z(R_i)>&=&\sum_{\gamma=0}^8\sum_{k=-2}^2<\lambda^{\gamma}_1\lambda^{\gamma}_4 O^k_{14} P^{\sigma fc}_{36}> e^{-\alpha_e(R^2_i+R^2_j)} \sqrt{\frac{2}{\pi}}\frac{1}{b^3} \nonumber\\
&& e^{\frac{-(R^2_i+R^2_j)}{8 b^2}} <2,2,-k',k|2,-z><2,2,0,0|2,0> \nonumber\\
&&  4\pi  \sqrt{\frac{5}{4\pi}} \int i_2(\frac{r R_i}{2b^2}) i_2(\frac{r R_j}{2b^2}) e^{\frac{-r^2}{2 b^2}} V^T_{\gamma}(r) r^2 dr \nonumber\\
&=&-\sum_{\gamma=0}^8 <\lambda^{\gamma}_1\lambda^{\gamma}_4 O^{k'-z}_{14} P^{\sigma fc}_{36}>  e^{-\alpha_e(R^2_i+R^2_j)} \sqrt{\frac{2}{\pi}}\frac{1}{b^3} \nonumber\\
&& e^{\frac{-(R^2_i+R^2_j)}{8 b^2}} 2 \sqrt{\frac{10\pi}{7}}<2,2,k',z-k'|2,z> K^{t,\gamma}_{0,2}(R_i,R_j)\nonumber
\end{eqnarray}
\end{itemize}

All the matrix elements for the flavor-spin operator associated to the tensor potential are given in the Table \ref{TENSORtable}.
\\

\begin{table}[H]
\centering

\begin{tabular}{|c|c|c|c|c|c|c|}
\hline
\rule[-4mm]{0mm}{8mm} & $m=-1$ & $m=0$ & $m=0$ & $m=0$ & $m=1$\\
$O$  &$s'_z=0$ & $s'_z=-1$ & $s'_z=0$ & $s'_z=1$ & $s'_z=-1$ \\
\rule[-4mm]{0mm}{5mm} &$s_z=-1$ & $s_z=-1$ & $s_z=0$ & $s_z=1$ & $s_z=0$ \\
\hline
\hline

\rule[-3mm]{0mm}{8mm}$O_{12}^m \tau_1 \cdot \tau_2$                   
&    0 &  405   &  405 &  405 &    0 \\
\rule[-3mm]{0mm}{5mm}$O_{36}^m \tau_3 \cdot \tau_6$                   
&  150 & -300   &  375 &  -75 & -150 \\
\rule[-3mm]{0mm}{5mm}$O_{12}^m \tau_1 \cdot \tau_2\ P_{36}^{f \sigma}$                   
&   24 & -147   &  -39 & -111 &  -24 \\
\rule[-3mm]{0mm}{5mm}$O_{36}^m \tau_3 \cdot \tau_6\ P_{36}^{f \sigma}$                   
& -102 &  330   & -129 &  177 &  102 \\
\rule[-3mm]{0mm}{5mm}$O_{13}^m \tau_1 \cdot \tau_3\ P_{36}^{f \sigma}$                   
&  -12 &   51   &   -3 &   33 &   12 \\
\rule[-3mm]{0mm}{5mm}$O_{16}^m \tau_1 \cdot \tau_6\ P_{36}^{f \sigma}$                   
&  -12 &   51   &   -3 &   33 &   12 \\
\rule[-3mm]{0mm}{5mm}$O_{14}^m\ \tau_1 \cdot \tau_4\ P_{36}^{f \sigma}$                   
&    9 &  -9/2  &   36 &    9 &    9 \\
\rule[-3mm]{0mm}{5mm}$O_{12}^m \lambda_1^{f} \cdot \lambda_2^{f}$                          
&    0 &  378   &  378 &  378 &    0 \\
\rule[-3mm]{0mm}{5mm}$O_{36}^m \lambda_3^{f} \cdot \lambda_6^{f}$                          
&  144 & -288   &  360 &  -84 & -144 \\
\rule[-3mm]{0mm}{5mm}$O_{12}^m \lambda_1^{f} \cdot \lambda_2^{f}\ P_{36}^{f \sigma}$     
&   32 & -142   &    2 &  -94 &  -32 \\
\rule[-3mm]{0mm}{5mm}$O_{36}^m \lambda_3^{f} \cdot \lambda_6^{f}\ P_{36}^{f \sigma}$     
&  -80 &  316   &  -44 &  208 &   80 \\
\rule[-3mm]{0mm}{5mm}$O_{13}^m \lambda_1^{f} \cdot \lambda_3^{f}\ P_{36}^{f \sigma}$     
&  -16 &   50   &  -22 &   26 &   16 \\
\rule[-3mm]{0mm}{5mm}$O_{16}^m \lambda_1^{f} \cdot \lambda_6^{f}\ P_{36}^{f \sigma}$     
&  -16 &   50   &  -22 &   26 &   16 \\
\rule[-3mm]{0mm}{5mm}$O_{14}^m \lambda_1^{f} \cdot \lambda_4^{f}\ P_{36}^{f \sigma}$     
&    8 &   -1   &   35 &    9 &   -8 \\
\rule[-3mm]{0mm}{5mm}$O_{12}^m \lambda_1^{f,0} \cdot \lambda_2^{f,0}$                          
&    0 &  -54   &  -54 &  -54 &    0 \\
\rule[-3mm]{0mm}{5mm}$O_{36}^m \lambda_3^{f,0} \cdot \lambda_6^{f,0}$                          
&  -12 &   24   &  -30 &  -18 &   12 \\
\rule[-3mm]{0mm}{5mm}$O_{12}^m \lambda_1^{f,0} \cdot \lambda_2^{f,0}\ P_{36}^{f \sigma}$     
&   16 &   10   &   82 &   34 &  -16 \\
\rule[-3mm]{0mm}{5mm}$O_{36}^m \lambda_3^{f,0} \cdot \lambda_6^{f,0}\ P_{36}^{f \sigma}$     
&   44 &  -28   &  170 &   62 &  -44 \\
\rule[-3mm]{0mm}{5mm}$O_{13}^m \lambda_1^{f,0} \cdot \lambda_3^{f,0}\ P_{36}^{f \sigma}$     
&   -8 &   -2   &  -38 &  -14 &    8 \\
\rule[-3mm]{0mm}{5mm}$O_{16}^m \lambda_1^{f,0} \cdot \lambda_6^{f,0}\ P_{36}^{f \sigma}$     
&   -8 &   -2   &  -38 &  -14 &    8 \\
\rule[-3mm]{0mm}{5mm}$O_{14}^m \lambda_1^{f,0} \cdot \lambda_4^{f,0}\ P_{36}^{f \sigma}$     
&   -2 &    7   &   -2 &    0 &    2 \\
\hline
\rule[-4mm]{0mm}{10mm}factor & $\frac{\sqrt{15 \pi}}{1620}$ & $\frac{\sqrt{5 \pi}}{1620}$ & $\frac{\sqrt{5 \pi}}{1620}$ & $\frac{\sqrt{5 \pi}}{1620}$ & $\frac{\sqrt{15 \pi}}{1620}$ \\
\hline
\end{tabular}
\caption{ The matrix elements $\langle NN_{s'_z}|O|NN_{s_z}\rangle$ of the different tensor operators $O$ in the case $(S,I)$ = $(1,0)$.}\label{TENSORtable}

\end{table}


\tableofcontents




\end{document}